\documentclass[12pt,twoside,notitlepage]{articleGS}
\usepackage{makeidx,color,enumerate,hyperref}

\usepackage{exscale}
\usepackage{graphics}
\usepackage{graphicx}
\usepackage{epsf}
\usepackage[english]{babel} 
\usepackage{verbatim}
\usepackage{amssymb}
\usepackage[applemac]{inputenc}

\rmfamily
\ttfamily
\itshape

\newcount\TotalFigureNo
\newdimen\broaderOne
\newdimen\smallerOne

\newdimen\BoxBreadth
\newdimen\BoxBreadthCOL
\newdimen\BoxHeightCOL
\def\chapter{\section}

   \newwrite\lit\openout\lit=PrivateLabelWrite

\newdimen\interspacereduction
\global\def\interspacereduction{\ifCUP 0.00cm\else -0.0pt\fi}
\newdimen\ColEntryShiftHoriz
\global\def\ColEntryShiftHoriz{\ifCUP 0.25cm\else 5pt\fi}
\newdimen\ColEntryShiftVerti
\global\def\ColEntryShiftVerti{\ifCUP 0.025cm\else 1pt\fi}

\newcount\EnuncNo
\def\enunc#1#2#3#4{\global\advance\EnuncNo by 1\immediate\write\lit{\def\ #4{\thesection.\the\EnuncNo}}
                   \label{#4}%
\bigskip\bigskip\noindent%
\setbox1=\hbox{\bf#1}%
\setbox2=\hbox{\bf#2}%
\setbox3=\hbox{\it#3}%
\ifdim\wd2>0.1pt{\bf\thesection.\the\EnuncNo\enspace
                 \hypertarget{#4}{#2}}\fi
\ifdim\wd1>0.1pt{\bf#1}\fi
\ifdim\wd3>0.1pt\enspace({\it#3\/})\fi{\bf .}}

\def\Proof{\bigskip\noindent{\bf Proof\kern0.05cm:\ }} 
\def\Bewende{{\unskip\nobreak\hfil\penalty50
        \hskip2em\hbox{}\nobreak\hfil$\square$
        \parfillskip=0pt \finalhyphendemerits=0\par}\bigskip\bigskip}

\def\int{\setbox1=\hbox{\includegraphics[scale=0.13]{\UniversalSources/GrossIntegral}}\kern-0.2\wd1\lower1.5\wd1\box1\kern-0.01\wd1}
\def\intFOLIE{\setbox1=\hbox{\includegraphics[scale=0.250]{\UniversalSources/GrossIntegral}}\kern-0.3\wd1\lower1.2\wd1\box1\kern-0.01\wd1}

\def\TarskiNull{\mathfrak{n}}
\def\TarskiAdd{\mathfrak{A}}
\def\FraktM{\mathfrak{M}}
\def\FraktJ{\mathfrak{J}}

\def\I1{\mathord{\setbox0=\hbox{\rm1}\setbox1=\hbox{\rm l}
   \copy0\kern-.45\wd0\box1}}

\def\PowerRel{\zeta}

\def\ExistIm{\mathop{\strut\vartheta}}

\def\PowTWO#1{{\bf 2}\sp{#1}}

\def\StrictFork#1#2{\setbox1=\hbox{$\,\bigcirc\,$}%
    \mathop{\,(#1\copy1\kern-\wd1\hbox to\wd1{\hfil$<$\hfil}#2)\,}}
\def\StrictJoin#1#2{\setbox1=\hbox{$\,\bigcirc\,$}%
    \mathop{\,(#1\copy1\kern-\wd1\hbox to\wd1{\hfil$>$\hfil}#2)\,}}
\def\Kronecker#1#2{\setbox1=\hbox{$\,\bigcirc\,$}%
    \mathop{\,(#1\copy1\kern-\wd1\hbox to\wd1{\hfil$\times$\hfil}#2)\,}}

\def\liftedIdent{\raise0.2pt\hbox{${\cal I}$}}
\def\liftedBOT{\raise0.2pt\hbox{${\cal BOT}$}}
\def\liftedTOP{\raise0.2pt\hbox{${\cal TOP}$}}
\def\liftedNega{\raise0.2pt\hbox{${\cal N}$}}
\def\liftedMeet{\raise0.2pt\hbox{${\cal M}$}}
\def\liftedJoin{\raise0.2pt\hbox{${\cal J}$}}
\def\liftedCompo{\raise0.2pt\hbox{${\cal C}$}}
\def\liftedTransp{\raise0.2pt\hbox{${\cal T}$}}

\newcount\CaptionNo
\newcount\TotalFigureNo

\long\def\Caption#1#2#3{%
\global\advance\CaptionNo by 1%
\global\advance\TotalFigureNo by 1%
\immediate\write\lit{\def\ #3{\thesection.\the\CaptionNo}}%
\label{#3}%
\vbox{
   
\bigskip
\hbox to\textwidth{\hfil#1\hfil}%

\medskip
\hbox to\textwidth{\small\hfil{\bf Fig.~\thesection.\the\CaptionNo}\kern0.3cm #2\hfil}%
}%

\bigskip         
}

\def\RelToVect #1{\mathop {\tt vec}(#1)}
\def\VectToRel #1{\mathop {\tt rel}(#1)}

\def\RELand#1#2{#1\kern0.1ex\cap\kern0.1ex#2}
\def\RELandOP{\kern0.1ex\cap\kern0.1ex}

\def\RELorOP{\kern0.1ex\cup\kern0.1ex}
\def\RELenthOP{\mathrel{\subseteq}}         
\def\RELnenthOP{\setbox7=\hbox{\mathsurround0pt$\subset$}
                \setbox8=\hbox{\mathsurround0pt$\not=$}
                \mathrel{\;\raise0.23\baselineskip\copy7\kern-0.5\wd7\kern-0.5\wd8\lower0.23\baselineskip\copy8}}         
\def\RELcontainedUnequalOP{\hbox{\kern0.15\baselineskip\lower0.19\baselineskip
                      \vbox{\setbox0=\hbox{$\subset$}
                            \setbox1=\hbox{$\>\not=\>$}
                            \hbox to\wd1{\hfil\box0\hfil}
                            \kern-0.48\baselineskip
                            \hbox to\wd1{\hfil\box1\hfil}}}}

\def\RELaboveUnequalOP{\hbox{\kern0.15\baselineskip\lower0.19\baselineskip
                      \vbox{\setbox0=\hbox{$\supset$}
                            \setbox1=\hbox{$\>\not=\>$}
                            \hbox to\wd1{\hfil\box0\hfil}
                            \kern-0.48\baselineskip
                            \hbox to\wd1{\hfil\box1\hfil}}}}

\def\RELaboveOP{\mathrel{\supseteq}}    
\def\RELcomp#1#2{
                 #1{\mathbin{\raise0.3ex\hbox{\large{\emph{;}}}}}#2}

 \def\RELcompOP{
                \mathbin{\raise0.3ex\hbox{\kern-0.6ex\tiny{\emph{\char'73}}\kern-0.6ex}}}
\def\RELtop{{\setbox8=\hbox{\mathsurround=0pt$\top$}{\vbox to\ht8{\vfil  \hrule height0.05\ht8 width0.85\wd8
                                                                         \hbox to0.85\wd8{\hfil\vrule height0.89999\ht8\kern0.046cm\vrule\hfil}}}}}
\def\RELbot{{\setbox8=\hbox{\mathsurround=0pt$\bot$}{\vbox to\ht8{\vfil
                                                                         \hbox to0.85\wd8{\hfil\vrule height0.89999\ht8\kern0.046cm\vrule\hfil}
                                                                         \hrule height0.05\ht8 width0.85\wd8}}}}
\def\RELide{{\mathop{\setbox8=\hbox{\mathsurround=0pt$\top$}\kern0.05\wd8
                                                            \vbox to\ht8{\hrule height0.05\ht8 width0.55\wd8
                                                                         \vfil
                                                                         \hbox to0.55\wd8{\hfil\vrule height0.89999\ht8\kern0.046cm\vrule\hfil}
                                                                         \vfil
                                                                         \hrule height0.05\ht8 width0.55\wd8}%
                                                            \kern0.05\wd8}}}
\def\RELtraOP{\sp{\hbox{\tiny\sf  T}}}
\def\RELfromTO#1#2#3{#1:#2\longrightarrow#3}

\def\leftResi#1#2{#1\backslash #2}

\def\RELneg#1{\overline{#1}}

\def\invlim{\mathop{\hbox{\tentt lim\kern1.0pt}}}

\def\lsem{\setbox0=\hbox{\lbrack\kern-.06cm\lbrack}\hbox to\wd0{\lbrack\kern-.06cm\lbrack}}       
\def\rsem{\setbox0=\hbox{\lbrack\kern-.06cm\lbrack}\hbox to\wd0{\rbrack\kern-.06cm\rbrack}}

\def\ext#1{{\tt ext}(#1)}
\def\ext#1#2#3{\setbox3=\hbox{#2}\setbox4=\hbox{3}
\ifdim \wd3=0pt{\tt ext}\ifdim \wd4=0pt \else_{#3}\fi(#1)\else{\tt
ext}(#1,#2)\fi}

\newif\ifCUP
\CUPfalse

\def\sqr#1#2{{\vcenter{\vbox{\hrule height.#2pt
   \hbox{\vrule width.#2pt height#1pt \kern#1pt\kern-.4pt
      \vrule width.#2pt}
       \hrule height.#2pt}}}}
\def\square{\mathchoice\sqr54\sqr54\sqr{3.1}3\sqr{2.5}3}

\def\nukoef{\hbox to\matrixskip{\hfil\textbf 0\hfil}}
\def\alkoef{\hbox to\matrixskip{\hfil\textbf 1\hfil}}
\def\NOC  {\hbox to\matrixskip{\hfil\textbf X\hfil}}
\def\a{\alkoef}
\def\n{\nukoef}

\def\predand{\land}
\def\predor{\lor}

\def\AtomIde{\hbox{\mathsurround0pt$\,\vert$\kern-0.095cm{\rm J}$\,$}}
\def\StrictIde{\hbox{\mathsurround0pt$\,\vert$\kern-0.145cm{\rm S}$\,$}}

\def\Source{source}
\def\Target{target}

\newbox\matrixstrutbox
\def\matrixstrut{%
   \relax\ifmmode\copy\matrixstrutbox\else\unhcopy\matrixstrutbox\fi}
\newdimen\matrixskip
\newdimen\rowskip
\newdimen\columnskip

\newdimen\HeightOfRowNames

\def\definematrixskips#1 #2 {%
   \dimen1=#1
   \global\matrixskip=\dimen1
   \global\rowskip=\dimen1%
   \global\columnskip=#2%
   \global\setbox\matrixstrutbox=\hbox{%
       \vrule height.7\dimen1 depth.3\dimen1 width0pt}}%

\def\vstrichTWO{\kern-0.2pt\relax\matrixstrut\vrule width2pt\kern-0.2pt\relax}

\def\vstrich{\kern-0.2pt\relax\matrixstrut\vrule\kern-0.2pt\relax}
   \global\matrixskip=\dimen1
   \global\setbox\matrixstrutbox=\hbox{%
       \vrule height.7\dimen1 depth.3\dimen1 width0pt}%
   \setbox0=\hbox{\nukoef\alkoef}\dimen0=\dimen1%
   \advance\dimen0 by-\ht0\advance\dimen0 by-\dp0\divide\dimen0 by 2%
   \advance\dimen0 by-.3\dimen1\advance\dimen0 by\dp0%
   \global\def\hstrich{\noalign{\kern\dimen0%
      \kern-0.2pt\hrule\kern-\dimen0\kern-0.2pt}}
   \global\def\mstrich{\kern\dimen0%
         \kern-0.2pt\hrule\kern-\dimen0\kern-0.2pt}

\catcode`@=11

\def\smatrix#1{\null\vcenter{\offinterlineskip
   \baselineskip=\matrixskip\m@th%
   \ialign{\matrixstrut\hfil$##$\hfil&&\hfil$##$\hfil\crcr
   \mathstrut\crcr\noalign{\kern-\baselineskip}
   #1\crcr\mathstrut\crcr\noalign{\kern-\baselineskip}}}}

\def\spmatrix#1{\left(\kern-.2\matrixskip%
    \smatrix{#1}\kern-.2\matrixskip\right)}

\catcode`@=12

\def\e{\setbox7=\hbox{\a}\copy7\kern-\wd7\n}

\def\not#1{\setbox7=\hbox{\mathsurround0pt$\,#1\,$}\copy7\kern-\wd7%
   \hbox to\wd7{\hfil /\hfil}}

 \def\CoefTrue{{\color{red}\a}}
\def\CoefTruFal{\hbox to\matrixskip{\hfill}}

\newdimen\breadthTruFal
\def\TruFal{\setbox7=\hbox{$\circ$}\setbox8=\hbox{$1$}
\if\wd7\gt\wd8\breadthTruFal=\wd7
\else \breadthTruFal=\wd8\fi
\raise0.1\breadthTruFal\hbox to\breadthTruFal{\hfil $\circ$\hfil}\kern-\breadthTruFal\hbox to\breadthTruFal{\hfil $1$\hfil}}

\def\FatONE{\hbox to\matrixskip{\NoMoreColor{blue}\large\hfil\textbf 1\hfil}}
\def\FatZERO{\bf\NoMoreColor{blue} \lower1.1pt\hbox to\matrixskip{\small\hfil\bf O\hfil}}

\def\Boolean{Boolean}

\def\LeftResi#1#2{#1\backslash#2\,}
\def\RightResi#1#2{#1/#2}

\def\syq{\mathop{\hbox{\tt syq\kern1.2pt}}}
\def\syqq#1#2{\syq(#1,#2)}
\def\lub{\mathop{\hbox{\tt lub\kern1.2pt}}}
\def\lubR{\mathop{\hbox{\tt lubR\kern1.2pt}}}
\def\ubd{\mathop{\hbox{\tt ubd\kern1.2pt}}}
\def\lbd{\mathop{\hbox{\tt lbd\kern1.2pt}}}
\def\glb{\mathop{\hbox{\tt glb\kern1.2pt}}}
\def\glbR{\mathop{\hbox{\tt glbR\kern1.2pt}}}
\def\sup{\mathop{\hbox{\tt sup\kern1.2pt}}}
\def\inf{\mathop{\hbox{\tt inf\kern1.2pt}}}
\def\wp{\mathop{\hbox{\tt wp\kern1.2pt}}}
\def\Ma{\mathop{\hbox{\tt Ma\kern1.2pt}}}
\def\Mi{\mathop{\hbox{\tt Mi\kern1.2pt}}}
\def\ma{\mathop{\hbox{\tt ma\kern1.2pt}}}
\def\mi{\mathop{\hbox{\tt mi\kern1.2pt}}}
\def\min{\mathop{\hbox{\tt min\kern1.0pt}}}
\def\max{\mathop{\hbox{\tt max\kern1.0pt}}}

\def\upaRT#1{{\tt upa}\kern1.0pt(\kern-1.0pt#1\kern1.0pt)}
\def\mupRT#1{{\tt mup}\kern1.0pt(\kern-1.0pt#1\kern1.0pt)}
\def\lea{\mathop{\hbox{\tt lea\kern1.0pt}}}
\def\gre{\mathop{\hbox{\tt gre\kern1.0pt}}}
\def\greR{\mathop{\hbox{\tt greR\kern1.0pt}}}

\global\def\germanmonth{%
\ifnum\the\month=1 Januar\fi%
\ifnum\the\month=2 Februar\fi%
\ifnum\the\month=3 M\"arz\fi%
\ifnum\the\month=4 April\fi%
\ifnum\the\month=5 Mai\fi%
\ifnum\the\month=6 Juni\fi%
\ifnum\the\month=7 Juli\fi%
\ifnum\the\month=8 August\fi%
\ifnum\the\month=9 September\fi%
\ifnum\the\month=10 Oktober\fi%
\ifnum\the\month=11 November\fi%
\ifnum\the\month=12 Dezember\fi}%
\def\datum{\the\day.~\germanmonth\ \the\year}

\global\def\englishmonth{%
\ifnum\the\month=1 January\fi%
\ifnum\the\month=2 February\fi%
\ifnum\the\month=3 March\fi%
\ifnum\the\month=4 April\fi%
\ifnum\the\month=5 May\fi%
\ifnum\the\month=6 June\fi%
\ifnum\the\month=7 July\fi%
\ifnum\the\month=8 August\fi%
\ifnum\the\month=9 September\fi%
\ifnum\the\month=10 October\fi%
\ifnum\the\month=11 November\fi%
\ifnum\the\month=12 December\fi}%
\def\engldatum{\englishmonth\ \the\day, \the\year}

\long\def\ignoriere#1{}

\def\Acut{\accent"13}

\def\Euro{\strut\kern-0cm\lower0.018cm\hbox{\includegraphics[scale=0.045]{/Users/guntherschmidt/Dropbox/Documents/SuSTeX/SuSTeXEIN/EuroLetter}\kern0.05cm}}

\newdimen\RestNebenFoto
\long\def\CaptionFotoBottom#1#2{\newpage
\setbox7=\hbox{#1}
\RestNebenFoto=\textheight
\advance\RestNebenFoto by-\ht7
\vbox to\textheight{\vfil\vfil\vfil%
\hbox to\textwidth{\hfil\hbox{#1}\hfil}
\vfil
\hbox to\textwidth{\hfil\strut\rm #2\hfil}
\vfil\vfil\vfil\vfil\hbox to\textwidth{\small\quad\the\value{page}\hfil{\color{blue}\hyperlink{zumIndex}{$\rightarrow$ Index}\quad}}\kern0.2cm}}

\newdimen\RestNebenFoto
\long\def\CaptionFotoLeft#1#2#3#4{\newpage
\RestNebenFoto=0pt%
\setbox7=\hbox{\strut\rm #2}
\ifdim\wd7>\RestNebenFoto\RestNebenFoto=\wd7\fi%
\setbox7=\hbox{\strut\rm #3}
\ifdim\wd7>\RestNebenFoto\RestNebenFoto=\wd7\fi%
\setbox7=\hbox{\strut\rm #4}
\ifdim\wd7>\RestNebenFoto\RestNebenFoto=\wd7\fi%
\vbox to\textheight{\vfil%
\hbox to\textwidth{\hfil$\vcenter{\hbox{#1}}\hfil\vcenter{\hbox to\RestNebenFoto{\hfil\strut\rm #2\hfil}%
\hbox to\RestNebenFoto{\hfil\strut\rm #3\hfil}\hbox to\RestNebenFoto{\hfil\strut\rm #4\hfil}}$\hfil}
\vfil\hbox to\textwidth{\small\quad\the\value{page}\hfil{\color{blue}\hyperlink{zumIndex}{$\rightarrow$ Index}\quad}}\kern0.2cm}}

\long\def\CaptionFotoRight#1#2#3#4{\newpage
\RestNebenFoto=0pt%
\setbox7=\hbox{\strut\rm #2}
\ifdim\wd7>\RestNebenFoto\RestNebenFoto=\wd7\fi%
\setbox7=\hbox{\strut\rm #3}
\ifdim\wd7>\RestNebenFoto\RestNebenFoto=\wd7\fi%
\setbox7=\hbox{\strut\rm #4}
\ifdim\wd7>\RestNebenFoto\RestNebenFoto=\wd7\fi%
\vbox to\textheight{\vfil%
\hbox to\textwidth{\hfil$\vcenter{\hbox to\RestNebenFoto{\hfil\strut\rm #2\hfil}%
\hbox to\RestNebenFoto{\hfil\strut\rm #3\hfil}\hbox to\RestNebenFoto{\hfil\strut\rm #4\hfil}}\hfil\vcenter{\hbox{#1}}$\hfil}
\vfil\hbox to\textwidth{\small\quad\the\value{page}\hfil{\color{blue}\hyperlink{zumIndex}{$\rightarrow$ Index}\quad}}\kern0.2cm}}

\def\NoMoreColor#1#2{#2}

\newfont{\HUGESF}{cminch scaled 580}
\newfont{\LARGESF}{cminch scaled 340}
\newfont{\BIGSF}{cminch scaled 170}

\def\PropSyqSurjFct{3.6}

\def\FigAtomSingle{4.1}

\def\PropExImagProps{5.2}

\def\PropRelVec{6.1}

\def\PropTranspositionPointfree{6.4}

\def\PropPiKrhoEquGeneral{7.2}
\def\PropForkMapKron{7.3}

\def\PropForkMapKronZierer{7.4}

\def\PropResiFork{7.6}
\def\FigSumPowToPowProdGeneral{7.4}
\def\AdditionSyqGeneralized{7.7}
\def\FigSumPowToPowProd{7.5}
\def\PropSumPowToPowProd{7.8}
\def\PropNegBinOp{8.1}

\def\DefBinOp{8.2}

\def\FigAssocLaw{8.2}

\def\FigHasInverses{8.3}

\def\FigHasRightNeutralUgly{8.6}
\def\FigTwoNeutral{8.7}

\def\DefDistr{8.8}
\def\FigSubsetElement{9.1}
\def\FigNegaDisjoint{9.2}
\def\FigSumPowToPowProdOnX{9.3}
\def\CorSumPowToPowProd{9.1}
\def\FigSumPowToPowProdMatrix{9.4}
\def\PropPowerMeet{9.2}

\def\PropPiAndRhoAndMeetSurj{9.4}

\def\PropAbsorbAssoc{9.5}

\newdimen\interspacereduction
\global\def\interspacereduction{-1pt}

\def\liftedJoin{\,\FraktJ\,}
\def\liftedMeet{\,\FraktM\,}

\newcount\ExerciseNo
\newcount\CaptionNo
\newcount\TotalFigureNo
\newcount\TotalExerciseNo
\title{Relational Mathematics Continued}

\author{{\sc Gunther Schmidt} \hskip3cm {\sc Michael Winter}\\[4ex]
{\normalsize Fakultät für Informatik, Universität der Bundeswehr München}
\\
{\small e-Mail: \texttt{Gunther.Schmidt@UniBw.DE}}
\\
\\
{\normalsize Brock University, St. Catharines, Ontario}
\\
{\small e-Mail: \texttt{Michael.Winter@BrockU.CA}}
}

\date{\today}

\makeindex
\begin{document}

\newpage

\titlepage
\vbox to\textheight{ 
\kern1cm

\hbox to\textwidth{\hfil\hfil\includegraphics[scale=0.5]{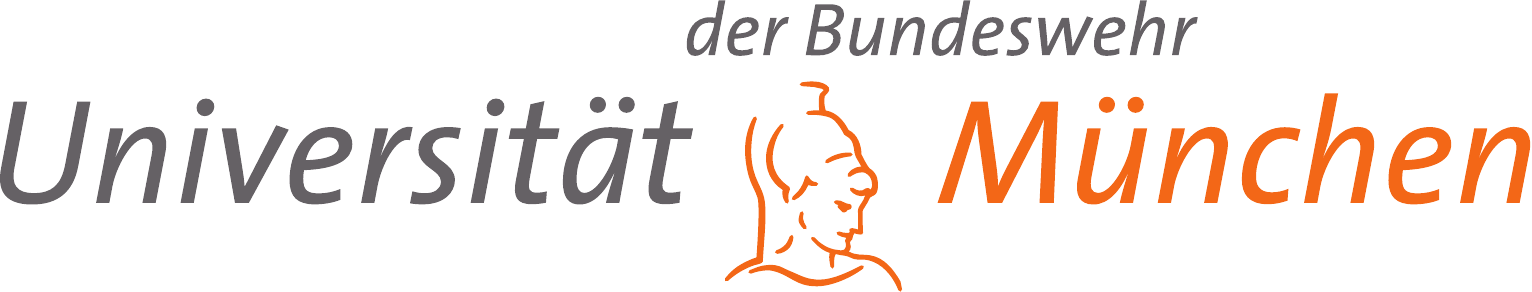}\hfil}

\kern0.1cm
\hbox to\textwidth{\hfil\hfil\sc Fakultät für Informatik\hfil}

\kern5cm
\hbox to\textwidth{\hfil{\LARGE\sc\bf Relational Mathematics Continued}\hfil}

\kern1cm
\hbox to\textwidth{\hfil\large\sc Gunther Schmidt\hfil Michael Winter\hfil}

\kern3cm
\hbox to\textwidth{\hfil\sc Bericht 2014-01\hfil}
\kern0.2cm
\hbox to\textwidth{\hfil\sc April 2014\hfil}

\vfill
\hbox to\textwidth{\hfil\sc Werner-Heisenberg-Weg 39 $\bullet$ 85577 Neubiberg $\bullet$  Germany\hfil}

\kern0.5cm}

\titlepage

\ 
\newpage

\maketitle
\pagenumbering{arabic}
\thispagestyle{empty}

\begin{center}
{\large{Abstract}}

\bigskip
\begin{minipage}{130mm}
\noindent
This is in some sense an addendum to \cite{RelaMath2010}. It originated from work on diverse other topics during which a lot of purely relational results with broad applicability have been produced. 
These include results on domain construction with novel formulae for existential and inverse image, a relational calculus for binary mappings, and the development of a formally derived relational calculus of Kronecker-, strict fork-, and strict join-operators. The many 
visualizations
in this report make it also a scrap- and picture book for examples.
\end{minipage}
\end{center}

\bigskip
\noindent
{\bf Keywords } relational mathematics, relation algebra, domain construction, vectorization, binary mapping, Kronecker-, fork-, and join-operator, products, 
existential and inverse image


\chapter{Introduction}
\ExerciseNo=0
\EnuncNo=0
\CaptionNo=0


In this report, several definitions, propositions and constructions are collected that already would have been incorporated in the book \cite{RelaMath2010} when they had been available at that time. This work is completely based on relation-algebraic methods. Nevertheless, we often use terms such as {\it set, powerset, etc.} to give intuition for the  concepts intended.

\bigskip
\noindent
Included is in Chapt.~\ref{ChaptPrerequisites}, what has to be mentioned from known relational methods to make the article self-contained. In addition, several new ideas of this kind are elaborated. Then follows a further study of the membership relation in Chapt.~\ref{ChapMemb} and, based on it, a presentation of novel insights on existential and inverse images in Chapt.~\ref{ChapPowerOps}. 
To underpin the often quite intuitive formulae with rigorous relation-algebraic proofs for the first time turned out to be an unexpectedly difficult task.

\bigskip
\noindent
The categorical product is again studied in Chapt.~\ref{SectRelAsPoint} and in 
Chapt.~\ref{ChapCateg}, working also with vectorization. Therewith, a relation may be seen in different incarnations, a relation as a rectangular \Boolean\ matrix or a \Boolean\ vector along the powerset, offering intricate interdependencies.

\bigskip
\noindent
While relations lend themselves mainly to being studied with linear concepts, it also possible to approach binary mappings or operations via relational mathematics as in Chapt.~\ref{ChapBinOp}. Application of such concepts allows to study \Boolean\ algebra from quite a different perspective in 
Chapt.~\ref{SectBoolAlg}.

\bigskip
\noindent
A slight generalization has taken place: It is known that relational mathematics
admits also non-representable relation algebras as models --- in case the Point Axiom should not have been postulated.

\bigskip
\noindent
The presentation via computer-generated examples allows a very detailed view. They have been generated with the language {\sc TituRel} (see \cite{Schmidt-2004a}), that directly interprets relational terms and formulae. So one can be sure to see the results of the explanations in the text directly mirrored.
Accumulating such a multitude of rules and formulae follows the idea of Ren\Acut e Descartes\index{Descartes, Ren\Acut e}, who is told to have said: \glqq Jedes Problem, das ich gelöst hatte, wurde zu einer Regel, mit deren Hilfe später weitere Probleme gelöst werden konnten.\grqq


\chapter{Prerequisites}\label{ChaptPrerequisites}
\ExerciseNo=0
\EnuncNo=0
\CaptionNo=0


The prerequisites presented routinely for relational work are fairly well-known: 
We will work with heterogeneous relations and provide a general reference to \cite{RelaMath2010}, but also to the earlier \cite{RelaGraDEUTSCH,RelaGraENGLISCH,SchmidtHattenspergerWinter1997}. Our operations are, thus, binary union \lq\lq$\RELorOP$\rq\rq, 
intersection \lq\lq$\RELandOP$\rq\rq, composition \lq\lq$\,\,\RELcompOP\,\,$\rq\rq, unary negation \lq\lq$\,\RELneg{\phantom{m}}\,$\rq\rq, transposition or conversion \lq\lq$\phantom{n}\RELtraOP\,$\rq\rq, together with zero-ary null relations \lq\lq$\RELbot$\rq\rq, universal relations \lq\lq$\,\RELtop\,$\rq\rq, and identities \lq\lq$\,\RELide\,$\rq\rq. A {\it heterogeneous relation algebra}

\begin{itemize}
\item{}is a category wrt.~composition \lq\lq$\;\RELcompOP\;$\rq\rq\ and identities $\RELide$,

\item{}has as morphism sets complete atomic boolean lattices with  
$\RELorOP,\RELandOP ,\RELneg{\phantom{n}}, \RELbot , \RELtop,\RELenthOP$,

\item{}obeys rules for transposition $\RELtraOP$ in connection with the latter two concepts
that may be stated in either one of the following two ways:

\smallskip
\noindent
{\it Dedekind rule:\index{Dedekind rule}}

\quad$R\RELcompOP S\RELandOP  Q
\RELenthOP(R\RELandOP  Q\RELcompOP S\RELtraOP)\RELcompOP\,
(S\RELandOP  R\RELtraOP\RELcompOP  Q)$

\smallskip
\noindent
{\it Schr\"oder equivalences:}\index{Schroeder equivalences@Schr\"oder equivalences}

\quad$A\RELcompOP B\RELenthOP C\quad\iff\quad 
A\RELtraOP\RELcompOP\RELneg{C}\RELenthOP\RELneg{B}\quad
\iff\quad\RELneg{C}\RELcompOP B\RELtraOP\RELenthOP\RELneg{A}$
\end{itemize}

\bigskip
\noindent
The two rules are equivalent in the context mentioned. Many rules follow out of this setting; not least everything for the concepts of a function, mapping, or ordering; e.g., that mappings $f$ may be {\it shunted}\index{shunting}, i.e., that $A\RELcompOP f\RELenthOP B\iff A\RELenthOP B\RELcompOP f\RELtraOP$. The rule $(A\RELandOP B\RELcompOP g\RELtraOP)\RELcompOP g=A\RELcompOP g\RELandOP B$ for univalent $g$ is also frequently applied and sometimes referred to as {\it destroy and append\/};  Prop.~5.4 of \cite{RelaMath2010}.

\bigskip
\noindent
A new and widely useful rule serves to negate the left-composition with a partial identity:

\enunc{}{Proposition}{}{PropNegDiag} $\RELneg{(\RELide\RELandOP\Delta)\RELcompOP\RELtop}=(\RELide\RELandOP\RELneg{\Delta})\RELcompOP\RELtop$\quad for an arbitrary homogeneous relation $\Delta$.

\Proof $\RELtop
=
\RELide\RELcompOP\RELtop
=
\big\lbrack\RELide\RELandOP(\Delta\RELorOP\RELneg{\Delta})\big\rbrack\RELcompOP\RELtop
=
\big\lbrack(\RELide\RELandOP\Delta)\RELorOP(\RELide\RELandOP\RELneg{\Delta})\big\rbrack\RELcompOP\RELtop
=
(\RELide\RELandOP\Delta)\RELcompOP\RELtop\RELorOP(\RELide\RELandOP\RELneg{\Delta})\RELcompOP\RELtop
$ implies

$\RELneg{(\RELide\RELandOP\Delta)\RELcompOP\RELtop}\RELenthOP(\RELide\RELandOP\RELneg{\Delta})\RELcompOP\RELtop$,\quad thus proving direction \lq\lq$\RELenthOP$\rq\rq.

\bigskip
\noindent
For \lq\lq$\RELaboveOP$\rq\rq, we use that $\RELide\RELandOP\RELneg{\Delta}\RELenthOP\RELide$ is univalent, prior to applying the Schröder rule:

\smallskip
$
(\RELide\RELandOP\RELneg{\Delta})\RELtraOP\RELcompOP(\RELide\RELandOP\Delta)\RELcompOP\RELtop
=
\lbrack(\RELide\RELandOP\RELneg{\Delta})\RELcompOP\RELide\RELandOP(\RELide\RELandOP\RELneg{\Delta})\RELcompOP\Delta\rbrack\RELcompOP\RELtop
\RELenthOP
\lbrack\RELide\RELandOP\RELneg{\Delta}\RELandOP\Delta\rbrack\RELcompOP\RELtop
=
\RELbot
$
\Bewende

\noindent
It is relatively hard to see: this  specializes Prop.~5.6 in \cite{RelaMath2010} for a homogeneous relation $\Delta$:

\smallskip
$\RELneg{(\RELide\RELandOP\Delta)\RELcompOP R}=
(\RELide\RELandOP\RELneg{\Delta})\RELcompOP\RELtop\RELorOP(\RELide\RELandOP\Delta)\RELcompOP\RELneg{R}$

\bigskip
\noindent
Another rule that sometimes proves helpful is the following:

\enunc{}{Proposition}{}{PropTwoMaps} For any two mappings $\RELfromTO{f,g}{X}{Y}$, this rule holds:

\smallskip
$
(\RELneg{f}\RELandOP g)\RELcompOP\RELtop
=
(f\RELandOP\RELneg{g})\RELcompOP\RELtop
$

\Proof $
(\RELneg{f}\RELandOP g)\RELcompOP\RELtop
=
(f\RELcompOP\RELneg{\RELide}\RELandOP g)\RELcompOP\RELtop
\RELenthOP
(f\RELandOP g\RELcompOP\RELneg{\RELide})\RELcompOP(\RELneg{\RELide}\RELandOP f\RELtraOP\RELcompOP g)\RELcompOP\RELtop
=
(f\RELandOP\RELneg{g})\RELcompOP(\RELneg{\RELide}\RELandOP f\RELtraOP\RELcompOP g)\RELcompOP\RELtop
\RELenthOP
(f\RELandOP\RELneg{g})\RELcompOP\RELtop
$
\Bewende

\noindent
There exist two resp.~three versions of an interpretation. The first one takes two mappings $f,g$ which never assign the same value. In this case both sides result in $\RELtop$. Then there may be two mappings with one or more values identical. In this case, precisely the respective arguments lead to $\n$-rows; they may even lead to $\RELbot$ when $f=g$.


\chapter{Symmetric quotient and membership\label{ChapMemb}}
\EnuncNo=0
\CaptionNo=0


\noindent
When a non-commutative composition is available, one usually looks for the left and the right residual\index{residual}, defined via 

\smallskip
$A\RELcompOP B\RELenthOP C\;\Longleftrightarrow\;
A \RELenthOP\RELneg{\RELneg{C}\RELcompOP B\RELtraOP} =:C/B$\quad and\quad $A\RELcompOP B\RELenthOP C\;\Longleftrightarrow\;
B \RELenthOP\RELneg{A\RELtraOP\RELcompOP\RELneg{C}} =:A\backslash C$.  

\smallskip
\noindent
Residuations have been studied intensively, not least in the context of Heyting algebras. 
We prove some rules for residuals:

\enunc{}{Proposition}{Residue cancellation}{PropResiRules}\index{cancellation} The following formulae hold for arbitrary relations $Q,R,T$ --- provided typing is correct\index{residue cancellation}: 

\begin{enumerate}[i)]
\item $\RightResi{(\LeftResi{Q}{R})}{T}=\LeftResi{Q}{(\RightResi{R}{T})}$
\item $\LeftResi{Q}{Q}=\RightResi{(\LeftResi{Q}{Q})}{(\LeftResi{Q}{Q})}$
\item $\RightResi{Q}{(R\RELcompOP U)}\RELenthOP\RightResi{(Q\RELcompOP U\RELtraOP)}{R}$\quad if $U$ is total
\end{enumerate}

\Proof i) $\RightResi{(\LeftResi{Q}{R})}{T}
=\RELneg{\RELneg{\RELneg{Q\RELtraOP\RELcompOP\RELneg{R}}}\RELcompOP T\RELtraOP}
=\RELneg{Q\RELtraOP\RELcompOP\RELneg{R}\RELcompOP T\RELtraOP}
$ and symmetrically to the other side.

\bigskip
\noindent
ii) $\LeftResi{Q}{Q}=\RELneg{Q\RELtraOP\RELcompOP\RELneg{Q}}$ implies that
$\RightResi{(\LeftResi{Q}{Q})}{(\LeftResi{Q}{Q})}
=
\RELneg{\RELneg{\RELneg{Q\RELtraOP\RELcompOP\RELneg{Q}}}\RELcompOP\RELneg{Q\RELtraOP\RELcompOP\RELneg{Q}}\RELtraOP}
=
\RELneg{Q\RELtraOP\RELcompOP\RELneg{Q}\RELcompOP\RELneg{\RELneg{Q}\RELtraOP\RELcompOP Q}}
=
\RELneg{Q\RELtraOP\RELcompOP\RELneg{Q}}
$, since $\RELneg{Q}\RELcompOP\RELneg{\RELneg{Q}\RELtraOP\RELcompOP Q}=\RELneg{Q}$, which in turn follows from Schröder's rule and reflexivity of $\RELneg{\RELneg{Q}\RELtraOP\RELcompOP Q}$.

\bigskip
\noindent
iii) $\RELtop=\RELtop\RELcompOP U\RELtraOP
=
(Q\RELorOP\RELneg{Q})\RELcompOP U\RELtraOP
=
Q\RELcompOP U\RELtraOP\RELorOP\RELneg{Q}\RELcompOP U\RELtraOP
\quad\iff\quad
\RELneg{Q\RELcompOP U\RELtraOP}\RELenthOP\RELneg{Q}\RELcompOP U\RELtraOP
$

$\Longrightarrow\quad
\RELneg{Q\RELcompOP U\RELtraOP}\RELcompOP R\RELtraOP\RELenthOP\RELneg{Q}\RELcompOP U\RELtraOP\RELcompOP R\RELtraOP
\quad\iff\quad
\RELneg{\RELneg{Q}\RELcompOP U\RELtraOP\RELcompOP R\RELtraOP}\RELenthOP\RELneg{\RELneg{Q\RELcompOP U\RELtraOP}\RELcompOP R\RELtraOP}
$
\Bewende

\newpage
\noindent
Intersecting such residuals in $\syqq{R}{S}:= \RELneg{R\RELtraOP\RELcompOP \RELneg{S}} \RELandOP  \RELneg{\RELneg{R}\RELtraOP\RELcompOP S}$,
\label{SyqDef}
the {\it symmetric quotient\/} $\RELfromTO{\syq(R,S)}{W}{Z}$ of two relations 
$\RELfromTO{R}{V}{W}$ and $\RELfromTO{S}{V}{Z}$ is defined. Symmetric quotients\index{symmetric quotient} serve the purpose of \lq column comparison\rq:

\smallskip
$\big\lbrack\syq(R,S)\big\rbrack_{wz}=\forall v\in V: R_{vw}\longleftrightarrow S_{vz}$.

\smallskip
\noindent
The following result may easily be understood. If a column of $A$ and the corresponding one of $B$ are equal to some column of $C$, then also their intersection and union will be equal. 

\enunc{}{Proposition}{}{PropSyqAndOr} For arbitrary relations $A,B,C$ with all the same source always

\smallskip
$\syqq{A}{C}\RELandOP\syqq{B}{C}\RELenthOP\syqq{A\RELandOP B}{C}\RELandOP\syqq{A\RELorOP B}{C}$.

\Proof For inclusion in the first term, we expand the symmetric quotients and negate to obtain

\smallskip
$\RELneg{A\RELandOP B}\RELtraOP\RELcompOP C\RELorOP(A\RELandOP B)\RELtraOP\RELcompOP\RELneg{C}
\RELenthOP
\RELneg{A}\RELtraOP\RELcompOP C\RELorOP A\RELtraOP\RELcompOP\RELneg{C}
\;\RELorOP\;
\RELneg{B}\RELtraOP\RELcompOP C\RELorOP B\RELtraOP\RELcompOP\RELneg{C}
$,

\smallskip
\noindent
which is obviously satisfied. This is then used to prove the other part.

\smallskip
$\syqq{A\RELorOP B}{C}
=
\syqq{\RELneg{A\RELorOP B}}{\RELneg{C}}
=
\syqq{\RELneg{A}\RELandOP\RELneg{B}}{\RELneg{C}}
$\quad now applying the former

$\RELaboveOP
\syqq{\RELneg{A}}{\RELneg{C}}\RELandOP\syqq{\RELneg{B}}{\RELneg{C}}
=
\syqq{A}{C}\RELandOP\syqq{B}{C}
$
\Bewende

\kern-0.1cm
\noindent
The illustration of the symmetric quotient is as follows:

\Caption{$\vcenter{\hbox{$
{\footnotesize%
\BoxBreadth=0pt%
\setbox7=\hbox{US}%
\ifdim\wd7>\BoxBreadth\BoxBreadth=\wd7\fi%
\setbox7=\hbox{French}%
\ifdim\wd7>\BoxBreadth\BoxBreadth=\wd7\fi%
\setbox7=\hbox{German}%
\ifdim\wd7>\BoxBreadth\BoxBreadth=\wd7\fi%
\setbox7=\hbox{British}%
\ifdim\wd7>\BoxBreadth\BoxBreadth=\wd7\fi%
\setbox7=\hbox{Spanish}%
\ifdim\wd7>\BoxBreadth\BoxBreadth=\wd7\fi%
\def\RowNames{\vcenter{\offinterlineskip\baselineskip=\matrixskip%
\hbox to\BoxBreadth{\strut\hfil US}\kern-1pt%
\hbox to\BoxBreadth{\strut\hfil French}\kern-1pt%
\hbox to\BoxBreadth{\strut\hfil German}\kern-1pt%
\hbox to\BoxBreadth{\strut\hfil British}\kern-1pt%
\hbox to\BoxBreadth{\strut\hfil Spanish}}}%
\BoxBreadthCOL=0pt%
\BoxHeightCOL=0pt%
\setbox7=\vbox{\hbox{\strut A}}%
\ifdim\wd7>\BoxBreadthCOL\BoxBreadthCOL=\wd7\fi%
\ifdim\ht7>\BoxHeightCOL\BoxHeightCOL=\ht7\fi\setbox7=\vbox{\hbox{\strut K}}%
\ifdim\wd7>\BoxBreadthCOL\BoxBreadthCOL=\wd7\fi%
\ifdim\ht7>\BoxHeightCOL\BoxHeightCOL=\ht7\fi\setbox7=\vbox{\hbox{\strut Q}}%
\ifdim\wd7>\BoxBreadthCOL\BoxBreadthCOL=\wd7\fi%
\ifdim\ht7>\BoxHeightCOL\BoxHeightCOL=\ht7\fi\setbox7=\vbox{\hbox{\strut J}}%
\ifdim\wd7>\BoxBreadthCOL\BoxBreadthCOL=\wd7\fi%
\ifdim\ht7>\BoxHeightCOL\BoxHeightCOL=\ht7\fi\setbox7=\vbox{\hbox{\strut 10}}%
\ifdim\wd7>\BoxBreadthCOL\BoxBreadthCOL=\wd7\fi%
\ifdim\ht7>\BoxHeightCOL\BoxHeightCOL=\ht7\fi\setbox7=\vbox{\hbox{\strut 9}}%
\ifdim\wd7>\BoxBreadthCOL\BoxBreadthCOL=\wd7\fi%
\ifdim\ht7>\BoxHeightCOL\BoxHeightCOL=\ht7\fi\setbox7=\vbox{\hbox{\strut 8}}%
\ifdim\wd7>\BoxBreadthCOL\BoxBreadthCOL=\wd7\fi%
\ifdim\ht7>\BoxHeightCOL\BoxHeightCOL=\ht7\fi\setbox7=\vbox{\hbox{\strut 7}}%
\ifdim\wd7>\BoxBreadthCOL\BoxBreadthCOL=\wd7\fi%
\ifdim\ht7>\BoxHeightCOL\BoxHeightCOL=\ht7\fi\setbox7=\vbox{\hbox{\strut 6}}%
\ifdim\wd7>\BoxBreadthCOL\BoxBreadthCOL=\wd7\fi%
\ifdim\ht7>\BoxHeightCOL\BoxHeightCOL=\ht7\fi\setbox7=\vbox{\hbox{\strut 5}}%
\ifdim\wd7>\BoxBreadthCOL\BoxBreadthCOL=\wd7\fi%
\ifdim\ht7>\BoxHeightCOL\BoxHeightCOL=\ht7\fi\setbox7=\vbox{\hbox{\strut 4}}%
\ifdim\wd7>\BoxBreadthCOL\BoxBreadthCOL=\wd7\fi%
\ifdim\ht7>\BoxHeightCOL\BoxHeightCOL=\ht7\fi\setbox7=\vbox{\hbox{\strut 3}}%
\ifdim\wd7>\BoxBreadthCOL\BoxBreadthCOL=\wd7\fi%
\ifdim\ht7>\BoxHeightCOL\BoxHeightCOL=\ht7\fi\setbox7=\vbox{\hbox{\strut 2}}%
\ifdim\wd7>\BoxBreadthCOL\BoxBreadthCOL=\wd7\fi%
\ifdim\ht7>\BoxHeightCOL\BoxHeightCOL=\ht7\fi%
\def\ColNames{\rotatebox{90}{\strut A}\kern-1pt%
\rotatebox{90}{\strut K}\kern-1pt%
\rotatebox{90}{\strut Q}\kern-1pt%
\rotatebox{90}{\strut J}\kern-1pt%
\rotatebox{90}{\strut 10}\kern-1pt%
\rotatebox{90}{\strut 9}\kern-1pt%
\rotatebox{90}{\strut 8}\kern-1pt%
\rotatebox{90}{\strut 7}\kern-1pt%
\rotatebox{90}{\strut 6}\kern-1pt%
\rotatebox{90}{\strut 5}\kern-1pt%
\rotatebox{90}{\strut 4}\kern-1pt%
\rotatebox{90}{\strut 3}\kern-1pt%
\rotatebox{90}{\strut 2}}%
\def\Matrix{\spmatrix{%
\noalign{\kern-2pt}
 \n&\n&\n&\n&\n&\n&\n&\n&\n&\n&\n&\n&{\CoefTrue}\cr
 \n&{\CoefTrue}&\n&\n&\n&\n&\n&{\CoefTrue}&\n&\n&\n&\n&\n\cr
 \n&\n&{\CoefTrue}&\n&\n&\n&{\CoefTrue}&{\CoefTrue}&\n&{\CoefTrue}&\n&{\CoefTrue}&\n\cr
 \n&{\CoefTrue}&{\CoefTrue}&\n&\n&\n&\n&{\CoefTrue}&\n&\n&\n&\n&{\CoefTrue}\cr
 \n&\n&\n&{\CoefTrue}&\n&\n&{\CoefTrue}&\n&\n&{\CoefTrue}&{\CoefTrue}&\n&{\CoefTrue}\cr
\noalign{\kern-2pt}}}
R= \vbox{\setbox8=\hbox{$\RowNames\Matrix$}
\hbox to\wd8{\hfil$\ColNames$\kern\ColEntryShiftHoriz}\kern\ColEntryShiftVerti
\box8}}
$}%
\hbox{$
{\footnotesize%
\BoxBreadth=0pt%
\setbox7=\hbox{US}%
\ifdim\wd7>\BoxBreadth\BoxBreadth=\wd7\fi%
\setbox7=\hbox{French}%
\ifdim\wd7>\BoxBreadth\BoxBreadth=\wd7\fi%
\setbox7=\hbox{German}%
\ifdim\wd7>\BoxBreadth\BoxBreadth=\wd7\fi%
\setbox7=\hbox{British}%
\ifdim\wd7>\BoxBreadth\BoxBreadth=\wd7\fi%
\setbox7=\hbox{Spanish}%
\ifdim\wd7>\BoxBreadth\BoxBreadth=\wd7\fi%
\def\RowNames{\vcenter{\offinterlineskip\baselineskip=\matrixskip%
\hbox to\BoxBreadth{\strut\hfil US}\kern-1pt%
\hbox to\BoxBreadth{\strut\hfil French}\kern-1pt%
\hbox to\BoxBreadth{\strut\hfil German}\kern-1pt%
\hbox to\BoxBreadth{\strut\hfil British}\kern-1pt%
\hbox to\BoxBreadth{\strut\hfil Spanish}}}%
\BoxBreadthCOL=0pt%
\BoxHeightCOL=0pt%
\setbox7=\vbox{\hbox{\strut Jan}}%
\ifdim\wd7>\BoxBreadthCOL\BoxBreadthCOL=\wd7\fi%
\ifdim\ht7>\BoxHeightCOL\BoxHeightCOL=\ht7\fi\setbox7=\vbox{\hbox{\strut Feb}}%
\ifdim\wd7>\BoxBreadthCOL\BoxBreadthCOL=\wd7\fi%
\ifdim\ht7>\BoxHeightCOL\BoxHeightCOL=\ht7\fi\setbox7=\vbox{\hbox{\strut Mar}}%
\ifdim\wd7>\BoxBreadthCOL\BoxBreadthCOL=\wd7\fi%
\ifdim\ht7>\BoxHeightCOL\BoxHeightCOL=\ht7\fi\setbox7=\vbox{\hbox{\strut Apr}}%
\ifdim\wd7>\BoxBreadthCOL\BoxBreadthCOL=\wd7\fi%
\ifdim\ht7>\BoxHeightCOL\BoxHeightCOL=\ht7\fi\setbox7=\vbox{\hbox{\strut May}}%
\ifdim\wd7>\BoxBreadthCOL\BoxBreadthCOL=\wd7\fi%
\ifdim\ht7>\BoxHeightCOL\BoxHeightCOL=\ht7\fi\setbox7=\vbox{\hbox{\strut Jun}}%
\ifdim\wd7>\BoxBreadthCOL\BoxBreadthCOL=\wd7\fi%
\ifdim\ht7>\BoxHeightCOL\BoxHeightCOL=\ht7\fi\setbox7=\vbox{\hbox{\strut Jul}}%
\ifdim\wd7>\BoxBreadthCOL\BoxBreadthCOL=\wd7\fi%
\ifdim\ht7>\BoxHeightCOL\BoxHeightCOL=\ht7\fi\setbox7=\vbox{\hbox{\strut Aug}}%
\ifdim\wd7>\BoxBreadthCOL\BoxBreadthCOL=\wd7\fi%
\ifdim\ht7>\BoxHeightCOL\BoxHeightCOL=\ht7\fi\setbox7=\vbox{\hbox{\strut Sep}}%
\ifdim\wd7>\BoxBreadthCOL\BoxBreadthCOL=\wd7\fi%
\ifdim\ht7>\BoxHeightCOL\BoxHeightCOL=\ht7\fi\setbox7=\vbox{\hbox{\strut Oct}}%
\ifdim\wd7>\BoxBreadthCOL\BoxBreadthCOL=\wd7\fi%
\ifdim\ht7>\BoxHeightCOL\BoxHeightCOL=\ht7\fi\setbox7=\vbox{\hbox{\strut Nov}}%
\ifdim\wd7>\BoxBreadthCOL\BoxBreadthCOL=\wd7\fi%
\ifdim\ht7>\BoxHeightCOL\BoxHeightCOL=\ht7\fi\setbox7=\vbox{\hbox{\strut Dec}}%
\ifdim\wd7>\BoxBreadthCOL\BoxBreadthCOL=\wd7\fi%
\ifdim\ht7>\BoxHeightCOL\BoxHeightCOL=\ht7\fi%
\def\ColNames{\rotatebox{90}{\strut Jan}\kern-1pt%
\rotatebox{90}{\strut Feb}\kern-1pt%
\rotatebox{90}{\strut Mar}\kern-1pt%
\rotatebox{90}{\strut Apr}\kern-1pt%
\rotatebox{90}{\strut May}\kern-1pt%
\rotatebox{90}{\strut Jun}\kern-1pt%
\rotatebox{90}{\strut Jul}\kern-1pt%
\rotatebox{90}{\strut Aug}\kern-1pt%
\rotatebox{90}{\strut Sep}\kern-1pt%
\rotatebox{90}{\strut Oct}\kern-1pt%
\rotatebox{90}{\strut Nov}\kern-1pt%
\rotatebox{90}{\strut Dec}}%
\def\Matrix{\spmatrix{%
\noalign{\kern-2pt}
 \n&\n&\n&{\CoefTrue}&\n&{\CoefTrue}&{\CoefTrue}&{\CoefTrue}&\n&{\CoefTrue}&\n&\n\cr
 {\CoefTrue}&\n&\n&{\CoefTrue}&\n&\n&{\CoefTrue}&\n&\n&{\CoefTrue}&\n&\n\cr
 {\CoefTrue}&{\CoefTrue}&\n&\n&{\CoefTrue}&{\CoefTrue}&\n&{\CoefTrue}&\n&\n&\n&{\CoefTrue}\cr
 {\CoefTrue}&{\CoefTrue}&\n&\n&\n&\n&{\CoefTrue}&\n&{\CoefTrue}&\n&{\CoefTrue}&{\CoefTrue}\cr
 \n&\n&\n&{\CoefTrue}&\n&{\CoefTrue}&{\CoefTrue}&{\CoefTrue}&\n&\n&\n&\n\cr
\noalign{\kern-2pt}}}
S= \vbox{\setbox8=\hbox{$\RowNames\Matrix$}
\hbox to\wd8{\hfil$\ColNames$\kern\ColEntryShiftHoriz}\kern\ColEntryShiftVerti
\box8}}
$}
}
{\footnotesize%
\BoxBreadth=0pt%
\setbox7=\hbox{A}%
\ifdim\wd7>\BoxBreadth\BoxBreadth=\wd7\fi%
\setbox7=\hbox{K}%
\ifdim\wd7>\BoxBreadth\BoxBreadth=\wd7\fi%
\setbox7=\hbox{Q}%
\ifdim\wd7>\BoxBreadth\BoxBreadth=\wd7\fi%
\setbox7=\hbox{J}%
\ifdim\wd7>\BoxBreadth\BoxBreadth=\wd7\fi%
\setbox7=\hbox{10}%
\ifdim\wd7>\BoxBreadth\BoxBreadth=\wd7\fi%
\setbox7=\hbox{9}%
\ifdim\wd7>\BoxBreadth\BoxBreadth=\wd7\fi%
\setbox7=\hbox{8}%
\ifdim\wd7>\BoxBreadth\BoxBreadth=\wd7\fi%
\setbox7=\hbox{7}%
\ifdim\wd7>\BoxBreadth\BoxBreadth=\wd7\fi%
\setbox7=\hbox{6}%
\ifdim\wd7>\BoxBreadth\BoxBreadth=\wd7\fi%
\setbox7=\hbox{5}%
\ifdim\wd7>\BoxBreadth\BoxBreadth=\wd7\fi%
\setbox7=\hbox{4}%
\ifdim\wd7>\BoxBreadth\BoxBreadth=\wd7\fi%
\setbox7=\hbox{3}%
\ifdim\wd7>\BoxBreadth\BoxBreadth=\wd7\fi%
\setbox7=\hbox{2}%
\ifdim\wd7>\BoxBreadth\BoxBreadth=\wd7\fi%
\def\RowNames{\vcenter{\offinterlineskip\baselineskip=\matrixskip%
\hbox to\BoxBreadth{\strut\hfil A}\kern-1.000099pt%
\hbox to\BoxBreadth{\strut\hfil K}\kern-1.000099pt%
\hbox to\BoxBreadth{\strut\hfil Q}\kern-1.000099pt%
\hbox to\BoxBreadth{\strut\hfil J}\kern-1.000099pt%
\hbox to\BoxBreadth{\strut\hfil 10}\kern-1.000099pt%
\hbox to\BoxBreadth{\strut\hfil 9}\kern-1.000099pt%
\hbox to\BoxBreadth{\strut\hfil 8}\kern-1.000099pt%
\hbox to\BoxBreadth{\strut\hfil 7}\kern-1.000099pt%
\hbox to\BoxBreadth{\strut\hfil 6}\kern-1.000099pt%
\hbox to\BoxBreadth{\strut\hfil 5}\kern-1.000099pt%
\hbox to\BoxBreadth{\strut\hfil 4}\kern-1.000099pt%
\hbox to\BoxBreadth{\strut\hfil 3}\kern-1.000099pt%
\hbox to\BoxBreadth{\strut\hfil 2}}}%
\BoxBreadthCOL=0pt%
\BoxHeightCOL=0pt%
\setbox7=\vbox{\hbox{\strut Jan}}%
\ifdim\wd7>\BoxBreadthCOL\BoxBreadthCOL=\wd7\fi%
\ifdim\ht7>\BoxHeightCOL\BoxHeightCOL=\ht7\fi\setbox7=\vbox{\hbox{\strut Feb}}%
\ifdim\wd7>\BoxBreadthCOL\BoxBreadthCOL=\wd7\fi%
\ifdim\ht7>\BoxHeightCOL\BoxHeightCOL=\ht7\fi\setbox7=\vbox{\hbox{\strut Mar}}%
\ifdim\wd7>\BoxBreadthCOL\BoxBreadthCOL=\wd7\fi%
\ifdim\ht7>\BoxHeightCOL\BoxHeightCOL=\ht7\fi\setbox7=\vbox{\hbox{\strut Apr}}%
\ifdim\wd7>\BoxBreadthCOL\BoxBreadthCOL=\wd7\fi%
\ifdim\ht7>\BoxHeightCOL\BoxHeightCOL=\ht7\fi\setbox7=\vbox{\hbox{\strut May}}%
\ifdim\wd7>\BoxBreadthCOL\BoxBreadthCOL=\wd7\fi%
\ifdim\ht7>\BoxHeightCOL\BoxHeightCOL=\ht7\fi\setbox7=\vbox{\hbox{\strut Jun}}%
\ifdim\wd7>\BoxBreadthCOL\BoxBreadthCOL=\wd7\fi%
\ifdim\ht7>\BoxHeightCOL\BoxHeightCOL=\ht7\fi\setbox7=\vbox{\hbox{\strut Jul}}%
\ifdim\wd7>\BoxBreadthCOL\BoxBreadthCOL=\wd7\fi%
\ifdim\ht7>\BoxHeightCOL\BoxHeightCOL=\ht7\fi\setbox7=\vbox{\hbox{\strut Aug}}%
\ifdim\wd7>\BoxBreadthCOL\BoxBreadthCOL=\wd7\fi%
\ifdim\ht7>\BoxHeightCOL\BoxHeightCOL=\ht7\fi\setbox7=\vbox{\hbox{\strut Sep}}%
\ifdim\wd7>\BoxBreadthCOL\BoxBreadthCOL=\wd7\fi%
\ifdim\ht7>\BoxHeightCOL\BoxHeightCOL=\ht7\fi\setbox7=\vbox{\hbox{\strut Oct}}%
\ifdim\wd7>\BoxBreadthCOL\BoxBreadthCOL=\wd7\fi%
\ifdim\ht7>\BoxHeightCOL\BoxHeightCOL=\ht7\fi\setbox7=\vbox{\hbox{\strut Nov}}%
\ifdim\wd7>\BoxBreadthCOL\BoxBreadthCOL=\wd7\fi%
\ifdim\ht7>\BoxHeightCOL\BoxHeightCOL=\ht7\fi\setbox7=\vbox{\hbox{\strut Dec}}%
\ifdim\wd7>\BoxBreadthCOL\BoxBreadthCOL=\wd7\fi%
\ifdim\ht7>\BoxHeightCOL\BoxHeightCOL=\ht7\fi%
\def\ColNames{\rotatebox{90}{\strut Jan}\kern-1pt%
\rotatebox{90}{\strut Feb}\kern-1pt%
\rotatebox{90}{\strut Mar}\kern-1pt%
\rotatebox{90}{\strut Apr}\kern-1pt%
\rotatebox{90}{\strut May}\kern-1pt%
\rotatebox{90}{\strut Jun}\kern-1pt%
\rotatebox{90}{\strut Jul}\kern-1pt%
\rotatebox{90}{\strut Aug}\kern-1pt%
\rotatebox{90}{\strut Sep}\kern-1pt%
\rotatebox{90}{\strut Oct}\kern-1pt%
\rotatebox{90}{\strut Nov}\kern-1pt%
\rotatebox{90}{\strut Dec}}%
\def\Matrix{\spmatrix{%
\noalign{\kern-2pt}
 \n&\n&{\CoefTrue}&\n&\n&\n&\n&\n&\n&\n&\n&\n\cr
 \n&\n&\n&\n&\n&\n&\n&\n&\n&\n&\n&\n\cr
 \n&{\CoefTrue}&\n&\n&\n&\n&\n&\n&\n&\n&\n&{\CoefTrue}\cr
 \n&\n&\n&\n&\n&\n&\n&\n&\n&\n&\n&\n\cr
 \n&\n&{\CoefTrue}&\n&\n&\n&\n&\n&\n&\n&\n&\n\cr
 \n&\n&{\CoefTrue}&\n&\n&\n&\n&\n&\n&\n&\n&\n\cr
 \n&\n&\n&\n&\n&\n&\n&\n&\n&\n&\n&\n\cr
 {\CoefTrue}&\n&\n&\n&\n&\n&\n&\n&\n&\n&\n&\n\cr
 \n&\n&{\CoefTrue}&\n&\n&\n&\n&\n&\n&\n&\n&\n\cr
 \n&\n&\n&\n&\n&\n&\n&\n&\n&\n&\n&\n\cr
 \n&\n&\n&\n&\n&\n&\n&\n&\n&\n&\n&\n\cr
 \n&\n&\n&\n&{\CoefTrue}&\n&\n&\n&\n&\n&\n&\n\cr
 \n&\n&\n&\n&\n&\n&\n&\n&\n&\n&\n&\n\cr
\noalign{\kern-2pt}}}
\quad\vcenter{\setbox8=\hbox{$\RowNames\Matrix$}
\hbox to\wd8{\hfil$\ColNames$\kern0.6\matrixskip}
\box8}}
$}
{$R,S$ and $\syqq{R}{S}$}{FigSyq}

\noindent
The symmetric quotient shows which columns of the left are equal to columns of the right relation in $\syq(R,S)$, with $S$ conceived as the denominator.

\bigskip
\noindent
It is extremely helpful that the symmetric quotient enjoys certain cancellation\index{cancellation} properties. These are far from being broadly known. Just minor side conditions have to be observed. In any of the following propositions correct typing is assumed. What is more important is that one may calculate with the symmetric quotient in a fairly traditional algebraic way. Proofs may be found in \cite{RelaMath2010}.

\enunc{}{Theorem}{}{ThmCancelA} Arbitrary relations $A,B$ satisfy in analogy to {\large$\; a\cdot{b\over a}=b$}:
\begin{enumerate}
\item[i)] $A\RELcompOP\syq (A, B)=B \RELandOP \RELtop\RELcompOP\syq (A, B)$,
\item[ii)] $\syq (A, B)$ surjective $\;\;\Longrightarrow\;\;A\RELcompOP\syq (A, B) = B$.
\Bewende
\end{enumerate}

\kern-\baselineskip

\newpage

\enunc{}{Theorem}{}{ThmCancelB} Arbitrary relations $A,B,C$ satisfy in analogy to {\large$\;{b\over a}\cdot{c\over b}={c\over a}$}
\setbox1=\hbox{$\!\!\syq (A,B)\RELcompOP\syq(B,C)$}
\begin{enumerate}
\item[i)]\ \hbox{\copy1\ $=\syq (A,C)\RELandOP\syq(A,B)\RELcompOP\RELtop
$}
\item[]\ \hbox{\ $\kern\wd1=\syq (A, C) \RELandOP \RELtop\RELcompOP\syq (B, C)
$}
\item[ii)] If  $\syq (A, B)$  is total, or if $\syq (B, C)$  is surjective, then          

\smallskip
$\syq (A,B)\RELcompOP\syq (B, C)=\syq (A,C).$
\Bewende
\end{enumerate}

\kern-\baselineskip

\enunc{}{Theorem}{}{ThmCancelC}
Assuming arbitrary relations $X,Y,Z$, always in analogy to {\large${z\over x} \;:\;{y\over x}={z\over y}$}
\begin{itemize}
\item $\LeftResi{\syqq{X}{Y}\;}{\syqq{Z}{X}}\RELaboveOP\syqq{Z}{Y}$
\item $\syqq{\syqq{X}{Y}}{\syqq{X}{Z}} \RELaboveOP\syqq{Y}{Z}$
\item $\syqq{\syqq{X}{Y}}{\syqq{X}{Z}}=\syqq{Y}{Z}$\quad\hbox{if $\syqq{X}{Y}$ and \/ $\syqq{X}{Z}$ are surjective}
\Bewende
\end{itemize}
\kern-\baselineskip

\noindent
Here is another basic rule:

\enunc{}{Proposition}{}{PropSyqSurjFct} For a surjective mapping $f$ always\quad
$\syqq{X}{f\RELcompOP Y}\RELenthOP\syqq{f\RELtraOP\RELcompOP X}{Y}$.

\Proof $\iff\quad
\RELneg{\RELneg{X\RELtraOP}\RELcompOP f\RELcompOP Y}\RELandOP\RELneg{X\RELtraOP\RELcompOP\RELneg{f\RELcompOP Y}}
\RELenthOP
\RELneg{\RELneg{X\RELtraOP\RELcompOP f}\RELcompOP Y}\RELandOP\RELneg{X\RELtraOP\RELcompOP f\RELcompOP\RELneg{Y}}
$

\smallskip
\noindent
Above, the second terms are equal since $f$ is a mapping. Containment of the first ones:

\smallskip
$\iff\quad\RELneg{X\RELtraOP\RELcompOP f}\RELcompOP Y
\RELenthOP
\RELneg{X\RELtraOP}\RELcompOP f\RELcompOP Y
\quad\Longleftarrow\quad
\RELneg{X\RELtraOP\RELcompOP f}
\RELenthOP
\RELneg{X\RELtraOP}\RELcompOP f
\quad\iff\quad
\RELtop=X\RELtraOP\RELcompOP f\RELorOP\RELneg{X\RELtraOP}\RELcompOP f
=\RELtop\RELcompOP f
$
\Bewende


\chapter{Membership and singleton injection\label{ChapMembSing}}
\EnuncNo=0
\CaptionNo=0


\noindent
The symmetric quotient is used to introduce {\it membership relations\/}\index{membership relation} 
$\RELfromTO{\varepsilon}{V}{{\cal P}(V)}$ between a set $V$ and its powerset ${\cal P}(V)$ or $\PowTWO{V}$. These can be characterized algebraically up to isomorphism demanding $\syqq{\varepsilon}{\varepsilon}\RELenthOP\RELide$ and surjectivity of $\syqq{\varepsilon}{R}$ for all $R$. 
With a membership $\varepsilon$, the powerset ordering is easily described as $\Omega=\RELneg{\varepsilon\RELtraOP\RELcompOP\RELneg{\varepsilon}}$. Also least upper bounds with regard to $\Omega$ may be expressed via membership and symmetric quotient, making this a very powerful tool; see \cite{RelaMath2010}.

\enunc{}{Proposition}{}{PropSyqLub}  If $\varepsilon$ is the membership relation 
and $\Omega$ the corresponding powerset ordering, the following equations hold
for arbitrary relations $X$:
\begin{enumerate}
\item[i)] $\varepsilon\RELcompOP\RELneg{\varepsilon\RELtraOP\RELcompOP X}=\RELneg{X}\quad\hbox{and}\quad
\RELneg{\varepsilon}\RELcompOP\RELneg{\RELneg{\varepsilon}\RELtraOP\RELcompOP X}=\RELneg{X},$
\smallskip
\item[ii)] $\textstyle\lub_\Omega(X)=\syq(\varepsilon,\varepsilon\RELcompOP X).$
\Bewende
\end{enumerate}

\noindent
We also introduce {\bf singleton injection}\index{singleton injection} $\sigma:=\syqq{\RELide}{\varepsilon}$ and {\bf atoms}\index{atom} $a:=\sigma\RELtraOP\RELcompOP\sigma$.

\Caption{$\vcenter{\hbox{$
{\footnotesize%
\BoxBreadth=0pt%
\setbox7=\hbox{a}%
\ifdim\wd7>\BoxBreadth\BoxBreadth=\wd7\fi%
\setbox7=\hbox{b}%
\ifdim\wd7>\BoxBreadth\BoxBreadth=\wd7\fi%
\setbox7=\hbox{c}%
\ifdim\wd7>\BoxBreadth\BoxBreadth=\wd7\fi%
\setbox7=\hbox{d}%
\ifdim\wd7>\BoxBreadth\BoxBreadth=\wd7\fi%
\def\RowNames{\vcenter{\offinterlineskip\baselineskip=\matrixskip%
\hbox to\BoxBreadth{\strut\hfil a}\kern\interspacereduction%
\hbox to\BoxBreadth{\strut\hfil b}\kern\interspacereduction%
\hbox to\BoxBreadth{\strut\hfil c}\kern\interspacereduction%
\hbox to\BoxBreadth{\strut\hfil d}}}%
\def\ColNames{\hbox{\rotatebox{90}{\strut \{\}}\kern\interspacereduction%
\rotatebox{90}{\strut \{a\}}\kern\interspacereduction%
\rotatebox{90}{\strut \{b\}}\kern\interspacereduction%
\rotatebox{90}{\strut \{a,b\}}\kern\interspacereduction%
\rotatebox{90}{\strut \{c\}}\kern\interspacereduction%
\rotatebox{90}{\strut \{a,c\}}\kern\interspacereduction%
\rotatebox{90}{\strut \{b,c\}}\kern\interspacereduction%
\rotatebox{90}{\strut abc}\kern\interspacereduction%
\rotatebox{90}{\strut \{d\}}\kern\interspacereduction%
\rotatebox{90}{\strut \{a,d\}}\kern\interspacereduction%
\rotatebox{90}{\strut \{b,d\}}\kern\interspacereduction%
\rotatebox{90}{\strut abd}\kern\interspacereduction%
\rotatebox{90}{\strut \{c,d\}}\kern\interspacereduction%
\rotatebox{90}{\strut \{a,c,d\}}\kern\interspacereduction%
\rotatebox{90}{\strut \{b,c,d\}}\kern\interspacereduction%
\rotatebox{90}{\strut \{a,b,c,d\}}\kern\interspacereduction%
}}%
\def\Matrix{\spmatrix{%
\noalign{\kern-2pt}
 \n&{\CoefTrue}&\n&{\CoefTrue}&\n&{\CoefTrue}&\n&{\CoefTrue}&\n&{\CoefTrue}&\n&{\CoefTrue}&\n&{\CoefTrue}&\n&{\CoefTrue}\cr
 \n&\n&{\CoefTrue}&{\CoefTrue}&\n&\n&{\CoefTrue}&{\CoefTrue}&\n&\n&{\CoefTrue}&{\CoefTrue}&\n&\n&{\CoefTrue}&{\CoefTrue}\cr
 \n&\n&\n&\n&{\CoefTrue}&{\CoefTrue}&{\CoefTrue}&{\CoefTrue}&\n&\n&\n&\n&{\CoefTrue}&{\CoefTrue}&{\CoefTrue}&{\CoefTrue}\cr
 \n&\n&\n&\n&\n&\n&\n&\n&{\CoefTrue}&{\CoefTrue}&{\CoefTrue}&{\CoefTrue}&{\CoefTrue}&{\CoefTrue}&{\CoefTrue}&{\CoefTrue}\cr
\noalign{\kern-2pt}}}%
\varepsilon= \vbox{\setbox8=\hbox{$\RowNames\Matrix$}
\hbox to\wd8{\hfil$\ColNames$\kern\ColEntryShiftHoriz}\kern\ColEntryShiftVerti
\box8}}
$}%
\hbox{$
{\footnotesize%
\BoxBreadth=0pt%
\setbox7=\hbox{a}%
\ifdim\wd7>\BoxBreadth\BoxBreadth=\wd7\fi%
\setbox7=\hbox{b}%
\ifdim\wd7>\BoxBreadth\BoxBreadth=\wd7\fi%
\setbox7=\hbox{c}%
\ifdim\wd7>\BoxBreadth\BoxBreadth=\wd7\fi%
\setbox7=\hbox{d}%
\ifdim\wd7>\BoxBreadth\BoxBreadth=\wd7\fi%
\def\RowNames{\vcenter{\offinterlineskip\baselineskip=\matrixskip%
\hbox to\BoxBreadth{\strut\hfil a}\kern\interspacereduction%
\hbox to\BoxBreadth{\strut\hfil b}\kern\interspacereduction%
\hbox to\BoxBreadth{\strut\hfil c}\kern\interspacereduction%
\hbox to\BoxBreadth{\strut\hfil d}}}%
\def\ColNames{\hbox{\rotatebox{90}{\strut \{\}}\kern\interspacereduction%
\rotatebox{90}{\strut \{a\}}\kern\interspacereduction%
\rotatebox{90}{\strut \{b\}}\kern\interspacereduction%
\rotatebox{90}{\strut \{a,b\}}\kern\interspacereduction%
\rotatebox{90}{\strut \{c\}}\kern\interspacereduction%
\rotatebox{90}{\strut \{a,c\}}\kern\interspacereduction%
\rotatebox{90}{\strut \{b,c\}}\kern\interspacereduction%
\rotatebox{90}{\strut abc}\kern\interspacereduction%
\rotatebox{90}{\strut \{d\}}\kern\interspacereduction%
\rotatebox{90}{\strut \{a,d\}}\kern\interspacereduction%
\rotatebox{90}{\strut \{b,d\}}\kern\interspacereduction%
\rotatebox{90}{\strut abd}\kern\interspacereduction%
\rotatebox{90}{\strut \{c,d\}}\kern\interspacereduction%
\rotatebox{90}{\strut \{a,c,d\}}\kern\interspacereduction%
\rotatebox{90}{\strut \{b,c,d\}}\kern\interspacereduction%
\rotatebox{90}{\strut \{a,b,c,d\}}\kern\interspacereduction%
}}%
\def\Matrix{\spmatrix{%
\noalign{\kern-2pt}
 \n&{\CoefTrue}&\n&\n&\n&\n&\n&\n&\n&\n&\n&\n&\n&\n&\n&\n\cr
 \n&\n&{\CoefTrue}&\n&\n&\n&\n&\n&\n&\n&\n&\n&\n&\n&\n&\n\cr
 \n&\n&\n&\n&{\CoefTrue}&\n&\n&\n&\n&\n&\n&\n&\n&\n&\n&\n\cr
 \n&\n&\n&\n&\n&\n&\n&\n&{\CoefTrue}&\n&\n&\n&\n&\n&\n&\n\cr
\noalign{\kern-2pt}}}%
\sigma= \vbox{\setbox8=\hbox{$\RowNames\Matrix$}
\hbox to\wd8{\hfil$\ColNames$\kern\ColEntryShiftHoriz}\kern\ColEntryShiftVerti
\box8}}
$}
\hbox{\kern-0.45cm
$
{\footnotesize%
\BoxBreadth=0pt%
\setbox7=\hbox{a}%
\ifdim\wd7>\BoxBreadth\BoxBreadth=\wd7\fi%
\setbox7=\hbox{b}%
\ifdim\wd7>\BoxBreadth\BoxBreadth=\wd7\fi%
\setbox7=\hbox{c}%
\ifdim\wd7>\BoxBreadth\BoxBreadth=\wd7\fi%
\setbox7=\hbox{d}%
\ifdim\wd7>\BoxBreadth\BoxBreadth=\wd7\fi%
\def\RowNames{\vcenter{\offinterlineskip\baselineskip=\matrixskip%
\hbox to\BoxBreadth{\strut\hfil a}\kern\interspacereduction%
\hbox to\BoxBreadth{\strut\hfil b}\kern\interspacereduction%
\hbox to\BoxBreadth{\strut\hfil c}\kern\interspacereduction%
\hbox to\BoxBreadth{\strut\hfil d}}}%
\def\ColNames{\hbox{\rotatebox{90}{\strut \{\}}\kern\interspacereduction%
\rotatebox{90}{\strut \{a\}}\kern\interspacereduction%
\rotatebox{90}{\strut \{b\}}\kern\interspacereduction%
\rotatebox{90}{\strut \{a,b\}}\kern\interspacereduction%
\rotatebox{90}{\strut \{c\}}\kern\interspacereduction%
\rotatebox{90}{\strut \{a,c\}}\kern\interspacereduction%
\rotatebox{90}{\strut \{b,c\}}\kern\interspacereduction%
\rotatebox{90}{\strut \{a,b,c\}}\kern\interspacereduction%
\rotatebox{90}{\strut \{d\}}\kern\interspacereduction%
\rotatebox{90}{\strut \{a,d\}}\kern\interspacereduction%
\rotatebox{90}{\strut \{b,d\}}\kern\interspacereduction%
\rotatebox{90}{\strut abd}\kern\interspacereduction%
\rotatebox{90}{\strut \{c,d\}}\kern\interspacereduction%
\rotatebox{90}{\strut \{a,c,d\}}\kern\interspacereduction%
\rotatebox{90}{\strut \{b,c,d\}}\kern\interspacereduction%
\rotatebox{90}{\strut \{a,b,c,d\}}\kern\interspacereduction%
}}%
\def\Matrix{\spmatrix{%
\noalign{\kern-2pt}
 \n&{\CoefTrue}&{\CoefTrue}&\n&{\CoefTrue}&\n&\n&\n&{\CoefTrue}&\n&\n&\n&\n&\n&\n&\n\cr
\noalign{\kern-2pt}}}%
\RELtop\RELcompOP\sigma\RELtraOP\RELcompOP\sigma= \vbox{\setbox8=\hbox{$
\Matrix$}
\hbox to\wd8{\hfil
\kern\ColEntryShiftHoriz}\kern\ColEntryShiftVerti
\box8}}$}}
\vcenter{\footnotesize%
\BoxBreadth=0pt%
\setbox7=\hbox{\{\}}%
\ifdim\wd7>\BoxBreadth\BoxBreadth=\wd7\fi%
\setbox7=\hbox{\{a\}}%
\ifdim\wd7>\BoxBreadth\BoxBreadth=\wd7\fi%
\setbox7=\hbox{\{b\}}%
\ifdim\wd7>\BoxBreadth\BoxBreadth=\wd7\fi%
\setbox7=\hbox{\{a,b\}}%
\ifdim\wd7>\BoxBreadth\BoxBreadth=\wd7\fi%
\setbox7=\hbox{\{c\}}%
\ifdim\wd7>\BoxBreadth\BoxBreadth=\wd7\fi%
\setbox7=\hbox{\{a,c\}}%
\ifdim\wd7>\BoxBreadth\BoxBreadth=\wd7\fi%
\setbox7=\hbox{\{b,c\}}%
\ifdim\wd7>\BoxBreadth\BoxBreadth=\wd7\fi%
\setbox7=\hbox{\{a,b,c\}}%
\ifdim\wd7>\BoxBreadth\BoxBreadth=\wd7\fi%
\setbox7=\hbox{\{d\}}%
\ifdim\wd7>\BoxBreadth\BoxBreadth=\wd7\fi%
\setbox7=\hbox{\{a,d\}}%
\ifdim\wd7>\BoxBreadth\BoxBreadth=\wd7\fi%
\setbox7=\hbox{\{b,d\}}%
\ifdim\wd7>\BoxBreadth\BoxBreadth=\wd7\fi%
\setbox7=\hbox{abd}%
\ifdim\wd7>\BoxBreadth\BoxBreadth=\wd7\fi%
\setbox7=\hbox{\{c,d\}}%
\ifdim\wd7>\BoxBreadth\BoxBreadth=\wd7\fi%
\setbox7=\hbox{\{a,c,d\}}%
\ifdim\wd7>\BoxBreadth\BoxBreadth=\wd7\fi%
\setbox7=\hbox{\{b,c,d\}}%
\ifdim\wd7>\BoxBreadth\BoxBreadth=\wd7\fi%
\setbox7=\hbox{\{a,b,c,d\}}%
\ifdim\wd7>\BoxBreadth\BoxBreadth=\wd7\fi%
\def\RowNames{\vcenter{\offinterlineskip\baselineskip=\matrixskip%
\hbox to\BoxBreadth{\strut\hfil \{\}}\kern\interspacereduction%
\hbox to\BoxBreadth{\strut\hfil \{a\}}\kern\interspacereduction%
\hbox to\BoxBreadth{\strut\hfil \{b\}}\kern\interspacereduction%
\hbox to\BoxBreadth{\strut\hfil \{a,b\}}\kern\interspacereduction%
\hbox to\BoxBreadth{\strut\hfil \{c\}}\kern\interspacereduction%
\hbox to\BoxBreadth{\strut\hfil \{a,c\}}\kern\interspacereduction%
\hbox to\BoxBreadth{\strut\hfil \{b,c\}}\kern\interspacereduction%
\hbox to\BoxBreadth{\strut\hfil \{a,b,c\}}\kern\interspacereduction%
\hbox to\BoxBreadth{\strut\hfil \{d\}}\kern\interspacereduction%
\hbox to\BoxBreadth{\strut\hfil \{a,d\}}\kern\interspacereduction%
\hbox to\BoxBreadth{\strut\hfil \{b,d\}}\kern\interspacereduction%
\hbox to\BoxBreadth{\strut\hfil abd}\kern\interspacereduction%
\hbox to\BoxBreadth{\strut\hfil \{c,d\}}\kern\interspacereduction%
\hbox to\BoxBreadth{\strut\hfil \{a,c,d\}}\kern\interspacereduction%
\hbox to\BoxBreadth{\strut\hfil \{b,c,d\}}\kern\interspacereduction%
\hbox to\BoxBreadth{\strut\hfil \{a,b,c,d\}}}}%
\def\ColNames{\hbox{\rotatebox{90}{\strut \{\}}\kern\interspacereduction%
\rotatebox{90}{\strut \{a\}}\kern\interspacereduction%
\rotatebox{90}{\strut \{b\}}\kern\interspacereduction%
\rotatebox{90}{\strut \{a,b\}}\kern\interspacereduction%
\rotatebox{90}{\strut \{c\}}\kern\interspacereduction%
\rotatebox{90}{\strut \{a,c\}}\kern\interspacereduction%
\rotatebox{90}{\strut \{b,c\}}\kern\interspacereduction%
\rotatebox{90}{\strut \{a,b,c\}}\kern\interspacereduction%
\rotatebox{90}{\strut \{d\}}\kern\interspacereduction%
\rotatebox{90}{\strut \{a,d\}}\kern\interspacereduction%
\rotatebox{90}{\strut \{b,d\}}\kern\interspacereduction%
\rotatebox{90}{\strut abd}\kern\interspacereduction%
\rotatebox{90}{\strut \{c,d\}}\kern\interspacereduction%
\rotatebox{90}{\strut \{a,c,d\}}\kern\interspacereduction%
\rotatebox{90}{\strut \{b,c,d\}}\kern\interspacereduction%
\rotatebox{90}{\strut \{a,b,c,d\}}\kern\interspacereduction%
}}%
\def\Matrix{\spmatrix{%
\noalign{\kern-2pt}
 \n&\n&\n&\n&\n&\n&\n&\n&\n&\n&\n&\n&\n&\n&\n&\n\cr
 \n&{\CoefTrue}&\n&\n&\n&\n&\n&\n&\n&\n&\n&\n&\n&\n&\n&\n\cr
 \n&\n&{\CoefTrue}&\n&\n&\n&\n&\n&\n&\n&\n&\n&\n&\n&\n&\n\cr
 \n&\n&\n&\n&\n&\n&\n&\n&\n&\n&\n&\n&\n&\n&\n&\n\cr
 \n&\n&\n&\n&{\CoefTrue}&\n&\n&\n&\n&\n&\n&\n&\n&\n&\n&\n\cr
 \n&\n&\n&\n&\n&\n&\n&\n&\n&\n&\n&\n&\n&\n&\n&\n\cr
 \n&\n&\n&\n&\n&\n&\n&\n&\n&\n&\n&\n&\n&\n&\n&\n\cr
 \n&\n&\n&\n&\n&\n&\n&\n&\n&\n&\n&\n&\n&\n&\n&\n\cr
 \n&\n&\n&\n&\n&\n&\n&\n&{\CoefTrue}&\n&\n&\n&\n&\n&\n&\n\cr
 \n&\n&\n&\n&\n&\n&\n&\n&\n&\n&\n&\n&\n&\n&\n&\n\cr
 \n&\n&\n&\n&\n&\n&\n&\n&\n&\n&\n&\n&\n&\n&\n&\n\cr
 \n&\n&\n&\n&\n&\n&\n&\n&\n&\n&\n&\n&\n&\n&\n&\n\cr
 \n&\n&\n&\n&\n&\n&\n&\n&\n&\n&\n&\n&\n&\n&\n&\n\cr
 \n&\n&\n&\n&\n&\n&\n&\n&\n&\n&\n&\n&\n&\n&\n&\n\cr
 \n&\n&\n&\n&\n&\n&\n&\n&\n&\n&\n&\n&\n&\n&\n&\n\cr
 \n&\n&\n&\n&\n&\n&\n&\n&\n&\n&\n&\n&\n&\n&\n&\n\cr
\noalign{\kern-2pt}}}%
\vbox{\setbox8=\hbox{$\RowNames\Matrix$}
\hbox to\wd8{\hfil$\ColNames$\kern\ColEntryShiftHoriz}\kern\ColEntryShiftVerti
\box8}}
$}
{$\varepsilon$, singleton injection $\sigma:=\syqq{\RELide}{\varepsilon}$ and atoms as vector $\sigma\RELtraOP\RELcompOP\sigma\RELcompOP\RELtop$ as well as diagonal
$\sigma\RELtraOP\RELcompOP\sigma$}{FigAtomSingle}

\noindent
The following results correspond to the lowest level of element-is-contained-in-set considerations. They are fairly intuitive and easy to understand from Fig.~\FigAtomSingle. The basic purpose of these statements is to make these tiny set arguments work together with more advanced algebraic mechanisms.

\enunc{}{Lemma}{}{LemmaEpsSigma} 
i) $\sigma\RELcompOP\varepsilon\RELtraOP=\RELide$

\bigskip
\noindent
\phantom{vi}ii) $\RELneg{\RELneg{\RELide}\RELcompOP\varepsilon}=\sigma\RELorOP\RELneg{\RELtop\RELcompOP\varepsilon}$

\bigskip
\noindent
\phantom{v}iii) $\sigma\RELcompOP\Omega=\varepsilon
\qquad
\sigma\RELcompOP\Omega\RELtraOP=\sigma\RELorOP\RELneg{\RELtop\RELcompOP\varepsilon}$

\bigskip
\noindent
\phantom{ii}iv) $
\varepsilon=\sigma\RELorOP(\varepsilon\RELandOP\RELneg{\RELtop\RELcompOP\sigma})
$

\bigskip
\noindent
\phantom{iii}v) $\RELneg{\RELide}\RELcompOP\varepsilon\RELandOP\RELneg{\RELneg{\RELide}\RELcompOP\sigma}
=
\RELtop\RELcompOP\varepsilon\RELandOP\RELneg{\RELtop\RELcompOP\sigma}$

\bigskip
\noindent
\phantom{ii}vi) $\Omega\RELandOP\varepsilon\RELtraOP\RELcompOP\varepsilon=
\Omega\RELandOP\varepsilon\RELtraOP\RELcompOP\RELtop
$

\bigskip
\noindent
\phantom{i}vii) $(\Omega\RELandOP\varepsilon\RELtraOP\RELcompOP\varepsilon)\RELcompOP
\varepsilon\RELtraOP=\varepsilon\RELtraOP\RELcompOP\RELtop
$

\bigskip
\noindent
viii) $(\Omega\RELandOP\varepsilon\RELtraOP\RELcompOP\RELtop)\RELtraOP\RELcompOP
(\Omega\RELandOP\varepsilon\RELtraOP\RELcompOP\RELtop)
=
\varepsilon\RELtraOP\RELcompOP\varepsilon
$

\Proof i) $\sigma\RELcompOP\varepsilon\RELtraOP
=
\big\lbrack\varepsilon\RELcompOP\sigma\RELtraOP\big\rbrack\RELtraOP
=
\big\lbrack\varepsilon\RELcompOP\syqq{\varepsilon}{\RELide}\big\rbrack\RELtraOP
=
\RELide$

\bigskip
\noindent
ii) $\sigma=\syqq{\RELide}{\varepsilon}=\RELneg{\RELneg{\RELide}\RELcompOP\varepsilon}\RELandOP\varepsilon$ by definition and $\RELneg{\RELneg{\RELide}\RELcompOP\varepsilon}\RELaboveOP\RELneg{\RELtop\RELcompOP\varepsilon}$
result in \lq\lq$\RELaboveOP$\rq\rq.

\bigskip
\noindent
\lq\lq$\RELenthOP$\rq\rq\quad means\quad $\RELneg{\RELneg{\RELide}\RELcompOP\varepsilon}\RELenthOP\sigma\RELorOP\RELneg{\RELtop\RELcompOP\varepsilon}=
\big\lbrack\RELneg{\RELneg{\RELide}\RELcompOP\varepsilon}\RELandOP\varepsilon\big\rbrack\RELorOP\RELneg{\RELtop\RELcompOP\varepsilon}$

$\iff\quad
\RELtop
=
\RELneg{\RELide}\RELcompOP\varepsilon\RELorOP\big\lbrack\RELneg{\RELneg{\RELide}\RELcompOP\varepsilon}\RELandOP\varepsilon\big\rbrack\RELorOP\RELneg{\RELtop\RELcompOP\varepsilon}
=
\big\lbrack\RELneg{\RELide}\RELcompOP\varepsilon\RELorOP\RELneg{\RELneg{\RELide}\RELcompOP\varepsilon}\RELorOP\RELneg{\RELtop\RELcompOP\varepsilon}\big\rbrack\RELandOP\big\lbrack\RELneg{\RELide}\RELcompOP\varepsilon\RELorOP\varepsilon\RELorOP\RELneg{\RELtop\RELcompOP\varepsilon}\big\rbrack
$, which is true.

\bigskip
\noindent
iii) $\sigma\RELcompOP\Omega
=
\sigma\RELcompOP\RELneg{\varepsilon\RELtraOP\RELcompOP\RELneg{\varepsilon}}
=
\RELneg{\sigma\RELcompOP\varepsilon\RELtraOP\RELcompOP\RELneg{\varepsilon}}
=
\RELneg{\RELide\RELcompOP\RELneg{\varepsilon}}
=
\varepsilon
$, \qquad using (i)

\smallskip
$\sigma\RELcompOP\Omega\RELtraOP
=
\sigma\RELcompOP\RELneg{\RELneg{\varepsilon\RELtraOP}\RELcompOP\varepsilon}
=
\RELneg{\RELneg{\sigma\RELcompOP\varepsilon\RELtraOP}\RELcompOP\varepsilon}
=
\RELneg{\RELneg{\RELide}\RELcompOP\varepsilon}
=
\sigma\RELorOP\RELneg{\RELtop\RELcompOP\varepsilon}$, \qquad using (i,ii)

\bigskip
\noindent
iv) \lq\lq$\RELaboveOP$\rq\rq\ is obvious. For \lq\lq$\RELenthOP$\rq\rq, it suffices to prove
$\RELtop\RELcompOP\sigma\RELandOP\varepsilon\RELenthOP(\RELtop\RELandOP\varepsilon\RELcompOP\sigma\RELtraOP)\RELcompOP(\sigma\RELandOP\RELtop\RELcompOP\varepsilon)\RELenthOP\sigma$ using (i).

\bigskip
\noindent
v) $\RELtop\RELcompOP\varepsilon\RELandOP\RELneg{\RELtop\RELcompOP\sigma}
=
(\RELide\RELcompOP\varepsilon\RELorOP\RELneg{\RELide}\RELcompOP\varepsilon)\RELandOP\RELneg{\RELide\RELcompOP\sigma\RELorOP\RELneg{\RELide}\RELcompOP\sigma}
=
(\RELide\RELcompOP\varepsilon\RELorOP\RELneg{\RELide}\RELcompOP\varepsilon)\RELandOP\RELneg{\sigma}\RELandOP\RELneg{\RELneg{\RELide}\RELcompOP\sigma}
=
(\RELide\RELcompOP\varepsilon\RELorOP\RELneg{\RELide}\RELcompOP\varepsilon)\RELandOP(\RELneg{\RELide}\RELcompOP\varepsilon\RELorOP\RELneg{\varepsilon})\RELandOP\RELneg{\RELneg{\RELide}\RELcompOP\sigma}
$

$=
\big[\RELneg{\RELide}\RELcompOP\varepsilon\RELorOP(\varepsilon\RELandOP\RELneg{\varepsilon})\big]\RELandOP\RELneg{\RELneg{\RELide}\RELcompOP\sigma}
=
\RELneg{\RELide}\RELcompOP\varepsilon\RELandOP\RELneg{\RELneg{\RELide}\RELcompOP\sigma}
$

\bigskip
\noindent
vi) This follows with the Dedekind rule from 

\smallskip
$\varepsilon\RELtraOP\RELcompOP\RELtop\RELandOP\Omega
\RELenthOP
(\varepsilon\RELtraOP\RELandOP\Omega\RELcompOP\RELtop)\RELcompOP(\RELtop\RELandOP\varepsilon\RELcompOP\Omega)
\RELenthOP
\varepsilon\RELtraOP\RELcompOP\varepsilon\RELcompOP\Omega
=
\varepsilon\RELtraOP\RELcompOP\varepsilon
$.

\bigskip
\noindent
vii) We start with formally showing the intuitively clear fact $\Omega\RELcompOP
\varepsilon\RELtraOP=\RELtop$:

\smallskip
$\Omega\RELcompOP\varepsilon\RELtraOP
\RELaboveOP
(\Omega\RELandOP\RELneg{\RELtop\RELcompOP\RELneg{\varepsilon}})\RELcompOP(\varepsilon\RELtraOP\RELandOP\RELneg{\RELneg{\varepsilon}\RELtraOP\RELcompOP\RELtop})
=
\RELneg{\RELtop\RELcompOP\RELneg{\varepsilon}}\,\RELcompOP\,\RELneg{\RELneg{\varepsilon}\RELtraOP\RELcompOP\RELtop}
$\qquad trivial

$\RELaboveOP
\syqq{\RELtop}{\varepsilon}\RELcompOP\syqq{\varepsilon}{\RELtop}
=
\syqq{\RELtop}{\RELtop}
=
\RELtop$\quad \cite{RelaMath2010} Prop.~8.13.ii; $\syqq{\varepsilon}{\RELtop}$ is surjective

\smallskip
\noindent
which is used together with (vi) in the following
chain of reasoning

\smallskip
$(\Omega\RELandOP\varepsilon\RELtraOP\RELcompOP\varepsilon)\RELcompOP
\varepsilon\RELtraOP
=
(\Omega\RELandOP\varepsilon\RELtraOP\RELcompOP\RELtop)\RELcompOP
\varepsilon\RELtraOP
=
\Omega\RELcompOP
\varepsilon\RELtraOP\RELandOP\varepsilon\RELtraOP\RELcompOP\RELtop
=
\RELtop\RELandOP\varepsilon\RELtraOP\RELcompOP\RELtop
=
\varepsilon\RELtraOP\RELcompOP\RELtop
$.

\bigskip
\noindent
viii) We recall the definition of singleton injection $\sigma:=\syqq{\RELide}{\varepsilon}$ and use (i,iii):

\smallskip
$\varepsilon
=
\varepsilon\RELandOP\RELtop
=
\varepsilon\RELandOP\RELide\RELcompOP\RELtop
=
\sigma\RELcompOP\Omega\RELandOP\sigma\RELcompOP\varepsilon\RELtraOP\RELcompOP\RELtop
=
\sigma\RELcompOP(\Omega\RELandOP\varepsilon\RELtraOP\RELcompOP\RELtop)
$\quad since $\sigma$ is univalent 

\smallskip
\noindent
Therefore

\smallskip
$\varepsilon\RELtraOP\RELcompOP\varepsilon
=
(\Omega\RELandOP\varepsilon\RELtraOP\RELcompOP\RELtop)\RELtraOP\RELcompOP\sigma\RELtraOP
\RELcompOP
\sigma\RELcompOP(\Omega\RELandOP\varepsilon\RELtraOP\RELcompOP\RELtop)
\RELenthOP
(\Omega\RELandOP\varepsilon\RELtraOP\RELcompOP\RELtop)\RELtraOP\RELcompOP(\Omega\RELandOP\varepsilon\RELtraOP\RELcompOP\RELtop)
$\quad since $\sigma$ is univalent

\smallskip
\noindent
The other inclusion \lq\lq$\RELenthOP$\rq\rq\ follows with (vi) from

\smallskip
$(\Omega\RELtraOP\RELandOP\RELtop\RELcompOP\varepsilon)\RELcompOP
(\Omega\RELandOP\varepsilon\RELtraOP\RELcompOP\RELtop)
=
(\Omega\RELtraOP\RELandOP\varepsilon\RELtraOP\RELcompOP\varepsilon)\RELcompOP
(\Omega\RELandOP\varepsilon\RELtraOP\RELcompOP\varepsilon)
\RELenthOP
\varepsilon\RELtraOP\RELcompOP\varepsilon\RELcompOP\Omega
=
\varepsilon\RELtraOP\RELcompOP\varepsilon
$
\Bewende

\kern-\baselineskip

\Caption{\includegraphics[scale=0.45]{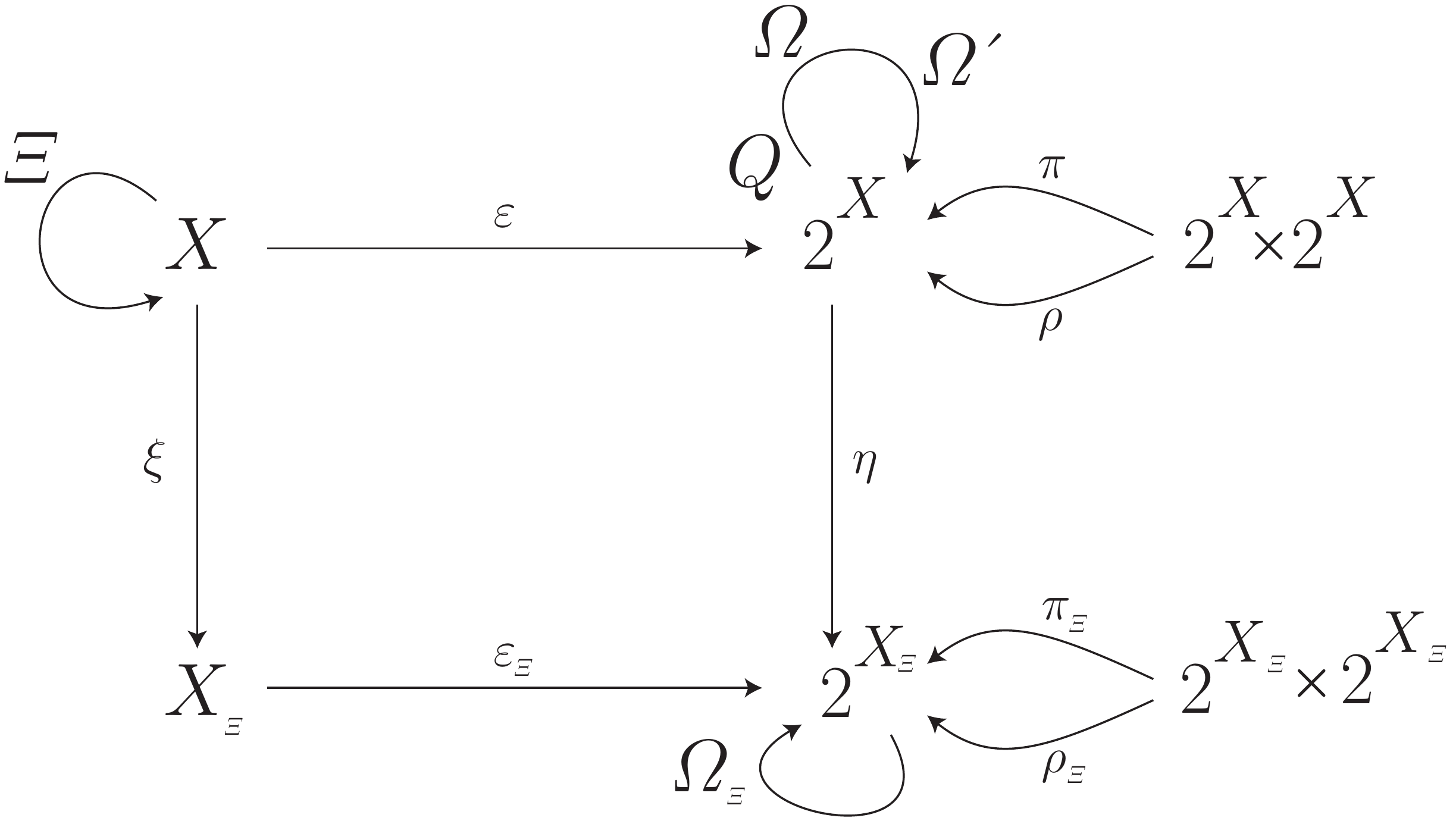}}
{Quotient of membership\index{quotient}}{FigQuotPow}

\kern-\baselineskip

\enunc{}{Proposition}{}{PropQuotPow}  Let be given the membership relation $\RELfromTO{\varepsilon}{X}{\PowTWO{X}}$ and an equivalence relation $\RELfromTO{\Xi}{X}{X}$ and its natural projection\index{natural projection} $\RELfromTO{\xi}{X}{X_\Xi}$
satisfying $\xi\RELcompOP\xi\RELtraOP=\Xi$. Then

\begin{enumerate}[i)]
\item
$\Omega':=\RELneg{\varepsilon\RELtraOP\RELcompOP\RELneg{\Xi\RELcompOP\varepsilon}}$\quad is a preorder.
\item $Q:=\syqq{\Xi\RELcompOP\varepsilon}{\Xi\RELcompOP\varepsilon}$ is an equivalence satisfying $Q=\Omega'\RELandOP{\Omega'}\RELtraOP$. 
\item[]Let $\RELfromTO{\eta}{\PowTWO{X}}{\PowTWO{X_\Xi}}$ denote its natural projection, i.e., the mapping that satisfies $Q=\eta\RELcompOP\eta\RELtraOP$.
\item $\varepsilon\RELcompOP Q=\Xi\RELcompOP \varepsilon$ 
\item $\varepsilon_\Xi:=\xi\RELtraOP\RELcompOP\varepsilon\RELcompOP\eta$\quad satisfies the properties of a membership relation.
\item $\xi\RELtraOP\RELcompOP\varepsilon=\varepsilon_\Xi\RELcompOP\eta\RELtraOP\qquad \Omega'\RELcompOP\eta=\eta\RELcompOP\Omega_\Xi$
\end{enumerate}

\Proof We recall that for an equivalence $\Psi$, in general $\Psi\RELcompOP\RELneg{\Psi\RELcompOP Y}=\RELneg{\Psi\RELcompOP Y}$ and $\RELneg{Z\RELcompOP\Psi}\RELcompOP\Psi=\RELneg{Z\RELcompOP\Psi}$, and that its natural projection is a surjective mapping.

\bigskip
\noindent
i) Reflexivity holds since $\RELide\RELenthOP\Omega=\RELneg{\varepsilon\RELtraOP\RELcompOP\RELneg{\varepsilon}}\RELenthOP\RELneg{\varepsilon\RELtraOP\RELcompOP\RELneg{\Xi\RELcompOP\varepsilon}}$, 
while transitivity follows from

\smallskip
$\RELneg{\varepsilon\RELtraOP\RELcompOP\Xi}\RELcompOP\Xi\RELcompOP\varepsilon
=
\RELneg{\varepsilon\RELtraOP\RELcompOP\Xi}\RELcompOP\varepsilon
\RELenthOP
\RELneg{\varepsilon\RELtraOP\RELcompOP\Xi}\RELcompOP\varepsilon
\quad\iff\quad
\RELneg{\Xi\RELcompOP\varepsilon}\RELcompOP\RELneg{\RELneg{\varepsilon\RELtraOP\RELcompOP\Xi}\RELcompOP\varepsilon}
\RELenthOP
\RELneg{\Xi\RELcompOP\varepsilon}
\quad\Longrightarrow\quad
\varepsilon\RELtraOP\RELcompOP\RELneg{\Xi\RELcompOP\varepsilon}\RELcompOP\RELneg{\RELneg{\varepsilon\RELtraOP\RELcompOP\Xi}\RELcompOP\varepsilon}
\RELenthOP
\varepsilon\RELtraOP\RELcompOP\RELneg{\Xi\RELcompOP\varepsilon}
$

$\iff\quad
\RELneg{\varepsilon\RELtraOP\RELcompOP\RELneg{\Xi\RELcompOP\varepsilon}}\RELcompOP\RELneg{\varepsilon\RELtraOP\RELcompOP\RELneg{\Xi\RELcompOP\varepsilon}}
\RELenthOP
\RELneg{\varepsilon\RELtraOP\RELcompOP\RELneg{\Xi\RELcompOP\varepsilon}}
$,\quad i.e.~$\Omega'\RELcompOP\Omega'\RELenthOP\Omega'$

\bigskip
\noindent
ii) A relation $\syqq{A}{A}$ is always an equivalence following \cite[Prop.~8.14.i]{RelaMath2010}; furthermore

\smallskip
$\Omega'\RELandOP{\Omega'}\RELtraOP
=
\RELneg{\varepsilon\RELtraOP\RELcompOP\RELneg{\Xi\RELcompOP\varepsilon}}\RELandOP\RELneg{\RELneg{\Xi\RELcompOP\varepsilon}\RELtraOP\RELcompOP\varepsilon}
=
\RELneg{\varepsilon\RELtraOP\RELcompOP\Xi\RELcompOP\RELneg{\Xi\RELcompOP\varepsilon}}\RELandOP\RELneg{\RELneg{\Xi\RELcompOP\varepsilon}\RELtraOP\RELcompOP\Xi\RELcompOP\varepsilon}
=
\syqq{\Xi\RELcompOP\varepsilon}{\Xi\RELcompOP\varepsilon}
=
Q
$

\bigskip
\noindent
iii) $
\Xi\RELcompOP\varepsilon
=
\varepsilon\RELcompOP\syqq{\varepsilon}{\Xi\RELcompOP\varepsilon}
$\quad since $\varepsilon$ is a membership

$\RELenthOP
\varepsilon\RELcompOP\syqq{\Xi\RELcompOP\varepsilon}{\Xi\RELcompOP\Xi\RELcompOP\varepsilon}
=\varepsilon\RELcompOP\syqq{\Xi\RELcompOP\varepsilon}{\Xi\RELcompOP\varepsilon}
=
\varepsilon\RELcompOP Q
$\quad \cite[Prop.~8.16.i]{RelaMath2010}

$\RELenthOP
\Xi\RELcompOP\varepsilon\RELcompOP \syqq{\Xi\RELcompOP\varepsilon}{\Xi\RELcompOP\varepsilon}
=
\Xi\RELcompOP\varepsilon
$\quad since always $A\RELcompOP\syqq{A}{A}=A$

\bigskip
\noindent
iv) $\syqq{\varepsilon_\Xi}{\varepsilon_\Xi}
=
\syqq{\xi\RELtraOP\RELcompOP\varepsilon\RELcompOP\eta}{\xi\RELtraOP\RELcompOP\varepsilon\RELcompOP\eta}
=
\eta\RELtraOP\RELcompOP\syqq{\xi\RELtraOP\RELcompOP\varepsilon}{\xi\RELtraOP\RELcompOP\varepsilon}\RELcompOP\eta
$\quad due to \cite[Prop.~8.18]{RelaMath2010}.

$=
\eta\RELtraOP\RELcompOP\syqq{\xi\RELcompOP\xi\RELtraOP\RELcompOP\varepsilon}{\xi\RELcompOP\xi\RELtraOP\RELcompOP\varepsilon}\RELcompOP\eta
$\quad due to \cite[Prop.~8.16.i]{RelaMath2010} since $\xi$ is a surjective mapping

$=
\eta\RELtraOP\RELcompOP\syqq{\Xi\RELcompOP\varepsilon}{\Xi\RELcompOP\varepsilon}\RELcompOP\eta
=
\eta\RELtraOP\RELcompOP Q\RELcompOP\eta
=
\eta\RELtraOP\RELcompOP\eta\RELcompOP\eta\RELtraOP\RELcompOP\eta
=
\RELide\RELcompOP\RELide
=
\RELide
$

\bigskip
\noindent
Assuming an arbitrary $X$ that is acceptable with regard to typing, 

$\RELtop\RELcompOP\syqq{\varepsilon_\Xi}{X}
=
\RELtop\RELcompOP\syqq{\xi\RELtraOP\RELcompOP\varepsilon\RELcompOP\eta}{X}
$\quad 

$=
\RELtop\RELcompOP\eta\RELtraOP\RELcompOP\syqq{\xi\RELtraOP\RELcompOP\varepsilon}{X}
$\quad due to \cite[Prop.~8.18]{RelaMath2010} since $\xi\RELtraOP\RELcompOP\varepsilon=\xi\RELtraOP\RELcompOP\varepsilon\RELcompOP Q
$, see above

$=
\RELtop\RELcompOP\syqq{\xi\RELtraOP\RELcompOP\varepsilon}{X}
$

$\RELaboveOP
\RELtop\RELcompOP\syqq{\varepsilon}{\xi\RELcompOP X}
$\quad Prop.~\PropSyqSurjFct

$=\RELtop$\quad because $\varepsilon$ is a membership.

\bigskip
\noindent
v) $\xi\RELtraOP\RELcompOP\varepsilon
=
\xi\RELtraOP\RELcompOP\xi\RELcompOP\xi\RELtraOP\RELcompOP\varepsilon
=
\xi\RELtraOP\RELcompOP\Xi\RELcompOP\varepsilon
=
\xi\RELtraOP\RELcompOP\varepsilon\RELcompOP Q
=
\xi\RELtraOP\RELcompOP\varepsilon\RELcompOP\eta\RELcompOP\eta\RELtraOP
=
\varepsilon_\Xi\RELcompOP\eta\RELtraOP
$

\smallskip
$\Omega'\RELcompOP\eta
=
\RELneg{\varepsilon\RELtraOP\RELcompOP\RELneg{\Xi\RELcompOP\varepsilon}}\RELcompOP\eta
=
\RELneg{\varepsilon\RELtraOP\RELcompOP\RELneg{\Xi\RELcompOP\Xi\RELcompOP\varepsilon}}\RELcompOP\eta
=
\RELneg{\varepsilon\RELtraOP\RELcompOP\RELneg{\Xi\RELcompOP\varepsilon\RELcompOP Q}}\RELcompOP\eta
=
\RELneg{\varepsilon\RELtraOP\RELcompOP\RELneg{\Xi\RELcompOP\varepsilon\RELcompOP\eta\RELcompOP\eta\RELtraOP}}\RELcompOP\eta
=
\RELneg{\varepsilon\RELtraOP\RELcompOP\RELneg{\Xi\RELcompOP\varepsilon\RELcompOP\eta}}\RELcompOP\eta\RELtraOP\RELcompOP\eta
=
\RELneg{\varepsilon\RELtraOP\RELcompOP\RELneg{\Xi\RELcompOP\varepsilon\RELcompOP\eta}}
$

$\qquad=
\RELneg{\varepsilon\RELtraOP\RELcompOP\Xi\RELcompOP\RELneg{\Xi\RELcompOP\varepsilon\RELcompOP\eta}}
=
\RELneg{Q\RELcompOP\varepsilon\RELtraOP\RELcompOP\RELneg{\Xi\RELcompOP\varepsilon\RELcompOP\eta}}
=
\RELneg{\eta\RELcompOP\eta\RELtraOP\RELcompOP\varepsilon\RELtraOP\RELcompOP\RELneg{\xi\RELcompOP\xi\RELtraOP\RELcompOP\varepsilon\RELcompOP\eta}}
=
\eta\RELcompOP\RELneg{\eta\RELtraOP\RELcompOP\varepsilon\RELtraOP\RELcompOP\xi\RELcompOP\RELneg{\xi\RELtraOP\RELcompOP\varepsilon\RELcompOP\eta}}
=
\eta\RELcompOP\RELneg{\varepsilon_\Xi\RELtraOP\RELcompOP\RELneg{\varepsilon_\Xi}}
=
\eta\RELcompOP\Omega_\Xi
$
\Bewende


\chapter{Power operations\label{ChapPowerOps}}
\EnuncNo=0
\CaptionNo=0


There is an interesting interrelationship from relations to their counterparts between the corresponding powersets.
It offers the possibility to work algebraically at situations where this has so far not been the classical approach; some has already been collected in \cite{RelaMath2010}.

\enunc{}{Definition}{}{DefExImInverseIm} Let any relation $\RELfromTO{R}{X}{Y}$ be given together with membership relations $\RELfromTO{\varepsilon}{X}{\PowTWO{X}},\RELfromTO{\varepsilon'}{Y}{\PowTWO{Y}}$. Then the corresponding {\bf existential image mapping}\index{existential image} is defined as $\ExistIm_R:=\syqq{R\RELtraOP\RELcompOP\varepsilon}{\varepsilon'}$. One may correspondingly study the {\bf inverse image mapping}\index{inverse image} defined as $\ExistIm_{R\RELtraOP}=\syqq{R\RELcompOP\varepsilon'}{\varepsilon}$.
\Bewende

\noindent
We recall an interesting fact concerning the existential image; see \cite{RelaMath2010}. Referring to \cite{deRoeverEngelhardtBuch}, the pair 
$\varepsilon,\varepsilon'$ constitutes an $L$-simulation\index{simulation} of $\ExistIm_R$ by $R$, and in addition, $\varepsilon\RELtraOP,{\varepsilon'}\RELtraOP$ show an $L\RELtraOP$-simulation of $R$ by $\ExistIm_R$.
In total, we have for an existential image the equality 

\smallskip
$\varepsilon\RELtraOP\RELcompOP R=\vartheta_R\RELcompOP{\varepsilon'}\RELtraOP$.

\smallskip
\noindent
Correspondingly, an application of this simulation rule to $R\RELtraOP$ instead of $R$ reads

\smallskip
${\varepsilon'}\RELtraOP\RELcompOP R\RELtraOP=\vartheta_{R\RELtraOP}\RELcompOP\varepsilon\RELtraOP$,
\qquad or else\qquad 
$R\RELcompOP\varepsilon'
=
\varepsilon\RELcompOP\vartheta_{R\RELtraOP}\RELtraOP
$.

\enunc{}{Proposition}{}{PropExImagProps} The existential image and the inverse image also satisfy formulae with respect to the powerset orderings:

\begin{enumerate}[i)]
\item $\Omega'\RELcompOP\vartheta_{f\RELtraOP}
\RELenthOP
\vartheta_{f\RELtraOP}\RELcompOP\Omega$\qquad if $f$ is a mapping,
\item $\Omega\RELcompOP\ExistIm_{f\RELtraOP}\RELtraOP=\ExistIm_{f}\,\RELcompOP\,\Omega'
$\qquad if $f$ is a mapping.
\end{enumerate}

\Proof i) Via shunting the claim is $\Omega'
\RELenthOP
\vartheta_{f\RELtraOP}\RELcompOP\Omega\RELcompOP\vartheta_{f\RELtraOP}\RELtraOP$, which we prove in negated form:

\smallskip
${\varepsilon'}\RELtraOP\RELcompOP\RELneg{\varepsilon'}
\RELaboveOP
{\varepsilon'}\RELtraOP\RELcompOP f\RELtraOP\RELcompOP f\RELcompOP\RELneg{\varepsilon'}
=
{\varepsilon'}\RELtraOP\RELcompOP f\RELtraOP\RELcompOP\RELneg{f\RELcompOP\varepsilon'}
=
{\varepsilon'}\RELtraOP\RELcompOP f\RELtraOP\RELcompOP\RELneg{\varepsilon\RELcompOP\vartheta_{f\RELtraOP}\RELtraOP}
=
\vartheta_{f\RELtraOP}\RELcompOP\varepsilon\RELtraOP\RELcompOP\RELneg{\varepsilon}\RELcompOP\vartheta_{f\RELtraOP}\RELtraOP
=
\vartheta_{f\RELtraOP}\RELcompOP\RELneg{\Omega}\RELcompOP\vartheta_{f\RELtraOP}\RELtraOP
=
\RELneg{\vartheta_{f\RELtraOP}\RELcompOP\Omega\RELcompOP\vartheta_{f\RELtraOP}\RELtraOP}
$

\bigskip
\noindent
ii) $\Omega\RELcompOP\ExistIm_{f\RELtraOP}\RELtraOP
=
\RELneg{\varepsilon\RELtraOP\RELcompOP\RELneg{\varepsilon}}\RELcompOP\ExistIm_{f\RELtraOP}\RELtraOP
=
\RELneg{\varepsilon\RELtraOP\RELcompOP\RELneg{\varepsilon\RELcompOP\ExistIm_{f\RELtraOP}\RELtraOP}}
=
\RELneg{\varepsilon\RELtraOP\RELcompOP\RELneg{f\RELcompOP\varepsilon'}}
=
\RELneg{\varepsilon\RELtraOP\RELcompOP f\RELcompOP\RELneg{\varepsilon'}}
=
\RELneg{\ExistIm_{f}\,\RELcompOP\,{\varepsilon'}\RELtraOP\RELcompOP\RELneg{\varepsilon'}}
=
\ExistIm_{f}\,\RELcompOP\,\RELneg{{\varepsilon'}\RELtraOP\RELcompOP\RELneg{\varepsilon'}}
=
\ExistIm_{f}\,\RELcompOP\,\Omega'
$
\Bewende

\noindent
Another rule combines the inverse image with the singleton injection.

\enunc{}{Proposition}{}{PropSingleInverse} i) Any relation $\RELfromTO{R}{X}{Y}$ with $\sigma_X,\sigma_Y$ the singleton injections satisfies

\smallskip

$\sigma_X\RELcompOP\ExistIm_{R\RELtraOP}\RELtraOP\RELcompOP\,\sigma_Y\RELtraOP\RELenthOP R$
\quad and \quad
$\varepsilon_X\RELcompOP\ExistIm_{R\RELtraOP}\RELtraOP\RELcompOP\,\sigma_Y\RELtraOP=R$.

\bigskip
\noindent
ii) When $f$ is a mapping, this sharpens to\quad  $\sigma_X\RELcompOP\ExistIm_f=f\RELcompOP\sigma_Y$.

\Proof i) $\sigma_X\RELcompOP\ExistIm_{R\RELtraOP}\RELtraOP\RELcompOP\,\sigma_Y\RELtraOP
\RELenthOP
\varepsilon_X\RELcompOP\ExistIm_{R\RELtraOP}\RELtraOP\RELcompOP\,\sigma_Y\RELtraOP
=
R\RELcompOP\varepsilon_Y\RELcompOP\sigma_Y\RELtraOP
=
R
$\quad Prop.~\PropExImagProps.i

\bigskip
\noindent
ii) $\sigma_X\RELcompOP\ExistIm_f
=
\sigma_X\RELcompOP\syqq{f\RELtraOP\RELcompOP\varepsilon_X}{\varepsilon_Y}
$\quad definition of symmetric quotient

$
=
\syqq{f\RELtraOP\RELcompOP\varepsilon_X\RELcompOP\sigma_X\RELtraOP}{\varepsilon_Y}
$\quad Prop.~8.16.ii of \cite{RelaMath2010}

$
=
\syqq{f\RELtraOP}{\varepsilon_Y}
=
\syqq{\RELide\RELcompOP f\RELtraOP}{\varepsilon_Y}
=
f\RELcompOP\syqq{\RELide}{\varepsilon_Y}
=
f\RELcompOP\sigma_Y
$\quad Prop.~\PropExImagProps.i
\Bewende

\noindent
The following rules are not unimportant when, in a forthcoming paper, continuity is studied in topology and transferred to a point-free relation-algebraic version.

\Caption{\vbox{%
\hbox to\textwidth{\hfil$\vcenter{\hbox{\includegraphics[scale=0.35]{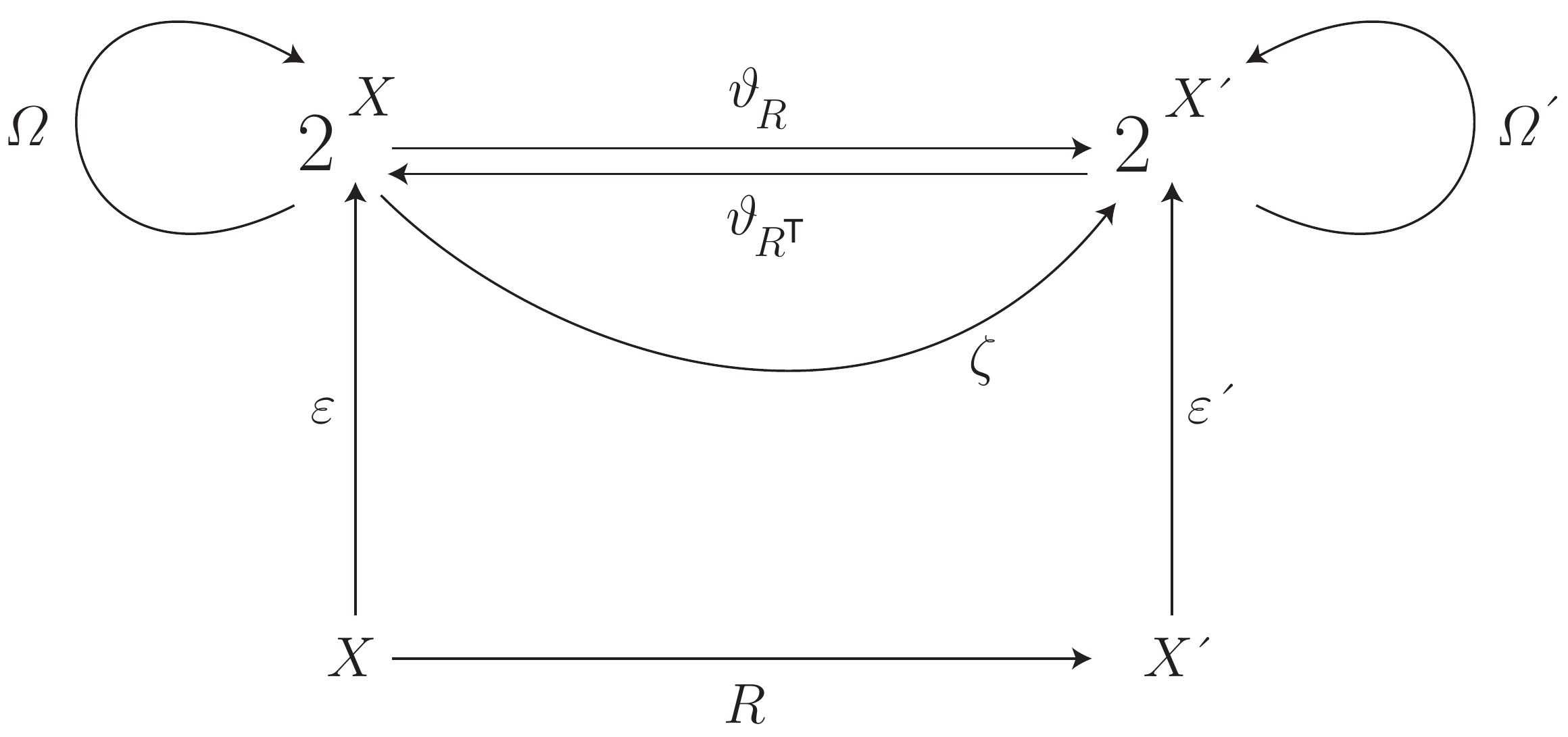}}
\kern2cm\hbox{\qquad\qquad$R$ =
{\footnotesize%
\BoxBreadth=0pt%
\setbox7=\hbox{1}%
\ifdim\wd7>\BoxBreadth\BoxBreadth=\wd7\fi%
\setbox7=\hbox{2}%
\ifdim\wd7>\BoxBreadth\BoxBreadth=\wd7\fi%
\setbox7=\hbox{3}%
\ifdim\wd7>\BoxBreadth\BoxBreadth=\wd7\fi%
\setbox7=\hbox{4}%
\ifdim\wd7>\BoxBreadth\BoxBreadth=\wd7\fi%
\setbox7=\hbox{5}%
\ifdim\wd7>\BoxBreadth\BoxBreadth=\wd7\fi%
\def\RowNames{\vcenter{\offinterlineskip\baselineskip=\matrixskip%
\hbox to\BoxBreadth{\strut\hfil 1}\kern\interspacereduction%
\hbox to\BoxBreadth{\strut\hfil 2}\kern\interspacereduction%
\hbox to\BoxBreadth{\strut\hfil 3}\kern\interspacereduction%
\hbox to\BoxBreadth{\strut\hfil 4}\kern\interspacereduction%
\hbox to\BoxBreadth{\strut\hfil 5}}}%
\def\ColNames{\hbox{\rotatebox{90}{\strut a}\kern\interspacereduction%
\rotatebox{90}{\strut b}\kern\interspacereduction%
\rotatebox{90}{\strut c}\kern\interspacereduction%
\rotatebox{90}{\strut d}\kern\interspacereduction%
}}%
\def\Matrix{\spmatrix{%
\noalign{\kern-2pt}
 \n&{\CoefTrue}&\n&{\CoefTrue}\cr
 {\CoefTrue}&\n&\n&\n\cr
 \n&\n&{\CoefTrue}&\n\cr
 \n&\n&\n&{\CoefTrue}\cr
 {\CoefTrue}&\n&{\CoefTrue}&\n\cr
\noalign{\kern-2pt}}}%
\vbox{\setbox8=\hbox{$\RowNames\Matrix$}
\hbox to\wd8{\hfil$\ColNames$\kern\ColEntryShiftHoriz}\kern\ColEntryShiftVerti
\box8}}}
}
\kern-1.2cm
\hfil
\vcenter{\hbox{$\vartheta_R=$
{\footnotesize%
\BoxBreadth=0pt%
\setbox7=\hbox{\{\}}%
\ifdim\wd7>\BoxBreadth\BoxBreadth=\wd7\fi%
\setbox7=\hbox{\{1\}}%
\ifdim\wd7>\BoxBreadth\BoxBreadth=\wd7\fi%
\setbox7=\hbox{\{2\}}%
\ifdim\wd7>\BoxBreadth\BoxBreadth=\wd7\fi%
\setbox7=\hbox{\{1,2\}}%
\ifdim\wd7>\BoxBreadth\BoxBreadth=\wd7\fi%
\setbox7=\hbox{\{3\}}%
\ifdim\wd7>\BoxBreadth\BoxBreadth=\wd7\fi%
\setbox7=\hbox{\{1,3\}}%
\ifdim\wd7>\BoxBreadth\BoxBreadth=\wd7\fi%
\setbox7=\hbox{\{2,3\}}%
\ifdim\wd7>\BoxBreadth\BoxBreadth=\wd7\fi%
\setbox7=\hbox{\{1,2,3\}}%
\ifdim\wd7>\BoxBreadth\BoxBreadth=\wd7\fi%
\setbox7=\hbox{\{4\}}%
\ifdim\wd7>\BoxBreadth\BoxBreadth=\wd7\fi%
\setbox7=\hbox{\{1,4\}}%
\ifdim\wd7>\BoxBreadth\BoxBreadth=\wd7\fi%
\setbox7=\hbox{\{2,4\}}%
\ifdim\wd7>\BoxBreadth\BoxBreadth=\wd7\fi%
\setbox7=\hbox{\{1,2,4\}}%
\ifdim\wd7>\BoxBreadth\BoxBreadth=\wd7\fi%
\setbox7=\hbox{\{3,4\}}%
\ifdim\wd7>\BoxBreadth\BoxBreadth=\wd7\fi%
\setbox7=\hbox{\{1,3,4\}}%
\ifdim\wd7>\BoxBreadth\BoxBreadth=\wd7\fi%
\setbox7=\hbox{\{2,3,4\}}%
\ifdim\wd7>\BoxBreadth\BoxBreadth=\wd7\fi%
\setbox7=\hbox{\{1,2,3,4\}}%
\ifdim\wd7>\BoxBreadth\BoxBreadth=\wd7\fi%
\setbox7=\hbox{\{5\}}%
\ifdim\wd7>\BoxBreadth\BoxBreadth=\wd7\fi%
\setbox7=\hbox{\{1,5\}}%
\ifdim\wd7>\BoxBreadth\BoxBreadth=\wd7\fi%
\setbox7=\hbox{\{2,5\}}%
\ifdim\wd7>\BoxBreadth\BoxBreadth=\wd7\fi%
\setbox7=\hbox{\{1,2,5\}}%
\ifdim\wd7>\BoxBreadth\BoxBreadth=\wd7\fi%
\setbox7=\hbox{\{3,5\}}%
\ifdim\wd7>\BoxBreadth\BoxBreadth=\wd7\fi%
\setbox7=\hbox{\{1,3,5\}}%
\ifdim\wd7>\BoxBreadth\BoxBreadth=\wd7\fi%
\setbox7=\hbox{\{2,3,5\}}%
\ifdim\wd7>\BoxBreadth\BoxBreadth=\wd7\fi%
\setbox7=\hbox{\{1,2,3,5\}}%
\ifdim\wd7>\BoxBreadth\BoxBreadth=\wd7\fi%
\setbox7=\hbox{\{4,5\}}%
\ifdim\wd7>\BoxBreadth\BoxBreadth=\wd7\fi%
\setbox7=\hbox{\{1,4,5\}}%
\ifdim\wd7>\BoxBreadth\BoxBreadth=\wd7\fi%
\setbox7=\hbox{\{2,4,5\}}%
\ifdim\wd7>\BoxBreadth\BoxBreadth=\wd7\fi%
\setbox7=\hbox{\{1,2,4,5\}}%
\ifdim\wd7>\BoxBreadth\BoxBreadth=\wd7\fi%
\setbox7=\hbox{\{3,4,5\}}%
\ifdim\wd7>\BoxBreadth\BoxBreadth=\wd7\fi%
\setbox7=\hbox{\{1,3,4,5\}}%
\ifdim\wd7>\BoxBreadth\BoxBreadth=\wd7\fi%
\setbox7=\hbox{\{2,3,4,5\}}%
\ifdim\wd7>\BoxBreadth\BoxBreadth=\wd7\fi%
\setbox7=\hbox{\{1,2,3,4,5\}}%
\ifdim\wd7>\BoxBreadth\BoxBreadth=\wd7\fi%
\def\RowNames{\vcenter{\offinterlineskip\baselineskip=\matrixskip%
\hbox to\BoxBreadth{\strut\hfil \{\}}\kern\interspacereduction%
\hbox to\BoxBreadth{\strut\hfil \{1\}}\kern\interspacereduction%
\hbox to\BoxBreadth{\strut\hfil \{2\}}\kern\interspacereduction%
\hbox to\BoxBreadth{\strut\hfil \{1,2\}}\kern\interspacereduction%
\hbox to\BoxBreadth{\strut\hfil \{3\}}\kern\interspacereduction%
\hbox to\BoxBreadth{\strut\hfil \{1,3\}}\kern\interspacereduction%
\hbox to\BoxBreadth{\strut\hfil \{2,3\}}\kern\interspacereduction%
\hbox to\BoxBreadth{\strut\hfil \{1,2,3\}}\kern\interspacereduction%
\hbox to\BoxBreadth{\strut\hfil \{4\}}\kern\interspacereduction%
\hbox to\BoxBreadth{\strut\hfil \{1,4\}}\kern\interspacereduction%
\hbox to\BoxBreadth{\strut\hfil \{2,4\}}\kern\interspacereduction%
\hbox to\BoxBreadth{\strut\hfil \{1,2,4\}}\kern\interspacereduction%
\hbox to\BoxBreadth{\strut\hfil \{3,4\}}\kern\interspacereduction%
\hbox to\BoxBreadth{\strut\hfil \{1,3,4\}}\kern\interspacereduction%
\hbox to\BoxBreadth{\strut\hfil \{2,3,4\}}\kern\interspacereduction%
\hbox to\BoxBreadth{\strut\hfil \{1,2,3,4\}}\kern\interspacereduction%
\hbox to\BoxBreadth{\strut\hfil \{5\}}\kern\interspacereduction%
\hbox to\BoxBreadth{\strut\hfil \{1,5\}}\kern\interspacereduction%
\hbox to\BoxBreadth{\strut\hfil \{2,5\}}\kern\interspacereduction%
\hbox to\BoxBreadth{\strut\hfil \{1,2,5\}}\kern\interspacereduction%
\hbox to\BoxBreadth{\strut\hfil \{3,5\}}\kern\interspacereduction%
\hbox to\BoxBreadth{\strut\hfil \{1,3,5\}}\kern\interspacereduction%
\hbox to\BoxBreadth{\strut\hfil \{2,3,5\}}\kern\interspacereduction%
\hbox to\BoxBreadth{\strut\hfil \{1,2,3,5\}}\kern\interspacereduction%
\hbox to\BoxBreadth{\strut\hfil \{4,5\}}\kern\interspacereduction%
\hbox to\BoxBreadth{\strut\hfil \{1,4,5\}}\kern\interspacereduction%
\hbox to\BoxBreadth{\strut\hfil \{2,4,5\}}\kern\interspacereduction%
\hbox to\BoxBreadth{\strut\hfil \{1,2,4,5\}}\kern\interspacereduction%
\hbox to\BoxBreadth{\strut\hfil \{3,4,5\}}\kern\interspacereduction%
\hbox to\BoxBreadth{\strut\hfil \{1,3,4,5\}}\kern\interspacereduction%
\hbox to\BoxBreadth{\strut\hfil \{2,3,4,5\}}\kern\interspacereduction%
\hbox to\BoxBreadth{\strut\hfil \{1,2,3,4,5\}}}}%
\def\ColNames{\hbox{\rotatebox{90}{\strut \{\}}\kern\interspacereduction%
\rotatebox{90}{\strut \{a\}}\kern\interspacereduction%
\rotatebox{90}{\strut \{b\}}\kern\interspacereduction%
\rotatebox{90}{\strut \{a,b\}}\kern\interspacereduction%
\rotatebox{90}{\strut \{c\}}\kern\interspacereduction%
\rotatebox{90}{\strut \{a,c\}}\kern\interspacereduction%
\rotatebox{90}{\strut \{b,c\}}\kern\interspacereduction%
\rotatebox{90}{\strut \{a,b,c\}}\kern\interspacereduction%
\rotatebox{90}{\strut \{d\}}\kern\interspacereduction%
\rotatebox{90}{\strut \{a,d\}}\kern\interspacereduction%
\rotatebox{90}{\strut \{b,d\}}\kern\interspacereduction%
\rotatebox{90}{\strut \{a,b,d\}}\kern\interspacereduction%
\rotatebox{90}{\strut \{c,d\}}\kern\interspacereduction%
\rotatebox{90}{\strut \{a,c,d\}}\kern\interspacereduction%
\rotatebox{90}{\strut \{b,c,d\}}\kern\interspacereduction%
\rotatebox{90}{\strut \{a,b,c,d\}}\kern\interspacereduction%
}}%
\def\Matrix{\spmatrix{%
\noalign{\kern-2pt}
 {\CoefTrue}&\n&\n&\n&\n&\n&\n&\n&\n&\n&\n&\n&\n&\n&\n&\n\cr
 \n&\n&\n&\n&\n&\n&\n&\n&\n&\n&{\CoefTrue}&\n&\n&\n&\n&\n\cr
 \n&{\CoefTrue}&\n&\n&\n&\n&\n&\n&\n&\n&\n&\n&\n&\n&\n&\n\cr
 \n&\n&\n&\n&\n&\n&\n&\n&\n&\n&\n&{\CoefTrue}&\n&\n&\n&\n\cr
 \n&\n&\n&\n&{\CoefTrue}&\n&\n&\n&\n&\n&\n&\n&\n&\n&\n&\n\cr
 \n&\n&\n&\n&\n&\n&\n&\n&\n&\n&\n&\n&\n&\n&{\CoefTrue}&\n\cr
 \n&\n&\n&\n&\n&{\CoefTrue}&\n&\n&\n&\n&\n&\n&\n&\n&\n&\n\cr
 \n&\n&\n&\n&\n&\n&\n&\n&\n&\n&\n&\n&\n&\n&\n&{\CoefTrue}\cr
 \n&\n&\n&\n&\n&\n&\n&\n&{\CoefTrue}&\n&\n&\n&\n&\n&\n&\n\cr
 \n&\n&\n&\n&\n&\n&\n&\n&\n&\n&{\CoefTrue}&\n&\n&\n&\n&\n\cr
 \n&\n&\n&\n&\n&\n&\n&\n&\n&{\CoefTrue}&\n&\n&\n&\n&\n&\n\cr
 \n&\n&\n&\n&\n&\n&\n&\n&\n&\n&\n&{\CoefTrue}&\n&\n&\n&\n\cr
 \n&\n&\n&\n&\n&\n&\n&\n&\n&\n&\n&\n&{\CoefTrue}&\n&\n&\n\cr
 \n&\n&\n&\n&\n&\n&\n&\n&\n&\n&\n&\n&\n&\n&{\CoefTrue}&\n\cr
 \n&\n&\n&\n&\n&\n&\n&\n&\n&\n&\n&\n&\n&{\CoefTrue}&\n&\n\cr
 \n&\n&\n&\n&\n&\n&\n&\n&\n&\n&\n&\n&\n&\n&\n&{\CoefTrue}\cr
 \n&\n&\n&\n&\n&{\CoefTrue}&\n&\n&\n&\n&\n&\n&\n&\n&\n&\n\cr
 \n&\n&\n&\n&\n&\n&\n&\n&\n&\n&\n&\n&\n&\n&\n&{\CoefTrue}\cr
 \n&\n&\n&\n&\n&{\CoefTrue}&\n&\n&\n&\n&\n&\n&\n&\n&\n&\n\cr
 \n&\n&\n&\n&\n&\n&\n&\n&\n&\n&\n&\n&\n&\n&\n&{\CoefTrue}\cr
 \n&\n&\n&\n&\n&{\CoefTrue}&\n&\n&\n&\n&\n&\n&\n&\n&\n&\n\cr
 \n&\n&\n&\n&\n&\n&\n&\n&\n&\n&\n&\n&\n&\n&\n&{\CoefTrue}\cr
 \n&\n&\n&\n&\n&{\CoefTrue}&\n&\n&\n&\n&\n&\n&\n&\n&\n&\n\cr
 \n&\n&\n&\n&\n&\n&\n&\n&\n&\n&\n&\n&\n&\n&\n&{\CoefTrue}\cr
 \n&\n&\n&\n&\n&\n&\n&\n&\n&\n&\n&\n&\n&{\CoefTrue}&\n&\n\cr
 \n&\n&\n&\n&\n&\n&\n&\n&\n&\n&\n&\n&\n&\n&\n&{\CoefTrue}\cr
 \n&\n&\n&\n&\n&\n&\n&\n&\n&\n&\n&\n&\n&{\CoefTrue}&\n&\n\cr
 \n&\n&\n&\n&\n&\n&\n&\n&\n&\n&\n&\n&\n&\n&\n&{\CoefTrue}\cr
 \n&\n&\n&\n&\n&\n&\n&\n&\n&\n&\n&\n&\n&{\CoefTrue}&\n&\n\cr
 \n&\n&\n&\n&\n&\n&\n&\n&\n&\n&\n&\n&\n&\n&\n&{\CoefTrue}\cr
 \n&\n&\n&\n&\n&\n&\n&\n&\n&\n&\n&\n&\n&{\CoefTrue}&\n&\n\cr
 \n&\n&\n&\n&\n&\n&\n&\n&\n&\n&\n&\n&\n&\n&\n&{\CoefTrue}\cr
\noalign{\kern-2pt}}}%
\vbox{\setbox8=\hbox{$\RowNames\Matrix$}
\hbox to\wd8{\hfil$\ColNames$\kern\ColEntryShiftHoriz}\kern\ColEntryShiftVerti
\box8}}}}$
\hfil}
\hbox to\textwidth{\hfil$\vartheta_{R\RELtraOP}=$
{\footnotesize%
\BoxBreadth=0pt%
\setbox7=\hbox{\{\}}%
\ifdim\wd7>\BoxBreadth\BoxBreadth=\wd7\fi%
\setbox7=\hbox{\{a\}}%
\ifdim\wd7>\BoxBreadth\BoxBreadth=\wd7\fi%
\setbox7=\hbox{\{b\}}%
\ifdim\wd7>\BoxBreadth\BoxBreadth=\wd7\fi%
\setbox7=\hbox{\{a,b\}}%
\ifdim\wd7>\BoxBreadth\BoxBreadth=\wd7\fi%
\setbox7=\hbox{\{c\}}%
\ifdim\wd7>\BoxBreadth\BoxBreadth=\wd7\fi%
\setbox7=\hbox{\{a,c\}}%
\ifdim\wd7>\BoxBreadth\BoxBreadth=\wd7\fi%
\setbox7=\hbox{\{b,c\}}%
\ifdim\wd7>\BoxBreadth\BoxBreadth=\wd7\fi%
\setbox7=\hbox{\{a,b,c\}}%
\ifdim\wd7>\BoxBreadth\BoxBreadth=\wd7\fi%
\setbox7=\hbox{\{d\}}%
\ifdim\wd7>\BoxBreadth\BoxBreadth=\wd7\fi%
\setbox7=\hbox{\{a,d\}}%
\ifdim\wd7>\BoxBreadth\BoxBreadth=\wd7\fi%
\setbox7=\hbox{\{b,d\}}%
\ifdim\wd7>\BoxBreadth\BoxBreadth=\wd7\fi%
\setbox7=\hbox{\{a,b,d\}}%
\ifdim\wd7>\BoxBreadth\BoxBreadth=\wd7\fi%
\setbox7=\hbox{\{c,d\}}%
\ifdim\wd7>\BoxBreadth\BoxBreadth=\wd7\fi%
\setbox7=\hbox{\{a,c,d\}}%
\ifdim\wd7>\BoxBreadth\BoxBreadth=\wd7\fi%
\setbox7=\hbox{\{b,c,d\}}%
\ifdim\wd7>\BoxBreadth\BoxBreadth=\wd7\fi%
\setbox7=\hbox{\{a,b,c,d\}}%
\ifdim\wd7>\BoxBreadth\BoxBreadth=\wd7\fi%
\def\RowNames{\vcenter{\offinterlineskip\baselineskip=\matrixskip%
\hbox to\BoxBreadth{\strut\hfil \{\}}\kern\interspacereduction%
\hbox to\BoxBreadth{\strut\hfil \{a\}}\kern\interspacereduction%
\hbox to\BoxBreadth{\strut\hfil \{b\}}\kern\interspacereduction%
\hbox to\BoxBreadth{\strut\hfil \{a,b\}}\kern\interspacereduction%
\hbox to\BoxBreadth{\strut\hfil \{c\}}\kern\interspacereduction%
\hbox to\BoxBreadth{\strut\hfil \{a,c\}}\kern\interspacereduction%
\hbox to\BoxBreadth{\strut\hfil \{b,c\}}\kern\interspacereduction%
\hbox to\BoxBreadth{\strut\hfil \{a,b,c\}}\kern\interspacereduction%
\hbox to\BoxBreadth{\strut\hfil \{d\}}\kern\interspacereduction%
\hbox to\BoxBreadth{\strut\hfil \{a,d\}}\kern\interspacereduction%
\hbox to\BoxBreadth{\strut\hfil \{b,d\}}\kern\interspacereduction%
\hbox to\BoxBreadth{\strut\hfil \{a,b,d\}}\kern\interspacereduction%
\hbox to\BoxBreadth{\strut\hfil \{c,d\}}\kern\interspacereduction%
\hbox to\BoxBreadth{\strut\hfil \{a,c,d\}}\kern\interspacereduction%
\hbox to\BoxBreadth{\strut\hfil \{b,c,d\}}\kern\interspacereduction%
\hbox to\BoxBreadth{\strut\hfil \{a,b,c,d\}}}}%
\def\ColNames{\hbox{\rotatebox{90}{\strut \{\}}\kern\interspacereduction%
\rotatebox{90}{\strut \{1\}}\kern\interspacereduction%
\rotatebox{90}{\strut \{2\}}\kern\interspacereduction%
\rotatebox{90}{\strut \{1,2\}}\kern\interspacereduction%
\rotatebox{90}{\strut \{3\}}\kern\interspacereduction%
\rotatebox{90}{\strut \{1,3\}}\kern\interspacereduction%
\rotatebox{90}{\strut \{2,3\}}\kern\interspacereduction%
\rotatebox{90}{\strut \{1,2,3\}}\kern\interspacereduction%
\rotatebox{90}{\strut \{4\}}\kern\interspacereduction%
\rotatebox{90}{\strut \{1,4\}}\kern\interspacereduction%
\rotatebox{90}{\strut \{2,4\}}\kern\interspacereduction%
\rotatebox{90}{\strut \{1,2,4\}}\kern\interspacereduction%
\rotatebox{90}{\strut \{3,4\}}\kern\interspacereduction%
\rotatebox{90}{\strut \{1,3,4\}}\kern\interspacereduction%
\rotatebox{90}{\strut \{2,3,4\}}\kern\interspacereduction%
\rotatebox{90}{\strut \{1,2,3,4\}}\kern\interspacereduction%
\rotatebox{90}{\strut \{5\}}\kern\interspacereduction%
\rotatebox{90}{\strut \{1,5\}}\kern\interspacereduction%
\rotatebox{90}{\strut \{2,5\}}\kern\interspacereduction%
\rotatebox{90}{\strut \{1,2,5\}}\kern\interspacereduction%
\rotatebox{90}{\strut \{3,5\}}\kern\interspacereduction%
\rotatebox{90}{\strut \{1,3,5\}}\kern\interspacereduction%
\rotatebox{90}{\strut \{2,3,5\}}\kern\interspacereduction%
\rotatebox{90}{\strut \{1,2,3,5\}}\kern\interspacereduction%
\rotatebox{90}{\strut \{4,5\}}\kern\interspacereduction%
\rotatebox{90}{\strut \{1,4,5\}}\kern\interspacereduction%
\rotatebox{90}{\strut \{2,4,5\}}\kern\interspacereduction%
\rotatebox{90}{\strut \{1,2,4,5\}}\kern\interspacereduction%
\rotatebox{90}{\strut \{3,4,5\}}\kern\interspacereduction%
\rotatebox{90}{\strut \{1,3,4,5\}}\kern\interspacereduction%
\rotatebox{90}{\strut \{2,3,4,5\}}\kern\interspacereduction%
\rotatebox{90}{\strut \{1,2,3,4,5\}}\kern\interspacereduction%
}}%
\def\Matrix{\spmatrix{%
\noalign{\kern-2pt}
 {\CoefTrue}&\n&\n&\n&\n&\n&\n&\n&\n&\n&\n&\n&\n&\n&\n&\n&\n&\n&\n&\n&\n&\n&\n&\n&\n&\n&\n&\n&\n&\n&\n&\n\cr
 \n&\n&\n&\n&\n&\n&\n&\n&\n&\n&\n&\n&\n&\n&\n&\n&\n&\n&{\CoefTrue}&\n&\n&\n&\n&\n&\n&\n&\n&\n&\n&\n&\n&\n\cr
 \n&{\CoefTrue}&\n&\n&\n&\n&\n&\n&\n&\n&\n&\n&\n&\n&\n&\n&\n&\n&\n&\n&\n&\n&\n&\n&\n&\n&\n&\n&\n&\n&\n&\n\cr
 \n&\n&\n&\n&\n&\n&\n&\n&\n&\n&\n&\n&\n&\n&\n&\n&\n&\n&\n&{\CoefTrue}&\n&\n&\n&\n&\n&\n&\n&\n&\n&\n&\n&\n\cr
 \n&\n&\n&\n&\n&\n&\n&\n&\n&\n&\n&\n&\n&\n&\n&\n&\n&\n&\n&\n&{\CoefTrue}&\n&\n&\n&\n&\n&\n&\n&\n&\n&\n&\n\cr
 \n&\n&\n&\n&\n&\n&\n&\n&\n&\n&\n&\n&\n&\n&\n&\n&\n&\n&\n&\n&\n&\n&{\CoefTrue}&\n&\n&\n&\n&\n&\n&\n&\n&\n\cr
 \n&\n&\n&\n&\n&\n&\n&\n&\n&\n&\n&\n&\n&\n&\n&\n&\n&\n&\n&\n&\n&{\CoefTrue}&\n&\n&\n&\n&\n&\n&\n&\n&\n&\n\cr
 \n&\n&\n&\n&\n&\n&\n&\n&\n&\n&\n&\n&\n&\n&\n&\n&\n&\n&\n&\n&\n&\n&\n&{\CoefTrue}&\n&\n&\n&\n&\n&\n&\n&\n\cr
 \n&\n&\n&\n&\n&\n&\n&\n&\n&{\CoefTrue}&\n&\n&\n&\n&\n&\n&\n&\n&\n&\n&\n&\n&\n&\n&\n&\n&\n&\n&\n&\n&\n&\n\cr
 \n&\n&\n&\n&\n&\n&\n&\n&\n&\n&\n&\n&\n&\n&\n&\n&\n&\n&\n&\n&\n&\n&\n&\n&\n&\n&\n&{\CoefTrue}&\n&\n&\n&\n\cr
 \n&\n&\n&\n&\n&\n&\n&\n&\n&{\CoefTrue}&\n&\n&\n&\n&\n&\n&\n&\n&\n&\n&\n&\n&\n&\n&\n&\n&\n&\n&\n&\n&\n&\n\cr
 \n&\n&\n&\n&\n&\n&\n&\n&\n&\n&\n&\n&\n&\n&\n&\n&\n&\n&\n&\n&\n&\n&\n&\n&\n&\n&\n&{\CoefTrue}&\n&\n&\n&\n\cr
 \n&\n&\n&\n&\n&\n&\n&\n&\n&\n&\n&\n&\n&\n&\n&\n&\n&\n&\n&\n&\n&\n&\n&\n&\n&\n&\n&\n&\n&{\CoefTrue}&\n&\n\cr
 \n&\n&\n&\n&\n&\n&\n&\n&\n&\n&\n&\n&\n&\n&\n&\n&\n&\n&\n&\n&\n&\n&\n&\n&\n&\n&\n&\n&\n&\n&\n&{\CoefTrue}\cr
 \n&\n&\n&\n&\n&\n&\n&\n&\n&\n&\n&\n&\n&\n&\n&\n&\n&\n&\n&\n&\n&\n&\n&\n&\n&\n&\n&\n&\n&{\CoefTrue}&\n&\n\cr
 \n&\n&\n&\n&\n&\n&\n&\n&\n&\n&\n&\n&\n&\n&\n&\n&\n&\n&\n&\n&\n&\n&\n&\n&\n&\n&\n&\n&\n&\n&\n&{\CoefTrue}\cr
\noalign{\kern-2pt}}}%
\vbox{\setbox8=\hbox{$\RowNames\Matrix$}
\hbox to\wd8{\hfil$\ColNames$\kern\ColEntryShiftHoriz}\kern\ColEntryShiftVerti
\box8}}\hfil}}}
{Existential and inverse images}{FigExImage} 

\enunc{}{Proposition}{}{PropThetaInverseTheta} Let $\RELfromTO{f}{X}{Y}$ be an arbitrary mapping. Then
\begin{enumerate}[i)]
\item $\ExistIm_{f\RELtraOP}\RELtraOP\RELcompOP\ExistIm_{f\RELtraOP}\RELcompOP\ExistIm_{f}
=
\ExistIm_{f\RELtraOP}\RELtraOP\RELcompOP\ExistIm_{f}\RELtraOP\RELcompOP\ExistIm_{f}
$,

\item $\ExistIm_{f\RELtraOP}\RELtraOP\RELcompOP\,\RELtop\RELandOP\ExistIm_{f}
=
\ExistIm_{f\RELtraOP}\RELtraOP\RELandOP\,\RELtop\RELcompOP\ExistIm_{f}$,

\item $\ExistIm_{f\RELtraOP}\RELtraOP\RELenthOP\ExistIm_{f}$\quad or\quad $\syqq{\varepsilon_X}{f\RELcompOP\varepsilon_Y}
\RELenthOP
\syqq{f\RELtraOP\RELcompOP\varepsilon_X}{\varepsilon_Y}$\quad when $f$ is surjective.
\end{enumerate}

\Proof i) and ii) are proved together, starting with

\smallskip
$\ExistIm_{f\RELtraOP}\RELtraOP\RELcompOP\RELtop\RELandOP\ExistIm_{f}
\RELenthOP
(\ExistIm_{f\RELtraOP}\RELtraOP\RELandOP\ExistIm_{f}\RELcompOP\RELtop)\RELcompOP(\RELtop\RELandOP\ExistIm_{f\RELtraOP}\RELcompOP\ExistIm_{f})
$\quad  Dedekind rule

$=
\ExistIm_{f\RELtraOP}\RELtraOP\RELcompOP\ExistIm_{f\RELtraOP}\RELcompOP\ExistIm_{f}
$\quad since existential images are total

$
\RELenthOP
\ExistIm_{f\RELtraOP}\RELtraOP\RELcompOP\RELtop\RELandOP\ExistIm_{f}
$\quad since existential images are univalent

\smallskip
\noindent
resulting in equality in between. Similarly

\smallskip
$\ExistIm_{f\RELtraOP}\RELtraOP\RELandOP\,\RELtop\RELcompOP\ExistIm_{f}
\RELenthOP
(\RELtop\RELandOP\ExistIm_{f\RELtraOP}\RELtraOP\RELcompOP\ExistIm_{f}\RELtraOP)\RELcompOP(\ExistIm_{f}\RELandOP\,\RELtop\RELcompOP\ExistIm_{f\RELtraOP}\RELtraOP)
$

$
=
\ExistIm_{f\RELtraOP}\RELtraOP\RELcompOP\ExistIm_{f}\RELtraOP\RELcompOP\ExistIm_{f}
$

$\RELenthOP
\ExistIm_{f\RELtraOP}\RELtraOP\RELandOP\,\RELtop\RELcompOP\ExistIm_{f}
$

\bigskip
\noindent 
Thus, (i,ii) mean the same. For the remaining proof we start from

\smallskip

$
\ExistIm_{f}\RELtraOP\RELcompOP\,(\ExistIm_{f\RELtraOP}\RELtraOP\RELandOP\,\RELtop\RELcompOP\ExistIm_{f})
=
\RELtop\RELcompOP\ExistIm_{f}\RELandOP\ExistIm_{f}\RELtraOP\RELcompOP\ExistIm_{f\RELtraOP}\RELtraOP
$\quad masking

$\RELenthOP
(\RELtop\RELandOP\ExistIm_{f}\RELtraOP\RELcompOP\ExistIm_{f\RELtraOP}\RELtraOP\RELcompOP\ExistIm_{f}\RELtraOP)\RELcompOP(\ExistIm_{f}\RELandOP\,\RELtop\RELcompOP\ExistIm_{f}\RELtraOP\RELcompOP\ExistIm_{f\RELtraOP}\RELtraOP)$\quad Dedekind rule

$\RELenthOP
\ExistIm_{f}\RELtraOP\RELcompOP\ExistIm_{f\RELtraOP}\RELtraOP\RELcompOP\ExistIm_{f}\RELtraOP\RELcompOP\,\ExistIm_{f}
$\quad

$=
\ExistIm_{f\,\,\RELcompOP\, f\RELtraOP\,\RELcompOP\, f}\RELtraOP\,\RELcompOP\,\ExistIm_{f}
$\quad existential images are multiplicative

$=
\ExistIm_{f}\RELtraOP\RELcompOP\,\ExistIm_{f}
\RELenthOP
\RELide
$\quad since $f\RELcompOP f\RELtraOP\RELcompOP f = f$ for a mapping

\smallskip
\noindent
Now shunting gives the needed result $
\ExistIm_{f\RELtraOP}\RELtraOP\RELandOP\,\RELtop\RELcompOP\ExistIm_{f}
\RELenthOP
\ExistIm_{f}
$.

\bigskip
\noindent 
The following is proved mutatis mutandis:

\smallskip
$(\ExistIm_{f\RELtraOP}\RELtraOP\RELcompOP\RELtop\RELandOP\ExistIm_{f})\RELcompOP\ExistIm_{f\RELtraOP}
=
\ExistIm_{f\RELtraOP}\RELtraOP\RELcompOP\RELtop\RELandOP\ExistIm_{f}\RELcompOP\ExistIm_{f\RELtraOP}
\RELenthOP
(\ExistIm_{f\RELtraOP}\RELtraOP\RELandOP\ExistIm_{f}\RELcompOP\ExistIm_{f\RELtraOP}\RELcompOP\RELtop)\RELcompOP(\RELtop\RELandOP\ExistIm_{f\RELtraOP}\RELcompOP\ExistIm_{f}\RELcompOP\ExistIm_{f\RELtraOP})
$

$=
(\ExistIm_{f\RELtraOP}\RELtraOP\RELandOP\ExistIm_{f}\RELcompOP\ExistIm_{f\RELtraOP}\RELcompOP\RELtop)\RELcompOP\ExistIm_{f\RELtraOP}\RELcompOP\ExistIm_{f}\RELcompOP\ExistIm_{f\RELtraOP}
\RELenthOP
\ExistIm_{f\RELtraOP}\RELtraOP\RELcompOP\ExistIm_{f\RELtraOP}\RELcompOP\ExistIm_{f}\RELcompOP\ExistIm_{f\RELtraOP}
$

$=
\ExistIm_{f\RELtraOP}\RELtraOP\RELcompOP\ExistIm_{f\RELtraOP\,\RELcompOP\, f\,\,\RELcompOP\, f\RELtraOP}
=
\ExistIm_{f\RELtraOP}\RELtraOP\RELcompOP\ExistIm_{f\RELtraOP}
\RELenthOP
\RELide$

\bigskip
\noindent 
iii) results simply from an application of Prop.~\PropSyqSurjFct.
\Bewende

\noindent
These results imply not least that $\ExistIm_{f\RELtraOP}\RELtraOP$ is univalent, or a partial function, when $f$ is surjective. With (ii), we have then also $\ExistIm_{f}\RELandOP\ExistIm_{f\RELtraOP}\RELtraOP\RELcompOP\RELtop=\ExistIm_{f\RELtraOP}\RELtraOP$.

\Caption{\vbox{\hbox to \textwidth{%
{\footnotesize%
\BoxBreadth=0pt%
\setbox7=\hbox{1}%
\ifdim\wd7>\BoxBreadth\BoxBreadth=\wd7\fi%
\setbox7=\hbox{2}%
\ifdim\wd7>\BoxBreadth\BoxBreadth=\wd7\fi%
\setbox7=\hbox{3}%
\ifdim\wd7>\BoxBreadth\BoxBreadth=\wd7\fi%
\setbox7=\hbox{4}%
\ifdim\wd7>\BoxBreadth\BoxBreadth=\wd7\fi%
\setbox7=\hbox{5}%
\ifdim\wd7>\BoxBreadth\BoxBreadth=\wd7\fi%
\def\RowNames{\vcenter{\offinterlineskip\baselineskip=\matrixskip%
\hbox to\BoxBreadth{\strut\hfil 1}\kern\interspacereduction%
\hbox to\BoxBreadth{\strut\hfil 2}\kern\interspacereduction%
\hbox to\BoxBreadth{\strut\hfil 3}\kern\interspacereduction%
\hbox to\BoxBreadth{\strut\hfil 4}\kern\interspacereduction%
\hbox to\BoxBreadth{\strut\hfil 5}}}%
\def\ColNames{\hbox{\rotatebox{90}{\strut a}\kern\interspacereduction%
\rotatebox{90}{\strut b}\kern\interspacereduction%
\rotatebox{90}{\strut c}\kern\interspacereduction%
\rotatebox{90}{\strut d}\kern\interspacereduction%
}}%
\def\Matrix{\spmatrix{%
\noalign{\kern-2pt}
 \n&{\CoefTrue}&\n&\n\cr
 {\CoefTrue}&\n&\n&\n\cr
 \n&\n&{\CoefTrue}&\n\cr
 \n&\n&\n&{\CoefTrue}\cr
 \n&\n&{\CoefTrue}&\n\cr
\noalign{\kern-2pt}}}%
\vbox{\setbox8=\hbox{$\RowNames\Matrix$}
\hbox to\wd8{\hfil$\ColNames$\kern\ColEntryShiftHoriz}\kern\ColEntryShiftVerti
\box8}}%
\kern-0.35cm\hfil%
{\footnotesize%
\BoxBreadth=0pt%
\setbox7=\hbox{\{\}}%
\ifdim\wd7>\BoxBreadth\BoxBreadth=\wd7\fi%
\setbox7=\hbox{\{1\}}%
\ifdim\wd7>\BoxBreadth\BoxBreadth=\wd7\fi%
\setbox7=\hbox{\{2\}}%
\ifdim\wd7>\BoxBreadth\BoxBreadth=\wd7\fi%
\setbox7=\hbox{\{1,2\}}%
\ifdim\wd7>\BoxBreadth\BoxBreadth=\wd7\fi%
\setbox7=\hbox{\{3\}}%
\ifdim\wd7>\BoxBreadth\BoxBreadth=\wd7\fi%
\setbox7=\hbox{\{1,3\}}%
\ifdim\wd7>\BoxBreadth\BoxBreadth=\wd7\fi%
\setbox7=\hbox{\{2,3\}}%
\ifdim\wd7>\BoxBreadth\BoxBreadth=\wd7\fi%
\setbox7=\hbox{\{1,2,3\}}%
\ifdim\wd7>\BoxBreadth\BoxBreadth=\wd7\fi%
\setbox7=\hbox{\{4\}}%
\ifdim\wd7>\BoxBreadth\BoxBreadth=\wd7\fi%
\setbox7=\hbox{\{1,4\}}%
\ifdim\wd7>\BoxBreadth\BoxBreadth=\wd7\fi%
\setbox7=\hbox{\{2,4\}}%
\ifdim\wd7>\BoxBreadth\BoxBreadth=\wd7\fi%
\setbox7=\hbox{\{1,2,4\}}%
\ifdim\wd7>\BoxBreadth\BoxBreadth=\wd7\fi%
\setbox7=\hbox{\{3,4\}}%
\ifdim\wd7>\BoxBreadth\BoxBreadth=\wd7\fi%
\setbox7=\hbox{\{1,3,4\}}%
\ifdim\wd7>\BoxBreadth\BoxBreadth=\wd7\fi%
\setbox7=\hbox{\{2,3,4\}}%
\ifdim\wd7>\BoxBreadth\BoxBreadth=\wd7\fi%
\setbox7=\hbox{\{1,2,3,4\}}%
\ifdim\wd7>\BoxBreadth\BoxBreadth=\wd7\fi%
\setbox7=\hbox{\{5\}}%
\ifdim\wd7>\BoxBreadth\BoxBreadth=\wd7\fi%
\setbox7=\hbox{\{1,5\}}%
\ifdim\wd7>\BoxBreadth\BoxBreadth=\wd7\fi%
\setbox7=\hbox{\{2,5\}}%
\ifdim\wd7>\BoxBreadth\BoxBreadth=\wd7\fi%
\setbox7=\hbox{\{1,2,5\}}%
\ifdim\wd7>\BoxBreadth\BoxBreadth=\wd7\fi%
\setbox7=\hbox{\{3,5\}}%
\ifdim\wd7>\BoxBreadth\BoxBreadth=\wd7\fi%
\setbox7=\hbox{\{1,3,5\}}%
\ifdim\wd7>\BoxBreadth\BoxBreadth=\wd7\fi%
\setbox7=\hbox{\{2,3,5\}}%
\ifdim\wd7>\BoxBreadth\BoxBreadth=\wd7\fi%
\setbox7=\hbox{\{1,2,3,5\}}%
\ifdim\wd7>\BoxBreadth\BoxBreadth=\wd7\fi%
\setbox7=\hbox{\{4,5\}}%
\ifdim\wd7>\BoxBreadth\BoxBreadth=\wd7\fi%
\setbox7=\hbox{\{1,4,5\}}%
\ifdim\wd7>\BoxBreadth\BoxBreadth=\wd7\fi%
\setbox7=\hbox{\{2,4,5\}}%
\ifdim\wd7>\BoxBreadth\BoxBreadth=\wd7\fi%
\setbox7=\hbox{\{1,2,4,5\}}%
\ifdim\wd7>\BoxBreadth\BoxBreadth=\wd7\fi%
\setbox7=\hbox{\{3,4,5\}}%
\ifdim\wd7>\BoxBreadth\BoxBreadth=\wd7\fi%
\setbox7=\hbox{\{1,3,4,5\}}%
\ifdim\wd7>\BoxBreadth\BoxBreadth=\wd7\fi%
\setbox7=\hbox{\{2,3,4,5\}}%
\ifdim\wd7>\BoxBreadth\BoxBreadth=\wd7\fi%
\setbox7=\hbox{\{1,2,3,4,5\}}%
\ifdim\wd7>\BoxBreadth\BoxBreadth=\wd7\fi%
\def\RowNames{\vcenter{\offinterlineskip\baselineskip=\matrixskip%
\hbox to\BoxBreadth{\strut\hfil \{\}}\kern\interspacereduction%
\hbox to\BoxBreadth{\strut\hfil \{1\}}\kern\interspacereduction%
\hbox to\BoxBreadth{\strut\hfil \{2\}}\kern\interspacereduction%
\hbox to\BoxBreadth{\strut\hfil \{1,2\}}\kern\interspacereduction%
\hbox to\BoxBreadth{\strut\hfil \{3\}}\kern\interspacereduction%
\hbox to\BoxBreadth{\strut\hfil \{1,3\}}\kern\interspacereduction%
\hbox to\BoxBreadth{\strut\hfil \{2,3\}}\kern\interspacereduction%
\hbox to\BoxBreadth{\strut\hfil \{1,2,3\}}\kern\interspacereduction%
\hbox to\BoxBreadth{\strut\hfil \{4\}}\kern\interspacereduction%
\hbox to\BoxBreadth{\strut\hfil \{1,4\}}\kern\interspacereduction%
\hbox to\BoxBreadth{\strut\hfil \{2,4\}}\kern\interspacereduction%
\hbox to\BoxBreadth{\strut\hfil \{1,2,4\}}\kern\interspacereduction%
\hbox to\BoxBreadth{\strut\hfil \{3,4\}}\kern\interspacereduction%
\hbox to\BoxBreadth{\strut\hfil \{1,3,4\}}\kern\interspacereduction%
\hbox to\BoxBreadth{\strut\hfil \{2,3,4\}}\kern\interspacereduction%
\hbox to\BoxBreadth{\strut\hfil \{1,2,3,4\}}\kern\interspacereduction%
\hbox to\BoxBreadth{\strut\hfil \{5\}}\kern\interspacereduction%
\hbox to\BoxBreadth{\strut\hfil \{1,5\}}\kern\interspacereduction%
\hbox to\BoxBreadth{\strut\hfil \{2,5\}}\kern\interspacereduction%
\hbox to\BoxBreadth{\strut\hfil \{1,2,5\}}\kern\interspacereduction%
\hbox to\BoxBreadth{\strut\hfil \{3,5\}}\kern\interspacereduction%
\hbox to\BoxBreadth{\strut\hfil \{1,3,5\}}\kern\interspacereduction%
\hbox to\BoxBreadth{\strut\hfil \{2,3,5\}}\kern\interspacereduction%
\hbox to\BoxBreadth{\strut\hfil \{1,2,3,5\}}\kern\interspacereduction%
\hbox to\BoxBreadth{\strut\hfil \{4,5\}}\kern\interspacereduction%
\hbox to\BoxBreadth{\strut\hfil \{1,4,5\}}\kern\interspacereduction%
\hbox to\BoxBreadth{\strut\hfil \{2,4,5\}}\kern\interspacereduction%
\hbox to\BoxBreadth{\strut\hfil \{1,2,4,5\}}\kern\interspacereduction%
\hbox to\BoxBreadth{\strut\hfil \{3,4,5\}}\kern\interspacereduction%
\hbox to\BoxBreadth{\strut\hfil \{1,3,4,5\}}\kern\interspacereduction%
\hbox to\BoxBreadth{\strut\hfil \{2,3,4,5\}}\kern\interspacereduction%
\hbox to\BoxBreadth{\strut\hfil \{1,2,3,4,5\}}}}%
\def\ColNames{\hbox{\rotatebox{90}{\strut \{\}}\kern\interspacereduction%
\rotatebox{90}{\strut \{a\}}\kern\interspacereduction%
\rotatebox{90}{\strut \{b\}}\kern\interspacereduction%
\rotatebox{90}{\strut \{a,b\}}\kern\interspacereduction%
\rotatebox{90}{\strut \{c\}}\kern\interspacereduction%
\rotatebox{90}{\strut \{a,c\}}\kern\interspacereduction%
\rotatebox{90}{\strut \{b,c\}}\kern\interspacereduction%
\rotatebox{90}{\strut \{a,b,c\}}\kern\interspacereduction%
\rotatebox{90}{\strut \{d\}}\kern\interspacereduction%
\rotatebox{90}{\strut \{a,d\}}\kern\interspacereduction%
\rotatebox{90}{\strut \{b,d\}}\kern\interspacereduction%
\rotatebox{90}{\strut \{a,b,d\}}\kern\interspacereduction%
\rotatebox{90}{\strut \{c,d\}}\kern\interspacereduction%
\rotatebox{90}{\strut \{a,c,d\}}\kern\interspacereduction%
\rotatebox{90}{\strut \{b,c,d\}}\kern\interspacereduction%
\rotatebox{90}{\strut \{a,b,c,d\}}\kern\interspacereduction%
}}%
\def\Matrix{\spmatrix{%
\noalign{\kern-2pt}
 {\CoefTrue}&\n&\n&\n&\n&\n&\n&\n&\n&\n&\n&\n&\n&\n&\n&\n\cr
 \n&\n&{\CoefTrue}&\n&\n&\n&\n&\n&\n&\n&\n&\n&\n&\n&\n&\n\cr
 \n&{\CoefTrue}&\n&\n&\n&\n&\n&\n&\n&\n&\n&\n&\n&\n&\n&\n\cr
 \n&\n&\n&{\CoefTrue}&\n&\n&\n&\n&\n&\n&\n&\n&\n&\n&\n&\n\cr
 \n&\n&\n&\n&{\CoefTrue}&\n&\n&\n&\n&\n&\n&\n&\n&\n&\n&\n\cr
 \n&\n&\n&\n&\n&\n&{\CoefTrue}&\n&\n&\n&\n&\n&\n&\n&\n&\n\cr
 \n&\n&\n&\n&\n&{\CoefTrue}&\n&\n&\n&\n&\n&\n&\n&\n&\n&\n\cr
 \n&\n&\n&\n&\n&\n&\n&{\CoefTrue}&\n&\n&\n&\n&\n&\n&\n&\n\cr
 \n&\n&\n&\n&\n&\n&\n&\n&{\CoefTrue}&\n&\n&\n&\n&\n&\n&\n\cr
 \n&\n&\n&\n&\n&\n&\n&\n&\n&\n&{\CoefTrue}&\n&\n&\n&\n&\n\cr
 \n&\n&\n&\n&\n&\n&\n&\n&\n&{\CoefTrue}&\n&\n&\n&\n&\n&\n\cr
 \n&\n&\n&\n&\n&\n&\n&\n&\n&\n&\n&{\CoefTrue}&\n&\n&\n&\n\cr
 \n&\n&\n&\n&\n&\n&\n&\n&\n&\n&\n&\n&{\CoefTrue}&\n&\n&\n\cr
 \n&\n&\n&\n&\n&\n&\n&\n&\n&\n&\n&\n&\n&\n&{\CoefTrue}&\n\cr
 \n&\n&\n&\n&\n&\n&\n&\n&\n&\n&\n&\n&\n&{\CoefTrue}&\n&\n\cr
 \n&\n&\n&\n&\n&\n&\n&\n&\n&\n&\n&\n&\n&\n&\n&{\CoefTrue}\cr
 \n&\n&\n&\n&{\CoefTrue}&\n&\n&\n&\n&\n&\n&\n&\n&\n&\n&\n\cr
 \n&\n&\n&\n&\n&\n&{\CoefTrue}&\n&\n&\n&\n&\n&\n&\n&\n&\n\cr
 \n&\n&\n&\n&\n&{\CoefTrue}&\n&\n&\n&\n&\n&\n&\n&\n&\n&\n\cr
 \n&\n&\n&\n&\n&\n&\n&{\CoefTrue}&\n&\n&\n&\n&\n&\n&\n&\n\cr
 \n&\n&\n&\n&{\CoefTrue}&\n&\n&\n&\n&\n&\n&\n&\n&\n&\n&\n\cr
 \n&\n&\n&\n&\n&\n&{\CoefTrue}&\n&\n&\n&\n&\n&\n&\n&\n&\n\cr
 \n&\n&\n&\n&\n&{\CoefTrue}&\n&\n&\n&\n&\n&\n&\n&\n&\n&\n\cr
 \n&\n&\n&\n&\n&\n&\n&{\CoefTrue}&\n&\n&\n&\n&\n&\n&\n&\n\cr
 \n&\n&\n&\n&\n&\n&\n&\n&\n&\n&\n&\n&{\CoefTrue}&\n&\n&\n\cr
 \n&\n&\n&\n&\n&\n&\n&\n&\n&\n&\n&\n&\n&\n&{\CoefTrue}&\n\cr
 \n&\n&\n&\n&\n&\n&\n&\n&\n&\n&\n&\n&\n&{\CoefTrue}&\n&\n\cr
 \n&\n&\n&\n&\n&\n&\n&\n&\n&\n&\n&\n&\n&\n&\n&{\CoefTrue}\cr
 \n&\n&\n&\n&\n&\n&\n&\n&\n&\n&\n&\n&{\CoefTrue}&\n&\n&\n\cr
 \n&\n&\n&\n&\n&\n&\n&\n&\n&\n&\n&\n&\n&\n&{\CoefTrue}&\n\cr
 \n&\n&\n&\n&\n&\n&\n&\n&\n&\n&\n&\n&\n&{\CoefTrue}&\n&\n\cr
 \n&\n&\n&\n&\n&\n&\n&\n&\n&\n&\n&\n&\n&\n&\n&{\CoefTrue}\cr
\noalign{\kern-2pt}}}%
\vbox{\setbox8=\hbox{$\RowNames\Matrix$}
\hbox to\wd8{\hfil$\ColNames$\kern\ColEntryShiftHoriz}\kern\ColEntryShiftVerti
\box8}}
\hfil\kern-0.2cm
{\footnotesize%
\BoxBreadth=0pt%
\setbox7=\hbox{\{\}}%
\ifdim\wd7>\BoxBreadth\BoxBreadth=\wd7\fi%
\setbox7=\hbox{\{1\}}%
\ifdim\wd7>\BoxBreadth\BoxBreadth=\wd7\fi%
\setbox7=\hbox{\{2\}}%
\ifdim\wd7>\BoxBreadth\BoxBreadth=\wd7\fi%
\setbox7=\hbox{\{1,2\}}%
\ifdim\wd7>\BoxBreadth\BoxBreadth=\wd7\fi%
\setbox7=\hbox{\{3\}}%
\ifdim\wd7>\BoxBreadth\BoxBreadth=\wd7\fi%
\setbox7=\hbox{\{1,3\}}%
\ifdim\wd7>\BoxBreadth\BoxBreadth=\wd7\fi%
\setbox7=\hbox{\{2,3\}}%
\ifdim\wd7>\BoxBreadth\BoxBreadth=\wd7\fi%
\setbox7=\hbox{\{1,2,3\}}%
\ifdim\wd7>\BoxBreadth\BoxBreadth=\wd7\fi%
\setbox7=\hbox{\{4\}}%
\ifdim\wd7>\BoxBreadth\BoxBreadth=\wd7\fi%
\setbox7=\hbox{\{1,4\}}%
\ifdim\wd7>\BoxBreadth\BoxBreadth=\wd7\fi%
\setbox7=\hbox{\{2,4\}}%
\ifdim\wd7>\BoxBreadth\BoxBreadth=\wd7\fi%
\setbox7=\hbox{\{1,2,4\}}%
\ifdim\wd7>\BoxBreadth\BoxBreadth=\wd7\fi%
\setbox7=\hbox{\{3,4\}}%
\ifdim\wd7>\BoxBreadth\BoxBreadth=\wd7\fi%
\setbox7=\hbox{\{1,3,4\}}%
\ifdim\wd7>\BoxBreadth\BoxBreadth=\wd7\fi%
\setbox7=\hbox{\{2,3,4\}}%
\ifdim\wd7>\BoxBreadth\BoxBreadth=\wd7\fi%
\setbox7=\hbox{\{1,2,3,4\}}%
\ifdim\wd7>\BoxBreadth\BoxBreadth=\wd7\fi%
\setbox7=\hbox{\{5\}}%
\ifdim\wd7>\BoxBreadth\BoxBreadth=\wd7\fi%
\setbox7=\hbox{\{1,5\}}%
\ifdim\wd7>\BoxBreadth\BoxBreadth=\wd7\fi%
\setbox7=\hbox{\{2,5\}}%
\ifdim\wd7>\BoxBreadth\BoxBreadth=\wd7\fi%
\setbox7=\hbox{\{1,2,5\}}%
\ifdim\wd7>\BoxBreadth\BoxBreadth=\wd7\fi%
\setbox7=\hbox{\{3,5\}}%
\ifdim\wd7>\BoxBreadth\BoxBreadth=\wd7\fi%
\setbox7=\hbox{\{1,3,5\}}%
\ifdim\wd7>\BoxBreadth\BoxBreadth=\wd7\fi%
\setbox7=\hbox{\{2,3,5\}}%
\ifdim\wd7>\BoxBreadth\BoxBreadth=\wd7\fi%
\setbox7=\hbox{\{1,2,3,5\}}%
\ifdim\wd7>\BoxBreadth\BoxBreadth=\wd7\fi%
\setbox7=\hbox{\{4,5\}}%
\ifdim\wd7>\BoxBreadth\BoxBreadth=\wd7\fi%
\setbox7=\hbox{\{1,4,5\}}%
\ifdim\wd7>\BoxBreadth\BoxBreadth=\wd7\fi%
\setbox7=\hbox{\{2,4,5\}}%
\ifdim\wd7>\BoxBreadth\BoxBreadth=\wd7\fi%
\setbox7=\hbox{\{1,2,4,5\}}%
\ifdim\wd7>\BoxBreadth\BoxBreadth=\wd7\fi%
\setbox7=\hbox{\{3,4,5\}}%
\ifdim\wd7>\BoxBreadth\BoxBreadth=\wd7\fi%
\setbox7=\hbox{\{1,3,4,5\}}%
\ifdim\wd7>\BoxBreadth\BoxBreadth=\wd7\fi%
\setbox7=\hbox{\{2,3,4,5\}}%
\ifdim\wd7>\BoxBreadth\BoxBreadth=\wd7\fi%
\setbox7=\hbox{\{1,2,3,4,5\}}%
\ifdim\wd7>\BoxBreadth\BoxBreadth=\wd7\fi%
\def\RowNames{\vcenter{\offinterlineskip\baselineskip=\matrixskip%
\hbox to\BoxBreadth{\strut\hfil \{\}}\kern\interspacereduction%
\hbox to\BoxBreadth{\strut\hfil \{1\}}\kern\interspacereduction%
\hbox to\BoxBreadth{\strut\hfil \{2\}}\kern\interspacereduction%
\hbox to\BoxBreadth{\strut\hfil \{1,2\}}\kern\interspacereduction%
\hbox to\BoxBreadth{\strut\hfil \{3\}}\kern\interspacereduction%
\hbox to\BoxBreadth{\strut\hfil \{1,3\}}\kern\interspacereduction%
\hbox to\BoxBreadth{\strut\hfil \{2,3\}}\kern\interspacereduction%
\hbox to\BoxBreadth{\strut\hfil \{1,2,3\}}\kern\interspacereduction%
\hbox to\BoxBreadth{\strut\hfil \{4\}}\kern\interspacereduction%
\hbox to\BoxBreadth{\strut\hfil \{1,4\}}\kern\interspacereduction%
\hbox to\BoxBreadth{\strut\hfil \{2,4\}}\kern\interspacereduction%
\hbox to\BoxBreadth{\strut\hfil \{1,2,4\}}\kern\interspacereduction%
\hbox to\BoxBreadth{\strut\hfil \{3,4\}}\kern\interspacereduction%
\hbox to\BoxBreadth{\strut\hfil \{1,3,4\}}\kern\interspacereduction%
\hbox to\BoxBreadth{\strut\hfil \{2,3,4\}}\kern\interspacereduction%
\hbox to\BoxBreadth{\strut\hfil \{1,2,3,4\}}\kern\interspacereduction%
\hbox to\BoxBreadth{\strut\hfil \{5\}}\kern\interspacereduction%
\hbox to\BoxBreadth{\strut\hfil \{1,5\}}\kern\interspacereduction%
\hbox to\BoxBreadth{\strut\hfil \{2,5\}}\kern\interspacereduction%
\hbox to\BoxBreadth{\strut\hfil \{1,2,5\}}\kern\interspacereduction%
\hbox to\BoxBreadth{\strut\hfil \{3,5\}}\kern\interspacereduction%
\hbox to\BoxBreadth{\strut\hfil \{1,3,5\}}\kern\interspacereduction%
\hbox to\BoxBreadth{\strut\hfil \{2,3,5\}}\kern\interspacereduction%
\hbox to\BoxBreadth{\strut\hfil \{1,2,3,5\}}\kern\interspacereduction%
\hbox to\BoxBreadth{\strut\hfil \{4,5\}}\kern\interspacereduction%
\hbox to\BoxBreadth{\strut\hfil \{1,4,5\}}\kern\interspacereduction%
\hbox to\BoxBreadth{\strut\hfil \{2,4,5\}}\kern\interspacereduction%
\hbox to\BoxBreadth{\strut\hfil \{1,2,4,5\}}\kern\interspacereduction%
\hbox to\BoxBreadth{\strut\hfil \{3,4,5\}}\kern\interspacereduction%
\hbox to\BoxBreadth{\strut\hfil \{1,3,4,5\}}\kern\interspacereduction%
\hbox to\BoxBreadth{\strut\hfil \{2,3,4,5\}}\kern\interspacereduction%
\hbox to\BoxBreadth{\strut\hfil \{1,2,3,4,5\}}}}%
\def\ColNames{\hbox{\rotatebox{90}{\strut \{\}}\kern\interspacereduction%
\rotatebox{90}{\strut \{a\}}\kern\interspacereduction%
\rotatebox{90}{\strut \{b\}}\kern\interspacereduction%
\rotatebox{90}{\strut \{a,b\}}\kern\interspacereduction%
\rotatebox{90}{\strut \{c\}}\kern\interspacereduction%
\rotatebox{90}{\strut \{a,c\}}\kern\interspacereduction%
\rotatebox{90}{\strut \{b,c\}}\kern\interspacereduction%
\rotatebox{90}{\strut \{a,b,c\}}\kern\interspacereduction%
\rotatebox{90}{\strut \{d\}}\kern\interspacereduction%
\rotatebox{90}{\strut \{a,d\}}\kern\interspacereduction%
\rotatebox{90}{\strut \{b,d\}}\kern\interspacereduction%
\rotatebox{90}{\strut \{a,b,d\}}\kern\interspacereduction%
\rotatebox{90}{\strut \{c,d\}}\kern\interspacereduction%
\rotatebox{90}{\strut \{a,c,d\}}\kern\interspacereduction%
\rotatebox{90}{\strut \{b,c,d\}}\kern\interspacereduction%
\rotatebox{90}{\strut \{a,b,c,d\}}\kern\interspacereduction%
}}%
\def\Matrix{\spmatrix{%
\noalign{\kern-2pt}
 {\CoefTrue}&\n&\n&\n&\n&\n&\n&\n&\n&\n&\n&\n&\n&\n&\n&\n\cr
 \n&\n&{\CoefTrue}&\n&\n&\n&\n&\n&\n&\n&\n&\n&\n&\n&\n&\n\cr
 \n&{\CoefTrue}&\n&\n&\n&\n&\n&\n&\n&\n&\n&\n&\n&\n&\n&\n\cr
 \n&\n&\n&{\CoefTrue}&\n&\n&\n&\n&\n&\n&\n&\n&\n&\n&\n&\n\cr
 \n&\n&\n&\n&\n&\n&\n&\n&\n&\n&\n&\n&\n&\n&\n&\n\cr
 \n&\n&\n&\n&\n&\n&\n&\n&\n&\n&\n&\n&\n&\n&\n&\n\cr
 \n&\n&\n&\n&\n&\n&\n&\n&\n&\n&\n&\n&\n&\n&\n&\n\cr
 \n&\n&\n&\n&\n&\n&\n&\n&\n&\n&\n&\n&\n&\n&\n&\n\cr
 \n&\n&\n&\n&\n&\n&\n&\n&{\CoefTrue}&\n&\n&\n&\n&\n&\n&\n\cr
 \n&\n&\n&\n&\n&\n&\n&\n&\n&\n&{\CoefTrue}&\n&\n&\n&\n&\n\cr
 \n&\n&\n&\n&\n&\n&\n&\n&\n&{\CoefTrue}&\n&\n&\n&\n&\n&\n\cr
 \n&\n&\n&\n&\n&\n&\n&\n&\n&\n&\n&{\CoefTrue}&\n&\n&\n&\n\cr
 \n&\n&\n&\n&\n&\n&\n&\n&\n&\n&\n&\n&\n&\n&\n&\n\cr
 \n&\n&\n&\n&\n&\n&\n&\n&\n&\n&\n&\n&\n&\n&\n&\n\cr
 \n&\n&\n&\n&\n&\n&\n&\n&\n&\n&\n&\n&\n&\n&\n&\n\cr
 \n&\n&\n&\n&\n&\n&\n&\n&\n&\n&\n&\n&\n&\n&\n&\n\cr
 \n&\n&\n&\n&\n&\n&\n&\n&\n&\n&\n&\n&\n&\n&\n&\n\cr
 \n&\n&\n&\n&\n&\n&\n&\n&\n&\n&\n&\n&\n&\n&\n&\n\cr
 \n&\n&\n&\n&\n&\n&\n&\n&\n&\n&\n&\n&\n&\n&\n&\n\cr
 \n&\n&\n&\n&\n&\n&\n&\n&\n&\n&\n&\n&\n&\n&\n&\n\cr
 \n&\n&\n&\n&{\CoefTrue}&\n&\n&\n&\n&\n&\n&\n&\n&\n&\n&\n\cr
 \n&\n&\n&\n&\n&\n&{\CoefTrue}&\n&\n&\n&\n&\n&\n&\n&\n&\n\cr
 \n&\n&\n&\n&\n&{\CoefTrue}&\n&\n&\n&\n&\n&\n&\n&\n&\n&\n\cr
 \n&\n&\n&\n&\n&\n&\n&{\CoefTrue}&\n&\n&\n&\n&\n&\n&\n&\n\cr
 \n&\n&\n&\n&\n&\n&\n&\n&\n&\n&\n&\n&\n&\n&\n&\n\cr
 \n&\n&\n&\n&\n&\n&\n&\n&\n&\n&\n&\n&\n&\n&\n&\n\cr
 \n&\n&\n&\n&\n&\n&\n&\n&\n&\n&\n&\n&\n&\n&\n&\n\cr
 \n&\n&\n&\n&\n&\n&\n&\n&\n&\n&\n&\n&\n&\n&\n&\n\cr
 \n&\n&\n&\n&\n&\n&\n&\n&\n&\n&\n&\n&{\CoefTrue}&\n&\n&\n\cr
 \n&\n&\n&\n&\n&\n&\n&\n&\n&\n&\n&\n&\n&\n&{\CoefTrue}&\n\cr
 \n&\n&\n&\n&\n&\n&\n&\n&\n&\n&\n&\n&\n&{\CoefTrue}&\n&\n\cr
 \n&\n&\n&\n&\n&\n&\n&\n&\n&\n&\n&\n&\n&\n&\n&{\CoefTrue}\cr
\noalign{\kern-2pt}}}%
\vbox{\setbox8=\hbox{$
\Matrix$}
\hbox to\wd8{\hfil$\ColNames$\kern\ColEntryShiftHoriz}\kern\ColEntryShiftVerti
\box8}}}%
\hbox to \textwidth{\qquad$\;f\qquad\hfil\vartheta_f\hfil\qquad\qquad\qquad\vartheta_{f\RELtraOP}\RELtraOP$\hfil}
}}
{Existential and inverse image for a surjective mapping}{FigExImageSurjective} 

\Caption{\vbox{\hbox to \textwidth{%
{\footnotesize%
\BoxBreadth=0pt%
\setbox7=\hbox{1}%
\ifdim\wd7>\BoxBreadth\BoxBreadth=\wd7\fi%
\setbox7=\hbox{2}%
\ifdim\wd7>\BoxBreadth\BoxBreadth=\wd7\fi%
\setbox7=\hbox{3}%
\ifdim\wd7>\BoxBreadth\BoxBreadth=\wd7\fi%
\def\RowNames{\vcenter{\offinterlineskip\baselineskip=\matrixskip%
\hbox to\BoxBreadth{\strut\hfil 1}\kern\interspacereduction%
\hbox to\BoxBreadth{\strut\hfil 2}\kern\interspacereduction%
\hbox to\BoxBreadth{\strut\hfil 3}}}%
\def\ColNames{\hbox{\rotatebox{90}{\strut a}\kern\interspacereduction%
\rotatebox{90}{\strut b}\kern\interspacereduction%
\rotatebox{90}{\strut c}\kern\interspacereduction%
\rotatebox{90}{\strut d}\kern\interspacereduction%
}}%
\def\Matrix{\spmatrix{%
\noalign{\kern-2pt}
 \n&\n&{\CoefTrue}&\n\cr
 \n&{\CoefTrue}&\n&\n\cr
 {\CoefTrue}&\n&\n&\n\cr
\noalign{\kern-2pt}}}%
\vbox{\setbox8=\hbox{$\RowNames\Matrix$}
\hbox to\wd8{\hfil$\ColNames$\kern\ColEntryShiftHoriz}\kern\ColEntryShiftVerti
\box8}}%
\kern-0.2cm\hfil
{\footnotesize%
\BoxBreadth=0pt%
\setbox7=\hbox{\{\}}%
\ifdim\wd7>\BoxBreadth\BoxBreadth=\wd7\fi%
\setbox7=\hbox{\{1\}}%
\ifdim\wd7>\BoxBreadth\BoxBreadth=\wd7\fi%
\setbox7=\hbox{\{2\}}%
\ifdim\wd7>\BoxBreadth\BoxBreadth=\wd7\fi%
\setbox7=\hbox{\{1,2\}}%
\ifdim\wd7>\BoxBreadth\BoxBreadth=\wd7\fi%
\setbox7=\hbox{\{3\}}%
\ifdim\wd7>\BoxBreadth\BoxBreadth=\wd7\fi%
\setbox7=\hbox{\{1,3\}}%
\ifdim\wd7>\BoxBreadth\BoxBreadth=\wd7\fi%
\setbox7=\hbox{\{2,3\}}%
\ifdim\wd7>\BoxBreadth\BoxBreadth=\wd7\fi%
\setbox7=\hbox{\{1,2,3\}}%
\ifdim\wd7>\BoxBreadth\BoxBreadth=\wd7\fi%
\def\RowNames{\vcenter{\offinterlineskip\baselineskip=\matrixskip%
\hbox to\BoxBreadth{\strut\hfil \{\}}\kern\interspacereduction%
\hbox to\BoxBreadth{\strut\hfil \{1\}}\kern\interspacereduction%
\hbox to\BoxBreadth{\strut\hfil \{2\}}\kern\interspacereduction%
\hbox to\BoxBreadth{\strut\hfil \{1,2\}}\kern\interspacereduction%
\hbox to\BoxBreadth{\strut\hfil \{3\}}\kern\interspacereduction%
\hbox to\BoxBreadth{\strut\hfil \{1,3\}}\kern\interspacereduction%
\hbox to\BoxBreadth{\strut\hfil \{2,3\}}\kern\interspacereduction%
\hbox to\BoxBreadth{\strut\hfil \{1,2,3\}}}}%
\def\ColNames{\hbox{\rotatebox{90}{\strut \{\}}\kern\interspacereduction%
\rotatebox{90}{\strut \{a\}}\kern\interspacereduction%
\rotatebox{90}{\strut \{b\}}\kern\interspacereduction%
\rotatebox{90}{\strut \{a,b\}}\kern\interspacereduction%
\rotatebox{90}{\strut \{c\}}\kern\interspacereduction%
\rotatebox{90}{\strut \{a,c\}}\kern\interspacereduction%
\rotatebox{90}{\strut \{b,c\}}\kern\interspacereduction%
\rotatebox{90}{\strut \{a,b,c\}}\kern\interspacereduction%
\rotatebox{90}{\strut \{d\}}\kern\interspacereduction%
\rotatebox{90}{\strut \{a,d\}}\kern\interspacereduction%
\rotatebox{90}{\strut \{b,d\}}\kern\interspacereduction%
\rotatebox{90}{\strut \{a,b,d\}}\kern\interspacereduction%
\rotatebox{90}{\strut \{c,d\}}\kern\interspacereduction%
\rotatebox{90}{\strut \{a,c,d\}}\kern\interspacereduction%
\rotatebox{90}{\strut \{b,c,d\}}\kern\interspacereduction%
\rotatebox{90}{\strut \{a,b,c,d\}}\kern\interspacereduction%
}}%
\def\Matrix{\spmatrix{%
\noalign{\kern-2pt}
 {\CoefTrue}&\n&\n&\n&\n&\n&\n&\n&\n&\n&\n&\n&\n&\n&\n&\n\cr
 \n&\n&\n&\n&{\CoefTrue}&\n&\n&\n&\n&\n&\n&\n&\n&\n&\n&\n\cr
 \n&\n&{\CoefTrue}&\n&\n&\n&\n&\n&\n&\n&\n&\n&\n&\n&\n&\n\cr
 \n&\n&\n&\n&\n&\n&{\CoefTrue}&\n&\n&\n&\n&\n&\n&\n&\n&\n\cr
 \n&{\CoefTrue}&\n&\n&\n&\n&\n&\n&\n&\n&\n&\n&\n&\n&\n&\n\cr
 \n&\n&\n&\n&\n&{\CoefTrue}&\n&\n&\n&\n&\n&\n&\n&\n&\n&\n\cr
 \n&\n&\n&{\CoefTrue}&\n&\n&\n&\n&\n&\n&\n&\n&\n&\n&\n&\n\cr
 \n&\n&\n&\n&\n&\n&\n&{\CoefTrue}&\n&\n&\n&\n&\n&\n&\n&\n\cr
\noalign{\kern-2pt}}}%
\vbox{\setbox8=\hbox{$\RowNames\Matrix$}
\hbox to\wd8{\hfil$\ColNames$\kern\ColEntryShiftHoriz}\kern\ColEntryShiftVerti
\box8}}
\hfil
{\footnotesize%
\BoxBreadth=0pt%
\setbox7=\hbox{\{\}}%
\ifdim\wd7>\BoxBreadth\BoxBreadth=\wd7\fi%
\setbox7=\hbox{\{1\}}%
\ifdim\wd7>\BoxBreadth\BoxBreadth=\wd7\fi%
\setbox7=\hbox{\{2\}}%
\ifdim\wd7>\BoxBreadth\BoxBreadth=\wd7\fi%
\setbox7=\hbox{\{1,2\}}%
\ifdim\wd7>\BoxBreadth\BoxBreadth=\wd7\fi%
\setbox7=\hbox{\{3\}}%
\ifdim\wd7>\BoxBreadth\BoxBreadth=\wd7\fi%
\setbox7=\hbox{\{1,3\}}%
\ifdim\wd7>\BoxBreadth\BoxBreadth=\wd7\fi%
\setbox7=\hbox{\{2,3\}}%
\ifdim\wd7>\BoxBreadth\BoxBreadth=\wd7\fi%
\setbox7=\hbox{\{1,2,3\}}%
\ifdim\wd7>\BoxBreadth\BoxBreadth=\wd7\fi%
\def\RowNames{\vcenter{\offinterlineskip\baselineskip=\matrixskip%
\hbox to\BoxBreadth{\strut\hfil \{\}}\kern\interspacereduction%
\hbox to\BoxBreadth{\strut\hfil \{1\}}\kern\interspacereduction%
\hbox to\BoxBreadth{\strut\hfil \{2\}}\kern\interspacereduction%
\hbox to\BoxBreadth{\strut\hfil \{1,2\}}\kern\interspacereduction%
\hbox to\BoxBreadth{\strut\hfil \{3\}}\kern\interspacereduction%
\hbox to\BoxBreadth{\strut\hfil \{1,3\}}\kern\interspacereduction%
\hbox to\BoxBreadth{\strut\hfil \{2,3\}}\kern\interspacereduction%
\hbox to\BoxBreadth{\strut\hfil \{1,2,3\}}}}%
\def\ColNames{\hbox{\rotatebox{90}{\strut \{\}}\kern\interspacereduction%
\rotatebox{90}{\strut \{a\}}\kern\interspacereduction%
\rotatebox{90}{\strut \{b\}}\kern\interspacereduction%
\rotatebox{90}{\strut \{a,b\}}\kern\interspacereduction%
\rotatebox{90}{\strut \{c\}}\kern\interspacereduction%
\rotatebox{90}{\strut \{a,c\}}\kern\interspacereduction%
\rotatebox{90}{\strut \{b,c\}}\kern\interspacereduction%
\rotatebox{90}{\strut \{a,b,c\}}\kern\interspacereduction%
\rotatebox{90}{\strut \{d\}}\kern\interspacereduction%
\rotatebox{90}{\strut \{a,d\}}\kern\interspacereduction%
\rotatebox{90}{\strut \{b,d\}}\kern\interspacereduction%
\rotatebox{90}{\strut \{a,b,d\}}\kern\interspacereduction%
\rotatebox{90}{\strut \{c,d\}}\kern\interspacereduction%
\rotatebox{90}{\strut \{a,c,d\}}\kern\interspacereduction%
\rotatebox{90}{\strut \{b,c,d\}}\kern\interspacereduction%
\rotatebox{90}{\strut \{a,b,c,d\}}\kern\interspacereduction%
}}%
\def\Matrix{\spmatrix{%
\noalign{\kern-2pt}
 {\CoefTrue}&\n&\n&\n&\n&\n&\n&\n&{\CoefTrue}&\n&\n&\n&\n&\n&\n&\n\cr
 \n&\n&\n&\n&{\CoefTrue}&\n&\n&\n&\n&\n&\n&\n&{\CoefTrue}&\n&\n&\n\cr
 \n&\n&{\CoefTrue}&\n&\n&\n&\n&\n&\n&\n&{\CoefTrue}&\n&\n&\n&\n&\n\cr
 \n&\n&\n&\n&\n&\n&{\CoefTrue}&\n&\n&\n&\n&\n&\n&\n&{\CoefTrue}&\n\cr
 \n&{\CoefTrue}&\n&\n&\n&\n&\n&\n&\n&{\CoefTrue}&\n&\n&\n&\n&\n&\n\cr
 \n&\n&\n&\n&\n&{\CoefTrue}&\n&\n&\n&\n&\n&\n&\n&{\CoefTrue}&\n&\n\cr
 \n&\n&\n&{\CoefTrue}&\n&\n&\n&\n&\n&\n&\n&{\CoefTrue}&\n&\n&\n&\n\cr
 \n&\n&\n&\n&\n&\n&\n&{\CoefTrue}&\n&\n&\n&\n&\n&\n&\n&{\CoefTrue}\cr
\noalign{\kern-2pt}}}%
\vbox{\setbox8=\hbox{$
\Matrix$}
\hbox to\wd8{\hfil$\ColNames$\kern\ColEntryShiftHoriz}\kern\ColEntryShiftVerti
\box8}}}%
\hbox to \textwidth{\qquad$\;f\qquad\hfil\vartheta_f\hfil\qquad\qquad\qquad\vartheta_{f\RELtraOP}\RELtraOP$\hfil}
}}
{Existential and inverse image for a non-surjective mapping}{FigExImageNonSurjective} 

\enunc{}{Proposition}{}{PropPowerRelProps} Consider a relation $\RELfromTO{R}{X}{Y}$ as well as the corresponding power relator\index{power relator} $\RELfromTO{\PowerRel_R:=\big(\varepsilon\backslash (R \RELcompOP \varepsilon') \big)   \RELandOP
       \big((\varepsilon \RELtraOP \RELcompOP R)  /   {\varepsilon'}\RELtraOP \big)
       =\RELneg{\varepsilon\RELtraOP\RELcompOP\RELneg{R\RELcompOP \varepsilon'}}
\RELandOP
\RELneg{\RELneg{\varepsilon\RELtraOP\RELcompOP R}\RELcompOP\varepsilon'}}{\PowTWO{X}}{\PowTWO{Y}}$. Then

\begin{enumerate}[i)]
\item\ \hbox to2.3cm{$R$ univalent\hfil}\qquad$\Longrightarrow\qquad\PowerRel_R$ univalent
\item\ \hbox to2.3cm{$R$ surjective\hfil}\qquad$\Longrightarrow\qquad\PowerRel_R$ surjective
\item\ \hbox to2.3cm{$R$ total\hfil}\qquad$\Longrightarrow\qquad\PowerRel_R$ total
\item\ \hbox to2.3cm{$R$ injective\hfil}\qquad$\Longrightarrow\qquad\PowerRel_R$ injective
\item\ \hbox to2.3cm{$f$ mapping\hfil}\qquad$\Longrightarrow\qquad\PowerRel_f=\ExistIm_f$ 
\end{enumerate}

\Proof The proofs of (i,\dots,iv) follow all the same scheme using Prop.~19.11 of \cite{RelaMath2010}.

\smallskip
$\PowerRel_R\RELtraOP\RELcompOP\PowerRel_R
=
\PowerRel_{R\RELtraOP}\RELcompOP\PowerRel_R
=
\PowerRel_{R\RELtraOP\,\RELcompOP\, R}
=
\RELneg{{\varepsilon'}\RELtraOP\RELcompOP\RELneg{R\RELtraOP\RELcompOP R\RELcompOP \varepsilon'}}
\RELandOP
\RELneg{\RELneg{{\varepsilon'}\RELtraOP\RELcompOP R\RELtraOP\RELcompOP R}\RELcompOP\varepsilon'}
$

$\RELenthOP
\RELneg{{\varepsilon'}\RELtraOP\RELcompOP\RELneg{\RELide\RELcompOP \varepsilon'}}
\RELandOP
\RELneg{\RELneg{{\varepsilon'}\RELtraOP\RELcompOP \RELide}\RELcompOP\varepsilon'}
=
\RELneg{{\varepsilon'}\RELtraOP\RELcompOP\RELneg{\varepsilon'}}
\RELandOP
\RELneg{\RELneg{{\varepsilon'}\RELtraOP}\RELcompOP\varepsilon'}
=
\Omega'\RELandOP{\Omega'}\RELtraOP
=
\RELide
$

\smallskip
\noindent
v) $\PowerRel_f
=
\RELneg{\varepsilon\RELtraOP\RELcompOP\RELneg{f\RELcompOP \varepsilon'}}
\RELandOP
\RELneg{\RELneg{\varepsilon\RELtraOP\RELcompOP f}\RELcompOP\varepsilon'}
=
\RELneg{\varepsilon\RELtraOP\RELcompOP f\RELcompOP \RELneg{\varepsilon'}}
\RELandOP
\RELneg{\RELneg{\varepsilon\RELtraOP\RELcompOP f}\RELcompOP\varepsilon'}
=
\syqq{f\RELtraOP\RELcompOP\varepsilon}{\varepsilon'}
=
\ExistIm_f
$
\Bewende

\noindent
The construct $\PowerRel$ looks quite similar to a symmetric quotient, but it is not!


\chapter{Relations in varying representations}\label{SectRelAsPoint}
\EnuncNo=0
\CaptionNo=0


\noindent
When dealing with relations, we have --- in principle --- three incarnations of the same idea. A relation between sets $X,Y$ may, namely, be represented 

\begin{itemize}
\item as $\RELfromTO{R}{X}{Y}$ corresponding to a possibly non-square \Boolean\ matrix,
\item as $\RELfromTO{\mathfrak{r}}{X\times Y}{\I1}$ corresponding to a \Boolean\ vector characterizing a subset of pairs,
\item as $\RELfromTO{r}{\PowTWO{X\times Y}}{\I1}$ corresponding to a point in the powerset of the pair set.
\end{itemize}

\noindent
Their interrelationship using projections $\RELfromTO{\pi}{X\times Y}{X}$, resp.~$\RELfromTO{\rho}{X\times Y}{Y}$, and the membership relation 
$\RELfromTO{\varepsilon_\times}{X\times Y}{\PowTWO{X\times Y}}$ starting in the product is as follows:

\smallskip
$\mathfrak{r}=\varepsilon_\times\RELcompOP r
\qquad\!\!\!
r=\syqq{\varepsilon_\times}{\mathfrak{r}}
\qquad\!\!\!
R=\VectToRel{\mathfrak{r}}=\pi\RELtraOP\RELcompOP(\mathfrak{r}\RELcompOP\RELtop_{\I1,Y}\RELandOP\rho)
\qquad\!\!\!
\mathfrak{r}=\RelToVect{R}=(\pi\RELcompOP R\RELandOP \rho)\RELcompOP\RELtop_{Y,\I1}$

\smallskip
\noindent
The transition from $R$ to $\mathfrak{r}$ is a {\bf vectorization}, known also at other occasions in algebra. While it may be considered an easy construction, one should think of a $5000\times1000$-relation and its vectorization that may be much harder to handle in practice.

\enunc{}{Proposition}{}{PropRelVec} \quad$R=\VectToRel{\RelToVect{R}}$ \quad and \quad $\mathfrak{r}=\RelToVect{\VectToRel{\mathfrak{r}}}$\index{vectorization}

\Proof $R
=
\pi\RELtraOP\RELcompOP\rho\RELcompOP\RELide_Y\RELandOP R
$

$\RELenthOP
(\pi\RELtraOP\RELcompOP\rho\RELandOP R\RELcompOP\RELide_Y)\RELcompOP(\RELide_Y\RELandOP(\pi\RELtraOP\RELcompOP\rho)\RELtraOP\RELcompOP R)
$\quad Dedekind

$=
\pi\RELtraOP\RELcompOP(\rho\RELandOP \pi\RELcompOP R)\RELcompOP(\RELide_Y\RELandOP\RELtop_{Y,X}\RELcompOP R)
$

$=
\pi\RELtraOP\RELcompOP\lbrack(\rho\RELandOP \pi\RELcompOP R)\RELcompOP\RELide_Y\RELandOP(\rho\RELandOP \pi\RELcompOP R)\RELcompOP\RELtop_{Y,X}\RELcompOP R)\rbrack
$\quad since $(\rho\RELandOP \pi\RELcompOP R)$ is univalent

$=
\pi\RELtraOP\RELcompOP\lbrack\rho\RELandOP \pi\RELcompOP R\RELandOP(\rho\RELandOP \pi\RELcompOP R)\RELcompOP\RELtop_{Y,X}\RELcompOP R)\rbrack
$\quad 

$\RELenthOP
\pi\RELtraOP\RELcompOP\lbrack\rho\RELandOP(\rho\RELandOP \pi\RELcompOP R)\RELcompOP\RELtop_{Y,Y})\rbrack
=
\pi\RELtraOP\RELcompOP\lbrack(\pi\RELcompOP R\RELandOP \rho)\RELcompOP\RELtop_{Y,\I1}\RELcompOP\RELtop_{\I1,Y})\RELandOP\rho\rbrack
=\VectToRel{\RelToVect{R}}
$\quad 

$\RELenthOP
\pi\RELtraOP\RELcompOP((\pi\RELcompOP R\RELandOP \rho)\RELandOP\rho\RELcompOP\RELtop_{Y,Y})\RELcompOP(\RELtop_{Y,Y}\RELandOP(\pi\RELcompOP R\RELandOP \rho)\RELtraOP\RELcompOP\rho)
$\quad Dedekind rule

$\RELenthOP
\pi\RELtraOP\RELcompOP\pi\RELcompOP R\RELcompOP\rho\RELtraOP\RELcompOP\rho
=
R
$\quad

\bigskip
$\mathfrak{r}
=
(\RELtop_{X\times Y,\I1}\RELandOP\mathfrak{r})\RELcompOP\RELtop_{\I1,\I1}
=
(\rho\RELcompOP\RELtop_{Y,\I1}\RELandOP\mathfrak{r})\RELcompOP\RELtop_{\I1,\I1}
$

$\RELenthOP
(\rho\RELandOP\mathfrak{r}\RELcompOP\RELtop_{\I1,Y})\RELcompOP(\RELtop_{Y,\I1}\RELandOP\rho\RELtraOP\RELcompOP\mathfrak{r})\RELcompOP\RELtop_{\I1,\I1}
$

$\RELenthOP
(\rho\RELandOP\mathfrak{r}\RELcompOP\RELtop_{\I1,Y})\RELcompOP\RELtop_{Y,\I1}
$

$=
((\mathfrak{r}\RELcompOP\RELtop_{\I1,Y}\RELandOP\rho)\RELandOP\rho)\RELcompOP\RELtop_{Y,\I1}
$

$\RELenthOP
\big\lbrack\pi\RELcompOP\pi\RELtraOP\RELcompOP(\mathfrak{r}\RELcompOP\RELtop_{\I1,Y}\RELandOP\rho)\RELandOP\rho\big\rbrack\RELcompOP\RELtop_{Y,\I1}
=
\RelToVect{\VectToRel{\mathfrak{r}}}
$

$\RELenthOP
\big\lbrack\pi\RELcompOP\pi\RELtraOP\RELandOP\rho\RELcompOP(\mathfrak{r}\RELcompOP\RELtop_{\I1,Y}\RELandOP\rho)\RELtraOP\big\rbrack\RELcompOP
\big\lbrack(\mathfrak{r}\RELcompOP\RELtop_{\I1,Y}\RELandOP\rho)\RELandOP\pi\RELcompOP\pi\RELtraOP\RELcompOP\rho\big\rbrack\RELcompOP\RELtop_{Y,\I1}
$

$\RELenthOP
\big\lbrack\pi\RELcompOP\pi\RELtraOP\RELandOP\rho\RELcompOP\rho\RELtraOP\big\rbrack\RELcompOP
\mathfrak{r}\RELcompOP\RELtop_{\I1,Y}\RELcompOP\RELtop_{Y,\I1}
$

$=
\mathfrak{r}\RELcompOP\RELtop_{\I1,Y}\RELcompOP\RELtop_{Y,\I1}
=
\mathfrak{r}\RELcompOP\RELtop_{\I1,\I1}
=
\mathfrak{r}
$
\Bewende

\noindent
It should be made clear that the relations with standard abbreviation $\varepsilon_\times,\pi,\rho$ do not fall from heaven. Rather, they are defined generically as characterizations {\it up to isomorphism} using the techniques of domain construction developed in
\cite{RelaMath2010}. They allow to formulate via a language called {\sc TituRel}\index{Titurel@{\sc TituRel}}

\smallskip
$\pi\approx {\tt Pi\ \ \hbox{$X$}\ \hbox{$Y$}} 
\qquad\quad
\rho\approx {\tt Rho \ \hbox{$X$}\ \hbox{$Y$}}$
\qquad\qquad given that $X=\hbox{\tt src}(R)$ and $Y=\hbox{\tt tgt}(R)$

$\varepsilon_\times\approx {\tt ElemIn \ \hbox{\tt (DirPro $X\; Y$)}}$

\bigskip
\noindent
Following the idea of the threefold ways of denoting, the identity $\RELfromTO{\RELide}{X}{X}$ gives rise to the vector $\RELfromTO{{\tt vec}(\RELide)=(\pi\RELandOP\rho)\RELcompOP\RELtop}{X\times X}{\I1}$ and finally to the element
$\RELfromTO{\liftedIdent=\syqq{\varepsilon_\times}{(\pi\RELandOP\rho)\RELcompOP\RELtop}}{\PowTWO{X\times X}}{\I1}$ in the powerset of all pairs.

\enunc{}{Proposition}{}{PropIdentityPointfree} Consider a set $X$ together with the membership
$\RELfromTO{\varepsilon_\times}{X\times X}{\PowTWO{X\times X}}$ on the direct product of the set with itself and define the point

\smallskip
$\liftedIdent:=
\syqq{\varepsilon_\times}{(\pi\RELandOP\rho)\RELcompOP\RELtop}
=
\syqq{\varepsilon_\times}{\RelToVect{\RELide}}$.

\smallskip
\noindent
Then $\VectToRel{\varepsilon_\times\RELcompOP \liftedIdent}=\RELide$.

\Proof $\VectToRel{\varepsilon_\times\RELcompOP \liftedIdent}
=
\VectToRel{\varepsilon_\times\RELcompOP\syqq{\varepsilon_\times}{(\pi\RELandOP\rho)\RELcompOP\RELtop}}
=
\VectToRel{(\pi\RELandOP\rho)\RELcompOP\RELtop}$

$=
\pi\RELtraOP\RELcompOP\big\lbrack(\pi\RELandOP\rho)\RELcompOP\RELtop\RELandOP\rho\big\rbrack
$\quad expanded

$=
\pi\RELtraOP\RELcompOP(\pi\RELandOP\rho)
$\quad see below

$=
\RELide\RELandOP\pi\RELtraOP\RELcompOP\rho
=
\RELide\RELandOP\RELtop
=
\RELide
$

\smallskip
\noindent
Now the postponed transition is justified with a sequence of containments implying equality:

\smallskip
$(\pi\RELandOP\rho)\RELcompOP\RELtop\RELandOP\rho
\RELenthOP
\lbrack(\pi\RELandOP\rho)\RELandOP \rho\RELcompOP\RELtop\rbrack \RELcompOP
\lbrack\RELtop\RELandOP (\pi\RELandOP\rho)\RELtraOP\RELcompOP\rho \rbrack
=
(\pi\RELandOP\rho) \RELcompOP
(\pi \RELtraOP\RELandOP\rho \RELtraOP)\RELcompOP\rho
$\qquad $\rho$ is total

$=
(\pi\RELandOP\rho) \RELcompOP
(\pi \RELtraOP\RELcompOP\rho\RELandOP\RELide)
=
(\pi\RELandOP\rho) \RELcompOP
(\RELtop\RELandOP\RELide)
=
\pi\RELandOP\rho
\RELenthOP
(\pi\RELandOP\rho)\RELcompOP\RELtop\RELandOP\rho$
\Bewende

\noindent
Much in the same way as later for $\liftedMeet,\liftedJoin$, we show here that it is possible to express the least and the greatest relations as points

\smallskip
$\RELfromTO{\liftedBOT}{\PowTWO{X\times Y}}{\I1}$,
\qquad\qquad$\RELfromTO{\liftedTOP}{\PowTWO{X\times Y}}{\I1}$.

\enunc{}{Proposition}{}{PropBOTTOPPointfree} Consider sets $X,Y$ together with the membership
$\RELfromTO{\varepsilon_\times}{X\times Y}{\PowTWO{X\times Y}}$ on the direct product of the sets and define the point

\smallskip
$\liftedBOT:=
\syqq{\varepsilon_\times}{\RelToVect{\RELbot}}=
\syqq{\varepsilon_\times}{\RELbot}
$

$
\liftedTOP:=
\syqq{\varepsilon_\times}{\RelToVect{\RELtop}}=
\syqq{\varepsilon_\times}{\RELtop}
$.

\smallskip
\noindent
Then $\VectToRel{\varepsilon_\times\RELcompOP \liftedBOT}=\RELbot$
and $\VectToRel{\varepsilon_\times\RELcompOP \liftedTOP}=\RELtop$.

\Proof $\VectToRel{\varepsilon_\times\RELcompOP \liftedBOT}
=
\VectToRel{\varepsilon_\times\RELcompOP \syqq{\varepsilon_\times}{\RelToVect{\RELbot}}}
=
\VectToRel{\RelToVect{\RELbot}}
=
\RELbot
$
\Bewende

\noindent
The processes of transposition and negation

\smallskip
$\RELfromTO{\liftedTransp}{\PowTWO{X\times Y}}{\PowTWO{Y\times X}}
\qquad
\RELfromTO{N}{\PowTWO{X\times Y}}{\PowTWO{X\times Y}}$, 

\smallskip
\noindent
may also be conceived as bijective mappings, as well as the process of composition

\smallskip
$\RELfromTO{\liftedCompo}{\PowTWO{X\times Y}\times\PowTWO{Y\times Z}}{\PowTWO{X\times Z}}$, 

\smallskip
\noindent
as a binary mapping, i.e., all three in a pointfree fashion. While we omit discussing $\liftedCompo$, we refer for $N$ to Fig.~\FigSumPowToPowProdOnX. Here, we restrict to studying formally the interchange of components of a pair, which obviously determines a bijective mapping 

\smallskip
$\RELfromTO{\liftedTransp}{\PowTWO{X\times Y}}{\PowTWO{Y\times X}}$,

\smallskip
\noindent
satisfying certain rules.

\Caption{\includegraphics[scale=0.5]{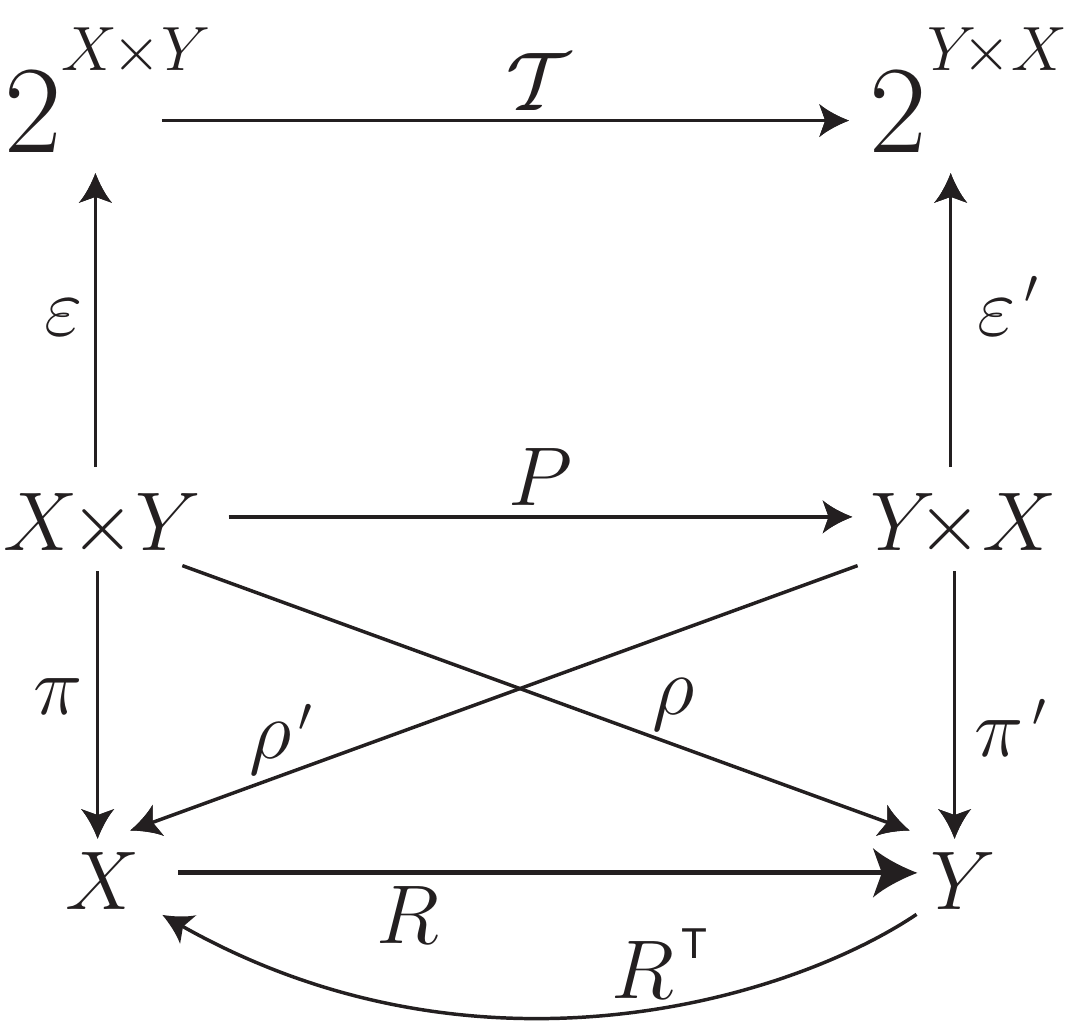}}
{Illustrating transposition as an operation on relations conceived as points}{FigTranspInPoints}

\enunc{}{Proposition}{}{PropTranspositionPointfree} Consider two sets $X,Y$ together with the memberships
$\RELfromTO{\varepsilon} {X\times Y}{\PowTWO{X\times Y}}$ and $ \RELfromTO{\varepsilon'}{Y\times X}{\PowTWO{Y\times X}}$ of both their direct products and define

\smallskip
$P:=\pi\RELcompOP{\rho'}\RELtraOP\RELandOP\rho\RELcompOP{\pi'}\RELtraOP
\qquad
\liftedTransp :=\syqq{P\RELtraOP\RELcompOP\varepsilon}{\varepsilon'}$.

\smallskip
\noindent
Then 

\begin{enumerate}[i)]
\item $P$ is a bijective mapping to be interpreted as sending $(x,y)$ to $(y,x)$.
\item $\liftedTransp$ is a bijective mapping resembling transposition.
\item $P\RELcompOP\rho'=\pi$\quad  \quad $P\RELcompOP\pi'=\rho$\quad  \quad $P \RELtraOP\RELcompOP\pi=\rho'$\quad  \quad $P \RELtraOP\RELcompOP\rho=\pi'$
\item $\liftedTransp \RELcompOP{\varepsilon'}\RELtraOP=\varepsilon\RELtraOP\RELcompOP P$,\quad i.e., $P$ and $\liftedTransp$ bisimulate one another via the membership relations.
\item $\VectToRel{P \RELtraOP\RELcompOP v}=
\big\lbrack\VectToRel{v}\big\rbrack\RELtraOP$
\item $\RelToVect{R \RELtraOP}=P\RELcompOP \RelToVect {R}$
\item $\VectToRel{\RELneg{v}}=\RELneg{\VectToRel{v}}$
\item $\RelToVect{\RELneg{R}}=\RELneg{\RelToVect{R}}$
\end{enumerate}

\Proof i) $P\RELtraOP\RELcompOP P=(\rho'\RELcompOP\pi\RELtraOP\RELandOP\pi'\RELcompOP\rho\RELtraOP) \RELcompOP(\pi\RELcompOP{\rho'}\RELtraOP\RELandOP\rho\RELcompOP{\pi'}\RELtraOP)
\RELenthOP
\rho'\ \RELcompOP\pi\RELtraOP \RELcompOP\pi \RELcompOP{\rho'}\RELtraOP
\RELandOP
\pi'\ \RELcompOP \rho\RELtraOP \RELcompOP \rho \RELcompOP{\pi'}\RELtraOP
=
\rho'\ \RELcompOP{\rho'}\RELtraOP
\RELandOP
\pi'\ \RELcompOP{\pi'}\RELtraOP
=
\RELide
$

\smallskip
\noindent
This shows univalency; analogously for injectivity. Therefore $P$ multiplies distributively over conjunction and we may proceed with

\smallskip
$P\RELcompOP P \RELtraOP
=
(\pi\RELcompOP{\rho'}\RELtraOP\RELandOP\rho\RELcompOP{\pi'}\RELtraOP)\RELcompOP
(\rho'\RELcompOP\pi\RELtraOP\RELandOP\pi'\RELcompOP\rho\RELtraOP)
=
(\pi\RELcompOP{\rho'}\RELtraOP\RELandOP\rho\RELcompOP{\pi'}\RELtraOP)\RELcompOP
\rho'\RELcompOP\pi\RELtraOP
\RELandOP
(\pi\RELcompOP{\rho'}\RELtraOP\RELandOP\rho\RELcompOP{\pi'}\RELtraOP) \RELcompOP\pi'\RELcompOP\rho\RELtraOP
$

$
=
(\pi\RELandOP\rho\RELcompOP{\pi'}\RELtraOP\RELcompOP
\rho')\RELcompOP\pi\RELtraOP
\RELandOP
(\pi\RELcompOP{\rho'}\RELtraOP\RELcompOP\pi'\RELandOP\rho) \RELcompOP\rho\RELtraOP
$

$
=
(\pi\RELandOP \RELtop)\RELcompOP\pi\RELtraOP
\RELandOP
(\RELtop\RELandOP\rho) \RELcompOP\rho\RELtraOP
=
\pi\RELcompOP\pi\RELtraOP
\RELandOP
\rho\RELcompOP\rho\RELtraOP
=
\RELide
$,

\smallskip
\noindent
giving totality, and in analogy also surjectivity.

\bigskip
\noindent
ii) $\liftedTransp $ is univalent, since $\liftedTransp \RELtraOP\RELcompOP\liftedTransp =
\syqq{\varepsilon'}{P\RELtraOP\RELcompOP\varepsilon}\RELcompOP\syqq{P\RELtraOP\RELcompOP\varepsilon}{\varepsilon'}
=
\syqq{\varepsilon'}{\varepsilon'}
=
\RELide$. It is total because $\liftedTransp \RELtraOP=\syqq{\varepsilon'}{P\RELtraOP\RELcompOP\varepsilon}$ is surjective by definition of the membership $\varepsilon'$.

\bigskip
\noindent
iii) is trivial.

\bigskip
\noindent
iv) $\liftedTransp \RELcompOP{\varepsilon'}\RELtraOP
=
\lbrack\varepsilon'\RELcompOP\liftedTransp  \RELtraOP\rbrack \RELtraOP
=
\lbrack\varepsilon'\RELcompOP\syqq {\varepsilon'}{P\RELtraOP\RELcompOP\varepsilon}\rbrack \RELtraOP
=
\lbrack P\RELtraOP\RELcompOP\varepsilon\rbrack \RELtraOP
=
\varepsilon\RELtraOP\RELcompOP P
$

\bigskip
\noindent
v) $\VectToRel{P \RELtraOP\RELcompOP v}
=
{\pi'}\RELtraOP\RELcompOP(P \RELtraOP\RELcompOP v\RELcompOP\RELtop\RELandOP\rho')
=
{\pi'}\RELtraOP\RELcompOP(P \RELtraOP\RELcompOP v\RELcompOP\RELtop\RELandOP P \RELtraOP\RELcompOP\pi)
=
{\pi'}\RELtraOP\RELcompOP P \RELtraOP\RELcompOP (v\RELcompOP\RELtop\RELandOP\pi)
=
\rho\RELtraOP\RELcompOP (v\RELcompOP\RELtop\RELandOP\pi)
$

$
=
\big\lbrack(v\RELcompOP\RELtop\RELandOP\pi)\RELtraOP\RELcompOP\rho \big\rbrack\RELtraOP
=
\big\lbrack(\RELtop\RELcompOP v\RELtraOP\RELandOP\pi\RELtraOP)\RELcompOP\rho \big\rbrack\RELtraOP
=
\big\lbrack\pi\RELtraOP(v\RELcompOP\RELtop\RELandOP\rho )\big\rbrack\RELtraOP
=
\big\lbrack\VectToRel{v}\big\rbrack\RELtraOP$

\bigskip
\noindent
vi) $\RelToVect {R\RELtraOP}
=
(\pi'\RELcompOP R \RELtraOP\RELandOP\rho')\RELcompOP\RELtop
=
(P \RELtraOP\RELcompOP\rho\RELcompOP R \RELtraOP\RELandOP P \RELtraOP\RELcompOP\pi)\RELcompOP\RELtop
=
P \RELtraOP\RELcompOP(\rho\RELcompOP R \RELtraOP\RELandOP\pi)\RELcompOP\RELtop
$

$
\RELenthOP
P \RELtraOP\RELcompOP (\rho\RELandOP\pi \RELcompOP R) \RELcompOP (R \RELtraOP\RELandOP\rho \RELtraOP\RELcompOP \pi)\RELcompOP\RELtop
=
P \RELtraOP\RELcompOP (\rho\RELandOP\pi \RELcompOP R) \RELcompOP (R \RELtraOP\RELandOP\RELtop)\RELcompOP\RELtop
\RELenthOP 
P \RELtraOP\RELcompOP (\rho\RELandOP\pi \RELcompOP R) \RELcompOP\RELtop
=
P \RELtraOP\RELcompOP (\pi \RELcompOP R \RELandOP\rho) \RELcompOP\RELtop
$

$
\RELenthOP
P \RELtraOP\RELcompOP (\pi \RELandOP \rho\RELcompOP R \RELtraOP) \RELcompOP
(R \RELandOP \pi \RELtraOP\rho) \RELcompOP\RELtop
\RELenthOP
P \RELtraOP\RELcompOP (\pi \RELandOP \rho\RELcompOP R \RELtraOP) \RELcompOP\RELtop
$\quad implying equality everywhere in between

$
=
P \RELtraOP\RELcompOP \RelToVect {R}
$

\bigskip
\noindent
vii) 
$\RELtop=\pi\RELtraOP\RELcompOP\rho
=
\pi\RELtraOP\RELcompOP(\rho\RELandOP \mathfrak{r}\RELcompOP\RELtop)\RELorOP\pi\RELtraOP\RELcompOP(\rho\RELandOP\RELneg{\mathfrak{r}\RELcompOP\RELtop})
\!\quad\Longrightarrow\quad\!
\RELneg{\VectToRel{\mathfrak{r}}}=\RELneg{\pi\RELtraOP\RELcompOP(\rho\RELandOP \mathfrak{r}\RELcompOP\RELtop)}
\RELenthOP
\pi\RELtraOP\RELcompOP(\rho\RELandOP\RELneg{\mathfrak{r}\RELcompOP\RELtop})=\VectToRel{\RELneg{\mathfrak{r}}}
$

$\pi\RELcompOP\pi\RELtraOP\RELcompOP(\rho\RELandOP\mathfrak{r}\RELcompOP\RELtop)\RELandOP\rho
\RELenthOP
(\pi\RELcompOP\pi\RELtraOP\RELandOP\rho\RELcompOP(\rho\RELandOP\mathfrak{r}\RELcompOP\RELtop)\RELtraOP)\RELcompOP(\rho\RELandOP\mathfrak{r}\RELcompOP\RELtop\RELandOP\pi\RELcompOP\pi\RELtraOP\RELcompOP\rho)
\RELenthOP
\RELide\RELcompOP\mathfrak{r}\RELcompOP\RELtop
=\mathfrak{r}\RELcompOP\RELtop
$\quad Dedekind rule

$\quad\Longrightarrow\quad
\pi\RELcompOP\pi\RELtraOP\RELcompOP(\rho\RELandOP\mathfrak{r}\RELcompOP\RELtop)
\RELenthOP
\RELneg{\rho}\RELorOP\mathfrak{r}\RELcompOP\RELtop
\quad\iff\quad
\VectToRel{\RELneg{\mathfrak{r}}}=\pi\RELtraOP\RELcompOP(\rho\RELandOP\RELneg{\mathfrak{r}\RELcompOP\RELtop})
\RELenthOP
\RELneg{\pi\RELtraOP\RELcompOP(\rho\RELandOP \mathfrak{r}\RELcompOP\RELtop)}
=\RELneg{\VectToRel{\mathfrak{r}}}
$

\smallskip

\bigskip
\noindent
viii) $\RelToVect{\RELneg{R}}
=
\RelToVect{\RELneg{\VectToRel{\RelToVect{R}}}}
$\quad Prop.~\PropRelVec

$
=
\RelToVect{\VectToRel{\RELneg{\RelToVect{R}}}}
$\quad according to (vii)

$
=
\RELneg{\RelToVect{R}}
$\quad Prop.~\PropRelVec
\Bewende


\chapter{Some categorical considerations\label{ChapCateg}}
\EnuncNo=0
\CaptionNo=0


\noindent
We here give relation-algebraic proofs of certain results we will use afterwards. Everything is fully based on the generic constructions of a direct sum, or product, etc. If any two heterogeneous relations $\pi,\rho$ with common \Source\ are given, they 
are said to form a {\bf direct product}\index{direct!product}\index{product, direct} if

\smallskip
$\pi\RELtraOP\RELcompOP\pi = \RELide,\quad\rho \RELtraOP\RELcompOP\rho = \RELide,\quad\pi\RELcompOP\pi \RELtraOP
\RELandOP  \rho\RELcompOP \rho  \RELtraOP  = \RELide,\quad\pi \RELtraOP\RELcompOP\rho 
= \RELtop.$

\smallskip
\noindent
Thus, the relations $\pi, \rho$ are mappings, 
usually called {\bf projections}\index{projection}. In a similar way, any two heterogeneous relations $\iota,\kappa$ with common \Target\  
are said to form the left, respectively right,
\textbf{injection\index{injection}} of a \textbf{direct sum\index{direct!sum}\index{sum, direct}} if

\smallskip
$\iota\RELcompOP\iota\RELtraOP
 = \RELide,\quad\kappa\RELcompOP\kappa\RELtraOP = \RELide,\quad\iota\RELtraOP\RELcompOP\iota
\RELorOP \kappa\RELtraOP\RELcompOP\kappa= \RELide,\quad\iota\RELcompOP\kappa\RELtraOP 
= \RELbot .$

\enunc{}{Definition}{}{DefKronForkJoin} Given any two direct products by projections

\smallskip
$\RELfromTO{\pi}{X\times Y}{X},\quad 
\RELfromTO{\rho}{X\times Y}{Y},\qquad
\RELfromTO{\pi'}{U\times V}{U},\quad
\RELfromTO{\rho'}{U\times V}{V}$,

\noindent
we define as binary operations on relations
\begin{enumerate}[i)]
\item $\RELfromTO{\Kronecker{A}{B}:=\pi\RELcompOP A\RELcompOP{\pi'}\RELtraOP\RELandOP\rho\,\RELcompOP B\RELcompOP{\rho'}\RELtraOP}{\quad X\times Y}{U\times V}$,\quad the {\bf Kronecker product}\index{Kronecker product},
\item $\RELfromTO{\StrictFork{C}{D}:=C\RELcompOP\pi\RELtraOP\RELandOP D\RELcompOP\rho\RELtraOP}{\quad Z}{X\times Y}$,\quad the {\bf fork-operator}\index{fork operator},
\item $\RELfromTO{\StrictJoin{E}{F}:=\pi\RELcompOP E\RELandOP\rho\,\RELcompOP F}{\quad X\times Y}{W}$,\quad the {\bf join-operator}\index{join operator}.
\Bewende
\end{enumerate}

\kern-\baselineskip

\Caption{\includegraphics[scale=0.4]{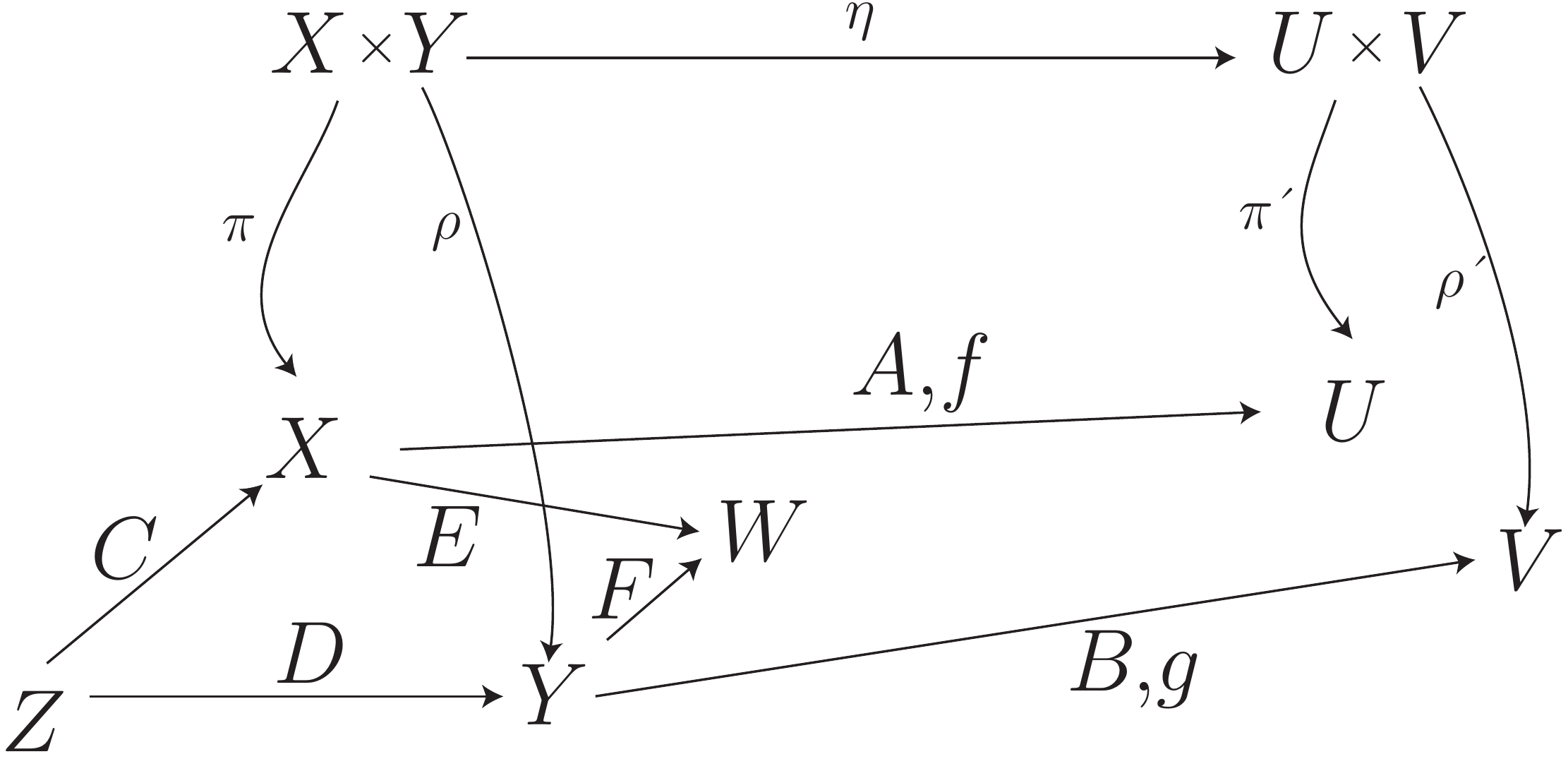}}
{Kronecker, fork-, and join-operators applied to relations and mappings}{FigKroneckerOfMaps}

\noindent
Obvious identities are $\Kronecker{A}{B}\RELtraOP=\Kronecker{A\RELtraOP}{B\RELtraOP}$ and $\StrictFork{C}{D}\RELtraOP=\StrictJoin{C\RELtraOP}{D\RELtraOP}$.
The next results are presented in some detail because they are very close to the \lq unsharpness\rq\index{unsharpness} situation where model problems arise: There exist relational formulae that hold in the classical interpretation, but cannot be derived in the axiomatization followed here; see \cite[Sect.~7.2]{RelaMath2010}. 

\enunc{}{Proposition}{}{PropPiKrhoEquGeneral} 
Let be given any two direct products by projections 

\smallskip
$\RELfromTO{\pi}{X\times Y}{X},\quad 
\RELfromTO{\rho}{X\times Y}{Y},\qquad
\RELfromTO{\pi'}{U\times V}{U},\quad
\RELfromTO{\rho'}{U\times V}{V}$ 

\smallskip
\noindent
together with relations $\RELfromTO{A}{X}{U}$ 
and $\RELfromTO{B}{Y}{V}$. Then

\begin{enumerate}[i)]
\item $\Kronecker{A}{B}\RELcompOP\,\pi'
=
\pi\RELcompOP A
\RELandOP 
\rho\RELcompOP B\RELcompOP\RELtop
\quad\quad\;
\Kronecker{A}{B}\RELcompOP\,\rho'
=
\pi\RELcompOP A\RELcompOP\RELtop
\RELandOP 
\rho\RELcompOP B
$\index{Kronecker operator}

\item $\Kronecker{A}{B}\RELcompOP\,\pi'
=
\pi\RELcompOP A$\quad in case $B$ is total

\item $\Kronecker{A}{B}\RELcompOP\,\rho'
=
\rho\RELcompOP B$\quad in case $A$ is total

\item $\Kronecker{A}{\RELide}\RELcompOP\,\pi'=\StrictJoin{A}{\RELtop}=\pi\RELcompOP A\quad\quad\;\Kronecker{\RELide}{B}\RELcompOP\,\rho'=\StrictJoin{\RELtop}{B}=\rho\RELcompOP B$.

\item[]$\StrictFork{A}{\RELide}\RELcompOP\,\pi'=A\quad\quad\;\StrictFork{\RELide}{B}\RELcompOP\,\rho'=B$.

\item If $A,B$ are both univalent, then so is $\Kronecker{A}{B}$.

\item If $A,B$ are both mappings, then so is $\Kronecker{A}{B}$.

\end{enumerate}

\Proof i) $\Kronecker{A}{B}\RELcompOP\,\pi'
=
(\pi\RELcompOP A\RELcompOP{\pi'}\RELtraOP
\RELandOP 
\rho\RELcompOP B\RELcompOP{\rho'}\RELtraOP)\RELcompOP\pi'$\quad by definition

$
=
\pi\RELcompOP A
\RELandOP 
\rho\RELcompOP B\RELcompOP{\rho'}\RELtraOP\RELcompOP\pi'
$\quad since $\pi'$ is univalent, \cite{RelaMath2010}, Prop.~5.4

$=
\pi\RELcompOP A
\RELandOP 
\rho\RELcompOP B\RELcompOP\RELtop
$\quad property of the direct product $\pi',\rho'$

\smallskip
\noindent
The second formula is derived analogously.

\bigskip
\noindent
ii) and (iii) are trivial consequences.

\bigskip
\noindent
iv) $\Kronecker{\RELide}{B}\RELcompOP\,\rho'
=
(\pi\RELcompOP{\pi'}\RELtraOP\RELandOP\rho\RELcompOP B\RELcompOP{\rho'}\RELtraOP)\RELcompOP\rho'
=
\pi\RELcompOP{\pi'}\RELtraOP\RELcompOP\rho'\RELandOP\rho\RELcompOP B
=
\pi\RELcompOP\RELtop\RELandOP\rho\RELcompOP B
=
\StrictJoin{\RELtop}{B}
=
\RELtop\RELandOP\rho\RELcompOP B
$

\bigskip
\noindent
v) $\Kronecker{A}{B}\RELtraOP\RELcompOP\Kronecker{A}{B}
=
(\pi'\RELcompOP A\RELtraOP\RELcompOP\pi\RELtraOP
\RELandOP 
\rho'\RELcompOP B\RELtraOP\RELcompOP\rho\RELtraOP)
\RELcompOP
(\pi\RELcompOP A\RELcompOP{\pi'}\RELtraOP
\RELandOP 
\rho\RELcompOP B\RELcompOP{\rho'}\RELtraOP)
$\quad by definition

$\RELenthOP
\pi'\RELcompOP A\RELtraOP\RELcompOP\pi\RELtraOP
\RELcompOP
\pi\RELcompOP A\RELcompOP{\pi'}\RELtraOP
\RELandOP 
\rho'\RELcompOP B\RELtraOP\RELcompOP\rho\RELtraOP
\RELcompOP
\rho\RELcompOP B\RELcompOP{\rho'}\RELtraOP
$\quad monotony

$\RELenthOP
\pi'\RELcompOP A\RELtraOP\RELcompOP A\RELcompOP{\pi'}\RELtraOP
\RELandOP 
\rho'\RELcompOP B\RELtraOP\RELcompOP B\RELcompOP{\rho'}\RELtraOP
$\quad since projections $\pi,\rho$ are univalent

$\RELenthOP
\pi'\RELcompOP {\pi'}\RELtraOP
\RELandOP 
\rho'\RELcompOP {\rho'}\RELtraOP
$\quad since $A,B$ are assumed to be univalent

$=
\RELide
$\quad by definition of a direct product

\bigskip
\noindent
vi) Univalency follows from (iv). 
$\Kronecker{A}{B}\RELcompOP\RELtop
\RELaboveOP
\Kronecker{A}{B}\RELcompOP\,\pi'\RELcompOP\RELtop
=
\pi\RELcompOP A\RELcompOP\RELtop
=
\pi\RELcompOP\RELtop
=
\RELtop
$
\Bewende

\noindent
The results above are more or less known. It was important to execute rigorous axiomatic proofs, i.e., not just based on Boolean matrices. Of course, analogous formulae hold in the converse situation.

\enunc{}{Proposition}{}{PropForkMapKron} Let be given the setting above.
\begin{enumerate}[i)]
\item $\Kronecker{R}{S}\RELcompOP\Kronecker{P}{Q}\RELenthOP\Kronecker{R\RELcompOP P}{S\RELcompOP Q}$
\item[]$\Kronecker{R}{S}\RELcompOP\StrictJoin{P}{Q}\RELenthOP\StrictJoin{R\RELcompOP P}{S\RELcompOP Q}\qquad
\StrictFork{R}{S}\RELcompOP\Kronecker{P}{Q}\RELenthOP\StrictFork{R\RELcompOP P}{S\RELcompOP Q}$

\item[]$\StrictFork{R}{S}\RELcompOP\StrictJoin{P}{Q}\RELenthOP R\RELcompOP P\RELandOP S\RELcompOP Q$

\item$\Kronecker{f}{g}\RELcompOP\Kronecker{A}{B}=\Kronecker{f\RELcompOP A}{g\RELcompOP B}$\quad provided $f,g$ are both univalent
\item $\Kronecker{f}{g}\RELcompOP\StrictJoin{A}{B}=\StrictJoin{f\RELcompOP A}{g\RELcompOP B}$\quad provided $f,g$ are both univalent
\item[]$\Kronecker{R}{S}\RELcompOP\StrictJoin{A}{B}=\StrictJoin{R\RELcompOP A}{S\RELcompOP B}$\quad provided $A,B$ are both injective
\item[]$\StrictFork{R}{S}\RELcompOP\StrictJoin{A}{B}=R\RELcompOP A\RELandOP S\RELcompOP B$\quad provided $A,B$ are both injective, or $R,S$ both univalent

\item $\Kronecker{R}{S}\RELcompOP\RELtop=\StrictJoin{R\RELcompOP\RELtop}{S\RELcompOP\RELtop}$
\item $\StrictFork{A}{B}\RELandOP\; C\RELcompOP\RELtop
=
\StrictFork{A\RELandOP C\RELcompOP\RELtop}{B\RELandOP C\RELcompOP\RELtop}
$
\item $\Kronecker{A}{B}\RELandOP
\StrictJoin{C\RELcompOP\RELtop}{D\RELcompOP\RELtop}
=
\Kronecker{A\RELandOP C\RELcompOP\RELtop}{B\RELandOP D\RELcompOP\RELtop}
$
\item $\StrictFork{A\RELandOP C}{B\RELandOP D}
=
\StrictFork{A}{B}\RELandOP\StrictFork{C}{D}
$
\item $\Kronecker{\LeftResi{R}{R}}{\LeftResi{S}{S}}\RELenthOP\LeftResi{\StrictFork{R}{S}}{\StrictFork{R}{S}}$
\item$C\RELcompOP\StrictFork{A}{B}=\StrictFork{C\RELcompOP A}{C\RELcompOP B}$\quad provided $C$ is univalent
\end{enumerate}

\Proof i) See \cite[Prop.~7.2.ii]{RelaMath2010}, where it is also mentioned that a pointfree proof of equality is impossible notwithstanding the fact that equality holds when the Point Axiom\index{Point Axiom} is demanded; i.e., not least for Boolean matrices. Indeed, there exist models where equality is violated.

\bigskip
\noindent
ii) According to Prop.~\PropPiKrhoEquGeneral.iv, $\Kronecker{f}{g}$ is univalent, so that we may reason

\smallskip
$\Kronecker{f}{g}\RELcompOP\Kronecker{A}{B}
=
\Kronecker{f}{g}\RELcompOP\,(\pi_2\RELcompOP A\RELcompOP\pi_3\RELtraOP\RELandOP\rho_2\RELcompOP B\RELcompOP\rho_3\RELtraOP)
$\quad by definition

$=
\Kronecker{f}{g}\RELcompOP\,\pi_2\RELcompOP A\RELcompOP\pi_3\RELtraOP\RELandOP\Kronecker{f}{g}\RELcompOP\,\rho_2\RELcompOP B\RELcompOP\rho_3\RELtraOP
$\quad univalency

$=
(\pi_1\RELcompOP f
\RELandOP 
\rho_1\RELcompOP g\RELcompOP\RELtop)\RELcompOP A\RELcompOP\pi_3\RELtraOP
\RELandOP
(\pi_1\RELcompOP f\RELcompOP\RELtop\RELandOP\rho_1\RELcompOP g)\RELcompOP B\RELcompOP\rho_3\RELtraOP
$\quad Prop.~\PropPiKrhoEquGeneral.i

$=
\pi_1\RELcompOP f\RELcompOP A\RELcompOP\pi_3\RELtraOP
\RELandOP 
\rho_1\RELcompOP g\RELcompOP\RELtop
\;\RELandOP\;  
\pi_1\RELcompOP f\RELcompOP\RELtop\RELandOP\rho_1\RELcompOP g\RELcompOP B\RELcompOP\rho_3\RELtraOP
$\quad masking

$=
\pi_1\RELcompOP f\RELcompOP A\RELcompOP\pi_3\RELtraOP
\RELandOP\rho_1\RELcompOP g\RELcompOP B\RELcompOP\rho_3\RELtraOP
$\quad trivial

$=
\Kronecker{f\RELcompOP A}{g\RELcompOP B}
$\quad by definition

\bigskip
\noindent
iii) is shown similar to (ii).

\bigskip
\noindent
iv) For clarity, we mention the ever changing typing of the universal relations explicitly:

\smallskip
$\Kronecker{R}{S}\RELcompOP\RELtop_{X'\times Y',Z}
=
(\pi\RELcompOP R\RELcompOP{\pi'}\RELtraOP\RELandOP\rho\RELcompOP S\RELcompOP{\rho'}\RELtraOP)\RELcompOP\RELtop_{X'\times Y',Z}
$\quad by definition

$=
(\pi\RELcompOP R\RELcompOP{\pi'}\RELtraOP\RELandOP\rho\RELcompOP S\RELcompOP{\rho'}\RELtraOP)\RELcompOP\pi'\RELcompOP\RELtop_{X',Z}
$\quad since  $\pi'$ is total

$=
(\pi\RELcompOP R\RELandOP\rho\RELcompOP S\RELcompOP{\rho'}\RELtraOP\RELcompOP\pi')\RELcompOP\RELtop_{X',Z}
$\quad since $\pi'$ is univalent, \cite{RelaMath2010}, Prop.~5.4

$=
(\pi\RELcompOP R\RELandOP\rho\RELcompOP S\RELcompOP\RELtop_{Y',X'})\RELcompOP\RELtop_{X',Z}
$\quad property of the direct product

$=
\pi\RELcompOP R\RELcompOP\RELtop_{X',Z}\RELandOP\rho\RELcompOP S\RELcompOP\RELtop_{Y',Z}
$\quad masking

$=\StrictJoin{R\RELcompOP\RELtop_{X',Z}}{S\RELcompOP\RELtop_{Y',Z}}
$\quad by definition

\bigskip
\noindent
v)
$\StrictFork{A\RELandOP C\RELcompOP\RELtop_{Z,X}}{B\RELandOP C\RELcompOP\RELtop_{Z,Y}}
=
(A\RELandOP C\RELcompOP\RELtop_{Z,X})\RELcompOP\pi\RELtraOP\RELandOP(B\RELandOP C\RELcompOP\RELtop_{Z,Y})\RELcompOP\rho\RELtraOP
$

$=
A\RELcompOP\pi\RELtraOP\RELandOP C\RELcompOP\RELtop_{Z,X}\RELcompOP\pi\RELtraOP\RELandOP B\RELcompOP\rho\RELtraOP\RELandOP C\RELcompOP\RELtop_{Z,Y}\RELcompOP\rho\RELtraOP
$

$=
A\RELcompOP\pi\RELtraOP\RELandOP C\RELcompOP\RELtop_{Z,X\times Y}\RELandOP B\RELcompOP\rho\RELtraOP\RELandOP C\RELcompOP\RELtop_{Z,X\times Y}
=
A\RELcompOP\pi\RELtraOP\RELandOP B\RELcompOP\rho\RELtraOP\RELandOP C\RELcompOP\RELtop_{Z,X\times Y}
$

$=
\StrictFork{A}{B}\RELandOP\; C\RELcompOP\RELtop_{Z,X\times Y}
$

\bigskip
\noindent
vi) Assume $\RELfromTO{A}{X}{Y},\RELfromTO{B}{U}{V}, \RELfromTO{C}{X}{Z}, \RELfromTO{D}{U}{W}$:

$\Kronecker{A}{B}\RELandOP
\StrictJoin{C\RELcompOP\RELtop_{Z,Y\times V}}{D\RELcompOP\RELtop_{W,Y\times V}}
=
\pi\RELcompOP A\RELcompOP{\pi'}\RELtraOP\RELandOP\rho\RELcompOP B\RELcompOP{\rho'}\RELtraOP\RELandOP\pi\RELcompOP C\RELcompOP\RELtop_{Z,Y\times V}\RELandOP\rho\RELcompOP D \RELcompOP\RELtop_{W,Y\times V}
$

$=
\pi\RELcompOP A\RELcompOP{\pi'}\RELtraOP\RELandOP\pi\RELcompOP C\RELcompOP\RELtop_{Z,Y\times V}\RELandOP\rho\RELcompOP B\RELcompOP{\rho'}\RELtraOP\RELandOP\rho\RELcompOP D \RELcompOP\RELtop_{W,Y\times V}
$\quad shuffled

$=
\pi\RELcompOP(A\RELcompOP{\pi'}\RELtraOP\RELandOP C\RELcompOP\RELtop_{Z,Y\times V})\RELandOP\rho\RELcompOP(B\RELcompOP{\rho'}\RELtraOP\RELandOP D \RELcompOP\RELtop_{W,Y\times V})
$\quad 

$=
\pi\RELcompOP(A\RELcompOP{\pi'}\RELtraOP\RELandOP C\RELcompOP\RELtop_{Z,Y}\RELcompOP{\pi'}\RELtraOP)\RELandOP\rho\RELcompOP(B\RELcompOP{\rho'}\RELtraOP\RELandOP D \RELcompOP\RELtop_{W,V}\RELcompOP{\rho'}\RELtraOP)
$\quad 

$=
\pi\RELcompOP(A\RELandOP C\RELcompOP\RELtop_{Z,Y})\RELcompOP{\pi'}\RELtraOP\RELandOP\rho\RELcompOP(B\RELandOP D \RELcompOP\RELtop_{W,V})\RELcompOP{\rho'}\RELtraOP
$\quad 

$=
\Kronecker{A\RELandOP C\RELcompOP\RELtop_{Z,Y}}{B\RELandOP D\RELcompOP\RELtop_{W,V}}
$

\bigskip
\noindent
vii) $\StrictFork{A\RELandOP C}{B\RELandOP D}
=
(A\RELandOP C)\RELcompOP\pi\RELtraOP\RELandOP
(B\RELandOP D)\RELcompOP\rho\RELtraOP
$

$=
A\RELcompOP\pi\RELtraOP\RELandOP C\RELcompOP\pi\RELtraOP\RELandOP
B\RELcompOP\rho\RELtraOP\RELandOP D\RELcompOP\rho\RELtraOP
$

$=
A\RELcompOP\pi\RELtraOP\RELandOP
B\RELcompOP\rho\RELtraOP\RELandOP C\RELcompOP\pi\RELtraOP\RELandOP D\RELcompOP\rho\RELtraOP
=
\StrictFork{A}{B}\RELandOP\StrictFork{C}{D}
$

\bigskip
\noindent
viii) $\LeftResi{\StrictFork{R}{S}}{\StrictFork{R}{S}}
=
\RELneg{\StrictFork{R}{S}\RELtraOP\RELcompOP\RELneg{\StrictFork{R}{S}}}
=
\RELneg{\StrictJoin{R\RELtraOP}{S\RELtraOP}\RELcompOP\RELneg{R\RELcompOP\pi\RELtraOP\RELandOP S\RELcompOP\rho\RELtraOP}}
$

$=
\RELneg{\StrictJoin{R\RELtraOP}{S\RELtraOP}\RELcompOP(\RELneg{R\RELcompOP\pi\RELtraOP}\RELorOP\RELneg{S\RELcompOP\rho\RELtraOP})}
=
\RELneg{\StrictJoin{R\RELtraOP}{S\RELtraOP}\RELcompOP\RELneg{R\RELcompOP\pi\RELtraOP}\RELorOP\StrictJoin{R\RELtraOP}{S\RELtraOP}\RELcompOP\RELneg{S\RELcompOP\rho\RELtraOP}}
$

$=
\RELneg{(\pi\RELcompOP R\RELtraOP\RELandOP\rho\RELcompOP S\RELtraOP)\RELcompOP\RELneg{R}\RELcompOP\pi\RELtraOP\RELorOP(\pi\RELcompOP R\RELtraOP\RELandOP\rho\RELcompOP S\RELtraOP)\RELcompOP\RELneg{S}\RELcompOP\rho\RELtraOP}
$

$\RELaboveOP
\RELneg{\pi\RELcompOP R\RELtraOP\RELcompOP\RELneg{R}\RELcompOP\pi\RELtraOP\RELorOP\rho\RELcompOP S\RELtraOP\RELcompOP\RELneg{S}\RELcompOP\rho\RELtraOP}
$

$=
\RELneg{\pi\RELcompOP R\RELtraOP\RELcompOP\RELneg{R}\RELcompOP\pi\RELtraOP}\RELandOP\RELneg{\rho\RELcompOP S\RELtraOP\RELcompOP\RELneg{S}\RELcompOP\rho\RELtraOP}
=
\pi\RELcompOP\RELneg{R\RELtraOP\RELcompOP\RELneg{R}}\RELcompOP\pi\RELtraOP\RELandOP
\rho\RELcompOP\RELneg{S\RELtraOP\RELcompOP\RELneg{S}}\RELcompOP\rho\RELtraOP
=
\Kronecker{\LeftResi{R}{R}}{\LeftResi{S}{S}}
$

\bigskip
\noindent
ix) trivial
\Bewende

\noindent
As mentioned, one must not demand {\it arbitrary\/} products to exist, because one will then run into model problems. To employ the Point Axiom is a requirement stronger than necessary. When here just two additional products are requested, this means some sort of an \lq\lq improved observability\rq\rq\index{observability} for the pairs in the product $A\times B$ via vectorization.

\Caption{\includegraphics[scale=0.4]{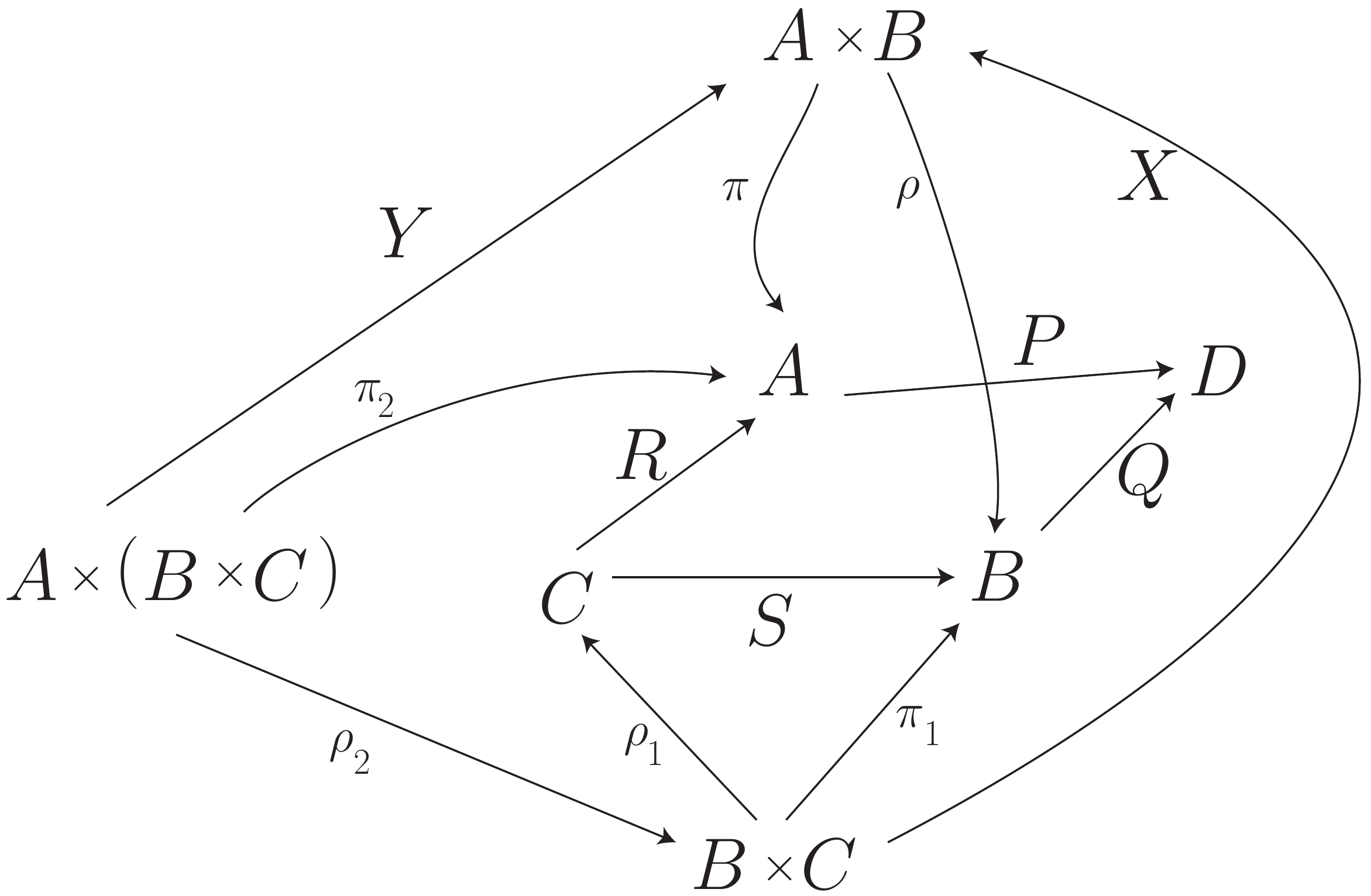}}
{The two additional products in the proof of $\StrictFork{R}{S}\RELcompOP\StrictJoin{P}{Q}= R\RELcompOP P\RELandOP S\RELcompOP Q$}{FigKroneckerOfMapsZierer}

\noindent
For better reference, we recall an important result by Hans Zierer with its difficult proof from \cite{ZiererDiss,Zierer1991}. It shows that when these additional products are available, there will hold equality in 
the third containment of Prop.~\PropForkMapKron.i.

\enunc{}{Proposition}{}{PropForkMapKronZierer} Let again be given the setting above. When products $\pi_1,\rho_1$ and $\pi_2,\rho_2$,

\smallskip
\ \hbox to5cm{$\RELfromTO{\pi_1}{B\times C}{B}$,\hfil}\quad $\RELfromTO{\rho_1}{B\times C}{C}$ 

\ \hbox to5cm{$\RELfromTO{\pi_2}{A\times(B\times C)}{A}$,\hfil}\quad $\RELfromTO{\rho_2}{A\times(B\times C)}{(B\times C)}$  

\smallskip
\noindent
exist, there will in addition to Prop.~\PropForkMapKron.i hold

\smallskip
$\StrictFork{R}{S}\RELcompOP\StrictJoin{P}{Q}= R\,\RELcompOP P\RELandOP S\RELcompOP Q$.

\Proof The intricate point is to define the following constructs

\smallskip
$X:=\rho_1\RELcompOP R\RELcompOP\pi\RELtraOP\RELandOP(\pi_1\RELandOP\rho_1\RELcompOP S)\RELcompOP\rho\RELtraOP
\qquad
Y:=(\pi_2\RELandOP\rho_2\RELcompOP\rho_1\RELcompOP R)\RELcompOP\pi\RELtraOP\RELandOP\rho_2\RELcompOP(\pi_1\RELandOP\rho_1\RELcompOP S)\RELcompOP\rho\RELtraOP$,

\smallskip
\noindent
of which $Y$ turns out to be univalent, and to show several rather simple consequences. These follow applying the destroy and append-rule for univalent relations repeatedly.

\smallskip
$\rho_1\RELtraOP\RELcompOP X=R\RELcompOP\pi\RELtraOP\RELandOP S\RELcompOP\rho\RELtraOP
$

$\rho_2\RELtraOP\RELcompOP Y=(\pi_1\RELandOP\rho_1\RELcompOP S)\RELcompOP\rho\RELtraOP\RELandOP\rho_1\RELcompOP R\RELcompOP\pi\RELtraOP=X
$

$ 
Y\RELcompOP\pi=(\pi_2\RELandOP\rho_2\RELcompOP\rho_1\RELcompOP R)\RELandOP\rho_2\RELcompOP(\pi_1\RELandOP\rho_1\RELcompOP S)\RELcompOP\RELtop
\qquad
Y\RELcompOP\rho=
(\pi_2\RELandOP\rho_2\RELcompOP\rho_1\RELcompOP R)\RELcompOP\RELtop\RELandOP\rho_2\RELcompOP(\pi_1\RELandOP\rho_1\RELcompOP S)
$

\smallskip
\noindent
Putting pieces together, we obtain

\smallskip
$\StrictFork{R}{S}\RELcompOP\StrictJoin{P}{Q}
=
(R\RELcompOP\pi\RELtraOP\RELandOP S\RELcompOP\rho\RELtraOP)\RELcompOP(\pi\RELcompOP P\RELandOP\rho\RELcompOP Q)
$

$=\rho_1\RELtraOP\RELcompOP\rho_2\RELtraOP\RELcompOP Y\RELcompOP(\pi\RELcompOP P\RELandOP\rho\RELcompOP Q)
$\quad see above

$=\rho_1\RELtraOP\RELcompOP\rho_2\RELtraOP\RELcompOP(Y\RELcompOP\pi\RELcompOP P\RELandOP Y\RELcompOP\rho\RELcompOP Q)
$\quad since $Y$ is univalent

$=\rho_1\RELtraOP\RELcompOP\rho_2\RELtraOP\RELcompOP\big[\{(\pi_2\RELandOP\rho_2\RELcompOP\rho_1\RELcompOP R)\RELandOP\rho_2\RELcompOP(\pi_1\RELandOP\rho_1\RELcompOP S)\RELcompOP\RELtop
\}\RELcompOP P$

$\qquad\qquad\RELandOP\,\{(\pi_2\RELandOP\rho_2\RELcompOP\rho_1\RELcompOP R)\RELcompOP\RELtop\RELandOP\rho_2\RELcompOP(\pi_1\RELandOP\rho_1\RELcompOP S)\}\RELcompOP Q\big]$\quad see above

$=\rho_1\RELtraOP\RELcompOP\rho_2\RELtraOP\RELcompOP\big[(\pi_2\RELandOP\rho_2\RELcompOP\rho_1\RELcompOP R)\RELcompOP P\RELandOP\rho_2\RELcompOP(\pi_1\RELandOP\rho_1\RELcompOP S)\RELcompOP\RELtop
$

$\qquad\qquad\RELandOP\,(\pi_2\RELandOP\rho_2\RELcompOP\rho_1\RELcompOP R)\RELcompOP\RELtop\RELandOP\rho_2\RELcompOP(\pi_1\RELandOP\rho_1\RELcompOP S)\RELcompOP Q\big]$\quad masking

$=\rho_1\RELtraOP\RELcompOP\rho_2\RELtraOP\RELcompOP\big[(\pi_2\RELandOP\rho_2\RELcompOP\rho_1\RELcompOP R)\RELcompOP P\RELandOP\rho_2\RELcompOP(\pi_1\RELandOP\rho_1\RELcompOP S)\RELcompOP Q\big]$\quad trivial

$=\rho_1\RELtraOP\RELcompOP\big[\rho_1\RELcompOP R\RELcompOP P\RELandOP(\pi_1\RELandOP\rho_1\RELcompOP S)\RELcompOP Q\big]$\quad destroy and append twice, $\rho_2\RELtraOP\RELcompOP\pi_2=\RELtop$

$=R\RELcompOP P\RELandOP S\RELcompOP Q$\quad destroy and append twice, $\rho_1\RELtraOP\RELcompOP\pi_1=\RELtop$
\Bewende

\noindent
That the two products requested are often met in practice may be seen from the discussion of associativity in Def.~\DefBinOp\ and Fig.~\FigAssocLaw.

\bigskip
\noindent
It is the merit of Jules Desharnais, to have sharpened the previous result in the important paper \cite{DesharnaisProducts}; also to be retrieved in 
\cite{WinterDiss}. Now just one of the relations $P,Q,R,S$ needs to possess a vectorization in order to obtain equality.

\enunc{}{Proposition}{}{PropForkMapKronJules} Let again be given the setting above. When the product $\pi',\rho'$ 

\smallskip
\ \hbox to5cm{$\RELfromTO{\pi'}{A\times C}{A}$,\hfil}\quad $\RELfromTO{\rho'}{A\times C}{C}$

\smallskip
\noindent
exists, there will in addition to Prop.~\PropForkMapKron.i hold

\smallskip
$\StrictFork{R}{S}\RELcompOP\StrictJoin{P}{Q}= R\RELcompOP P\RELandOP S\RELcompOP Q$.

\Proof Only one direction needs to be proved.

\smallskip
$R\RELcompOP P\RELandOP S\RELcompOP Q
=
\StrictFork{R}{\RELide}\RELcompOP\,\pi'\RELcompOP P\RELandOP S\RELcompOP Q
$\quad Prop.~\PropPiKrhoEquGeneral.iv

$\RELenthOP
\big[\!\StrictFork{R}{\RELide}\RELandOP\, S\RELcompOP Q\RELcompOP P\RELtraOP\RELcompOP{\pi'}\RELtraOP\big]\RELcompOP\big[\pi'\RELcompOP P\RELandOP\StrictFork{R}{\RELide}\RELtraOP\RELcompOP\, S\RELcompOP Q\big]
$\quad Dedekind rule 

$\RELenthOP
\StrictFork{R}{\RELide}\RELcompOP\big[\pi'\RELcompOP P\RELandOP\StrictJoin{R\RELtraOP}{\RELide}\RELcompOP\, S\RELcompOP Q\big]
$\quad monotony and transposition

$\RELenthOP
\StrictFork{R}{\RELide}\RELcompOP\big[\pi'\RELcompOP P\RELandOP\{\StrictJoin{R\RELtraOP}{\RELide}\RELcompOP\, S\RELcompOP Q\RELandOP\pi'\RELcompOP P\}\big]
$\quad trivial

$\RELenthOP
\StrictFork{R}{\RELide}\RELcompOP\,(\pi'\RELcompOP P\RELandOP\big\{\!\!\StrictJoin{R\RELtraOP}{\RELide}\RELcompOP\, S\RELandOP\pi'\RELcompOP P\RELcompOP Q\RELtraOP\big\}\RELcompOP Q)
$\quad Dedekind rule and monotony

$\RELenthOP
\StrictFork{R}{\RELide}\RELcompOP\,(\pi'\RELcompOP P\RELandOP\big\{\pi'\RELcompOP R\RELtraOP\RELcompOP S\RELandOP\rho'\RELcompOP S\RELandOP\pi'\RELcompOP P\RELcompOP Q\RELtraOP\big\}\RELcompOP Q)
$

$\RELenthOP
\StrictFork{R}{\RELide}\RELcompOP\,(\pi'\RELcompOP P\RELandOP\big\{\pi'\RELcompOP\RELtop\RELandOP\rho'\RELcompOP S\big\}\RELcompOP Q)
$\quad trivial

$\RELenthOP
\StrictFork{R}{\RELide}\RELcompOP\,(\pi'\RELcompOP P\RELandOP\StrictJoin{\RELtop}{S}
\RELcompOP\, Q)
$\quad definition of join

$\RELenthOP
\StrictFork{R}{\RELide}\RELcompOP\,(\pi'\RELcompOP P\RELandOP\Kronecker{\RELide}{S}\RELcompOP\,\rho\,\RELcompOP Q)
$\quad Prop.~\PropPiKrhoEquGeneral.iv

$\RELenthOP
\StrictFork{R}{\RELide}\RELcompOP\Kronecker{\RELide}{S}\RELcompOP\,(\rho\RELcompOP Q\RELandOP\Kronecker{\RELide}{S}\RELtraOP\RELcompOP\,\pi'\RELcompOP P)
$\quad Dedekind rule and monotony

$\RELenthOP
\StrictFork{R}{\RELide}\RELcompOP\Kronecker{\RELide}{S}\RELcompOP\,(\pi\RELcompOP P\RELandOP\rho\RELcompOP Q)
$\quad Prop.~\PropPiKrhoEquGeneral.i

$=
\StrictFork{R}{\RELide}\RELcompOP\Kronecker{\RELide}{S}\RELcompOP\StrictJoin{P}{Q}
$\quad definition of join

$=
\StrictFork{R}{S}\RELcompOP\StrictJoin{P}{Q}
$\quad according to \cite{RelaMath2010} Prop.~7.5
\Bewende

\kern-\baselineskip

\Caption{\includegraphics[scale=0.45]{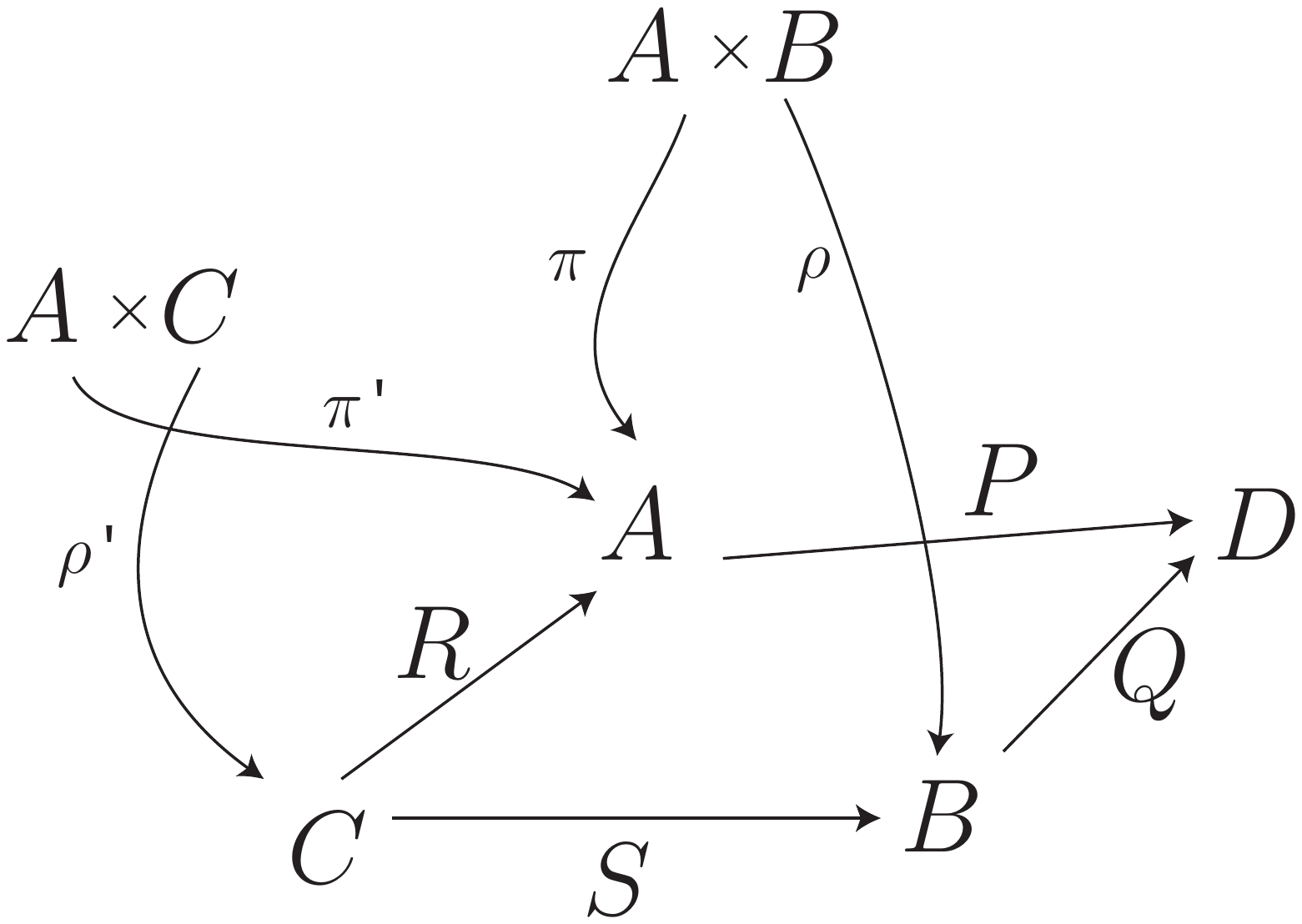}}
{The additional product in the proof of $\StrictFork{R}{S}\RELcompOP\StrictJoin{P}{Q}= R\RELcompOP P\RELandOP S\RELcompOP Q$}{FigKroneckerOfMapsJules}

\bigskip
\noindent 
The following proposition states that left residuation distributes over the strict fork.

\enunc{}{Proposition}{}{PropResiFork} For relations typed\quad $\RELfromTO{A}{W}{X},\quad\RELfromTO{B}{X}{Y},\quad\RELfromTO{C}{X}{Z}$ 

\smallskip
$\LeftResi{A}{\StrictFork{B}{C}}=\StrictFork{\LeftResi{A}{B}}{\LeftResi{A}{C}}
$

\Proof
$\LeftResi{A}{\StrictFork{B}{C}}
=
\RELneg{A\RELtraOP\RELcompOP\RELneg{\StrictFork{B}{C}}}
=
\RELneg{A\RELtraOP\RELcompOP\RELneg{B\RELcompOP\pi\RELtraOP\RELandOP C\RELcompOP\rho\RELtraOP}}
=
\RELneg{A\RELtraOP\RELcompOP(\RELneg{B\RELcompOP\pi\RELtraOP}\RELorOP\RELneg{C\RELcompOP\rho\RELtraOP})}
$

$=
\RELneg{A\RELtraOP\RELcompOP(\RELneg{B}\RELcompOP\pi\RELtraOP\RELorOP\RELneg{C}\RELcompOP\rho\RELtraOP)}
=
\RELneg{A\RELtraOP\RELcompOP\RELneg{B}\RELcompOP\pi\RELtraOP\RELorOP A\RELtraOP\RELcompOP\RELneg{C}\RELcompOP\rho\RELtraOP}
=
\RELneg{A\RELtraOP\RELcompOP\RELneg{B}\RELcompOP\pi\RELtraOP}\RELandOP\RELneg{A\RELtraOP\RELcompOP\RELneg{C}\RELcompOP\rho\RELtraOP}
=
\RELneg{A\RELtraOP\RELcompOP\RELneg{B}}\RELcompOP\pi\RELtraOP\RELandOP\RELneg{A\RELtraOP\RELcompOP\RELneg{C}}\RELcompOP\rho\RELtraOP
$
\Bewende

\kern-\baselineskip

\Caption{\includegraphics[scale=0.4]{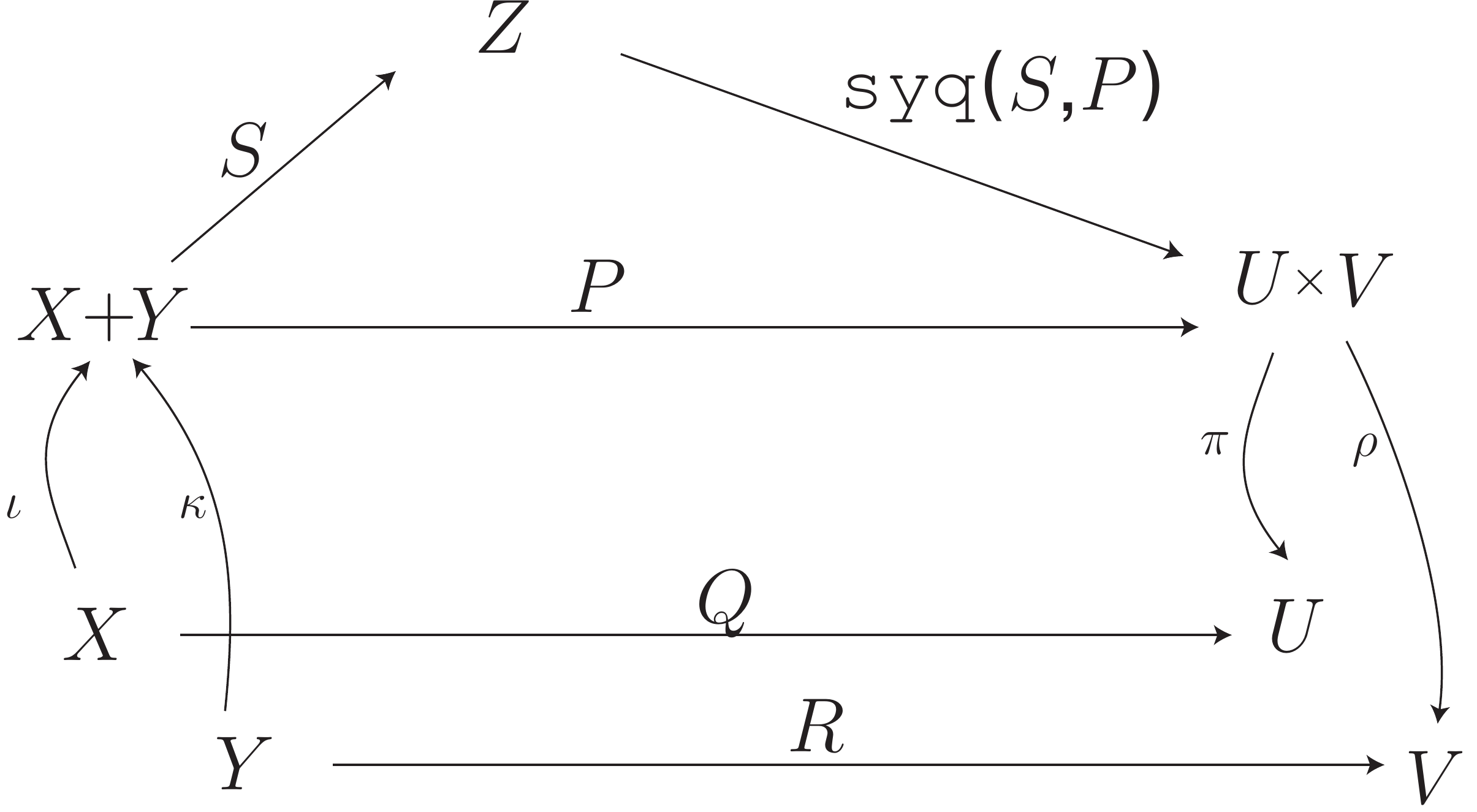}}
{Illustrating the addition theorem with $P:=\iota\RELtraOP\RELcompOP Q\RELcompOP\pi\RELtraOP\RELorOP\kappa\RELtraOP\RELcompOP R\RELcompOP\rho\RELtraOP$}{FigSumPowToPowProdGeneral}

\noindent
An addition theorem\index{addition theorem}, quite similar to the broadly known 

\smallskip
$ \sin ( x + y ) = \sin x\cdot\cos y + \cos x\cdot\sin y $, 

\smallskip
\noindent
holds for direct sum and direct product, cf.~Fig.~\FigSumPowToPowProdGeneral:

\enunc{}{Proposition}{}{AdditionSyqGeneralized} Let be given any three relations
$\RELfromTO{Q}{X}{U}$, $\RELfromTO{R}{Y}{V}$ and $\RELfromTO{S}{X+Y}{Z}$. In addition, we consider the injections
$\RELfromTO{\iota}{X}{X+Y}$, $\RELfromTO{\kappa}{Y}{X+Y}$ as well as the projections $\RELfromTO{\pi}{U\times V}{U}$, $\RELfromTO{\rho}{U\times V}{V}$, generically given. Then the following generalized addition theorem holds

\smallskip
$\syqq{S}{\iota\RELtraOP\RELcompOP Q\RELcompOP\pi\RELtraOP\RELorOP\kappa\RELtraOP\RELcompOP R\RELcompOP\rho\RELtraOP}
=\syqq{\iota\RELcompOP S}{Q}\RELcompOP\pi\RELtraOP\RELandOP\syqq{\kappa\RELcompOP S}{R}\RELcompOP\rho\RELtraOP$.

\smallskip
\noindent
In another notation, this looks as follows:

\smallskip
$\syqq{S}{\iota\RELtraOP\RELcompOP Q\RELcompOP\pi\RELtraOP\RELorOP\kappa\RELtraOP\RELcompOP R\RELcompOP\rho\RELtraOP}
=\StrictFork{\syqq{\iota\RELcompOP S}{Q}}{\syqq{\kappa\RELcompOP S}{R}}
$

\Proof In what follows, we abbreviate $P:=\iota\RELtraOP\RELcompOP Q\RELcompOP\pi\RELtraOP\RELorOP\kappa\RELtraOP\RELcompOP R\RELcompOP\rho\RELtraOP$.

\smallskip
$\syqq{\iota\RELcompOP S}{Q}\RELcompOP\pi\RELtraOP
=
\syqq{\iota\RELcompOP S}{Q\RELcompOP\pi\RELtraOP}
=
\syqq{\iota\RELcompOP S}{\iota\RELcompOP P}
$

$\syqq{\kappa\RELcompOP S}{R}\RELcompOP\rho\RELtraOP
=
\syqq{\kappa\RELcompOP S}{R\RELcompOP\rho\RELtraOP}
=
\syqq{\kappa\RELcompOP S}{\kappa\RELcompOP P}
$

\smallskip
$\syqq{\iota\RELcompOP S}{\iota\RELcompOP P}\RELandOP\syqq{\kappa\RELcompOP S}{\kappa\RELcompOP P}
=
\RELneg{\RELneg{S\RELtraOP\RELcompOP\iota\RELtraOP}\RELcompOP\iota\RELcompOP P}\RELandOP\RELneg{S\RELtraOP\RELcompOP\iota\RELtraOP\RELcompOP\RELneg{\iota\RELcompOP P}}\RELandOP\RELneg{\RELneg{S\RELtraOP\RELcompOP\kappa\RELtraOP}\RELcompOP\kappa\RELcompOP P}\RELandOP\RELneg{S\RELtraOP\RELcompOP\kappa\RELtraOP\RELcompOP\RELneg{\kappa\RELcompOP P}}
$

$=\RELneg{\RELneg{S\RELtraOP}\RELcompOP\iota\RELtraOP\RELcompOP\iota\RELcompOP P}\RELandOP\RELneg{S\RELtraOP\RELcompOP\iota\RELtraOP\RELcompOP\iota\RELcompOP\RELneg{P}}\RELandOP\RELneg{\RELneg{S\RELtraOP}\RELcompOP\kappa\RELtraOP\RELcompOP\kappa\RELcompOP P}\RELandOP\RELneg{S\RELtraOP\RELcompOP\kappa\RELtraOP\RELcompOP\kappa\RELcompOP\RELneg{P}}
$\quad $\iota,\kappa$ are mappings

$=\RELneg{\RELneg{S\RELtraOP}\RELcompOP\iota\RELtraOP\RELcompOP\iota\RELcompOP P}\RELandOP\RELneg{\RELneg{S\RELtraOP}\RELcompOP\kappa\RELtraOP\RELcompOP\kappa\RELcompOP P}\RELandOP\RELneg{S\RELtraOP\RELcompOP\iota\RELtraOP\RELcompOP\iota\RELcompOP\RELneg{P}}\RELandOP\RELneg{S\RELtraOP\RELcompOP\kappa\RELtraOP\RELcompOP\kappa\RELcompOP\RELneg{P}}
$\quad shuffled

$=\RELneg{\RELneg{S\RELtraOP}\RELcompOP(\iota\RELtraOP\RELcompOP\iota\RELorOP\kappa\RELtraOP\RELcompOP\kappa)\RELcompOP P}\RELandOP\RELneg{S\RELtraOP\RELcompOP(\iota\RELtraOP\RELcompOP\iota\RELorOP\kappa\RELtraOP\RELcompOP\kappa)\RELcompOP\RELneg{P}}
$\quad 

$=\RELneg{\RELneg{S\RELtraOP}\RELcompOP P}\RELandOP\RELneg{S\RELtraOP\RELcompOP\RELneg{P}}
=\syqq{S}{P}$\quad $\iota,\kappa$ form a direct sum
\Bewende

\noindent
Now we relate pairs of subsets of two sets $X,Y$ with subsets of the direct sum $X+Y$.

\Caption{\includegraphics[scale=0.4]{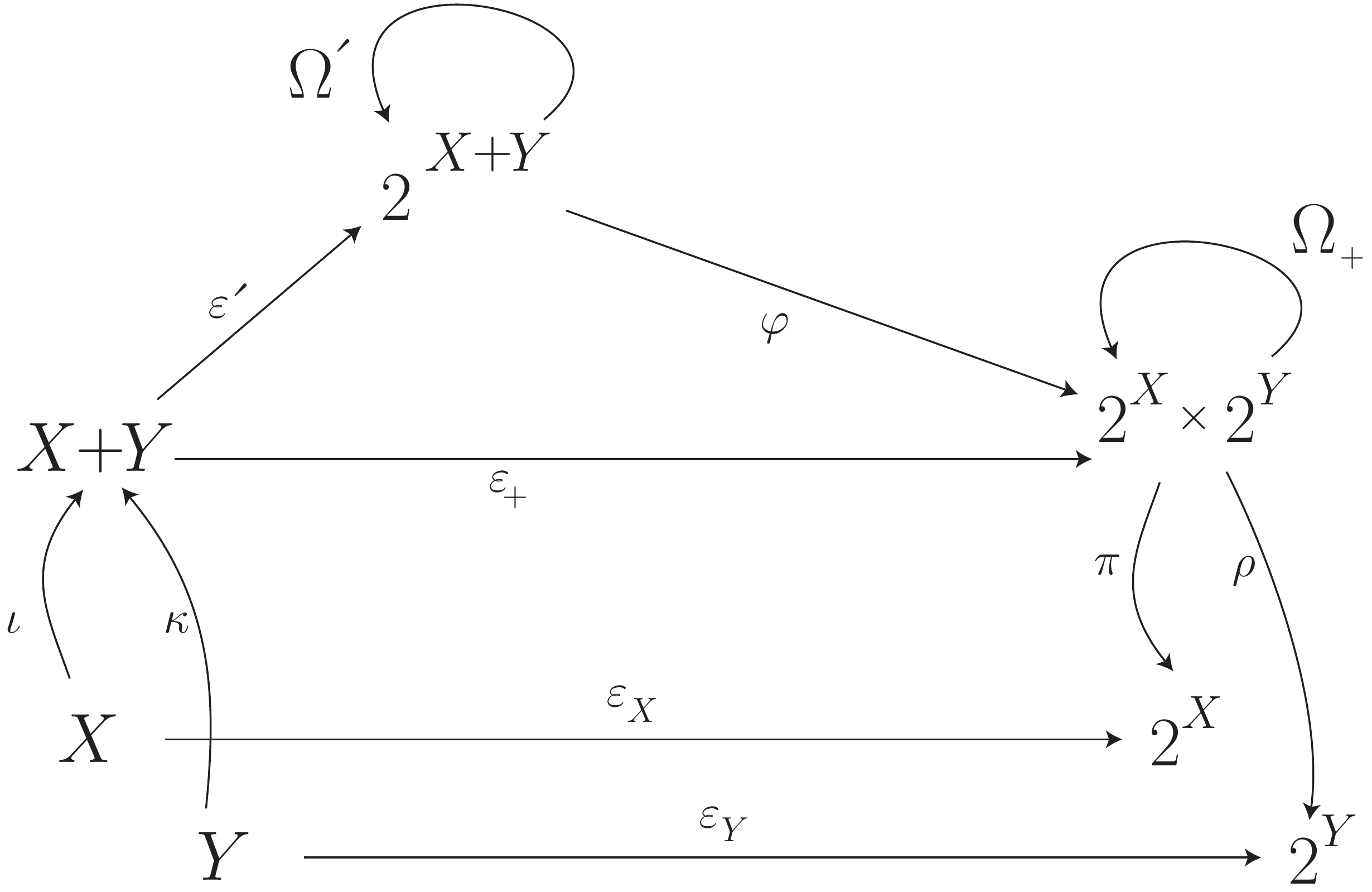}}
{Converting subsets of a sum to pairs of subsets}{FigSumPowToPowProd}

\noindent
In Prop.~\PropSumPowToPowProd, it is demonstrated that $\PowTWO{X+Y}$ is isomorphic to $\PowTWO{X}\times\PowTWO{Y}$.
In addition, it turns out that $\RELfromTO{\varepsilon_+}{X+Y}{\PowTWO{X}\times\PowTWO{Y}}$ satisfies the properties of a relational power, although it is constructed differently.
In largely the same sense, the mappings 

\smallskip
$\RELfromTO{\syqq{\iota\RELcompOP\varepsilon'}{\varepsilon_X}}{\PowTWO{X+Y}}{\PowTWO{X}}$
\quad and\quad 
$\RELfromTO{\syqq{\kappa\RELcompOP\varepsilon'}{\varepsilon_Y}}{\PowTWO{X+Y}}{\PowTWO{Y}}$ 

\smallskip
\noindent
establish $\PowTWO{X+Y}$ as another direct product of $\PowTWO{X}$ and $\PowTWO{Y}$. Since the direct product is uniquely determined up to isomorphism, however, we are able to prove the isomorphism via the bijective mapping $\varphi$.

\bigskip
\noindent
Earlier, we have been scrupulous with regard to the existence of products; we should maintain this here. Everything is fine, when $\varepsilon',\pi,\rho$ are available. We assume this to be the case. In another model of relation algebra, however, $\pi,\rho$ may not exist; then additional investigations are necessary.

\enunc{}{Proposition}{}{PropSumPowToPowProd} Let arbitrary sets $X,Y$ be given for which we consider their membership relations $\RELfromTO{\varepsilon_X}{X}{\PowTWO{X}}$, $\RELfromTO{\varepsilon_Y}{Y}{\PowTWO{Y}}$, the direct product $\PowTWO{X}\times\PowTWO{Y}$ of these powersets, as well as the direct sum $X+Y$ and its membership relation $\RELfromTO{\varepsilon'}{X+Y}{\PowTWO{X+Y}}$; see Fig.~\FigSumPowToPowProd. 
Then the following hold
\begin{enumerate}[i)]
\item for the construct\qquad$\varepsilon_+:=\iota\RELtraOP\RELcompOP\varepsilon_X\RELcompOP\pi\RELtraOP
\RELorOP
\kappa\RELtraOP\RELcompOP\varepsilon_Y\RELcompOP\rho\RELtraOP$
\begin{itemize}
\item $\iota\RELcompOP\varepsilon_+=\varepsilon_X\RELcompOP\pi\RELtraOP
\qquad
\iota\RELcompOP\varepsilon_+\RELcompOP\pi=\varepsilon_X
\qquad\qquad\kappa\RELcompOP\varepsilon_+=\varepsilon_Y\RELcompOP\rho\RELtraOP
\qquad
\kappa\RELcompOP\varepsilon_+\RELcompOP\rho=\varepsilon_Y
$
\item $\syqq{\iota\RELcompOP\varepsilon_+}{\varepsilon_X}=\pi
\qquad\qquad\qquad
\syqq{\kappa\RELcompOP\varepsilon_+}{\varepsilon_Y}=\rho
$
\end{itemize}

\item for the construct\qquad$\varphi:=\syqq{\varepsilon'}{\varepsilon_+}$
\begin{itemize}
\item $\varphi
= \syqq{\iota\RELcompOP\varepsilon'}{\varepsilon_X}\RELcompOP\pi\RELtraOP
\RELandOP
\syqq{\kappa\RELcompOP\varepsilon'}{\varepsilon_Y}\RELcompOP\rho\RELtraOP
$, \qquad i.e.~$\varphi$ satisfies an addition theorem
\item $\varphi\RELcompOP\pi=\syqq{\iota\RELcompOP\varepsilon'}{\varepsilon_X}\qquad
\varphi\RELcompOP\rho=\syqq{\kappa\RELcompOP\varepsilon'}{\varepsilon_Y}$
\item $\varphi$ is a bijective mapping
\item $\varepsilon'\RELcompOP\varphi=\varepsilon_+
\qquad
\varepsilon_+\RELcompOP\varphi\RELtraOP=\varepsilon'
$
\item $\syqq{\pi\RELcompOP\varepsilon_X\RELtraOP}{\varepsilon_+\RELtraOP}=\iota$
\qquad\qquad\qquad 
$\!\syqq{\rho\RELcompOP\varepsilon_Y\RELtraOP}{\varepsilon_+\RELtraOP}=\kappa$
\item $\varepsilon_+$\quad satisfies the relational requirements of a membership relation.
\end{itemize}

\newpage
\item for the construct\qquad$\Omega_{+}:=\RELneg{{\varepsilon_+}\RELtraOP\RELcompOP\RELneg{\varepsilon_+}}$ 

\begin{itemize}
\item is an ordering.
\item $\Omega_{+}=\Kronecker{\Omega_X}{\Omega_Y}$
\item $\varphi$ is an order isomorphism between the orderings

$\RELfromTO{\Omega':=\RELneg{{\varepsilon'}\RELtraOP\RELcompOP\RELneg{\varepsilon'}}}{\PowTWO{X+Y}}{\PowTWO{X+Y}}$
\quad and \quad 
$\RELfromTO{\Omega_{+}}{\PowTWO{X}\times\PowTWO{Y}}{\PowTWO{X}\times\PowTWO{Y}}$.
\end{itemize}
\smallskip
\noindent
\end{enumerate}

\kern-\baselineskip

\Proof i) We demonstrate the main sample cases:

\smallskip
$\iota\RELcompOP\varepsilon_+
=
\iota\RELcompOP(\iota\RELtraOP\RELcompOP\varepsilon_X\RELcompOP\pi\RELtraOP
\RELorOP
\kappa\RELtraOP\RELcompOP\varepsilon_Y\RELcompOP\rho\RELtraOP)
=
\iota\RELcompOP\iota\RELtraOP\RELcompOP\varepsilon_X\RELcompOP\pi\RELtraOP
\RELorOP
\iota\RELcompOP\kappa\RELtraOP\RELcompOP\varepsilon_Y\RELcompOP\rho\RELtraOP
=
\RELide\RELcompOP\varepsilon_X\RELcompOP\pi\RELtraOP
\RELorOP
\RELbot
=
\varepsilon_X\RELcompOP\pi\RELtraOP
$

$\iota\RELcompOP\varepsilon_+\RELcompOP\pi
=
\varepsilon_X\RELcompOP\pi\RELtraOP\RELcompOP\pi
=
\varepsilon_X
$

$\syqq{\iota\RELcompOP\varepsilon_+}{\varepsilon_X}
=
\syqq{\varepsilon_X\RELcompOP\pi\RELtraOP}{\varepsilon_X}
=
\pi\RELcompOP\syqq{\varepsilon_X}{\varepsilon_X}
=
\pi
$

\bigskip
\noindent
ii) The first formula is an immediate consequence of the addition theorem Prop.~\AdditionSyqGeneralized.
We have to obey some care: Only $\varepsilon_X,\varepsilon_Y,\varepsilon'$ have been introduced as membership relations; $\varepsilon_+$ is defined differently but denoted similarly, since it will soon turn out to be one also.

\bigskip
\noindent
Then we prove with the addition theorem

\smallskip
$\varphi\RELcompOP\pi
=
\big\lbrack\syqq{\iota\RELcompOP\varepsilon'}{\varepsilon_X}\RELcompOP\pi\RELtraOP
\RELandOP
\syqq{\kappa\RELcompOP\varepsilon'}{\varepsilon_Y}\RELcompOP\rho\RELtraOP\big\rbrack\RELcompOP\pi
$

$=
\syqq{\iota\RELcompOP\varepsilon'}{\varepsilon_X}
\RELandOP
\syqq{\kappa\RELcompOP\varepsilon'}{\varepsilon_Y}\RELcompOP\rho\RELtraOP\RELcompOP\pi
$

$=
\syqq{\iota\RELcompOP\varepsilon'}{\varepsilon_X}
\RELandOP
\syqq{\kappa\RELcompOP\varepsilon'}{\varepsilon_Y}\RELcompOP\RELtop
$\qquad since $\pi,\rho$ form a direct product

$=
\syqq{\iota\RELcompOP\varepsilon'}{\varepsilon_X}
\RELandOP
\RELtop
$\qquad since $\syqq{\varepsilon_Y}{\dots}$ is always surjective

$=
\syqq{\iota\RELcompOP\varepsilon'}{\varepsilon_X}
$

\bigskip
\noindent
Now, we convince ourselves that $\varphi$ is total, which follows with the preceding result from
 
\smallskip
$\varphi\RELcompOP\RELtop
=
\varphi\RELcompOP\pi\RELcompOP\RELtop
=
\syqq{\iota\RELcompOP\varepsilon'}{\varepsilon_X}\RELcompOP\RELtop
$\quad and the fact that $\varepsilon_X$ is a membership

\bigskip
\noindent
Univalency follows also with the addition theorem\index{addition theorem}

\smallskip
$\varphi\RELtraOP\RELcompOP\varphi
\RELenthOP
\big\lbrack\pi\RELcompOP\syqq{\varepsilon_X}{\iota\RELcompOP\varepsilon'}\RELandOP
\rho\RELcompOP\syqq{\varepsilon_Y}{\kappa\RELcompOP\varepsilon'}\big\rbrack\RELcompOP
\big\lbrack\syqq{\iota\RELcompOP\varepsilon'}{\varepsilon_X}\RELcompOP\pi\RELtraOP
\RELandOP
\syqq{\kappa\RELcompOP\varepsilon'}{\varepsilon_Y}\RELcompOP\rho\RELtraOP\big\rbrack
$

$\RELenthOP
\pi\RELcompOP\syqq{\varepsilon_X}{\iota\RELcompOP\varepsilon'}\RELcompOP\syqq{\iota\RELcompOP\varepsilon'}{\varepsilon_X}\RELcompOP\pi\RELtraOP\RELandOP
\rho\RELcompOP\syqq{\varepsilon_Y}{\kappa\RELcompOP\varepsilon'}\RELcompOP\syqq{\kappa\RELcompOP\varepsilon'}{\varepsilon_Y}\RELcompOP\rho\RELtraOP
$

$\RELenthOP
\pi\RELcompOP\syqq{\varepsilon_X}{\varepsilon_X}\RELcompOP\pi\RELtraOP\RELandOP
\rho\RELcompOP\syqq{\varepsilon_Y}{\varepsilon_Y}\RELcompOP\rho\RELtraOP
\RELenthOP
\pi\RELcompOP\pi\RELtraOP\RELandOP
\rho\RELcompOP\rho\RELtraOP
=\RELide
$

\bigskip
\noindent
Even simpler and without the addition theorem we get $\varphi\RELcompOP\varphi\RELtraOP\RELenthOP\syqq{\varepsilon'}{\varepsilon'}\RELenthOP\RELide$, so that $\varphi$ is injective. Finally, $\varphi$ is surjective since $\varepsilon'$ is a membership relation.

\bigskip
$\varepsilon'\RELcompOP\varphi
=
\varepsilon'\RELcompOP\syqq{\varepsilon'}{\varepsilon_+}=\varepsilon_+$\quad since $\varepsilon'$ is a membership relation

$\varepsilon_+\RELcompOP\varphi\RELtraOP
=
\varepsilon'\RELcompOP\varphi\RELcompOP\varphi\RELtraOP
=\varepsilon'
$\qquad since $\varphi$ is already established as a bijective mapping

\bigskip
$\syqq{\pi\RELcompOP\varepsilon_X\RELtraOP}{\varepsilon_+\RELtraOP}
=
\syqq{\pi\RELcompOP\varepsilon_X\RELtraOP}{\varphi\RELtraOP\RELcompOP{\varepsilon'}\RELtraOP}
=
\syqq{\varphi\RELcompOP\pi\RELcompOP\varepsilon_X\RELtraOP}{{\varepsilon'}\RELtraOP}
$\quad since $\varphi$ is a bijective mapping

$
=
\syqq{\syqq{\iota\RELcompOP\varepsilon'}{\varepsilon_X}\RELcompOP\varepsilon_X\RELtraOP}{{\varepsilon'}\RELtraOP}
$\quad see above

$
=
\syqq{{\varepsilon'}\RELtraOP\RELcompOP\iota\RELtraOP}{{\varepsilon'}\RELtraOP}
$

$
=
\iota\RELcompOP\syqq{{\varepsilon'}\RELtraOP}{{\varepsilon'}\RELtraOP}
$

$
=
\iota\RELcompOP(\RELneg{\RELneg{\varepsilon'}\RELcompOP{\varepsilon'}\RELtraOP}\RELandOP\RELneg{\varepsilon'\RELcompOP\RELneg{{\varepsilon'}\RELtraOP}})
$

$
=
\iota\RELcompOP({\Omega'}\RELtraOP\RELandOP\Omega')
=
\iota\RELcompOP\RELide
=
\iota
$

\bigskip
\noindent
It is relatively easy to prove that the differently constructed $\varepsilon_+$ is a membership relation:

\newpage
$\syqq{\varepsilon_+}{\varepsilon_+}
=
\syqq{\varepsilon'\RELcompOP\varphi}{\varepsilon'\RELcompOP\varphi}
=
\varphi\RELtraOP\RELcompOP\syqq{\varepsilon'}{\varepsilon'\RELcompOP\varphi}
=
\varphi\RELtraOP\RELcompOP\syqq{\varepsilon'}{\varepsilon'}\RELcompOP\varphi
=
\varphi\RELtraOP\RELcompOP\RELide\RELcompOP\varphi
=
\varphi\RELtraOP\RELcompOP\varphi
=
\RELide
$ 

$\syqq{\varepsilon_+}{U}
=
\syqq{\varepsilon'\RELcompOP\varphi}{U}
=
\varphi\RELtraOP\RELcompOP\syqq{\varepsilon'}{U}
$ is surjective since $\varepsilon'$ is a membership

\smallskip
\noindent
according to \cite{RelaMath2010} Prop.~8.18.

\bigskip
\noindent
iii) $\Omega_{+}$ is --- consequently --- indeed an ordering. It satisfies

\smallskip
$\Kronecker{\Omega_X}{\Omega_Y}
=
\pi\RELcompOP\RELneg{\varepsilon_X\RELtraOP\RELcompOP\RELneg{\varepsilon_X}}\RELcompOP\pi\RELtraOP
\RELandOP
\rho\RELcompOP\RELneg{\varepsilon_Y\RELtraOP\RELcompOP\RELneg{\varepsilon_Y}}\RELcompOP\rho\RELtraOP
=
\RELneg{\pi\RELcompOP\varepsilon_X\RELtraOP\RELcompOP\RELneg{\varepsilon_X\RELcompOP\pi\RELtraOP}}
\RELandOP
\RELneg{\rho\RELcompOP\varepsilon_Y\RELtraOP\RELcompOP\RELneg{\varepsilon_Y\RELcompOP\rho\RELtraOP}}
$

$=
\RELneg{\varepsilon_+\RELtraOP\RELcompOP\iota\RELtraOP\RELcompOP\RELneg{\iota\RELcompOP\varepsilon_+}}
\RELandOP
\RELneg{\varepsilon_+\RELtraOP\RELcompOP\kappa\RELtraOP\RELcompOP\RELneg{\kappa\RELcompOP\varepsilon_+}}
$\quad Prop.~\PropSumPowToPowProd.i

$=
\RELneg{\varepsilon_+\RELtraOP\RELcompOP\iota\RELtraOP\RELcompOP\iota\RELcompOP\RELneg{\varepsilon_+}}
\RELandOP
\RELneg{\varepsilon_+\RELtraOP\RELcompOP\kappa\RELtraOP\RELcompOP\kappa\RELcompOP\RELneg{\varepsilon_+}}
$\quad $\iota,\kappa$ are mappings

$=
\RELneg{\varepsilon_+\RELtraOP\RELcompOP\iota\RELtraOP\RELcompOP\iota\RELcompOP\RELneg{\varepsilon_+}\RELorOP
\varepsilon_+\RELtraOP\RELcompOP\kappa\RELtraOP\RELcompOP\kappa\RELcompOP\RELneg{\varepsilon_+}}
$

$=
\RELneg{\varepsilon_+\RELtraOP\RELcompOP(\iota\RELtraOP\RELcompOP\iota\RELorOP
\kappa\RELtraOP\RELcompOP\kappa)\RELcompOP\RELneg{\varepsilon_+}}
=
\RELneg{\varepsilon_+\RELtraOP\RELcompOP\RELneg{\varepsilon_+}}
$\quad $\iota,\kappa$ form a direct sum

$=
\Omega_{+}
$

\bigskip
\noindent
First direction of the isomorphism proposition, using that $\varphi$ is a bijective mapping:  

\smallskip
$\Omega'\RELcompOP\varphi
=
\RELneg{{\varepsilon'}\RELtraOP\RELcompOP\RELneg{\varepsilon'}}\RELcompOP\varphi
=
\RELneg{{\varepsilon'}\RELtraOP\RELcompOP\RELneg{\varepsilon'\RELcompOP\varphi}}
=
\RELneg{{\varepsilon'}\RELtraOP\RELcompOP\RELneg{\varepsilon_{+}}}
=
\RELneg{\varphi\RELcompOP\varepsilon_{+}\RELtraOP\RELcompOP\RELneg{\varepsilon_{+}}}
=
\varphi\RELcompOP\RELneg{\varepsilon_{+}\RELtraOP\RELcompOP\RELneg{\varepsilon_{+}}}
=
\varphi\RELcompOP\Omega_{+}
$

\bigskip
\noindent
Second direction: 

\smallskip
$\Omega_{+}\RELcompOP\varphi\RELtraOP
=
\RELneg{\varepsilon_{+}\RELtraOP\RELcompOP\RELneg{\varepsilon_{+}}}\RELcompOP\varphi\RELtraOP
=
\RELneg{\varepsilon_{+}\RELtraOP\RELcompOP\RELneg{\varepsilon_{+}\RELcompOP\varphi\RELtraOP}}
=
\RELneg{\varepsilon_{+}\RELtraOP\RELcompOP\RELneg{\varepsilon'}}
=
\RELneg{\varphi\RELtraOP\RELcompOP{\varepsilon'}\RELtraOP\RELcompOP\RELneg{\varepsilon'}}
=
\varphi\RELtraOP\RELcompOP\RELneg{{\varepsilon'}\RELtraOP\RELcompOP\RELneg{\varepsilon'}}
=
\varphi\RELtraOP\RELcompOP\Omega'
$
\Bewende


\chapter{Binary operations\label{ChapBinOp}}
\ExerciseNo=0
\EnuncNo=0
\CaptionNo=0


\def\TarskiSwitch{P}
\def\TarskiShuffle{T}
\def\TarskiNeutrL{n_l}
\def\TarskiNeutrR{n_r}
\def\TarskiDelta{\delta_r}
\def\TarskiInversR{i_r}

\noindent
We now attempt to study also binary operations\index{binary operation} on a set relationally. This will already allow a very basic look on group theory. It will turn out that such elements as the unit, e.g., will be points. A {\it point\/}\index{point} resembles the classic {\it element\/} of set theory. In the relational setting, a {\it point} is a row-constant, injective, and surjective relation $x$, i.e, it satisfies

\smallskip
$x\RELcompOP\RELtop=x,\qquad
x\RELcompOP x\RELtraOP\RELenthOP\RELide,\qquad
\RELtop\RELcompOP x=\RELtop
$.

\bigskip
\noindent
We assume a direct product with projections $\RELfromTO{\pi,\rho}{X\times X}{X}$ and in addition a binary mapping $\RELfromTO{\TarskiAdd}{X\times X}{X}$. A first preparatory observation concerns what one might consider as coretract or section in a category, here simply a left-inverse of the projection $\rho$.

\enunc{}{Proposition}{}{PropNegBinOp} If $x$ is any point, then $f:=(\rho\RELandOP\pi\RELcompOP x\RELcompOP\RELtop)\RELtraOP$ is a mapping. It satisfies $f\RELcompOP\rho=\RELide$ and $\rho\RELenthOP\leftResi{f}{\RELide}$.

\Proof Since $x$ is row-constant and injective, we have univalence

\smallskip
$f\RELtraOP\RELcompOP f=(\rho\RELandOP\pi\RELcompOP x\RELcompOP\RELtop)\RELcompOP(\rho\RELandOP\pi\RELcompOP x\RELcompOP\RELtop)\RELtraOP
\RELenthOP
\rho\RELcompOP\rho\RELtraOP\RELandOP\pi\RELcompOP x\RELcompOP\RELtop\RELcompOP\RELtop\RELtraOP\RELcompOP x\RELtraOP\RELcompOP\pi\RELtraOP
\RELenthOP
\rho\RELcompOP\rho\RELtraOP\RELandOP\pi\RELcompOP\pi\RELtraOP
=
\RELide
$

\smallskip
\noindent
as well as totality

\smallskip
$f\RELcompOP\RELtop
=
(\rho\RELtraOP\RELandOP\RELtop\RELtraOP\RELcompOP x\RELtraOP\RELcompOP\pi\RELtraOP)\RELcompOP\RELtop
=
\rho\RELtraOP\RELcompOP(\RELtop\RELandOP\pi\RELcompOP x\RELcompOP\RELtop)
=
\rho\RELtraOP\RELcompOP\pi\RELcompOP x\RELcompOP\RELtop
=
\RELtop\RELcompOP x\RELcompOP\RELtop
=
\RELtop
$\quad since $x$ is a point.

$f\RELcompOP\rho
=
(\rho\RELtraOP\RELandOP\RELtop\RELcompOP x\RELtraOP\RELcompOP \pi\RELtraOP)\RELcompOP\rho
=
\RELide\RELandOP\RELtop\RELcompOP x\RELtraOP\RELcompOP \pi\RELtraOP\RELcompOP\rho
=
\RELide\RELandOP\RELtop\RELcompOP x\RELtraOP\RELcompOP\RELtop
=
\RELide\RELandOP\RELtop
=
\RELide
$

$\rho\RELenthOP\leftResi{f}{\RELide}=\RELneg{f\RELtraOP\RELcompOP\RELneg{\RELide}}
\quad\iff\quad
f\RELtraOP\RELcompOP\RELneg{\RELide}\RELenthOP\RELneg{\rho}
\quad\iff\quad
f\RELcompOP\rho\RELenthOP\RELide
\quad\iff\quad
{\tt True}
$
\Bewende

\Caption{\vbox{\hbox to\textwidth{\hfil
\vbox{\hbox{\footnotesize%
\BoxBreadth=0pt%
\setbox7=\hbox{a}%
\ifdim\wd7>\BoxBreadth\BoxBreadth=\wd7\fi%
\setbox7=\hbox{b}%
\ifdim\wd7>\BoxBreadth\BoxBreadth=\wd7\fi%
\setbox7=\hbox{c}%
\ifdim\wd7>\BoxBreadth\BoxBreadth=\wd7\fi%
\setbox7=\hbox{d}%
\ifdim\wd7>\BoxBreadth\BoxBreadth=\wd7\fi%
\setbox7=\hbox{e}%
\ifdim\wd7>\BoxBreadth\BoxBreadth=\wd7\fi%
\setbox7=\hbox{f}%
\ifdim\wd7>\BoxBreadth\BoxBreadth=\wd7\fi%
\setbox7=\hbox{g}%
\ifdim\wd7>\BoxBreadth\BoxBreadth=\wd7\fi%
\def\RowNames{\def\scalable{0.5\baselineskip}
\vcenter{\offinterlineskip\baselineskip=\matrixskip%
\hbox to\BoxBreadth{\strut\hfil a}\kern\scalable%
\hbox to\BoxBreadth{\strut\hfil b}\kern\scalable%
\hbox to\BoxBreadth{\strut\hfil c}\kern\scalable%
\hbox to\BoxBreadth{\strut\hfil d}\kern\scalable%
\hbox to\BoxBreadth{\strut\hfil e}\kern\scalable%
\hbox to\BoxBreadth{\strut\hfil f}\kern\scalable%
\hbox to\BoxBreadth{\strut\hfil g}}}%
\def\ColNames{\def\scalable{0.18\baselineskip}
\hbox{\rotatebox{90}{\strut a}\kern\scalable%
\rotatebox{90}{\strut b}\kern\scalable%
\rotatebox{90}{\strut c}\kern\scalable%
\rotatebox{90}{\strut d}\kern\scalable%
\rotatebox{90}{\strut e}\kern\scalable%
\rotatebox{90}{\strut f}\kern\scalable%
\rotatebox{90}{\strut g}\kern\scalable%
\kern0pt
}}%
\def\Matrix{\spmatrix{%
\noalign{\kern-2pt}
 \hbox{\vbox to14pt{\vfil\hbox to14pt{\hfil 3\hfil}\vfil}}&\hbox{\vbox to14pt{\vfil\hbox to14pt{\hfil 2\hfil}\vfil}}&\hbox{\vbox to14pt{\vfil\hbox to14pt{\hfil 1\hfil}\vfil}}&\hbox{\vbox to14pt{\vfil\hbox to14pt{\hfil 4\hfil}\vfil}}&\hbox{\vbox to14pt{\vfil\hbox to14pt{\hfil 1\hfil}\vfil}}&\hbox{\vbox to14pt{\vfil\hbox to14pt{\hfil 6\hfil}\vfil}}&\hbox{\vbox to14pt{\vfil\hbox to14pt{\hfil 7\hfil}\vfil}}\cr
 \hbox{\vbox to14pt{\vfil\hbox to14pt{\hfil 1\hfil}\vfil}}&\hbox{\vbox to14pt{\vfil\hbox to14pt{\hfil 2\hfil}\vfil}}&\hbox{\vbox to14pt{\vfil\hbox to14pt{\hfil 5\hfil}\vfil}}&\hbox{\vbox to14pt{\vfil\hbox to14pt{\hfil 4\hfil}\vfil}}&\hbox{\vbox to14pt{\vfil\hbox to14pt{\hfil 3\hfil}\vfil}}&\hbox{\vbox to14pt{\vfil\hbox to14pt{\hfil 6\hfil}\vfil}}&\hbox{\vbox to14pt{\vfil\hbox to14pt{\hfil 7\hfil}\vfil}}\cr
 \hbox{\vbox to14pt{\vfil\hbox to14pt{\hfil 1\hfil}\vfil}}&\hbox{\vbox to14pt{\vfil\hbox to14pt{\hfil 3\hfil}\vfil}}&\hbox{\vbox to14pt{\vfil\hbox to14pt{\hfil 2\hfil}\vfil}}&\hbox{\vbox to14pt{\vfil\hbox to14pt{\hfil 4\hfil}\vfil}}&\hbox{\vbox to14pt{\vfil\hbox to14pt{\hfil 2\hfil}\vfil}}&\hbox{\vbox to14pt{\vfil\hbox to14pt{\hfil 6\hfil}\vfil}}&\hbox{\vbox to14pt{\vfil\hbox to14pt{\hfil 7\hfil}\vfil}}\cr
 \hbox{\vbox to14pt{\vfil\hbox to14pt{\hfil 1\hfil}\vfil}}&\hbox{\vbox to14pt{\vfil\hbox to14pt{\hfil 5\hfil}\vfil}}&\hbox{\vbox to14pt{\vfil\hbox to14pt{\hfil 4\hfil}\vfil}}&\hbox{\vbox to14pt{\vfil\hbox to14pt{\hfil 7\hfil}\vfil}}&\hbox{\vbox to14pt{\vfil\hbox to14pt{\hfil 4\hfil}\vfil}}&\hbox{\vbox to14pt{\vfil\hbox to14pt{\hfil 3\hfil}\vfil}}&\hbox{\vbox to14pt{\vfil\hbox to14pt{\hfil 6\hfil}\vfil}}\cr
 \hbox{\vbox to14pt{\vfil\hbox to14pt{\hfil 1\hfil}\vfil}}&\hbox{\vbox to14pt{\vfil\hbox to14pt{\hfil 2\hfil}\vfil}}&\hbox{\vbox to14pt{\vfil\hbox to14pt{\hfil 5\hfil}\vfil}}&\hbox{\vbox to14pt{\vfil\hbox to14pt{\hfil 4\hfil}\vfil}}&\hbox{\vbox to14pt{\vfil\hbox to14pt{\hfil 5\hfil}\vfil}}&\hbox{\vbox to14pt{\vfil\hbox to14pt{\hfil 6\hfil}\vfil}}&\hbox{\vbox to14pt{\vfil\hbox to14pt{\hfil 7\hfil}\vfil}}\cr
 \hbox{\vbox to14pt{\vfil\hbox to14pt{\hfil 1\hfil}\vfil}}&\hbox{\vbox to14pt{\vfil\hbox to14pt{\hfil 3\hfil}\vfil}}&\hbox{\vbox to14pt{\vfil\hbox to14pt{\hfil 6\hfil}\vfil}}&\hbox{\vbox to14pt{\vfil\hbox to14pt{\hfil 5\hfil}\vfil}}&\hbox{\vbox to14pt{\vfil\hbox to14pt{\hfil 2\hfil}\vfil}}&\hbox{\vbox to14pt{\vfil\hbox to14pt{\hfil 4\hfil}\vfil}}&\hbox{\vbox to14pt{\vfil\hbox to14pt{\hfil 7\hfil}\vfil}}\cr
 \hbox{\vbox to14pt{\vfil\hbox to14pt{\hfil 1\hfil}\vfil}}&\hbox{\vbox to14pt{\vfil\hbox to14pt{\hfil 2\hfil}\vfil}}&\hbox{\vbox to14pt{\vfil\hbox to14pt{\hfil 7\hfil}\vfil}}&\hbox{\vbox to14pt{\vfil\hbox to14pt{\hfil 4\hfil}\vfil}}&\hbox{\vbox to14pt{\vfil\hbox to14pt{\hfil 7\hfil}\vfil}}&\hbox{\vbox to14pt{\vfil\hbox to14pt{\hfil 6\hfil}\vfil}}&\hbox{\vbox to14pt{\vfil\hbox to14pt{\hfil 3\hfil}\vfil}}\cr
\noalign{\kern-2pt}}}
\vbox{\setbox8=\hbox{$\RowNames\Matrix$}
\hbox to\wd8{\hfil$\ColNames$\kern0.6\matrixskip}\kern0.1cm
\box8}}
\kern0.1cm
\hbox{$\;\;\;$\small number indicates row}
\kern0.8cm
\hbox{\quad point $c=$ {\footnotesize%
\BoxBreadth=0pt%
\setbox7=\hbox{a}%
\ifdim\wd7>\BoxBreadth\BoxBreadth=\wd7\fi%
\setbox7=\hbox{b}%
\ifdim\wd7>\BoxBreadth\BoxBreadth=\wd7\fi%
\setbox7=\hbox{c}%
\ifdim\wd7>\BoxBreadth\BoxBreadth=\wd7\fi%
\setbox7=\hbox{d}%
\ifdim\wd7>\BoxBreadth\BoxBreadth=\wd7\fi%
\setbox7=\hbox{e}%
\ifdim\wd7>\BoxBreadth\BoxBreadth=\wd7\fi%
\setbox7=\hbox{f}%
\ifdim\wd7>\BoxBreadth\BoxBreadth=\wd7\fi%
\setbox7=\hbox{g}%
\ifdim\wd7>\BoxBreadth\BoxBreadth=\wd7\fi%
\def\RowNames{\def\scalable{\interspacereduction}
\vcenter{\offinterlineskip\baselineskip=\matrixskip%
\hbox to\BoxBreadth{\strut\hfil a}\kern\scalable%
\hbox to\BoxBreadth{\strut\hfil b}\kern\scalable%
\hbox to\BoxBreadth{\strut\hfil c}\kern\scalable%
\hbox to\BoxBreadth{\strut\hfil d}\kern\scalable%
\hbox to\BoxBreadth{\strut\hfil e}\kern\scalable%
\hbox to\BoxBreadth{\strut\hfil f}\kern\scalable%
\hbox to\BoxBreadth{\strut\hfil g}}}%
\def\MarkVect{\setbox2=\hbox{\strut}\matrixskip=\ht2\spmatrix{\noalign{\kern-2pt}%
\n\cr
\n\cr
{\CoefTrue}\cr
\n\cr
\n\cr
\n\cr
\n\cr
\noalign{\kern-2pt}}%
}%
$\RowNames\MarkVect$}}
\kern-4cm}
\hfil
{\footnotesize%
\BoxBreadth=0pt%
\setbox7=\hbox{(a,a)}%
\ifdim\wd7>\BoxBreadth\BoxBreadth=\wd7\fi%
\setbox7=\hbox{(b,a)}%
\ifdim\wd7>\BoxBreadth\BoxBreadth=\wd7\fi%
\setbox7=\hbox{(a,b)}%
\ifdim\wd7>\BoxBreadth\BoxBreadth=\wd7\fi%
\setbox7=\hbox{(c,a)}%
\ifdim\wd7>\BoxBreadth\BoxBreadth=\wd7\fi%
\setbox7=\hbox{(b,b)}%
\ifdim\wd7>\BoxBreadth\BoxBreadth=\wd7\fi%
\setbox7=\hbox{(a,c)}%
\ifdim\wd7>\BoxBreadth\BoxBreadth=\wd7\fi%
\setbox7=\hbox{(d,a)}%
\ifdim\wd7>\BoxBreadth\BoxBreadth=\wd7\fi%
\setbox7=\hbox{(c,b)}%
\ifdim\wd7>\BoxBreadth\BoxBreadth=\wd7\fi%
\setbox7=\hbox{(b,c)}%
\ifdim\wd7>\BoxBreadth\BoxBreadth=\wd7\fi%
\setbox7=\hbox{(a,d)}%
\ifdim\wd7>\BoxBreadth\BoxBreadth=\wd7\fi%
\setbox7=\hbox{(e,a)}%
\ifdim\wd7>\BoxBreadth\BoxBreadth=\wd7\fi%
\setbox7=\hbox{(d,b)}%
\ifdim\wd7>\BoxBreadth\BoxBreadth=\wd7\fi%
\setbox7=\hbox{(c,c)}%
\ifdim\wd7>\BoxBreadth\BoxBreadth=\wd7\fi%
\setbox7=\hbox{(b,d)}%
\ifdim\wd7>\BoxBreadth\BoxBreadth=\wd7\fi%
\setbox7=\hbox{(a,e)}%
\ifdim\wd7>\BoxBreadth\BoxBreadth=\wd7\fi%
\setbox7=\hbox{(f,a)}%
\ifdim\wd7>\BoxBreadth\BoxBreadth=\wd7\fi%
\setbox7=\hbox{(e,b)}%
\ifdim\wd7>\BoxBreadth\BoxBreadth=\wd7\fi%
\setbox7=\hbox{(d,c)}%
\ifdim\wd7>\BoxBreadth\BoxBreadth=\wd7\fi%
\setbox7=\hbox{(c,d)}%
\ifdim\wd7>\BoxBreadth\BoxBreadth=\wd7\fi%
\setbox7=\hbox{(b,e)}%
\ifdim\wd7>\BoxBreadth\BoxBreadth=\wd7\fi%
\setbox7=\hbox{(a,f)}%
\ifdim\wd7>\BoxBreadth\BoxBreadth=\wd7\fi%
\setbox7=\hbox{(g,a)}%
\ifdim\wd7>\BoxBreadth\BoxBreadth=\wd7\fi%
\setbox7=\hbox{(f,b)}%
\ifdim\wd7>\BoxBreadth\BoxBreadth=\wd7\fi%
\setbox7=\hbox{(e,c)}%
\ifdim\wd7>\BoxBreadth\BoxBreadth=\wd7\fi%
\setbox7=\hbox{(d,d)}%
\ifdim\wd7>\BoxBreadth\BoxBreadth=\wd7\fi%
\setbox7=\hbox{(c,e)}%
\ifdim\wd7>\BoxBreadth\BoxBreadth=\wd7\fi%
\setbox7=\hbox{(b,f)}%
\ifdim\wd7>\BoxBreadth\BoxBreadth=\wd7\fi%
\setbox7=\hbox{(a,g)}%
\ifdim\wd7>\BoxBreadth\BoxBreadth=\wd7\fi%
\setbox7=\hbox{(g,b)}%
\ifdim\wd7>\BoxBreadth\BoxBreadth=\wd7\fi%
\setbox7=\hbox{(f,c)}%
\ifdim\wd7>\BoxBreadth\BoxBreadth=\wd7\fi%
\setbox7=\hbox{(e,d)}%
\ifdim\wd7>\BoxBreadth\BoxBreadth=\wd7\fi%
\setbox7=\hbox{(d,e)}%
\ifdim\wd7>\BoxBreadth\BoxBreadth=\wd7\fi%
\setbox7=\hbox{(c,f)}%
\ifdim\wd7>\BoxBreadth\BoxBreadth=\wd7\fi%
\setbox7=\hbox{(b,g)}%
\ifdim\wd7>\BoxBreadth\BoxBreadth=\wd7\fi%
\setbox7=\hbox{(g,c)}%
\ifdim\wd7>\BoxBreadth\BoxBreadth=\wd7\fi%
\setbox7=\hbox{(f,d)}%
\ifdim\wd7>\BoxBreadth\BoxBreadth=\wd7\fi%
\setbox7=\hbox{(e,e)}%
\ifdim\wd7>\BoxBreadth\BoxBreadth=\wd7\fi%
\setbox7=\hbox{(d,f)}%
\ifdim\wd7>\BoxBreadth\BoxBreadth=\wd7\fi%
\setbox7=\hbox{(c,g)}%
\ifdim\wd7>\BoxBreadth\BoxBreadth=\wd7\fi%
\setbox7=\hbox{(g,d)}%
\ifdim\wd7>\BoxBreadth\BoxBreadth=\wd7\fi%
\setbox7=\hbox{(f,e)}%
\ifdim\wd7>\BoxBreadth\BoxBreadth=\wd7\fi%
\setbox7=\hbox{(e,f)}%
\ifdim\wd7>\BoxBreadth\BoxBreadth=\wd7\fi%
\setbox7=\hbox{(d,g)}%
\ifdim\wd7>\BoxBreadth\BoxBreadth=\wd7\fi%
\setbox7=\hbox{(g,e)}%
\ifdim\wd7>\BoxBreadth\BoxBreadth=\wd7\fi%
\setbox7=\hbox{(f,f)}%
\ifdim\wd7>\BoxBreadth\BoxBreadth=\wd7\fi%
\setbox7=\hbox{(e,g)}%
\ifdim\wd7>\BoxBreadth\BoxBreadth=\wd7\fi%
\setbox7=\hbox{(g,f)}%
\ifdim\wd7>\BoxBreadth\BoxBreadth=\wd7\fi%
\setbox7=\hbox{(f,g)}%
\ifdim\wd7>\BoxBreadth\BoxBreadth=\wd7\fi%
\setbox7=\hbox{(g,g)}%
\ifdim\wd7>\BoxBreadth\BoxBreadth=\wd7\fi%
\def\RowNames{\def\scalable{\interspacereduction}
\vcenter{\offinterlineskip\baselineskip=\matrixskip%
\hbox to\BoxBreadth{\strut\hfil (a,a)}\kern\scalable%
\hbox to\BoxBreadth{\strut\hfil (b,a)}\kern\scalable%
\hbox to\BoxBreadth{\strut\hfil (a,b)}\kern\scalable%
\hbox to\BoxBreadth{\strut\hfil (c,a)}\kern\scalable%
\hbox to\BoxBreadth{\strut\hfil (b,b)}\kern\scalable%
\hbox to\BoxBreadth{\strut\hfil (a,c)}\kern\scalable%
\hbox to\BoxBreadth{\strut\hfil (d,a)}\kern\scalable%
\hbox to\BoxBreadth{\strut\hfil (c,b)}\kern\scalable%
\hbox to\BoxBreadth{\strut\hfil (b,c)}\kern\scalable%
\hbox to\BoxBreadth{\strut\hfil (a,d)}\kern\scalable%
\hbox to\BoxBreadth{\strut\hfil (e,a)}\kern\scalable%
\hbox to\BoxBreadth{\strut\hfil (d,b)}\kern\scalable%
\hbox to\BoxBreadth{\strut\hfil (c,c)}\kern\scalable%
\hbox to\BoxBreadth{\strut\hfil (b,d)}\kern\scalable%
\hbox to\BoxBreadth{\strut\hfil (a,e)}\kern\scalable%
\hbox to\BoxBreadth{\strut\hfil (f,a)}\kern\scalable%
\hbox to\BoxBreadth{\strut\hfil (e,b)}\kern\scalable%
\hbox to\BoxBreadth{\strut\hfil (d,c)}\kern\scalable%
\hbox to\BoxBreadth{\strut\hfil (c,d)}\kern\scalable%
\hbox to\BoxBreadth{\strut\hfil (b,e)}\kern\scalable%
\hbox to\BoxBreadth{\strut\hfil (a,f)}\kern\scalable%
\hbox to\BoxBreadth{\strut\hfil (g,a)}\kern\scalable%
\hbox to\BoxBreadth{\strut\hfil (f,b)}\kern\scalable%
\hbox to\BoxBreadth{\strut\hfil (e,c)}\kern\scalable%
\hbox to\BoxBreadth{\strut\hfil (d,d)}\kern\scalable%
\hbox to\BoxBreadth{\strut\hfil (c,e)}\kern\scalable%
\hbox to\BoxBreadth{\strut\hfil (b,f)}\kern\scalable%
\hbox to\BoxBreadth{\strut\hfil (a,g)}\kern\scalable%
\hbox to\BoxBreadth{\strut\hfil (g,b)}\kern\scalable%
\hbox to\BoxBreadth{\strut\hfil (f,c)}\kern\scalable%
\hbox to\BoxBreadth{\strut\hfil (e,d)}\kern\scalable%
\hbox to\BoxBreadth{\strut\hfil (d,e)}\kern\scalable%
\hbox to\BoxBreadth{\strut\hfil (c,f)}\kern\scalable%
\hbox to\BoxBreadth{\strut\hfil (b,g)}\kern\scalable%
\hbox to\BoxBreadth{\strut\hfil (g,c)}\kern\scalable%
\hbox to\BoxBreadth{\strut\hfil (f,d)}\kern\scalable%
\hbox to\BoxBreadth{\strut\hfil (e,e)}\kern\scalable%
\hbox to\BoxBreadth{\strut\hfil (d,f)}\kern\scalable%
\hbox to\BoxBreadth{\strut\hfil (c,g)}\kern\scalable%
\hbox to\BoxBreadth{\strut\hfil (g,d)}\kern\scalable%
\hbox to\BoxBreadth{\strut\hfil (f,e)}\kern\scalable%
\hbox to\BoxBreadth{\strut\hfil (e,f)}\kern\scalable%
\hbox to\BoxBreadth{\strut\hfil (d,g)}\kern\scalable%
\hbox to\BoxBreadth{\strut\hfil (g,e)}\kern\scalable%
\hbox to\BoxBreadth{\strut\hfil (f,f)}\kern\scalable%
\hbox to\BoxBreadth{\strut\hfil (e,g)}\kern\scalable%
\hbox to\BoxBreadth{\strut\hfil (g,f)}\kern\scalable%
\hbox to\BoxBreadth{\strut\hfil (f,g)}\kern\scalable%
\hbox to\BoxBreadth{\strut\hfil (g,g)}}}%
\def\ColNames{\def\scalable{\interspacereduction}
\hbox{\rotatebox{90}{\strut a}\kern\scalable%
\rotatebox{90}{\strut b}\kern\scalable%
\rotatebox{90}{\strut c}\kern\scalable%
\rotatebox{90}{\strut d}\kern\scalable%
\rotatebox{90}{\strut e}\kern\scalable%
\rotatebox{90}{\strut f}\kern\scalable%
\rotatebox{90}{\strut g}\kern\scalable%
\kern0pt
}}%
\def\Matrix{\spmatrix{%
\noalign{\kern-2pt}
 \n&\n&{\CoefTrue}&\n&\n&\n&\n\cr
 {\CoefTrue}&\n&\n&\n&\n&\n&\n\cr
 \n&{\CoefTrue}&\n&\n&\n&\n&\n\cr
 {\CoefTrue}&\n&\n&\n&\n&\n&\n\cr
 \n&\n&{\CoefTrue}&\n&\n&\n&\n\cr
 {\CoefTrue}&\n&\n&\n&\n&\n&\n\cr
 {\CoefTrue}&\n&\n&\n&\n&\n&\n\cr
 \n&{\CoefTrue}&\n&\n&\n&\n&\n\cr
 \n&{\CoefTrue}&\n&\n&\n&\n&\n\cr
 \n&\n&\n&{\CoefTrue}&\n&\n&\n\cr
 {\CoefTrue}&\n&\n&\n&\n&\n&\n\cr
 \n&\n&\n&\n&{\CoefTrue}&\n&\n\cr
 \n&\n&{\CoefTrue}&\n&\n&\n&\n\cr
 \n&\n&\n&{\CoefTrue}&\n&\n&\n\cr
 {\CoefTrue}&\n&\n&\n&\n&\n&\n\cr
 {\CoefTrue}&\n&\n&\n&\n&\n&\n\cr
 \n&{\CoefTrue}&\n&\n&\n&\n&\n\cr
 \n&\n&\n&{\CoefTrue}&\n&\n&\n\cr
 \n&\n&\n&{\CoefTrue}&\n&\n&\n\cr
 \n&{\CoefTrue}&\n&\n&\n&\n&\n\cr
 \n&\n&\n&\n&\n&{\CoefTrue}&\n\cr
 {\CoefTrue}&\n&\n&\n&\n&\n&\n\cr
 \n&\n&{\CoefTrue}&\n&\n&\n&\n\cr
 \n&\n&\n&\n&{\CoefTrue}&\n&\n\cr
 \n&\n&\n&\n&\n&\n&{\CoefTrue}\cr
 \n&\n&{\CoefTrue}&\n&\n&\n&\n\cr
 \n&\n&\n&\n&\n&{\CoefTrue}&\n\cr
 \n&\n&\n&\n&\n&\n&{\CoefTrue}\cr
 \n&{\CoefTrue}&\n&\n&\n&\n&\n\cr
 \n&\n&\n&\n&\n&{\CoefTrue}&\n\cr
 \n&\n&\n&{\CoefTrue}&\n&\n&\n\cr
 \n&\n&\n&{\CoefTrue}&\n&\n&\n\cr
 \n&\n&\n&\n&\n&{\CoefTrue}&\n\cr
 \n&\n&\n&\n&\n&\n&{\CoefTrue}\cr
 \n&\n&\n&\n&\n&\n&{\CoefTrue}\cr
 \n&\n&\n&\n&{\CoefTrue}&\n&\n\cr
 \n&\n&\n&\n&{\CoefTrue}&\n&\n\cr
 \n&\n&{\CoefTrue}&\n&\n&\n&\n\cr
 \n&\n&\n&\n&\n&\n&{\CoefTrue}\cr
 \n&\n&\n&{\CoefTrue}&\n&\n&\n\cr
 \n&\n&\n&\n&\n&{\CoefTrue}&\n\cr
 \n&\n&\n&\n&\n&{\CoefTrue}&\n\cr
 \n&\n&\n&\n&\n&{\CoefTrue}&\n\cr
 \n&\n&\n&\n&\n&\n&{\CoefTrue}\cr
 \n&\n&\n&{\CoefTrue}&\n&\n&\n\cr
 \n&\n&\n&\n&\n&\n&{\CoefTrue}\cr
 \n&\n&\n&\n&\n&{\CoefTrue}&\n\cr
 \n&\n&\n&\n&\n&\n&{\CoefTrue}\cr
 \n&\n&{\CoefTrue}&\n&\n&\n&\n\cr
\noalign{\kern-2pt}}}%
\vbox{\setbox8=\hbox{$\RowNames\Matrix$}
\hbox to\wd8{\hfil$\ColNames$\kern\ColEntryShiftHoriz}\kern\ColEntryShiftVerti
\box8}}
\hfil
{\footnotesize%
\BoxBreadth=0pt%
\setbox7=\hbox{(a,a)}%
\ifdim\wd7>\BoxBreadth\BoxBreadth=\wd7\fi%
\setbox7=\hbox{(b,a)}%
\ifdim\wd7>\BoxBreadth\BoxBreadth=\wd7\fi%
\setbox7=\hbox{(a,b)}%
\ifdim\wd7>\BoxBreadth\BoxBreadth=\wd7\fi%
\setbox7=\hbox{(c,a)}%
\ifdim\wd7>\BoxBreadth\BoxBreadth=\wd7\fi%
\setbox7=\hbox{(b,b)}%
\ifdim\wd7>\BoxBreadth\BoxBreadth=\wd7\fi%
\setbox7=\hbox{(a,c)}%
\ifdim\wd7>\BoxBreadth\BoxBreadth=\wd7\fi%
\setbox7=\hbox{(d,a)}%
\ifdim\wd7>\BoxBreadth\BoxBreadth=\wd7\fi%
\setbox7=\hbox{(c,b)}%
\ifdim\wd7>\BoxBreadth\BoxBreadth=\wd7\fi%
\setbox7=\hbox{(b,c)}%
\ifdim\wd7>\BoxBreadth\BoxBreadth=\wd7\fi%
\setbox7=\hbox{(a,d)}%
\ifdim\wd7>\BoxBreadth\BoxBreadth=\wd7\fi%
\setbox7=\hbox{(e,a)}%
\ifdim\wd7>\BoxBreadth\BoxBreadth=\wd7\fi%
\setbox7=\hbox{(d,b)}%
\ifdim\wd7>\BoxBreadth\BoxBreadth=\wd7\fi%
\setbox7=\hbox{(c,c)}%
\ifdim\wd7>\BoxBreadth\BoxBreadth=\wd7\fi%
\setbox7=\hbox{(b,d)}%
\ifdim\wd7>\BoxBreadth\BoxBreadth=\wd7\fi%
\setbox7=\hbox{(a,e)}%
\ifdim\wd7>\BoxBreadth\BoxBreadth=\wd7\fi%
\setbox7=\hbox{(f,a)}%
\ifdim\wd7>\BoxBreadth\BoxBreadth=\wd7\fi%
\setbox7=\hbox{(e,b)}%
\ifdim\wd7>\BoxBreadth\BoxBreadth=\wd7\fi%
\setbox7=\hbox{(d,c)}%
\ifdim\wd7>\BoxBreadth\BoxBreadth=\wd7\fi%
\setbox7=\hbox{(c,d)}%
\ifdim\wd7>\BoxBreadth\BoxBreadth=\wd7\fi%
\setbox7=\hbox{(b,e)}%
\ifdim\wd7>\BoxBreadth\BoxBreadth=\wd7\fi%
\setbox7=\hbox{(a,f)}%
\ifdim\wd7>\BoxBreadth\BoxBreadth=\wd7\fi%
\setbox7=\hbox{(g,a)}%
\ifdim\wd7>\BoxBreadth\BoxBreadth=\wd7\fi%
\setbox7=\hbox{(f,b)}%
\ifdim\wd7>\BoxBreadth\BoxBreadth=\wd7\fi%
\setbox7=\hbox{(e,c)}%
\ifdim\wd7>\BoxBreadth\BoxBreadth=\wd7\fi%
\setbox7=\hbox{(d,d)}%
\ifdim\wd7>\BoxBreadth\BoxBreadth=\wd7\fi%
\setbox7=\hbox{(c,e)}%
\ifdim\wd7>\BoxBreadth\BoxBreadth=\wd7\fi%
\setbox7=\hbox{(b,f)}%
\ifdim\wd7>\BoxBreadth\BoxBreadth=\wd7\fi%
\setbox7=\hbox{(a,g)}%
\ifdim\wd7>\BoxBreadth\BoxBreadth=\wd7\fi%
\setbox7=\hbox{(g,b)}%
\ifdim\wd7>\BoxBreadth\BoxBreadth=\wd7\fi%
\setbox7=\hbox{(f,c)}%
\ifdim\wd7>\BoxBreadth\BoxBreadth=\wd7\fi%
\setbox7=\hbox{(e,d)}%
\ifdim\wd7>\BoxBreadth\BoxBreadth=\wd7\fi%
\setbox7=\hbox{(d,e)}%
\ifdim\wd7>\BoxBreadth\BoxBreadth=\wd7\fi%
\setbox7=\hbox{(c,f)}%
\ifdim\wd7>\BoxBreadth\BoxBreadth=\wd7\fi%
\setbox7=\hbox{(b,g)}%
\ifdim\wd7>\BoxBreadth\BoxBreadth=\wd7\fi%
\setbox7=\hbox{(g,c)}%
\ifdim\wd7>\BoxBreadth\BoxBreadth=\wd7\fi%
\setbox7=\hbox{(f,d)}%
\ifdim\wd7>\BoxBreadth\BoxBreadth=\wd7\fi%
\setbox7=\hbox{(e,e)}%
\ifdim\wd7>\BoxBreadth\BoxBreadth=\wd7\fi%
\setbox7=\hbox{(d,f)}%
\ifdim\wd7>\BoxBreadth\BoxBreadth=\wd7\fi%
\setbox7=\hbox{(c,g)}%
\ifdim\wd7>\BoxBreadth\BoxBreadth=\wd7\fi%
\setbox7=\hbox{(g,d)}%
\ifdim\wd7>\BoxBreadth\BoxBreadth=\wd7\fi%
\setbox7=\hbox{(f,e)}%
\ifdim\wd7>\BoxBreadth\BoxBreadth=\wd7\fi%
\setbox7=\hbox{(e,f)}%
\ifdim\wd7>\BoxBreadth\BoxBreadth=\wd7\fi%
\setbox7=\hbox{(d,g)}%
\ifdim\wd7>\BoxBreadth\BoxBreadth=\wd7\fi%
\setbox7=\hbox{(g,e)}%
\ifdim\wd7>\BoxBreadth\BoxBreadth=\wd7\fi%
\setbox7=\hbox{(f,f)}%
\ifdim\wd7>\BoxBreadth\BoxBreadth=\wd7\fi%
\setbox7=\hbox{(e,g)}%
\ifdim\wd7>\BoxBreadth\BoxBreadth=\wd7\fi%
\setbox7=\hbox{(g,f)}%
\ifdim\wd7>\BoxBreadth\BoxBreadth=\wd7\fi%
\setbox7=\hbox{(f,g)}%
\ifdim\wd7>\BoxBreadth\BoxBreadth=\wd7\fi%
\setbox7=\hbox{(g,g)}%
\ifdim\wd7>\BoxBreadth\BoxBreadth=\wd7\fi%
\def\RowNames{\def\scalable{\interspacereduction}
\vcenter{\offinterlineskip\baselineskip=\matrixskip%
\hbox to\BoxBreadth{\strut\hfil (a,a)}\kern\scalable%
\hbox to\BoxBreadth{\strut\hfil (b,a)}\kern\scalable%
\hbox to\BoxBreadth{\strut\hfil (a,b)}\kern\scalable%
\hbox to\BoxBreadth{\strut\hfil (c,a)}\kern\scalable%
\hbox to\BoxBreadth{\strut\hfil (b,b)}\kern\scalable%
\hbox to\BoxBreadth{\strut\hfil (a,c)}\kern\scalable%
\hbox to\BoxBreadth{\strut\hfil (d,a)}\kern\scalable%
\hbox to\BoxBreadth{\strut\hfil (c,b)}\kern\scalable%
\hbox to\BoxBreadth{\strut\hfil (b,c)}\kern\scalable%
\hbox to\BoxBreadth{\strut\hfil (a,d)}\kern\scalable%
\hbox to\BoxBreadth{\strut\hfil (e,a)}\kern\scalable%
\hbox to\BoxBreadth{\strut\hfil (d,b)}\kern\scalable%
\hbox to\BoxBreadth{\strut\hfil (c,c)}\kern\scalable%
\hbox to\BoxBreadth{\strut\hfil (b,d)}\kern\scalable%
\hbox to\BoxBreadth{\strut\hfil (a,e)}\kern\scalable%
\hbox to\BoxBreadth{\strut\hfil (f,a)}\kern\scalable%
\hbox to\BoxBreadth{\strut\hfil (e,b)}\kern\scalable%
\hbox to\BoxBreadth{\strut\hfil (d,c)}\kern\scalable%
\hbox to\BoxBreadth{\strut\hfil (c,d)}\kern\scalable%
\hbox to\BoxBreadth{\strut\hfil (b,e)}\kern\scalable%
\hbox to\BoxBreadth{\strut\hfil (a,f)}\kern\scalable%
\hbox to\BoxBreadth{\strut\hfil (g,a)}\kern\scalable%
\hbox to\BoxBreadth{\strut\hfil (f,b)}\kern\scalable%
\hbox to\BoxBreadth{\strut\hfil (e,c)}\kern\scalable%
\hbox to\BoxBreadth{\strut\hfil (d,d)}\kern\scalable%
\hbox to\BoxBreadth{\strut\hfil (c,e)}\kern\scalable%
\hbox to\BoxBreadth{\strut\hfil (b,f)}\kern\scalable%
\hbox to\BoxBreadth{\strut\hfil (a,g)}\kern\scalable%
\hbox to\BoxBreadth{\strut\hfil (g,b)}\kern\scalable%
\hbox to\BoxBreadth{\strut\hfil (f,c)}\kern\scalable%
\hbox to\BoxBreadth{\strut\hfil (e,d)}\kern\scalable%
\hbox to\BoxBreadth{\strut\hfil (d,e)}\kern\scalable%
\hbox to\BoxBreadth{\strut\hfil (c,f)}\kern\scalable%
\hbox to\BoxBreadth{\strut\hfil (b,g)}\kern\scalable%
\hbox to\BoxBreadth{\strut\hfil (g,c)}\kern\scalable%
\hbox to\BoxBreadth{\strut\hfil (f,d)}\kern\scalable%
\hbox to\BoxBreadth{\strut\hfil (e,e)}\kern\scalable%
\hbox to\BoxBreadth{\strut\hfil (d,f)}\kern\scalable%
\hbox to\BoxBreadth{\strut\hfil (c,g)}\kern\scalable%
\hbox to\BoxBreadth{\strut\hfil (g,d)}\kern\scalable%
\hbox to\BoxBreadth{\strut\hfil (f,e)}\kern\scalable%
\hbox to\BoxBreadth{\strut\hfil (e,f)}\kern\scalable%
\hbox to\BoxBreadth{\strut\hfil (d,g)}\kern\scalable%
\hbox to\BoxBreadth{\strut\hfil (g,e)}\kern\scalable%
\hbox to\BoxBreadth{\strut\hfil (f,f)}\kern\scalable%
\hbox to\BoxBreadth{\strut\hfil (e,g)}\kern\scalable%
\hbox to\BoxBreadth{\strut\hfil (g,f)}\kern\scalable%
\hbox to\BoxBreadth{\strut\hfil (f,g)}\kern\scalable%
\hbox to\BoxBreadth{\strut\hfil (g,g)}}}%
\def\ColNames{\def\scalable{\interspacereduction}
\hbox{\rotatebox{90}{\strut a}\kern\scalable%
\rotatebox{90}{\strut b}\kern\scalable%
\rotatebox{90}{\strut c}\kern\scalable%
\rotatebox{90}{\strut d}\kern\scalable%
\rotatebox{90}{\strut e}\kern\scalable%
\rotatebox{90}{\strut f}\kern\scalable%
\rotatebox{90}{\strut g}\kern\scalable%
\kern0pt
}}%
\def\Matrix{\spmatrix{%
\noalign{\kern-2pt}
 {\CoefTrue}&\n&\n&\n&\n&\n&\n\cr
 {\CoefTrue}&\n&\n&\n&\n&\n&\n\cr
 \n&{\CoefTrue}&\n&\n&\n&\n&\n\cr
 {\CoefTrue}&\n&\n&\n&\n&\n&\n\cr
 \n&{\CoefTrue}&\n&\n&\n&\n&\n\cr
 \n&\n&{\CoefTrue}&\n&\n&\n&\n\cr
 {\CoefTrue}&\n&\n&\n&\n&\n&\n\cr
 \n&{\CoefTrue}&\n&\n&\n&\n&\n\cr
 \n&\n&{\CoefTrue}&\n&\n&\n&\n\cr
 \n&\n&\n&{\CoefTrue}&\n&\n&\n\cr
 {\CoefTrue}&\n&\n&\n&\n&\n&\n\cr
 \n&{\CoefTrue}&\n&\n&\n&\n&\n\cr
 \n&\n&{\CoefTrue}&\n&\n&\n&\n\cr
 \n&\n&\n&{\CoefTrue}&\n&\n&\n\cr
 \n&\n&\n&\n&{\CoefTrue}&\n&\n\cr
 {\CoefTrue}&\n&\n&\n&\n&\n&\n\cr
 \n&{\CoefTrue}&\n&\n&\n&\n&\n\cr
 \n&\n&{\CoefTrue}&\n&\n&\n&\n\cr
 \n&\n&\n&{\CoefTrue}&\n&\n&\n\cr
 \n&\n&\n&\n&{\CoefTrue}&\n&\n\cr
 \n&\n&\n&\n&\n&{\CoefTrue}&\n\cr
 {\CoefTrue}&\n&\n&\n&\n&\n&\n\cr
 \n&{\CoefTrue}&\n&\n&\n&\n&\n\cr
 \n&\n&{\CoefTrue}&\n&\n&\n&\n\cr
 \n&\n&\n&{\CoefTrue}&\n&\n&\n\cr
 \n&\n&\n&\n&{\CoefTrue}&\n&\n\cr
 \n&\n&\n&\n&\n&{\CoefTrue}&\n\cr
 \n&\n&\n&\n&\n&\n&{\CoefTrue}\cr
 \n&{\CoefTrue}&\n&\n&\n&\n&\n\cr
 \n&\n&{\CoefTrue}&\n&\n&\n&\n\cr
 \n&\n&\n&{\CoefTrue}&\n&\n&\n\cr
 \n&\n&\n&\n&{\CoefTrue}&\n&\n\cr
 \n&\n&\n&\n&\n&{\CoefTrue}&\n\cr
 \n&\n&\n&\n&\n&\n&{\CoefTrue}\cr
 \n&\n&{\CoefTrue}&\n&\n&\n&\n\cr
 \n&\n&\n&{\CoefTrue}&\n&\n&\n\cr
 \n&\n&\n&\n&{\CoefTrue}&\n&\n\cr
 \n&\n&\n&\n&\n&{\CoefTrue}&\n\cr
 \n&\n&\n&\n&\n&\n&{\CoefTrue}\cr
 \n&\n&\n&{\CoefTrue}&\n&\n&\n\cr
 \n&\n&\n&\n&{\CoefTrue}&\n&\n\cr
 \n&\n&\n&\n&\n&{\CoefTrue}&\n\cr
 \n&\n&\n&\n&\n&\n&{\CoefTrue}\cr
 \n&\n&\n&\n&{\CoefTrue}&\n&\n\cr
 \n&\n&\n&\n&\n&{\CoefTrue}&\n\cr
 \n&\n&\n&\n&\n&\n&{\CoefTrue}\cr
 \n&\n&\n&\n&\n&{\CoefTrue}&\n\cr
 \n&\n&\n&\n&\n&\n&{\CoefTrue}\cr
 \n&\n&\n&\n&\n&\n&{\CoefTrue}\cr
\noalign{\kern-2pt}}}%
\vbox{\setbox8=\hbox{$\RowNames\Matrix$}
\hbox to\wd8{\hfil$\ColNames$\kern\ColEntryShiftHoriz}\kern\ColEntryShiftVerti
\box8}}
\hfil
{\footnotesize%
\BoxBreadth=0pt%
\setbox7=\hbox{(a,a)}%
\ifdim\wd7>\BoxBreadth\BoxBreadth=\wd7\fi%
\setbox7=\hbox{(b,a)}%
\ifdim\wd7>\BoxBreadth\BoxBreadth=\wd7\fi%
\setbox7=\hbox{(a,b)}%
\ifdim\wd7>\BoxBreadth\BoxBreadth=\wd7\fi%
\setbox7=\hbox{(c,a)}%
\ifdim\wd7>\BoxBreadth\BoxBreadth=\wd7\fi%
\setbox7=\hbox{(b,b)}%
\ifdim\wd7>\BoxBreadth\BoxBreadth=\wd7\fi%
\setbox7=\hbox{(a,c)}%
\ifdim\wd7>\BoxBreadth\BoxBreadth=\wd7\fi%
\setbox7=\hbox{(d,a)}%
\ifdim\wd7>\BoxBreadth\BoxBreadth=\wd7\fi%
\setbox7=\hbox{(c,b)}%
\ifdim\wd7>\BoxBreadth\BoxBreadth=\wd7\fi%
\setbox7=\hbox{(b,c)}%
\ifdim\wd7>\BoxBreadth\BoxBreadth=\wd7\fi%
\setbox7=\hbox{(a,d)}%
\ifdim\wd7>\BoxBreadth\BoxBreadth=\wd7\fi%
\setbox7=\hbox{(e,a)}%
\ifdim\wd7>\BoxBreadth\BoxBreadth=\wd7\fi%
\setbox7=\hbox{(d,b)}%
\ifdim\wd7>\BoxBreadth\BoxBreadth=\wd7\fi%
\setbox7=\hbox{(c,c)}%
\ifdim\wd7>\BoxBreadth\BoxBreadth=\wd7\fi%
\setbox7=\hbox{(b,d)}%
\ifdim\wd7>\BoxBreadth\BoxBreadth=\wd7\fi%
\setbox7=\hbox{(a,e)}%
\ifdim\wd7>\BoxBreadth\BoxBreadth=\wd7\fi%
\setbox7=\hbox{(f,a)}%
\ifdim\wd7>\BoxBreadth\BoxBreadth=\wd7\fi%
\setbox7=\hbox{(e,b)}%
\ifdim\wd7>\BoxBreadth\BoxBreadth=\wd7\fi%
\setbox7=\hbox{(d,c)}%
\ifdim\wd7>\BoxBreadth\BoxBreadth=\wd7\fi%
\setbox7=\hbox{(c,d)}%
\ifdim\wd7>\BoxBreadth\BoxBreadth=\wd7\fi%
\setbox7=\hbox{(b,e)}%
\ifdim\wd7>\BoxBreadth\BoxBreadth=\wd7\fi%
\setbox7=\hbox{(a,f)}%
\ifdim\wd7>\BoxBreadth\BoxBreadth=\wd7\fi%
\setbox7=\hbox{(g,a)}%
\ifdim\wd7>\BoxBreadth\BoxBreadth=\wd7\fi%
\setbox7=\hbox{(f,b)}%
\ifdim\wd7>\BoxBreadth\BoxBreadth=\wd7\fi%
\setbox7=\hbox{(e,c)}%
\ifdim\wd7>\BoxBreadth\BoxBreadth=\wd7\fi%
\setbox7=\hbox{(d,d)}%
\ifdim\wd7>\BoxBreadth\BoxBreadth=\wd7\fi%
\setbox7=\hbox{(c,e)}%
\ifdim\wd7>\BoxBreadth\BoxBreadth=\wd7\fi%
\setbox7=\hbox{(b,f)}%
\ifdim\wd7>\BoxBreadth\BoxBreadth=\wd7\fi%
\setbox7=\hbox{(a,g)}%
\ifdim\wd7>\BoxBreadth\BoxBreadth=\wd7\fi%
\setbox7=\hbox{(g,b)}%
\ifdim\wd7>\BoxBreadth\BoxBreadth=\wd7\fi%
\setbox7=\hbox{(f,c)}%
\ifdim\wd7>\BoxBreadth\BoxBreadth=\wd7\fi%
\setbox7=\hbox{(e,d)}%
\ifdim\wd7>\BoxBreadth\BoxBreadth=\wd7\fi%
\setbox7=\hbox{(d,e)}%
\ifdim\wd7>\BoxBreadth\BoxBreadth=\wd7\fi%
\setbox7=\hbox{(c,f)}%
\ifdim\wd7>\BoxBreadth\BoxBreadth=\wd7\fi%
\setbox7=\hbox{(b,g)}%
\ifdim\wd7>\BoxBreadth\BoxBreadth=\wd7\fi%
\setbox7=\hbox{(g,c)}%
\ifdim\wd7>\BoxBreadth\BoxBreadth=\wd7\fi%
\setbox7=\hbox{(f,d)}%
\ifdim\wd7>\BoxBreadth\BoxBreadth=\wd7\fi%
\setbox7=\hbox{(e,e)}%
\ifdim\wd7>\BoxBreadth\BoxBreadth=\wd7\fi%
\setbox7=\hbox{(d,f)}%
\ifdim\wd7>\BoxBreadth\BoxBreadth=\wd7\fi%
\setbox7=\hbox{(c,g)}%
\ifdim\wd7>\BoxBreadth\BoxBreadth=\wd7\fi%
\setbox7=\hbox{(g,d)}%
\ifdim\wd7>\BoxBreadth\BoxBreadth=\wd7\fi%
\setbox7=\hbox{(f,e)}%
\ifdim\wd7>\BoxBreadth\BoxBreadth=\wd7\fi%
\setbox7=\hbox{(e,f)}%
\ifdim\wd7>\BoxBreadth\BoxBreadth=\wd7\fi%
\setbox7=\hbox{(d,g)}%
\ifdim\wd7>\BoxBreadth\BoxBreadth=\wd7\fi%
\setbox7=\hbox{(g,e)}%
\ifdim\wd7>\BoxBreadth\BoxBreadth=\wd7\fi%
\setbox7=\hbox{(f,f)}%
\ifdim\wd7>\BoxBreadth\BoxBreadth=\wd7\fi%
\setbox7=\hbox{(e,g)}%
\ifdim\wd7>\BoxBreadth\BoxBreadth=\wd7\fi%
\setbox7=\hbox{(g,f)}%
\ifdim\wd7>\BoxBreadth\BoxBreadth=\wd7\fi%
\setbox7=\hbox{(f,g)}%
\ifdim\wd7>\BoxBreadth\BoxBreadth=\wd7\fi%
\setbox7=\hbox{(g,g)}%
\ifdim\wd7>\BoxBreadth\BoxBreadth=\wd7\fi%
\def\RowNames{\def\scalable{\interspacereduction}
\vcenter{\offinterlineskip\baselineskip=\matrixskip%
\hbox to\BoxBreadth{\strut\hfil (a,a)}\kern\scalable%
\hbox to\BoxBreadth{\strut\hfil (b,a)}\kern\scalable%
\hbox to\BoxBreadth{\strut\hfil (a,b)}\kern\scalable%
\hbox to\BoxBreadth{\strut\hfil (c,a)}\kern\scalable%
\hbox to\BoxBreadth{\strut\hfil (b,b)}\kern\scalable%
\hbox to\BoxBreadth{\strut\hfil (a,c)}\kern\scalable%
\hbox to\BoxBreadth{\strut\hfil (d,a)}\kern\scalable%
\hbox to\BoxBreadth{\strut\hfil (c,b)}\kern\scalable%
\hbox to\BoxBreadth{\strut\hfil (b,c)}\kern\scalable%
\hbox to\BoxBreadth{\strut\hfil (a,d)}\kern\scalable%
\hbox to\BoxBreadth{\strut\hfil (e,a)}\kern\scalable%
\hbox to\BoxBreadth{\strut\hfil (d,b)}\kern\scalable%
\hbox to\BoxBreadth{\strut\hfil (c,c)}\kern\scalable%
\hbox to\BoxBreadth{\strut\hfil (b,d)}\kern\scalable%
\hbox to\BoxBreadth{\strut\hfil (a,e)}\kern\scalable%
\hbox to\BoxBreadth{\strut\hfil (f,a)}\kern\scalable%
\hbox to\BoxBreadth{\strut\hfil (e,b)}\kern\scalable%
\hbox to\BoxBreadth{\strut\hfil (d,c)}\kern\scalable%
\hbox to\BoxBreadth{\strut\hfil (c,d)}\kern\scalable%
\hbox to\BoxBreadth{\strut\hfil (b,e)}\kern\scalable%
\hbox to\BoxBreadth{\strut\hfil (a,f)}\kern\scalable%
\hbox to\BoxBreadth{\strut\hfil (g,a)}\kern\scalable%
\hbox to\BoxBreadth{\strut\hfil (f,b)}\kern\scalable%
\hbox to\BoxBreadth{\strut\hfil (e,c)}\kern\scalable%
\hbox to\BoxBreadth{\strut\hfil (d,d)}\kern\scalable%
\hbox to\BoxBreadth{\strut\hfil (c,e)}\kern\scalable%
\hbox to\BoxBreadth{\strut\hfil (b,f)}\kern\scalable%
\hbox to\BoxBreadth{\strut\hfil (a,g)}\kern\scalable%
\hbox to\BoxBreadth{\strut\hfil (g,b)}\kern\scalable%
\hbox to\BoxBreadth{\strut\hfil (f,c)}\kern\scalable%
\hbox to\BoxBreadth{\strut\hfil (e,d)}\kern\scalable%
\hbox to\BoxBreadth{\strut\hfil (d,e)}\kern\scalable%
\hbox to\BoxBreadth{\strut\hfil (c,f)}\kern\scalable%
\hbox to\BoxBreadth{\strut\hfil (b,g)}\kern\scalable%
\hbox to\BoxBreadth{\strut\hfil (g,c)}\kern\scalable%
\hbox to\BoxBreadth{\strut\hfil (f,d)}\kern\scalable%
\hbox to\BoxBreadth{\strut\hfil (e,e)}\kern\scalable%
\hbox to\BoxBreadth{\strut\hfil (d,f)}\kern\scalable%
\hbox to\BoxBreadth{\strut\hfil (c,g)}\kern\scalable%
\hbox to\BoxBreadth{\strut\hfil (g,d)}\kern\scalable%
\hbox to\BoxBreadth{\strut\hfil (f,e)}\kern\scalable%
\hbox to\BoxBreadth{\strut\hfil (e,f)}\kern\scalable%
\hbox to\BoxBreadth{\strut\hfil (d,g)}\kern\scalable%
\hbox to\BoxBreadth{\strut\hfil (g,e)}\kern\scalable%
\hbox to\BoxBreadth{\strut\hfil (f,f)}\kern\scalable%
\hbox to\BoxBreadth{\strut\hfil (e,g)}\kern\scalable%
\hbox to\BoxBreadth{\strut\hfil (g,f)}\kern\scalable%
\hbox to\BoxBreadth{\strut\hfil (f,g)}\kern\scalable%
\hbox to\BoxBreadth{\strut\hfil (g,g)}}}%
\def\ColNames{\def\scalable{\interspacereduction}
\hbox{\rotatebox{90}{\strut a}\kern\scalable%
\rotatebox{90}{\strut b}\kern\scalable%
\rotatebox{90}{\strut c}\kern\scalable%
\rotatebox{90}{\strut d}\kern\scalable%
\rotatebox{90}{\strut e}\kern\scalable%
\rotatebox{90}{\strut f}\kern\scalable%
\rotatebox{90}{\strut g}\kern\scalable%
\kern0pt
}}%
\def\Matrix{\spmatrix{%
\noalign{\kern-2pt}
 \n&\n&\n&\n&\n&\n&\n\cr
 \n&\n&\n&\n&\n&\n&\n\cr
 \n&\n&\n&\n&\n&\n&\n\cr
 {\CoefTrue}&\n&\n&\n&\n&\n&\n\cr
 \n&\n&\n&\n&\n&\n&\n\cr
 \n&\n&\n&\n&\n&\n&\n\cr
 \n&\n&\n&\n&\n&\n&\n\cr
 \n&{\CoefTrue}&\n&\n&\n&\n&\n\cr
 \n&\n&\n&\n&\n&\n&\n\cr
 \n&\n&\n&\n&\n&\n&\n\cr
 \n&\n&\n&\n&\n&\n&\n\cr
 \n&\n&\n&\n&\n&\n&\n\cr
 \n&\n&{\CoefTrue}&\n&\n&\n&\n\cr
 \n&\n&\n&\n&\n&\n&\n\cr
 \n&\n&\n&\n&\n&\n&\n\cr
 \n&\n&\n&\n&\n&\n&\n\cr
 \n&\n&\n&\n&\n&\n&\n\cr
 \n&\n&\n&\n&\n&\n&\n\cr
 \n&\n&\n&{\CoefTrue}&\n&\n&\n\cr
 \n&\n&\n&\n&\n&\n&\n\cr
 \n&\n&\n&\n&\n&\n&\n\cr
 \n&\n&\n&\n&\n&\n&\n\cr
 \n&\n&\n&\n&\n&\n&\n\cr
 \n&\n&\n&\n&\n&\n&\n\cr
 \n&\n&\n&\n&\n&\n&\n\cr
 \n&\n&\n&\n&{\CoefTrue}&\n&\n\cr
 \n&\n&\n&\n&\n&\n&\n\cr
 \n&\n&\n&\n&\n&\n&\n\cr
 \n&\n&\n&\n&\n&\n&\n\cr
 \n&\n&\n&\n&\n&\n&\n\cr
 \n&\n&\n&\n&\n&\n&\n\cr
 \n&\n&\n&\n&\n&\n&\n\cr
 \n&\n&\n&\n&\n&{\CoefTrue}&\n\cr
 \n&\n&\n&\n&\n&\n&\n\cr
 \n&\n&\n&\n&\n&\n&\n\cr
 \n&\n&\n&\n&\n&\n&\n\cr
 \n&\n&\n&\n&\n&\n&\n\cr
 \n&\n&\n&\n&\n&\n&\n\cr
 \n&\n&\n&\n&\n&\n&{\CoefTrue}\cr
 \n&\n&\n&\n&\n&\n&\n\cr
 \n&\n&\n&\n&\n&\n&\n\cr
 \n&\n&\n&\n&\n&\n&\n\cr
 \n&\n&\n&\n&\n&\n&\n\cr
 \n&\n&\n&\n&\n&\n&\n\cr
 \n&\n&\n&\n&\n&\n&\n\cr
 \n&\n&\n&\n&\n&\n&\n\cr
 \n&\n&\n&\n&\n&\n&\n\cr
 \n&\n&\n&\n&\n&\n&\n\cr
 \n&\n&\n&\n&\n&\n&\n\cr
\noalign{\kern-2pt}}}%
\vbox{\setbox8=\hbox{$\RowNames\Matrix$}
\hbox to\wd8{\hfil$\ColNames$\kern\ColEntryShiftHoriz}\kern\ColEntryShiftVerti
\box8}}\hfil}
\hbox to\textwidth{\hfil table $\TarskiAdd$\hfil relation $\TarskiAdd$\hfil$\rho$\hfil$f\RELtraOP$\hfil}}}
{Binary map as table and as relation, projection $\rho$ and transposed mapping $f$ for point $c$}{FigBinMap}

\noindent
Given $x$, we map with $f$ every $y$ to the pair $(x,y)$. By symmetry, $g:=(\pi\RELandOP\rho\RELcompOP x\RELcompOP\RELtop)\RELtraOP$ is also a mapping; it satisfies $g\RELcompOP\pi=\RELide$ and $\pi=\leftResi{g}{\RELide}$.

\enunc{}{Definition}{}{DefBinOp} Given this setting, we define as follows:
\begin{enumerate}[i)]
\item $\TarskiSwitch:=\pi\RELcompOP{\rho}\RELtraOP\RELandOP\rho\RELcompOP{\pi}\RELtraOP$ flips components of a pair.
\item $\TarskiAdd$ {\bf commutative}\index{commutative}\quad $:\Longleftrightarrow\quad\TarskiSwitch\RELcompOP\TarskiAdd=\TarskiAdd$.
\item The shuffling for the associative\index{associative} law is achieved by one version of the following
\item[]$\;\TarskiShuffle:=
\pi'\RELcompOP\pi\RELcompOP\pi_1\RELtraOP\RELandOP\pi'\RELcompOP\rho\RELcompOP\pi\RELtraOP\RELcompOP\rho_1\RELtraOP\RELandOP\rho'\RELcompOP\rho\RELtraOP\RELcompOP\rho_1\RELtraOP$\quad or, grouped suitably,
\item[]\qquad=
$\pi'\RELcompOP\pi\RELcompOP\pi_1\RELtraOP\RELandOP(\pi'\RELcompOP\rho\RELcompOP\pi\RELtraOP\RELandOP\rho'\RELcompOP\rho\RELtraOP)\RELcompOP\rho_1\RELtraOP=
\StrictFork{\pi'\RELcompOP\pi}{\Kronecker{\rho}{\RELide}}$
\item[]\qquad=
$\pi'\RELcompOP(\pi\RELcompOP\pi_1\RELtraOP\RELandOP\rho\RELcompOP\pi\RELtraOP\RELcompOP\rho_1\RELtraOP)\RELandOP\rho'\RELcompOP\rho\RELtraOP\RELcompOP\rho_1\RELtraOP
=
\StrictJoin{\Kronecker{\RELide}{\pi\RELtraOP}}{\rho\RELtraOP\RELcompOP\rho_1\RELtraOP}$
\item $\TarskiAdd$ {\bf associative}\quad $:\Longleftrightarrow \quad
\Kronecker{\TarskiAdd}{\RELide_X}\RELcompOP\,\TarskiAdd
=\TarskiShuffle\RELcompOP\Kronecker{\RELide_X}{\TarskiAdd}\RELcompOP\,\TarskiAdd$
\Bewende
\end{enumerate}

\noindent
The associativity condition is here given in an acceptably concise form; written down without sufficient care, it appears considerably longer.

\enunc{}{Lemma}{}{LemSwap} Several identities for $\TarskiSwitch,\TarskiShuffle$ --- correct typing assumed.

\begin{enumerate}[i)]
\item $\TarskiSwitch,\TarskiShuffle$ are bijective mappings.
\item $\TarskiSwitch\RELtraOP=\TarskiSwitch$
\item $\TarskiSwitch\RELcompOP\Kronecker{R}{S}=\Kronecker{S}{R}\RELcompOP\,\TarskiSwitch'$
\item $\TarskiSwitch\RELcompOP\StrictJoin{R}{S}=\StrictJoin{S}{R}\qquad\StrictFork{R}{S}\RELcompOP\,\TarskiSwitch'=\StrictFork{S}{R}$
\item $\,\TarskiShuffle\RELcompOP\Kronecker{Q}{\!\Kronecker{R}{S}}=\Kronecker{\Kronecker{Q}{R}\!}{S}\RELcompOP\,\,\TarskiShuffle'$
\item $\!\StrictFork{Q}{\!\StrictFork{R}{S}}=\StrictFork{\StrictFork{Q}{R}\!}{S}\RELcompOP\,\,\TarskiShuffle'$
\end{enumerate}

\Proof iii) $\TarskiSwitch\RELcompOP\Kronecker{R}{S}
=
(\pi\RELcompOP{\rho}\RELtraOP\RELandOP\rho\RELcompOP{\pi}\RELtraOP)\RELcompOP(\pi\RELcompOP R\RELcompOP{\pi'}\RELtraOP\RELandOP\rho\RELcompOP S\RELcompOP{\rho'}\RELtraOP)
$\quad by definition

$=
(\pi\RELcompOP{\rho}\RELtraOP\RELandOP\rho\RELcompOP{\pi}\RELtraOP)\RELcompOP\pi\RELcompOP R\RELcompOP{\pi'}\RELtraOP\RELandOP(\pi\RELcompOP{\rho}\RELtraOP\RELandOP\rho\RELcompOP{\pi}\RELtraOP)\RELcompOP\rho\RELcompOP S\RELcompOP{\rho'}\RELtraOP
$\quad since $P$ is univalent

$=
(\pi\RELcompOP{\rho}\RELtraOP\RELcompOP\pi\RELandOP\rho)\RELcompOP R\RELcompOP{\pi'}\RELtraOP\RELandOP(\pi\RELandOP\rho\RELcompOP{\pi}\RELtraOP\RELcompOP\rho)\RELcompOP S\RELcompOP{\rho'}\RELtraOP
$\quad Prop.~\PropPiKrhoEquGeneral.ii

$=
\rho\RELcompOP R\RELcompOP{\pi'}\RELtraOP\RELandOP\pi\RELcompOP S\RELcompOP{\rho'}\RELtraOP
=
\pi\RELcompOP S\RELcompOP{\rho'}\RELtraOP\RELandOP\rho\RELcompOP R\RELcompOP{\pi'}\RELtraOP
$\quad $\pi,\rho$ are projections

\bigskip
\noindent
Similarly from the other side:

$\Kronecker{S}{R}\RELcompOP\,\TarskiSwitch'
$

$=(\pi\RELcompOP S\RELcompOP{\pi'}\RELtraOP\RELandOP\rho\RELcompOP R\RELcompOP{\rho'}\RELtraOP)
\RELcompOP
(\pi'\RELcompOP{\rho'}\RELtraOP\RELandOP\rho'\RELcompOP{\pi'}\RELtraOP)
$

$=
\pi\RELcompOP S\RELcompOP{\pi'}\RELtraOP\RELcompOP
(\pi'\RELcompOP{\rho'}\RELtraOP\RELandOP\rho'\RELcompOP{\pi'}\RELtraOP)\RELandOP\rho\RELcompOP R\RELcompOP{\rho'}\RELtraOP
\RELcompOP
(\pi'\RELcompOP{\rho'}\RELtraOP\RELandOP\rho'\RELcompOP{\pi'}\RELtraOP)
$

$=
\pi\RELcompOP S\RELcompOP
({\rho'}\RELtraOP\RELandOP{\pi'}\RELtraOP\RELcompOP\rho'\RELcompOP{\pi'}\RELtraOP)\RELandOP\rho\RELcompOP R\RELcompOP
({\rho'}\RELtraOP\RELcompOP\pi'\RELcompOP{\rho'}\RELtraOP\RELandOP{\pi'}\RELtraOP)
$

$=
\pi\RELcompOP S\RELcompOP
{\rho'}\RELtraOP\RELandOP\rho\RELcompOP R\RELcompOP
{\pi'}\RELtraOP
$

\bigskip
\noindent
iv) $\TarskiSwitch\RELcompOP\StrictJoin{R}{S}
=
(\pi\RELcompOP{\rho}\RELtraOP\RELandOP\rho\RELcompOP{\pi}\RELtraOP)\RELcompOP(\pi\RELcompOP R\RELandOP\rho\RELcompOP S)
$\quad by definition

$=
(\pi\RELcompOP{\rho}\RELtraOP\RELandOP\rho\RELcompOP{\pi}\RELtraOP)\RELcompOP\pi\RELcompOP R\RELandOP(\pi\RELcompOP{\rho}\RELtraOP\RELandOP\rho\RELcompOP{\pi}\RELtraOP)\RELcompOP\rho\RELcompOP S
$\quad since $P$ is univalent

$=
(\pi\RELcompOP{\rho}\RELtraOP\RELcompOP\pi\RELandOP\rho)\RELcompOP R\RELandOP(\pi\RELandOP\rho\RELcompOP{\pi}\RELtraOP\RELcompOP\rho)\RELcompOP S
$\quad Prop.~\PropPiKrhoEquGeneral.ii

$=
\rho\RELcompOP R\RELandOP\pi\RELcompOP S
=
\pi\RELcompOP S\RELandOP\rho\RELcompOP R
$\quad $\pi,\rho$ are projections

\Caption{$\vcenter{\hbox to\textwidth{\hfil\includegraphics[scale=0.45]{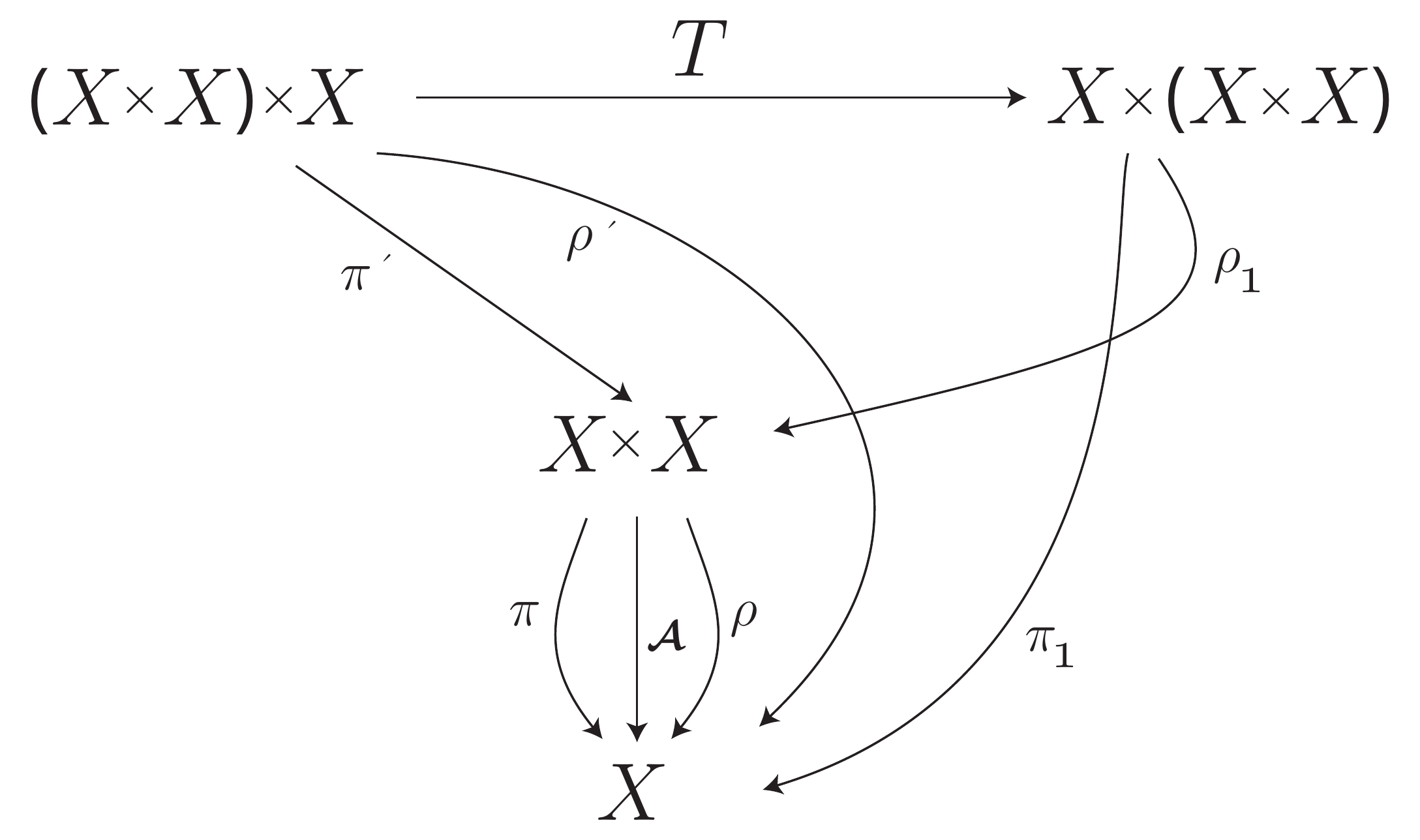}\hfil}
\kern\baselineskip
\hbox to\textwidth{\hfil{\large$\TarskiShuffle=$} {\footnotesize%
\BoxBreadth=0pt%
\setbox7=\hbox{((1,1),1)}%
\ifdim\wd7>\BoxBreadth\BoxBreadth=\wd7\fi%
\setbox7=\hbox{((2,1),1)}%
\ifdim\wd7>\BoxBreadth\BoxBreadth=\wd7\fi%
\setbox7=\hbox{((1,1),2)}%
\ifdim\wd7>\BoxBreadth\BoxBreadth=\wd7\fi%
\setbox7=\hbox{((1,2),1)}%
\ifdim\wd7>\BoxBreadth\BoxBreadth=\wd7\fi%
\setbox7=\hbox{((2,1),2)}%
\ifdim\wd7>\BoxBreadth\BoxBreadth=\wd7\fi%
\setbox7=\hbox{((1,1),3)}%
\ifdim\wd7>\BoxBreadth\BoxBreadth=\wd7\fi%
\setbox7=\hbox{((3,1),1)}%
\ifdim\wd7>\BoxBreadth\BoxBreadth=\wd7\fi%
\setbox7=\hbox{((1,2),2)}%
\ifdim\wd7>\BoxBreadth\BoxBreadth=\wd7\fi%
\setbox7=\hbox{((2,1),3)}%
\ifdim\wd7>\BoxBreadth\BoxBreadth=\wd7\fi%
\setbox7=\hbox{((2,2),1)}%
\ifdim\wd7>\BoxBreadth\BoxBreadth=\wd7\fi%
\setbox7=\hbox{((3,1),2)}%
\ifdim\wd7>\BoxBreadth\BoxBreadth=\wd7\fi%
\setbox7=\hbox{((1,2),3)}%
\ifdim\wd7>\BoxBreadth\BoxBreadth=\wd7\fi%
\setbox7=\hbox{((1,3),1)}%
\ifdim\wd7>\BoxBreadth\BoxBreadth=\wd7\fi%
\setbox7=\hbox{((2,2),2)}%
\ifdim\wd7>\BoxBreadth\BoxBreadth=\wd7\fi%
\setbox7=\hbox{((3,1),3)}%
\ifdim\wd7>\BoxBreadth\BoxBreadth=\wd7\fi%
\setbox7=\hbox{((3,2),1)}%
\ifdim\wd7>\BoxBreadth\BoxBreadth=\wd7\fi%
\setbox7=\hbox{((1,3),2)}%
\ifdim\wd7>\BoxBreadth\BoxBreadth=\wd7\fi%
\setbox7=\hbox{((2,2),3)}%
\ifdim\wd7>\BoxBreadth\BoxBreadth=\wd7\fi%
\setbox7=\hbox{((2,3),1)}%
\ifdim\wd7>\BoxBreadth\BoxBreadth=\wd7\fi%
\setbox7=\hbox{((3,2),2)}%
\ifdim\wd7>\BoxBreadth\BoxBreadth=\wd7\fi%
\setbox7=\hbox{((1,3),3)}%
\ifdim\wd7>\BoxBreadth\BoxBreadth=\wd7\fi%
\setbox7=\hbox{((3,3),1)}%
\ifdim\wd7>\BoxBreadth\BoxBreadth=\wd7\fi%
\setbox7=\hbox{((2,3),2)}%
\ifdim\wd7>\BoxBreadth\BoxBreadth=\wd7\fi%
\setbox7=\hbox{((3,2),3)}%
\ifdim\wd7>\BoxBreadth\BoxBreadth=\wd7\fi%
\setbox7=\hbox{((3,3),2)}%
\ifdim\wd7>\BoxBreadth\BoxBreadth=\wd7\fi%
\setbox7=\hbox{((2,3),3)}%
\ifdim\wd7>\BoxBreadth\BoxBreadth=\wd7\fi%
\setbox7=\hbox{((3,3),3)}%
\ifdim\wd7>\BoxBreadth\BoxBreadth=\wd7\fi%
\def\RowNames{\vcenter{\offinterlineskip\baselineskip=\matrixskip%
\hbox to\BoxBreadth{\strut\hfil ((1,1),1)}\kern\interspacereduction%
\hbox to\BoxBreadth{\strut\hfil ((2,1),1)}\kern\interspacereduction%
\hbox to\BoxBreadth{\strut\hfil ((1,1),2)}\kern\interspacereduction%
\hbox to\BoxBreadth{\strut\hfil ((1,2),1)}\kern\interspacereduction%
\hbox to\BoxBreadth{\strut\hfil ((2,1),2)}\kern\interspacereduction%
\hbox to\BoxBreadth{\strut\hfil ((1,1),3)}\kern\interspacereduction%
\hbox to\BoxBreadth{\strut\hfil ((3,1),1)}\kern\interspacereduction%
\hbox to\BoxBreadth{\strut\hfil ((1,2),2)}\kern\interspacereduction%
\hbox to\BoxBreadth{\strut\hfil ((2,1),3)}\kern\interspacereduction%
\hbox to\BoxBreadth{\strut\hfil ((2,2),1)}\kern\interspacereduction%
\hbox to\BoxBreadth{\strut\hfil ((3,1),2)}\kern\interspacereduction%
\hbox to\BoxBreadth{\strut\hfil ((1,2),3)}\kern\interspacereduction%
\hbox to\BoxBreadth{\strut\hfil ((1,3),1)}\kern\interspacereduction%
\hbox to\BoxBreadth{\strut\hfil ((2,2),2)}\kern\interspacereduction%
\hbox to\BoxBreadth{\strut\hfil ((3,1),3)}\kern\interspacereduction%
\hbox to\BoxBreadth{\strut\hfil ((3,2),1)}\kern\interspacereduction%
\hbox to\BoxBreadth{\strut\hfil ((1,3),2)}\kern\interspacereduction%
\hbox to\BoxBreadth{\strut\hfil ((2,2),3)}\kern\interspacereduction%
\hbox to\BoxBreadth{\strut\hfil ((2,3),1)}\kern\interspacereduction%
\hbox to\BoxBreadth{\strut\hfil ((3,2),2)}\kern\interspacereduction%
\hbox to\BoxBreadth{\strut\hfil ((1,3),3)}\kern\interspacereduction%
\hbox to\BoxBreadth{\strut\hfil ((3,3),1)}\kern\interspacereduction%
\hbox to\BoxBreadth{\strut\hfil ((2,3),2)}\kern\interspacereduction%
\hbox to\BoxBreadth{\strut\hfil ((3,2),3)}\kern\interspacereduction%
\hbox to\BoxBreadth{\strut\hfil ((3,3),2)}\kern\interspacereduction%
\hbox to\BoxBreadth{\strut\hfil ((2,3),3)}\kern\interspacereduction%
\hbox to\BoxBreadth{\strut\hfil ((3,3),3)}}}%
\def\ColNames{\hbox{\rotatebox{90}{\strut (1,(1,1))}\kern\interspacereduction%
\rotatebox{90}{\strut (2,(1,1))}\kern\interspacereduction%
\rotatebox{90}{\strut (1,(2,1))}\kern\interspacereduction%
\rotatebox{90}{\strut (3,(1,1))}\kern\interspacereduction%
\rotatebox{90}{\strut (2,(2,1))}\kern\interspacereduction%
\rotatebox{90}{\strut (1,(1,2))}\kern\interspacereduction%
\rotatebox{90}{\strut (3,(2,1))}\kern\interspacereduction%
\rotatebox{90}{\strut (2,(1,2))}\kern\interspacereduction%
\rotatebox{90}{\strut (1,(3,1))}\kern\interspacereduction%
\rotatebox{90}{\strut (3,(1,2))}\kern\interspacereduction%
\rotatebox{90}{\strut (2,(3,1))}\kern\interspacereduction%
\rotatebox{90}{\strut (1,(2,2))}\kern\interspacereduction%
\rotatebox{90}{\strut (3,(3,1))}\kern\interspacereduction%
\rotatebox{90}{\strut (2,(2,2))}\kern\interspacereduction%
\rotatebox{90}{\strut (1,(1,3))}\kern\interspacereduction%
\rotatebox{90}{\strut (3,(2,2))}\kern\interspacereduction%
\rotatebox{90}{\strut (2,(1,3))}\kern\interspacereduction%
\rotatebox{90}{\strut (1,(3,2))}\kern\interspacereduction%
\rotatebox{90}{\strut (3,(1,3))}\kern\interspacereduction%
\rotatebox{90}{\strut (2,(3,2))}\kern\interspacereduction%
\rotatebox{90}{\strut (1,(2,3))}\kern\interspacereduction%
\rotatebox{90}{\strut (3,(3,2))}\kern\interspacereduction%
\rotatebox{90}{\strut (2,(2,3))}\kern\interspacereduction%
\rotatebox{90}{\strut (1,(3,3))}\kern\interspacereduction%
\rotatebox{90}{\strut (3,(2,3))}\kern\interspacereduction%
\rotatebox{90}{\strut (2,(3,3))}\kern\interspacereduction%
\rotatebox{90}{\strut (3,(3,3))}\kern\interspacereduction%
}}%
\def\Matrix{\spmatrix{%
\noalign{\kern-2pt}
 {\CoefTrue}&\n&\n&\n&\n&\n&\n&\n&\n&\n&\n&\n&\n&\n&\n&\n&\n&\n&\n&\n&\n&\n&\n&\n&\n&\n&\n\cr
 \n&{\CoefTrue}&\n&\n&\n&\n&\n&\n&\n&\n&\n&\n&\n&\n&\n&\n&\n&\n&\n&\n&\n&\n&\n&\n&\n&\n&\n\cr
 \n&\n&\n&\n&\n&{\CoefTrue}&\n&\n&\n&\n&\n&\n&\n&\n&\n&\n&\n&\n&\n&\n&\n&\n&\n&\n&\n&\n&\n\cr
 \n&\n&{\CoefTrue}&\n&\n&\n&\n&\n&\n&\n&\n&\n&\n&\n&\n&\n&\n&\n&\n&\n&\n&\n&\n&\n&\n&\n&\n\cr
 \n&\n&\n&\n&\n&\n&\n&{\CoefTrue}&\n&\n&\n&\n&\n&\n&\n&\n&\n&\n&\n&\n&\n&\n&\n&\n&\n&\n&\n\cr
 \n&\n&\n&\n&\n&\n&\n&\n&\n&\n&\n&\n&\n&\n&{\CoefTrue}&\n&\n&\n&\n&\n&\n&\n&\n&\n&\n&\n&\n\cr
 \n&\n&\n&{\CoefTrue}&\n&\n&\n&\n&\n&\n&\n&\n&\n&\n&\n&\n&\n&\n&\n&\n&\n&\n&\n&\n&\n&\n&\n\cr
 \n&\n&\n&\n&\n&\n&\n&\n&\n&\n&\n&{\CoefTrue}&\n&\n&\n&\n&\n&\n&\n&\n&\n&\n&\n&\n&\n&\n&\n\cr
 \n&\n&\n&\n&\n&\n&\n&\n&\n&\n&\n&\n&\n&\n&\n&\n&{\CoefTrue}&\n&\n&\n&\n&\n&\n&\n&\n&\n&\n\cr
 \n&\n&\n&\n&{\CoefTrue}&\n&\n&\n&\n&\n&\n&\n&\n&\n&\n&\n&\n&\n&\n&\n&\n&\n&\n&\n&\n&\n&\n\cr
 \n&\n&\n&\n&\n&\n&\n&\n&\n&{\CoefTrue}&\n&\n&\n&\n&\n&\n&\n&\n&\n&\n&\n&\n&\n&\n&\n&\n&\n\cr
 \n&\n&\n&\n&\n&\n&\n&\n&\n&\n&\n&\n&\n&\n&\n&\n&\n&\n&\n&\n&{\CoefTrue}&\n&\n&\n&\n&\n&\n\cr
 \n&\n&\n&\n&\n&\n&\n&\n&{\CoefTrue}&\n&\n&\n&\n&\n&\n&\n&\n&\n&\n&\n&\n&\n&\n&\n&\n&\n&\n\cr
 \n&\n&\n&\n&\n&\n&\n&\n&\n&\n&\n&\n&\n&{\CoefTrue}&\n&\n&\n&\n&\n&\n&\n&\n&\n&\n&\n&\n&\n\cr
 \n&\n&\n&\n&\n&\n&\n&\n&\n&\n&\n&\n&\n&\n&\n&\n&\n&\n&{\CoefTrue}&\n&\n&\n&\n&\n&\n&\n&\n\cr
 \n&\n&\n&\n&\n&\n&{\CoefTrue}&\n&\n&\n&\n&\n&\n&\n&\n&\n&\n&\n&\n&\n&\n&\n&\n&\n&\n&\n&\n\cr
 \n&\n&\n&\n&\n&\n&\n&\n&\n&\n&\n&\n&\n&\n&\n&\n&\n&{\CoefTrue}&\n&\n&\n&\n&\n&\n&\n&\n&\n\cr
 \n&\n&\n&\n&\n&\n&\n&\n&\n&\n&\n&\n&\n&\n&\n&\n&\n&\n&\n&\n&\n&\n&{\CoefTrue}&\n&\n&\n&\n\cr
 \n&\n&\n&\n&\n&\n&\n&\n&\n&\n&{\CoefTrue}&\n&\n&\n&\n&\n&\n&\n&\n&\n&\n&\n&\n&\n&\n&\n&\n\cr
 \n&\n&\n&\n&\n&\n&\n&\n&\n&\n&\n&\n&\n&\n&\n&{\CoefTrue}&\n&\n&\n&\n&\n&\n&\n&\n&\n&\n&\n\cr
 \n&\n&\n&\n&\n&\n&\n&\n&\n&\n&\n&\n&\n&\n&\n&\n&\n&\n&\n&\n&\n&\n&\n&{\CoefTrue}&\n&\n&\n\cr
 \n&\n&\n&\n&\n&\n&\n&\n&\n&\n&\n&\n&{\CoefTrue}&\n&\n&\n&\n&\n&\n&\n&\n&\n&\n&\n&\n&\n&\n\cr
 \n&\n&\n&\n&\n&\n&\n&\n&\n&\n&\n&\n&\n&\n&\n&\n&\n&\n&\n&{\CoefTrue}&\n&\n&\n&\n&\n&\n&\n\cr
 \n&\n&\n&\n&\n&\n&\n&\n&\n&\n&\n&\n&\n&\n&\n&\n&\n&\n&\n&\n&\n&\n&\n&\n&{\CoefTrue}&\n&\n\cr
 \n&\n&\n&\n&\n&\n&\n&\n&\n&\n&\n&\n&\n&\n&\n&\n&\n&\n&\n&\n&\n&{\CoefTrue}&\n&\n&\n&\n&\n\cr
 \n&\n&\n&\n&\n&\n&\n&\n&\n&\n&\n&\n&\n&\n&\n&\n&\n&\n&\n&\n&\n&\n&\n&\n&\n&{\CoefTrue}&\n\cr
 \n&\n&\n&\n&\n&\n&\n&\n&\n&\n&\n&\n&\n&\n&\n&\n&\n&\n&\n&\n&\n&\n&\n&\n&\n&\n&{\CoefTrue}\cr
\noalign{\kern-2pt}}}%
\vbox{\setbox8=\hbox{$\RowNames\Matrix$}
\hbox to\wd8{\hfil$\ColNames$\kern\ColEntryShiftHoriz}\kern\ColEntryShiftVerti
\box8}}\hfil}}$}
{Illustrating the associative shuffling}{FigAssocLaw}

\noindent
v) similar to (iii) and (vi)

\bigskip
\noindent
vi) $\StrictFork{\StrictFork{Q}{R}\!}{S}\RELcompOP\,\,\TarskiShuffle
=
\StrictFork{\StrictFork{Q}{R}\!}{S}\RELcompOP\StrictJoin{\Kronecker{\RELide}{\pi\RELtraOP}}{\rho\RELtraOP\RELcompOP\rho_1\RELtraOP}$\quad by definition of $\TarskiShuffle$

$=
\StrictFork{Q}{R}\RELcompOP\Kronecker{\RELide}{\pi\RELtraOP}\RELandOP\, S\RELcompOP\rho\RELtraOP\RELcompOP\rho_1\RELtraOP
$\quad Prop.~\PropForkMapKron.iii since $\Kronecker{\RELide}{\pi\RELtraOP}$ and $\rho\RELtraOP\RELcompOP\rho_1\RELtraOP$ are both injective

$=
\StrictFork{Q\RELcompOP\RELide}{R\RELcompOP\pi\RELtraOP}\RELandOP\, S\RELcompOP\rho\RELtraOP\RELcompOP\rho_1\RELtraOP
$

$=
Q\RELcompOP\pi_1\RELtraOP\RELandOP R\RELcompOP\pi\RELtraOP\RELcompOP\rho_1\RELtraOP\RELandOP\, S\RELcompOP\rho\RELtraOP\RELcompOP\rho_1\RELtraOP
$

$=
Q\RELcompOP\pi_1\RELtraOP\RELandOP(R\RELcompOP\pi\RELtraOP\RELandOP\, S\RELcompOP\rho\RELtraOP)\RELcompOP\rho_1\RELtraOP
$

$=
Q\RELcompOP\pi_1\RELtraOP\RELandOP\StrictFork{R}{S}\RELcompOP\rho_1\RELtraOP
$

$=
\StrictFork{Q}{\StrictFork{R}{S}}
$
\Bewende

\noindent
Several identities are satisfied for $\TarskiShuffle$:

\smallskip
$\TarskiShuffle\RELcompOP\pi_1=\pi'\RELcompOP\pi,\qquad\qquad\qquad\qquad\qquad\qquad
\TarskiShuffle\RELcompOP\rho_1=\pi'\RELcompOP\rho\RELcompOP\pi\RELtraOP\RELandOP\rho'\RELcompOP\rho\RELtraOP
=
\Kronecker{\rho}{\RELide}
$

${\pi'}\RELtraOP\RELcompOP\TarskiShuffle=\pi\RELcompOP\pi_1\RELtraOP\RELandOP\rho\RELcompOP\pi\RELtraOP\RELcompOP\rho_1\RELtraOP
=
\Kronecker{\RELide}{\pi\RELtraOP}
,\qquad
{\rho'}\RELtraOP\RELcompOP\TarskiShuffle=\rho\RELtraOP\RELcompOP\rho_1\RELtraOP
$

\bigskip
\noindent
There follow characterizations of elements as being neutral, being inverses, etc. One will observe in (i), that the possibility of left-inversion of $x$, (i.e.~$\forall y: \exists p:\pi_{px}\predand\TarskiAdd_{py},\quad \forall y:\exists z:x+z=y$) is defined without mentioning the neutral element. A left-invertible element is characterized by the fact that the corresponding row of the composition table for $\TarskiAdd$ contains all the elements in some sequence.

\enunc{}{Definition}{}{DefInverse} Let be given the binary mapping $\TarskiAdd$ as before.
\begin{enumerate}[i)]
\item $\RELneg{\RELneg{\pi\RELtraOP\RELcompOP\TarskiAdd}\RELcompOP\RELtop}$\quad the set of elements that may be {\bf left-inverted}\index{left-invertible}, i.e., $\{x \mid \forall y :\exists p:\pi_{px}\predand\TarskiAdd_{py}\}$
\item$\RELneg{\RELneg{\rho\RELtraOP\RELcompOP\TarskiAdd}\RELcompOP\RELtop}$\quad the set of elements that may be {\bf right-inverted}\index{right-invertible}, i.e., $\{y \mid \forall x :\exists p:\rho_{py}\predand\TarskiAdd_{px}\}$
\item $\TarskiAdd$ allows {\bf left-inversion}\quad $:\Longleftrightarrow\quad\pi\RELtraOP\RELcompOP\TarskiAdd=\RELtop$
\item $\TarskiAdd$ allows {\bf right-inversion}\quad $:\Longleftrightarrow\quad\rho\RELtraOP\RELcompOP\TarskiAdd=\RELtop$
\Bewende
\end{enumerate}

\noindent
To identify a left-invertible point $e$ (i.e.~a transposed map) means via shunting also

\smallskip
$e\RELenthOP\RELneg{\RELneg{\pi\RELtraOP\RELcompOP\TarskiAdd}\RELcompOP\RELtop}
\quad\iff\quad
\RELtop
\RELenthOP\TarskiAdd\RELtraOP\RELcompOP\pi\RELcompOP e
\quad\iff\quad
e\RELcompOP\RELtop\RELenthOP\pi\RELtraOP\RELcompOP\TarskiAdd
$,

\smallskip
\noindent
and relates (i) with (iii). In Fig.~\FigHasInverses, for the element $a$, e.g., there is no element $x$ such that $\TarskiAdd_{ax}=e$, the fifth.

\Caption{\hbox{$\TarskiAdd=${\footnotesize%
\BoxBreadth=0pt%
\setbox7=\hbox{a}%
\ifdim\wd7>\BoxBreadth\BoxBreadth=\wd7\fi%
\setbox7=\hbox{b}%
\ifdim\wd7>\BoxBreadth\BoxBreadth=\wd7\fi%
\setbox7=\hbox{c}%
\ifdim\wd7>\BoxBreadth\BoxBreadth=\wd7\fi%
\setbox7=\hbox{d}%
\ifdim\wd7>\BoxBreadth\BoxBreadth=\wd7\fi%
\setbox7=\hbox{e}%
\ifdim\wd7>\BoxBreadth\BoxBreadth=\wd7\fi%
\setbox7=\hbox{f}%
\ifdim\wd7>\BoxBreadth\BoxBreadth=\wd7\fi%
\setbox7=\hbox{g}%
\ifdim\wd7>\BoxBreadth\BoxBreadth=\wd7\fi%
\def\RowNames{\def\scalable{0.5\baselineskip}
\vcenter{\offinterlineskip\baselineskip=\matrixskip%
\hbox to\BoxBreadth{\strut\hfil a}\kern\scalable%
\hbox to\BoxBreadth{\strut\hfil b}\kern\scalable%
\hbox to\BoxBreadth{\strut\hfil c}\kern\scalable%
\hbox to\BoxBreadth{\strut\hfil d}\kern\scalable%
\hbox to\BoxBreadth{\strut\hfil e}\kern\scalable%
\hbox to\BoxBreadth{\strut\hfil f}\kern\scalable%
\hbox to\BoxBreadth{\strut\hfil g}}}%
\def\ColNames{\def\scalable{0.18\baselineskip}
\hbox{\rotatebox{90}{\strut a}\kern\scalable%
\rotatebox{90}{\strut b}\kern\scalable%
\rotatebox{90}{\strut c}\kern\scalable%
\rotatebox{90}{\strut d}\kern\scalable%
\rotatebox{90}{\strut e}\kern\scalable%
\rotatebox{90}{\strut f}\kern\scalable%
\rotatebox{90}{\strut g}\kern\scalable%
\kern0pt
}}%
\def\Matrix{\spmatrix{%
\noalign{\kern-2pt}
 \hbox{\vbox to14pt{\vfil\hbox to14pt{\hfil 3\hfil}\vfil}}&\hbox{\vbox to14pt{\vfil\hbox to14pt{\hfil 2\hfil}\vfil}}&\hbox{\vbox to14pt{\vfil\hbox to14pt{\hfil 1\hfil}\vfil}}&\hbox{\vbox to14pt{\vfil\hbox to14pt{\hfil 4\hfil}\vfil}}&\hbox{\vbox to14pt{\vfil\hbox to14pt{\hfil 1\hfil}\vfil}}&\hbox{\vbox to14pt{\vfil\hbox to14pt{\hfil 6\hfil}\vfil}}&\hbox{\vbox to14pt{\vfil\hbox to14pt{\hfil 7\hfil}\vfil}}\cr
 \hbox{\vbox to14pt{\vfil\hbox to14pt{\hfil 1\hfil}\vfil}}&\hbox{\vbox to14pt{\vfil\hbox to14pt{\hfil 2\hfil}\vfil}}&\hbox{\vbox to14pt{\vfil\hbox to14pt{\hfil 5\hfil}\vfil}}&\hbox{\vbox to14pt{\vfil\hbox to14pt{\hfil 4\hfil}\vfil}}&\hbox{\vbox to14pt{\vfil\hbox to14pt{\hfil 3\hfil}\vfil}}&\hbox{\vbox to14pt{\vfil\hbox to14pt{\hfil 6\hfil}\vfil}}&\hbox{\vbox to14pt{\vfil\hbox to14pt{\hfil 7\hfil}\vfil}}\cr
 \hbox{\vbox to14pt{\vfil\hbox to14pt{\hfil 1\hfil}\vfil}}&\hbox{\vbox to14pt{\vfil\hbox to14pt{\hfil 3\hfil}\vfil}}&\hbox{\vbox to14pt{\vfil\hbox to14pt{\hfil 2\hfil}\vfil}}&\hbox{\vbox to14pt{\vfil\hbox to14pt{\hfil 4\hfil}\vfil}}&\hbox{\vbox to14pt{\vfil\hbox to14pt{\hfil 2\hfil}\vfil}}&\hbox{\vbox to14pt{\vfil\hbox to14pt{\hfil 6\hfil}\vfil}}&\hbox{\vbox to14pt{\vfil\hbox to14pt{\hfil 7\hfil}\vfil}}\cr
 \hbox{\vbox to14pt{\vfil\hbox to14pt{\hfil 1\hfil}\vfil}}&\hbox{\vbox to14pt{\vfil\hbox to14pt{\hfil 5\hfil}\vfil}}&\hbox{\vbox to14pt{\vfil\hbox to14pt{\hfil 4\hfil}\vfil}}&\hbox{\vbox to14pt{\vfil\hbox to14pt{\hfil 7\hfil}\vfil}}&\hbox{\vbox to14pt{\vfil\hbox to14pt{\hfil 4\hfil}\vfil}}&\hbox{\vbox to14pt{\vfil\hbox to14pt{\hfil 3\hfil}\vfil}}&\hbox{\vbox to14pt{\vfil\hbox to14pt{\hfil 6\hfil}\vfil}}\cr
 \hbox{\vbox to14pt{\vfil\hbox to14pt{\hfil 1\hfil}\vfil}}&\hbox{\vbox to14pt{\vfil\hbox to14pt{\hfil 2\hfil}\vfil}}&\hbox{\vbox to14pt{\vfil\hbox to14pt{\hfil 5\hfil}\vfil}}&\hbox{\vbox to14pt{\vfil\hbox to14pt{\hfil 4\hfil}\vfil}}&\hbox{\vbox to14pt{\vfil\hbox to14pt{\hfil 5\hfil}\vfil}}&\hbox{\vbox to14pt{\vfil\hbox to14pt{\hfil 6\hfil}\vfil}}&\hbox{\vbox to14pt{\vfil\hbox to14pt{\hfil 7\hfil}\vfil}}\cr
 \hbox{\vbox to14pt{\vfil\hbox to14pt{\hfil 1\hfil}\vfil}}&\hbox{\vbox to14pt{\vfil\hbox to14pt{\hfil 3\hfil}\vfil}}&\hbox{\vbox to14pt{\vfil\hbox to14pt{\hfil 6\hfil}\vfil}}&\hbox{\vbox to14pt{\vfil\hbox to14pt{\hfil 5\hfil}\vfil}}&\hbox{\vbox to14pt{\vfil\hbox to14pt{\hfil 2\hfil}\vfil}}&\hbox{\vbox to14pt{\vfil\hbox to14pt{\hfil 4\hfil}\vfil}}&\hbox{\vbox to14pt{\vfil\hbox to14pt{\hfil 7\hfil}\vfil}}\cr
 \hbox{\vbox to14pt{\vfil\hbox to14pt{\hfil 1\hfil}\vfil}}&\hbox{\vbox to14pt{\vfil\hbox to14pt{\hfil 2\hfil}\vfil}}&\hbox{\vbox to14pt{\vfil\hbox to14pt{\hfil 7\hfil}\vfil}}&\hbox{\vbox to14pt{\vfil\hbox to14pt{\hfil 4\hfil}\vfil}}&\hbox{\vbox to14pt{\vfil\hbox to14pt{\hfil 7\hfil}\vfil}}&\hbox{\vbox to14pt{\vfil\hbox to14pt{\hfil 6\hfil}\vfil}}&\hbox{\vbox to14pt{\vfil\hbox to14pt{\hfil 3\hfil}\vfil}}\cr
\noalign{\kern-2pt}}}
\vbox{\setbox8=\hbox{$\RowNames\Matrix$}
\hbox to\wd8{\hfil$\ColNames$\kern0.6\matrixskip}\kern0.1cm
\box8}}

\quad

{\footnotesize%
\BoxBreadth=0pt%
\setbox7=\hbox{a}%
\ifdim\wd7>\BoxBreadth\BoxBreadth=\wd7\fi%
\setbox7=\hbox{b}%
\ifdim\wd7>\BoxBreadth\BoxBreadth=\wd7\fi%
\setbox7=\hbox{c}%
\ifdim\wd7>\BoxBreadth\BoxBreadth=\wd7\fi%
\setbox7=\hbox{d}%
\ifdim\wd7>\BoxBreadth\BoxBreadth=\wd7\fi%
\setbox7=\hbox{e}%
\ifdim\wd7>\BoxBreadth\BoxBreadth=\wd7\fi%
\setbox7=\hbox{f}%
\ifdim\wd7>\BoxBreadth\BoxBreadth=\wd7\fi%
\setbox7=\hbox{g}%
\ifdim\wd7>\BoxBreadth\BoxBreadth=\wd7\fi%
\def\RowNames{\vcenter{\offinterlineskip\baselineskip=\matrixskip%
\hbox to11.00789pt{\strut\hfil a}\kern\interspacereduction%
\hbox to11.00789pt{\strut\hfil b}\kern\interspacereduction%
\hbox to11.00789pt{\strut\hfil c}\kern\interspacereduction%
\hbox to11.00789pt{\strut\hfil d}\kern\interspacereduction%
\hbox to11.00789pt{\strut\hfil e}\kern\interspacereduction%
\hbox to11.00789pt{\strut\hfil f}\kern\interspacereduction%
\hbox to11.00789pt{\strut\hfil g}}}%
\def\MarkVect{\setbox2=\hbox{\strut}\matrixskip=\ht2\spmatrix{\noalign{\kern-2pt}%
\n\cr
{\CoefTrue}\cr
\n\cr
\n\cr
\n\cr
{\CoefTrue}\cr
\n\cr
\noalign{\kern-2pt}}%
}%
{\normalsize$\RELneg{\RELneg{\pi\RELtraOP\RELcompOP\TarskiAdd}\RELcompOP\RELtop}=$}$\RowNames\MarkVect$}
\quad {\footnotesize%
\BoxBreadth=0pt%
\setbox7=\hbox{a}%
\ifdim\wd7>\BoxBreadth\BoxBreadth=\wd7\fi%
\setbox7=\hbox{b}%
\ifdim\wd7>\BoxBreadth\BoxBreadth=\wd7\fi%
\setbox7=\hbox{c}%
\ifdim\wd7>\BoxBreadth\BoxBreadth=\wd7\fi%
\setbox7=\hbox{d}%
\ifdim\wd7>\BoxBreadth\BoxBreadth=\wd7\fi%
\setbox7=\hbox{e}%
\ifdim\wd7>\BoxBreadth\BoxBreadth=\wd7\fi%
\setbox7=\hbox{f}%
\ifdim\wd7>\BoxBreadth\BoxBreadth=\wd7\fi%
\setbox7=\hbox{g}%
\ifdim\wd7>\BoxBreadth\BoxBreadth=\wd7\fi%
\def\RowNames{\def\scalable{\interspacereduction}
\vcenter{\offinterlineskip\baselineskip=\matrixskip%
\hbox to\BoxBreadth{\strut\hfil a}\kern\scalable%
\hbox to\BoxBreadth{\strut\hfil b}\kern\scalable%
\hbox to\BoxBreadth{\strut\hfil c}\kern\scalable%
\hbox to\BoxBreadth{\strut\hfil d}\kern\scalable%
\hbox to\BoxBreadth{\strut\hfil e}\kern\scalable%
\hbox to\BoxBreadth{\strut\hfil f}\kern\scalable%
\hbox to\BoxBreadth{\strut\hfil g}}}%
\def\ColNames{\def\scalable{\interspacereduction}
\hbox{\rotatebox{90}{\strut a}\kern\scalable%
\rotatebox{90}{\strut b}\kern\scalable%
\rotatebox{90}{\strut c}\kern\scalable%
\rotatebox{90}{\strut d}\kern\scalable%
\rotatebox{90}{\strut e}\kern\scalable%
\rotatebox{90}{\strut f}\kern\scalable%
\rotatebox{90}{\strut g}\kern\scalable%
\kern0pt
}}%
\def\Matrix{\spmatrix{%
\noalign{\kern-2pt}
 {\CoefTrue}&\n&\n&\n&\n&\n&\n\cr
 \n&\n&\n&\n&{\CoefTrue}&\n&\n\cr
 \n&{\CoefTrue}&\n&\n&\n&\n&\n\cr
 \n&\n&\n&\n&\n&{\CoefTrue}&\n\cr
 \n&\n&\n&{\CoefTrue}&\n&\n&\n\cr
 \n&\n&{\CoefTrue}&\n&\n&\n&\n\cr
 \n&\n&\n&\n&\n&\n&{\CoefTrue}\cr
\noalign{\kern-2pt}}}%
$\TarskiAdd\RELtraOP\RELcompOP(\rho\RELandOP\pi\RELcompOP f\RELcompOP\RELtop)=$
\vbox{\setbox8=\hbox{$\RowNames\Matrix$}
\hbox to\wd8{\hfil$\ColNames$\kern\ColEntryShiftHoriz}\kern\ColEntryShiftVerti
\box8}}}}
{A mapping with elements ${\rm b,f}$ posessing all left-inverses and left-division by ${\rm f}$}{FigHasInverses}

\noindent
Whenever one takes a point $i\RELenthOP\RELneg{\RELneg{\pi\RELtraOP\RELcompOP\TarskiAdd}\RELcompOP\RELtop}$, the construct $f:=(\rho\RELandOP\pi\RELcompOP i\RELcompOP\RELtop)\RELtraOP$ is a mapping, according to Prop.~\PropNegBinOp. As an example, left-division by ${\rm f}$ is shown as a mapping $\TarskiAdd\RELtraOP\RELcompOP(\rho\RELandOP\pi\RELcompOP f\RELcompOP\RELtop)$ on the right: \quad$\TarskiAdd_{{\rm f\,b}}=c\;\Longrightarrow\;{\rm f}\backslash {\rm c}={\rm b}$\quad or else\quad $\TarskiAdd_{{\rm f\,d}}={\rm e}\;\Longrightarrow\;{\rm f}\backslash {\rm e}={\rm d}$.

\bigskip

\bigskip
\noindent
Invariant elements commute with every other one. In the table representation, row and column concerning this element
are equal.

\enunc{}{Definition}{}{DefInvariant} Let be given the binary mapping $\TarskiAdd$ as before. Then

\smallskip
$\RELneg{\pi\RELtraOP\RELcompOP\RELneg{[\TarskiAdd\RELandOP P\RELcompOP\TarskiAdd]\RELcompOP\RELtop}}
=
\RELneg{\rho\RELtraOP\RELcompOP\RELneg{[\TarskiAdd\RELandOP P\RELcompOP\TarskiAdd]\RELcompOP\RELtop}}
=
\LeftResi{\pi}{([\TarskiAdd\RELandOP P\RELcompOP\TarskiAdd]\RELcompOP\RELtop)}
$ is the set of {\bf invariant elements}\index{invariant element}, 

\smallskip
\noindent
i.e., those $x$ with $\forall y: \TarskiAdd_{xy}=\TarskiAdd_{yx}$
\Bewende

\Caption{\hbox{$\TarskiAdd=$ {\footnotesize%
\BoxBreadth=0pt%
\setbox7=\hbox{a}%
\ifdim\wd7>\BoxBreadth\BoxBreadth=\wd7\fi%
\setbox7=\hbox{b}%
\ifdim\wd7>\BoxBreadth\BoxBreadth=\wd7\fi%
\setbox7=\hbox{c}%
\ifdim\wd7>\BoxBreadth\BoxBreadth=\wd7\fi%
\setbox7=\hbox{d}%
\ifdim\wd7>\BoxBreadth\BoxBreadth=\wd7\fi%
\setbox7=\hbox{e}%
\ifdim\wd7>\BoxBreadth\BoxBreadth=\wd7\fi%
\setbox7=\hbox{f}%
\ifdim\wd7>\BoxBreadth\BoxBreadth=\wd7\fi%
\setbox7=\hbox{g}%
\ifdim\wd7>\BoxBreadth\BoxBreadth=\wd7\fi%
\def\RowNames{\def\scalable{0.5\baselineskip}
\vcenter{\offinterlineskip\baselineskip=\matrixskip%
\hbox to\BoxBreadth{\strut\hfil a}\kern\scalable%
\hbox to\BoxBreadth{\strut\hfil b}\kern\scalable%
\hbox to\BoxBreadth{\strut\hfil c}\kern\scalable%
\hbox to\BoxBreadth{\strut\hfil d}\kern\scalable%
\hbox to\BoxBreadth{\strut\hfil e}\kern\scalable%
\hbox to\BoxBreadth{\strut\hfil f}\kern\scalable%
\hbox to\BoxBreadth{\strut\hfil g}}}%
\def\ColNames{\def\scalable{0.17\baselineskip}
\hbox{\rotatebox{90}{\strut a}\kern\scalable%
\rotatebox{90}{\strut b}\kern\scalable%
\rotatebox{90}{\strut c}\kern\scalable%
\rotatebox{90}{\strut d}\kern\scalable%
\rotatebox{90}{\strut e}\kern\scalable%
\rotatebox{90}{\strut f}\kern\scalable%
\rotatebox{90}{\strut g}\kern\scalable%
\kern0pt
}}%
\def\Matrix{\spmatrix{%
\noalign{\kern-2pt}
 \hbox{\vbox to14pt{\vfil\hbox to14pt{\hfil 3\hfil}\vfil}}&\hbox{\vbox to14pt{\vfil\hbox to14pt{\hfil 2\hfil}\vfil}}&\hbox{\vbox to14pt{\vfil\hbox to14pt{\hfil 1\hfil}\vfil}}&\hbox{\vbox to14pt{\vfil\hbox to14pt{\hfil 4\hfil}\vfil}}&\hbox{\vbox to14pt{\vfil\hbox to14pt{\hfil 5\hfil}\vfil}}&\hbox{\vbox to14pt{\vfil\hbox to14pt{\hfil 6\hfil}\vfil}}&\hbox{\vbox to14pt{\vfil\hbox to14pt{\hfil 7\hfil}\vfil}}\cr
 \hbox{\vbox to14pt{\vfil\hbox to14pt{\hfil 1\hfil}\vfil}}&\hbox{\vbox to14pt{\vfil\hbox to14pt{\hfil 3\hfil}\vfil}}&\hbox{\vbox to14pt{\vfil\hbox to14pt{\hfil 2\hfil}\vfil}}&\hbox{\vbox to14pt{\vfil\hbox to14pt{\hfil 4\hfil}\vfil}}&\hbox{\vbox to14pt{\vfil\hbox to14pt{\hfil 5\hfil}\vfil}}&\hbox{\vbox to14pt{\vfil\hbox to14pt{\hfil 6\hfil}\vfil}}&\hbox{\vbox to14pt{\vfil\hbox to14pt{\hfil 7\hfil}\vfil}}\cr
 \hbox{\vbox to14pt{\vfil\hbox to14pt{\hfil 1\hfil}\vfil}}&\hbox{\vbox to14pt{\vfil\hbox to14pt{\hfil 2\hfil}\vfil}}&\hbox{\vbox to14pt{\vfil\hbox to14pt{\hfil 3\hfil}\vfil}}&\hbox{\vbox to14pt{\vfil\hbox to14pt{\hfil 4\hfil}\vfil}}&\hbox{\vbox to14pt{\vfil\hbox to14pt{\hfil 5\hfil}\vfil}}&\hbox{\vbox to14pt{\vfil\hbox to14pt{\hfil 6\hfil}\vfil}}&\hbox{\vbox to14pt{\vfil\hbox to14pt{\hfil 7\hfil}\vfil}}\cr
 \hbox{\vbox to14pt{\vfil\hbox to14pt{\hfil 1\hfil}\vfil}}&\hbox{\vbox to14pt{\vfil\hbox to14pt{\hfil 5\hfil}\vfil}}&\hbox{\vbox to14pt{\vfil\hbox to14pt{\hfil 4\hfil}\vfil}}&\hbox{\vbox to14pt{\vfil\hbox to14pt{\hfil 7\hfil}\vfil}}&\hbox{\vbox to14pt{\vfil\hbox to14pt{\hfil 2\hfil}\vfil}}&\hbox{\vbox to14pt{\vfil\hbox to14pt{\hfil 3\hfil}\vfil}}&\hbox{\vbox to14pt{\vfil\hbox to14pt{\hfil 6\hfil}\vfil}}\cr
 \hbox{\vbox to14pt{\vfil\hbox to14pt{\hfil 1\hfil}\vfil}}&\hbox{\vbox to14pt{\vfil\hbox to14pt{\hfil 2\hfil}\vfil}}&\hbox{\vbox to14pt{\vfil\hbox to14pt{\hfil 5\hfil}\vfil}}&\hbox{\vbox to14pt{\vfil\hbox to14pt{\hfil 4\hfil}\vfil}}&\hbox{\vbox to14pt{\vfil\hbox to14pt{\hfil 3\hfil}\vfil}}&\hbox{\vbox to14pt{\vfil\hbox to14pt{\hfil 6\hfil}\vfil}}&\hbox{\vbox to14pt{\vfil\hbox to14pt{\hfil 7\hfil}\vfil}}\cr
 \hbox{\vbox to14pt{\vfil\hbox to14pt{\hfil 1\hfil}\vfil}}&\hbox{\vbox to14pt{\vfil\hbox to14pt{\hfil 3\hfil}\vfil}}&\hbox{\vbox to14pt{\vfil\hbox to14pt{\hfil 6\hfil}\vfil}}&\hbox{\vbox to14pt{\vfil\hbox to14pt{\hfil 5\hfil}\vfil}}&\hbox{\vbox to14pt{\vfil\hbox to14pt{\hfil 2\hfil}\vfil}}&\hbox{\vbox to14pt{\vfil\hbox to14pt{\hfil 4\hfil}\vfil}}&\hbox{\vbox to14pt{\vfil\hbox to14pt{\hfil 7\hfil}\vfil}}\cr
 \hbox{\vbox to14pt{\vfil\hbox to14pt{\hfil 1\hfil}\vfil}}&\hbox{\vbox to14pt{\vfil\hbox to14pt{\hfil 2\hfil}\vfil}}&\hbox{\vbox to14pt{\vfil\hbox to14pt{\hfil 7\hfil}\vfil}}&\hbox{\vbox to14pt{\vfil\hbox to14pt{\hfil 4\hfil}\vfil}}&\hbox{\vbox to14pt{\vfil\hbox to14pt{\hfil 5\hfil}\vfil}}&\hbox{\vbox to14pt{\vfil\hbox to14pt{\hfil 6\hfil}\vfil}}&\hbox{\vbox to14pt{\vfil\hbox to14pt{\hfil 3\hfil}\vfil}}\cr
\noalign{\kern-2pt}}}
\vbox{\setbox8=\hbox{$\RowNames\Matrix$}
\hbox to\wd8{\hfil$\ColNames$\kern0.6\matrixskip}\kern0.1cm
\box8}}

\quad

{\footnotesize%
\BoxBreadth=0pt%
\setbox7=\hbox{a}%
\ifdim\wd7>\BoxBreadth\BoxBreadth=\wd7\fi%
\setbox7=\hbox{b}%
\ifdim\wd7>\BoxBreadth\BoxBreadth=\wd7\fi%
\setbox7=\hbox{c}%
\ifdim\wd7>\BoxBreadth\BoxBreadth=\wd7\fi%
\setbox7=\hbox{d}%
\ifdim\wd7>\BoxBreadth\BoxBreadth=\wd7\fi%
\setbox7=\hbox{e}%
\ifdim\wd7>\BoxBreadth\BoxBreadth=\wd7\fi%
\setbox7=\hbox{f}%
\ifdim\wd7>\BoxBreadth\BoxBreadth=\wd7\fi%
\setbox7=\hbox{g}%
\ifdim\wd7>\BoxBreadth\BoxBreadth=\wd7\fi%
\def\RowNames{\vcenter{\offinterlineskip\baselineskip=\matrixskip%
\hbox to11.00789pt{\strut\hfil a}\kern\interspacereduction%
\hbox to11.00789pt{\strut\hfil b}\kern\interspacereduction%
\hbox to11.00789pt{\strut\hfil c}\kern\interspacereduction%
\hbox to11.00789pt{\strut\hfil d}\kern\interspacereduction%
\hbox to11.00789pt{\strut\hfil e}\kern\interspacereduction%
\hbox to11.00789pt{\strut\hfil f}\kern\interspacereduction%
\hbox to11.00789pt{\strut\hfil g}}}%
\def\MarkVect{\setbox2=\hbox{\strut}\matrixskip=\ht2\spmatrix{\noalign{\kern-2pt}%
{\CoefTrue}\cr
\n\cr
{\CoefTrue}\cr
\n\cr
\n\cr
\n\cr
\n\cr
\noalign{\kern-2pt}}%
}%
$\RowNames\MarkVect${\normalsize$=\RELneg{\pi\RELtraOP\RELcompOP\RELneg{[\TarskiAdd\RELandOP P\RELcompOP\TarskiAdd]\RELcompOP\RELtop}}$}}
}}
{A hardly interesting binary mapping and its invariant elements}{FigHasInvariants}

\noindent
In case $\TarskiAdd$ is a group operation, the invariant elements together form the center of the group.

\bigskip

\bigskip
\noindent
Next  interesting are left- resp.~right-neutral elements. The intention for a right-neutral element $\RELfromTO{\TarskiNeutrR}{X}{\I1}$ is that application of $\TarskiAdd$ to any pair $(x, \TarskiNeutrR)$ with $x$ chosen arbitrarily results in $x$. In the relational setting with points $x, \TarskiNeutrR $, this reads 

\smallskip
$\TarskiAdd\RELtraOP\RELcompOP\StrictJoin{x}{\TarskiNeutrR}=\TarskiAdd\RELtraOP\RELcompOP(\pi \RELcompOP x\RELandOP\rho\RELcompOP \TarskiNeutrR)\RELcompOP 
=
x
$.

\smallskip
\noindent
When working in a group theory environment, $\TarskiNeutrR $ is usually called  zero or unit element, depending on whether one works in an additive or multiplicative setting. 
A point-free formulation for all $x$ simultaneously is 

\smallskip
$\TarskiAdd\RELtraOP\RELcompOP\StrictJoin{\RELide}{\TarskiNeutrR\RELcompOP\RELtop_{\I1X}}=\TarskiAdd\RELtraOP\RELcompOP(\pi \RELcompOP\RELide_X\RELandOP\rho\RELcompOP \TarskiNeutrR \RELcompOP\RELtop_{\I1X})\RELcompOP 
=
\RELide_X
$.

\smallskip
\noindent
This is a condition $\TarskiNeutrR$ has to satisfy. Concentrating on \lq\lq$\RELenthOP$\rq\rq\ alone, the following equivalences make it more explicit:

\smallskip
$\TarskiAdd\RELtraOP\RELcompOP(\pi\RELcompOP\RELide\RELandOP\rho\RELcompOP\TarskiNeutrR \RELcompOP\RELtop) 
\RELenthOP
\RELide
\quad\iff\quad
\pi\RELandOP\rho\RELcompOP\TarskiNeutrR \RELcompOP\RELtop
\RELenthOP
\TarskiAdd
\quad\iff\quad
\rho\RELcompOP\TarskiNeutrR \RELcompOP\RELtop
\RELenthOP
\TarskiAdd\RELorOP\RELneg{\pi}
 $
 
 $\quad\iff\quad
\rho\RELtraOP\RELcompOP(\RELneg{\TarskiAdd}\RELandOP\pi)
\RELenthOP
\RELneg{\TarskiNeutrR \RELcompOP\RELtop}
\quad\iff\quad
\TarskiNeutrR \RELcompOP\RELtop
 \RELenthOP
 \RELneg{\rho\RELtraOP\RELcompOP(\RELneg{\TarskiAdd}\RELandOP\pi)}
 $

\smallskip
\noindent
The $\TarskiNeutrR$ thus characterized may in arbitrarily chosen cases uninterestingly be equal to $\RELbot$ for which $\rho\RELtraOP\RELcompOP\pi=\RELtop$ gives a hint. We assume, however, a point $e\RELenthOP\TarskiNeutrR$ and recall that according to Prop.~\PropNegBinOp\ $g:=(\pi\RELandOP\rho\RELcompOP e\RELcompOP\RELtop)\RELtraOP$ is a map.
From

\smallskip
$\TarskiAdd\RELtraOP\RELcompOP g\RELtraOP
=
\TarskiAdd\RELtraOP\RELcompOP(\pi\RELandOP\rho\RELcompOP e\RELcompOP\RELtop)
\RELenthOP
\TarskiAdd\RELtraOP\RELcompOP(\pi\RELandOP\rho\RELcompOP\TarskiNeutrR \RELcompOP\RELtop)
\RELenthOP
\RELide$

\smallskip
\noindent
we then derive equality: The mapping $g\RELcompOP\TarskiAdd$
contained in the mapping $\RELide$ means that they are equal.

\enunc{}{Definition}{}{DefNeutral} Let be given the binary mapping $\TarskiAdd$ as before. We call any point $e$ in
\begin{enumerate}[i)]
\item[]$\RELneg{\rho\RELtraOP\RELcompOP(\RELneg{\TarskiAdd}\RELandOP\pi)}
$\quad a {\bf right-neutral} element\index{right-neutral element},
\item[]$
\RELneg{\pi\RELtraOP\RELcompOP(\RELneg{\TarskiAdd}\RELandOP\rho)}$\quad a {\bf left-neutral} element\index{left-neutral element},
\item[]$\RELneg{\rho\RELtraOP\RELcompOP(\RELneg{\TarskiAdd}\RELandOP\pi)}\RELandOP\RELneg{\pi\RELtraOP\RELcompOP(\RELneg{\TarskiAdd}\RELandOP\rho)}$\quad a {\bf neutral} element\index{neutral element}.
\Bewende
\end{enumerate}

\bigskip
\noindent
In an alternative approach, we might have considered

\smallskip
$\RELfromTO{\TarskiDelta:=\RELide_{X\times X}\RELandOP\TarskiAdd\RELcompOP\pi\RELtraOP}{X\times X}{X\times X}$

\smallskip
\noindent
i.e., all the pairs with result and left component equal. Then one would look for points $e$ in

\smallskip
$\RELfromTO{\TarskiNeutrR:=\RELneg{\rho\RELtraOP\RELcompOP\RELneg{\TarskiDelta\RELcompOP\RELtop_{X\times X,X}}}}{X}{X}$,

\smallskip
\noindent
indicating right-neutral elements if any, and then giving rise to forming of right-inverses

\smallskip
$\RELfromTO{\TarskiInversR:=\pi\RELtraOP\RELcompOP(\TarskiAdd\RELcompOP e\RELcompOP\RELtop\RELandOP\rho)=\VectToRel{\TarskiAdd\RELcompOP e\RELcompOP\RELtop}}{X}{X}
$.

\bigskip
\noindent
With the standard methods, it is possible to prove

\smallskip
$\TarskiAdd\RELtraOP\RELcompOP\StrictJoin{\TarskiInversR}{\RELide}
=
\TarskiAdd\RELtraOP\RELcompOP(\pi \RELcompOP\TarskiInversR\RELandOP\rho)
\RELenthOP 
e\RELcompOP\RELtop
$

$\Longleftrightarrow\quad
\TarskiAdd\RELcompOP\RELneg{e\RELcompOP\RELtop}
\RELenthOP 
\RELneg{\pi\RELcompOP\TarskiInversR\RELandOP\rho}
=
\RELneg{\pi\RELcompOP\TarskiInversR}\RELorOP\RELneg{\rho}
$

$\Longleftrightarrow\quad
\RELtop
=
\RELneg{\TarskiAdd\RELcompOP\RELneg{e\RELcompOP\RELtop}}
\RELorOP
\RELneg{\pi\RELcompOP\TarskiInversR}\RELorOP\RELneg{\rho}
=
\TarskiAdd\RELcompOP e\RELcompOP\RELtop
\RELorOP
\RELneg{\pi\RELcompOP\TarskiInversR}\RELorOP\RELneg{\rho}
$\quad since $\TarskiAdd$ is a map

$\Longleftrightarrow\quad
\pi\RELcompOP\TarskiInversR
\RELenthOP
\TarskiAdd\RELcompOP e\RELcompOP\RELtop
\RELorOP\RELneg{\rho}
$

$\Longleftrightarrow\quad
\pi\RELtraOP\RELcompOP\RELneg{\TarskiAdd\RELcompOP e\RELcompOP\RELtop
\RELorOP\RELneg{\rho}}
\RELenthOP
\RELneg{\TarskiInversR}
$

$\Longleftrightarrow\quad
\TarskiInversR
\RELenthOP
\RELneg{\pi\RELtraOP\RELcompOP(\RELneg{\TarskiAdd\RELcompOP e\RELcompOP\RELtop}
\RELandOP\rho)}
=
\RELneg{\VectToRel{\RELneg{\TarskiAdd\RELcompOP e\RELcompOP\RELtop}}}
=
\VectToRel{\TarskiAdd\RELcompOP e\RELcompOP\RELtop}
$\quad due to Prop.~\PropTranspositionPointfree.vii

\smallskip
\noindent
We have to show equality $\TarskiAdd\RELtraOP\RELcompOP\StrictJoin{\TarskiInversR}{\RELide}
=
\TarskiAdd\RELtraOP\RELcompOP(\pi \RELcompOP\TarskiInversR\RELandOP\rho)
= 
e\RELcompOP\RELtop$ with a
separate argument, based on the fact that $e$ is a neutral point, or else, a transposed mapping. It suffices, according to Prop.~5.2.iii of \cite{RelaMath2010}, when 
$\TarskiAdd\RELtraOP\RELcompOP(\pi \RELcompOP\TarskiInversR\RELandOP\rho)$ turns out to be surjective

\smallskip
$\RELtop\RELcompOP\TarskiAdd\RELtraOP\RELcompOP(\pi \RELcompOP\TarskiInversR\RELandOP\rho)
=
\RELtop\RELcompOP(\pi \RELcompOP\TarskiInversR\RELandOP\rho)
=
\RELtop\RELcompOP\rho\RELtraOP\RELcompOP(\pi \RELcompOP\TarskiInversR\RELandOP\rho)
=
\RELtop\RELcompOP(\rho\RELtraOP\RELcompOP\pi \RELcompOP\TarskiInversR\RELandOP\RELide)
=
\RELtop\RELcompOP(\RELtop\RELcompOP\TarskiInversR\RELandOP\RELide)
=
\RELtop
$,\quad since

$\RELtop\RELcompOP\TarskiInversR
=
\RELtop\RELcompOP\pi\RELtraOP\RELcompOP(\TarskiAdd\RELcompOP e\RELcompOP\RELtop\RELandOP\rho)
=
\RELtop\RELcompOP(\TarskiAdd\RELcompOP e\RELcompOP\RELtop\RELandOP\rho)
= (\RELtop\RELandOP\RELtop\RELcompOP e\RELtraOP\RELcompOP \TarskiAdd\RELtraOP)\RELcompOP\rho
= \RELtop\RELcompOP e\RELtraOP\RELcompOP \TarskiAdd\RELtraOP\RELcompOP\rho
= \RELtop\RELcompOP e\RELtraOP\RELcompOP\RELtop
= \RELtop$

\smallskip
\noindent
when $\TarskiAdd$ allows right-inversion and $e$ is a point.

\bigskip
\noindent
As an example, we show the alternating group $A_3$ as well as a constant binary mapping.

\Caption{\vbox{\hbox to\textwidth{\hfil{\footnotesize%
\BoxBreadth=0pt%
\setbox7=\hbox{[1,2,3]}%
\ifdim\wd7>\BoxBreadth\BoxBreadth=\wd7\fi%
\setbox7=\hbox{[2,3,1]}%
\ifdim\wd7>\BoxBreadth\BoxBreadth=\wd7\fi%
\setbox7=\hbox{[3,1,2]}%
\ifdim\wd7>\BoxBreadth\BoxBreadth=\wd7\fi%
\def\RowNames{\def\scalable{0.45\baselineskip}
\vcenter{\offinterlineskip\baselineskip=\matrixskip%
\hbox to\BoxBreadth{\strut\hfil [1,2,3]}\kern\scalable%
\hbox to\BoxBreadth{\strut\hfil [2,3,1]}\kern\scalable%
\hbox to\BoxBreadth{\strut\hfil [3,1,2]}}}%
\def\ColNames{\def\scalable{0.2\baselineskip}
\hbox{\rotatebox{90}{\strut [1,2,3]}\kern\scalable%
\rotatebox{90}{\strut [2,3,1]}\kern\scalable%
\rotatebox{90}{\strut [3,1,2]}\kern\scalable%
\kern0pt
}}%
\def\Matrix{\spmatrix{%
\noalign{\kern-2pt}
 \hbox{\vbox to14pt{\vfil\hbox to14pt{\hfil 1\hfil}\vfil}}&\hbox{\vbox to14pt{\vfil\hbox to14pt{\hfil 2\hfil}\vfil}}&\hbox{\vbox to14pt{\vfil\hbox to14pt{\hfil 3\hfil}\vfil}}\cr
 \hbox{\vbox to14pt{\vfil\hbox to14pt{\hfil 2\hfil}\vfil}}&\hbox{\vbox to14pt{\vfil\hbox to14pt{\hfil 3\hfil}\vfil}}&\hbox{\vbox to14pt{\vfil\hbox to14pt{\hfil 1\hfil}\vfil}}\cr
 \hbox{\vbox to14pt{\vfil\hbox to14pt{\hfil 3\hfil}\vfil}}&\hbox{\vbox to14pt{\vfil\hbox to14pt{\hfil 1\hfil}\vfil}}&\hbox{\vbox to14pt{\vfil\hbox to14pt{\hfil 2\hfil}\vfil}}\cr
\noalign{\kern-2pt}}}
\vbox{\setbox8=\hbox{$\RowNames\Matrix$}
\hbox to\wd8{\hfil$\ColNames$\kern0.6\matrixskip}\kern0.1cm
\box8}}

\quad
{\footnotesize%
\BoxBreadth=0pt%
\setbox7=\hbox{([1,2,3],[1,2,3])}%
\ifdim\wd7>\BoxBreadth\BoxBreadth=\wd7\fi%
\setbox7=\hbox{([2,3,1],[1,2,3])}%
\ifdim\wd7>\BoxBreadth\BoxBreadth=\wd7\fi%
\setbox7=\hbox{([1,2,3],[2,3,1])}%
\ifdim\wd7>\BoxBreadth\BoxBreadth=\wd7\fi%
\setbox7=\hbox{([3,1,2],[1,2,3])}%
\ifdim\wd7>\BoxBreadth\BoxBreadth=\wd7\fi%
\setbox7=\hbox{([2,3,1],[2,3,1])}%
\ifdim\wd7>\BoxBreadth\BoxBreadth=\wd7\fi%
\setbox7=\hbox{([1,2,3],[3,1,2])}%
\ifdim\wd7>\BoxBreadth\BoxBreadth=\wd7\fi%
\setbox7=\hbox{([3,1,2],[2,3,1])}%
\ifdim\wd7>\BoxBreadth\BoxBreadth=\wd7\fi%
\setbox7=\hbox{([2,3,1],[3,1,2])}%
\ifdim\wd7>\BoxBreadth\BoxBreadth=\wd7\fi%
\setbox7=\hbox{([3,1,2],[3,1,2])}%
\ifdim\wd7>\BoxBreadth\BoxBreadth=\wd7\fi%
\def\RowNames{\def\scalable{\interspacereduction}
\vcenter{\offinterlineskip\baselineskip=\matrixskip%
\hbox to\BoxBreadth{\strut\hfil ([1,2,3],[1,2,3])}\kern\scalable%
\hbox to\BoxBreadth{\strut\hfil ([2,3,1],[1,2,3])}\kern\scalable%
\hbox to\BoxBreadth{\strut\hfil ([1,2,3],[2,3,1])}\kern\scalable%
\hbox to\BoxBreadth{\strut\hfil ([3,1,2],[1,2,3])}\kern\scalable%
\hbox to\BoxBreadth{\strut\hfil ([2,3,1],[2,3,1])}\kern\scalable%
\hbox to\BoxBreadth{\strut\hfil ([1,2,3],[3,1,2])}\kern\scalable%
\hbox to\BoxBreadth{\strut\hfil ([3,1,2],[2,3,1])}\kern\scalable%
\hbox to\BoxBreadth{\strut\hfil ([2,3,1],[3,1,2])}\kern\scalable%
\hbox to\BoxBreadth{\strut\hfil ([3,1,2],[3,1,2])}}}%
\def\ColNames{\def\scalable{\interspacereduction}
\hbox{\rotatebox{90}{\strut ([1,2,3],[1,2,3])}\kern\scalable%
\rotatebox{90}{\strut ([2,3,1],[1,2,3])}\kern\scalable%
\rotatebox{90}{\strut ([1,2,3],[2,3,1])}\kern\scalable%
\rotatebox{90}{\strut ([3,1,2],[1,2,3])}\kern\scalable%
\rotatebox{90}{\strut ([2,3,1],[2,3,1])}\kern\scalable%
\rotatebox{90}{\strut ([1,2,3],[3,1,2])}\kern\scalable%
\rotatebox{90}{\strut ([3,1,2],[2,3,1])}\kern\scalable%
\rotatebox{90}{\strut ([2,3,1],[3,1,2])}\kern\scalable%
\rotatebox{90}{\strut ([3,1,2],[3,1,2])}\kern\scalable%
\kern0pt
}}%
\def\Matrix{\spmatrix {%
\noalign{\kern-2pt}
 {\CoefTrue}&\n&\n&\n&\n&\n&\n&\n&\n\cr
 \n&{\CoefTrue}&\n&\n&\n&\n&\n&\n&\n\cr
 \n&\n&\n&\n&\n&\n&\n&\n&\n\cr
 \n&\n&\n&{\CoefTrue}&\n&\n&\n&\n&\n\cr
 \n&\n&\n&\n&\n&\n&\n&\n&\n\cr
 \n&\n&\n&\n&\n&\n&\n&\n&\n\cr
 \n&\n&\n&\n&\n&\n&\n&\n&\n\cr
 \n&\n&\n&\n&\n&\n&\n&\n&\n\cr
 \n&\n&\n&\n&\n&\n&\n&\n&\n\cr
\noalign{\kern-2pt}}}%
\vbox{\setbox8=\hbox{$\RowNames\Matrix$}
\hbox to\wd8{\hfil$\ColNames$\kern\ColEntryShiftHoriz}\kern\ColEntryShiftVerti
\box8}}
{\footnotesize%
\BoxBreadth=0pt%
\setbox7=\hbox{[1,2,3]}%
\ifdim\wd7>\BoxBreadth\BoxBreadth=\wd7\fi%
\setbox7=\hbox{[2,3,1]}%
\ifdim\wd7>\BoxBreadth\BoxBreadth=\wd7\fi%
\setbox7=\hbox{[3,1,2]}%
\ifdim\wd7>\BoxBreadth\BoxBreadth=\wd7\fi%
\def\RowNames{\def\scalable{\interspacereduction}
\vcenter{\offinterlineskip\baselineskip=\matrixskip%
\hbox to\BoxBreadth{\strut\hfil [1,2,3]}\kern\scalable%
\hbox to\BoxBreadth{\strut\hfil [2,3,1]}\kern\scalable%
\hbox to\BoxBreadth{\strut\hfil [3,1,2]}}}%
\def\ColNames{\def\scalable{\interspacereduction}
\hbox{\rotatebox{90}{\strut [1,2,3]}\kern\scalable%
\rotatebox{90}{\strut [2,3,1]}\kern\scalable%
\rotatebox{90}{\strut [3,1,2]}\kern\scalable%
\kern0pt
}}%
\def\Matrix{\spmatrix{%
\noalign{\kern-2pt}
 {\CoefTrue}&{\CoefTrue}&{\CoefTrue}\cr
 \n&\n&\n\cr
 \n&\n&\n\cr
\noalign{\kern-2pt}}}%
\vbox{\setbox8=\hbox{$\RowNames\Matrix$}
\hbox to\wd8{\hfil$\ColNames$\kern\ColEntryShiftHoriz}\kern\ColEntryShiftVerti
\box8}}
{\footnotesize%
\BoxBreadth=0pt%
\setbox7=\hbox{[1,2,3]}%
\ifdim\wd7>\BoxBreadth\BoxBreadth=\wd7\fi%
\setbox7=\hbox{[2,3,1]}%
\ifdim\wd7>\BoxBreadth\BoxBreadth=\wd7\fi%
\setbox7=\hbox{[3,1,2]}%
\ifdim\wd7>\BoxBreadth\BoxBreadth=\wd7\fi%
\def\RowNames{\def\scalable{\interspacereduction}
\vcenter{\offinterlineskip\baselineskip=\matrixskip%
\hbox to\BoxBreadth{\strut\hfil [1,2,3]}\kern\scalable%
\hbox to\BoxBreadth{\strut\hfil [2,3,1]}\kern\scalable%
\hbox to\BoxBreadth{\strut\hfil [3,1,2]}}}%
\def\ColNames{\def\scalable{\interspacereduction}
\hbox{\rotatebox{90}{\strut [1,2,3]}\kern\scalable%
\rotatebox{90}{\strut [2,3,1]}\kern\scalable%
\rotatebox{90}{\strut [3,1,2]}\kern\scalable%
\kern0pt
}}%
\def\Matrix{\spmatrix{%
\noalign{\kern-2pt}
 {\CoefTrue}&\n&\n\cr
 \n&\n&{\CoefTrue}\cr
 \n&{\CoefTrue}&\n\cr
\noalign{\kern-2pt}}}%
\vbox{\setbox8=\hbox{$\RowNames\Matrix$}
\hbox to\wd8{\hfil$\ColNames$\kern\ColEntryShiftHoriz}\kern\ColEntryShiftVerti
\box8}}
\hfil}
\hbox to\textwidth{\hfil\quad$\TarskiAdd\hfil\qquad\qquad\qquad\qquad\TarskiDelta\hfil\qquad\TarskiNeutrR\hfil\TarskiInversR\hfil$}
}}
{Existence of right-neutral elements}{FigHasRightNeutral}

\noindent
Here also the forming of inverses $\TarskiInversR$ is indicated. Since $\TarskiNeutrR$ in Fig.~\FigHasRightNeutralUgly\  is not row-constant, it cannot contain a point, so that there is no right-neutral element.

\Caption{\vbox{\hbox to\textwidth{\hfil{\footnotesize%
\BoxBreadth=0pt%
\setbox7=\hbox{a}%
\ifdim\wd7>\BoxBreadth\BoxBreadth=\wd7\fi%
\setbox7=\hbox{b}%
\ifdim\wd7>\BoxBreadth\BoxBreadth=\wd7\fi%
\setbox7=\hbox{c}%
\ifdim\wd7>\BoxBreadth\BoxBreadth=\wd7\fi%
\setbox7=\hbox{d}%
\ifdim\wd7>\BoxBreadth\BoxBreadth=\wd7\fi%
\def\RowNames{\def\scalable{0.45\baselineskip}
\vcenter{\offinterlineskip\baselineskip=\matrixskip%
\hbox to\BoxBreadth{\strut\hfil a}\kern\scalable%
\hbox to\BoxBreadth{\strut\hfil b}\kern\scalable%
\hbox to\BoxBreadth{\strut\hfil c}\kern\scalable%
\hbox to\BoxBreadth{\strut\hfil d}}}%
\def\ColNames{\def\scalable{0.18\baselineskip}
\hbox{\rotatebox{90}{\strut a}\kern\scalable%
\rotatebox{90}{\strut b}\kern\scalable%
\rotatebox{90}{\strut c}\kern\scalable%
\rotatebox{90}{\strut d}\kern\scalable%
\kern0pt
}}%
\def\Matrix{\spmatrix{%
\noalign{\kern-2pt}
 \hbox{\vbox to14pt{\vfil\hbox to14pt{\hfil 1\hfil}\vfil}}&\hbox{\vbox to14pt{\vfil\hbox to14pt{\hfil 1\hfil}\vfil}}&\hbox{\vbox to14pt{\vfil\hbox to14pt{\hfil 1\hfil}\vfil}}&\hbox{\vbox to14pt{\vfil\hbox to14pt{\hfil 1\hfil}\vfil}}\cr
 \hbox{\vbox to14pt{\vfil\hbox to14pt{\hfil 1\hfil}\vfil}}&\hbox{\vbox to14pt{\vfil\hbox to14pt{\hfil 1\hfil}\vfil}}&\hbox{\vbox to14pt{\vfil\hbox to14pt{\hfil 1\hfil}\vfil}}&\hbox{\vbox to14pt{\vfil\hbox to14pt{\hfil 1\hfil}\vfil}}\cr
 \hbox{\vbox to14pt{\vfil\hbox to14pt{\hfil 1\hfil}\vfil}}&\hbox{\vbox to14pt{\vfil\hbox to14pt{\hfil 1\hfil}\vfil}}&\hbox{\vbox to14pt{\vfil\hbox to14pt{\hfil 1\hfil}\vfil}}&\hbox{\vbox to14pt{\vfil\hbox to14pt{\hfil 1\hfil}\vfil}}\cr
 \hbox{\vbox to14pt{\vfil\hbox to14pt{\hfil 1\hfil}\vfil}}&\hbox{\vbox to14pt{\vfil\hbox to14pt{\hfil 1\hfil}\vfil}}&\hbox{\vbox to14pt{\vfil\hbox to14pt{\hfil 1\hfil}\vfil}}&\hbox{\vbox to14pt{\vfil\hbox to14pt{\hfil 1\hfil}\vfil}}\cr
\noalign{\kern-2pt}}}
\vbox{\setbox8=\hbox{$\RowNames\Matrix$}
\hbox to\wd8{\hfil$\ColNames$\kern0.6\matrixskip}\kern0.1cm
\box8}}

\quad
{\footnotesize%
\BoxBreadth=0pt%
\setbox7=\hbox{(a,a)}%
\ifdim\wd7>\BoxBreadth\BoxBreadth=\wd7\fi%
\setbox7=\hbox{(b,a)}%
\ifdim\wd7>\BoxBreadth\BoxBreadth=\wd7\fi%
\setbox7=\hbox{(a,b)}%
\ifdim\wd7>\BoxBreadth\BoxBreadth=\wd7\fi%
\setbox7=\hbox{(c,a)}%
\ifdim\wd7>\BoxBreadth\BoxBreadth=\wd7\fi%
\setbox7=\hbox{(b,b)}%
\ifdim\wd7>\BoxBreadth\BoxBreadth=\wd7\fi%
\setbox7=\hbox{(a,c)}%
\ifdim\wd7>\BoxBreadth\BoxBreadth=\wd7\fi%
\setbox7=\hbox{(d,a)}%
\ifdim\wd7>\BoxBreadth\BoxBreadth=\wd7\fi%
\setbox7=\hbox{(c,b)}%
\ifdim\wd7>\BoxBreadth\BoxBreadth=\wd7\fi%
\setbox7=\hbox{(b,c)}%
\ifdim\wd7>\BoxBreadth\BoxBreadth=\wd7\fi%
\setbox7=\hbox{(a,d)}%
\ifdim\wd7>\BoxBreadth\BoxBreadth=\wd7\fi%
\setbox7=\hbox{(d,b)}%
\ifdim\wd7>\BoxBreadth\BoxBreadth=\wd7\fi%
\setbox7=\hbox{(c,c)}%
\ifdim\wd7>\BoxBreadth\BoxBreadth=\wd7\fi%
\setbox7=\hbox{(b,d)}%
\ifdim\wd7>\BoxBreadth\BoxBreadth=\wd7\fi%
\setbox7=\hbox{(d,c)}%
\ifdim\wd7>\BoxBreadth\BoxBreadth=\wd7\fi%
\setbox7=\hbox{(c,d)}%
\ifdim\wd7>\BoxBreadth\BoxBreadth=\wd7\fi%
\setbox7=\hbox{(d,d)}%
\ifdim\wd7>\BoxBreadth\BoxBreadth=\wd7\fi%
\def\RowNames{\def\scalable{\interspacereduction}
\vcenter{\offinterlineskip\baselineskip=\matrixskip%
\hbox to\BoxBreadth{\strut\hfil (a,a)}\kern\scalable%
\hbox to\BoxBreadth{\strut\hfil (b,a)}\kern\scalable%
\hbox to\BoxBreadth{\strut\hfil (a,b)}\kern\scalable%
\hbox to\BoxBreadth{\strut\hfil (c,a)}\kern\scalable%
\hbox to\BoxBreadth{\strut\hfil (b,b)}\kern\scalable%
\hbox to\BoxBreadth{\strut\hfil (a,c)}\kern\scalable%
\hbox to\BoxBreadth{\strut\hfil (d,a)}\kern\scalable%
\hbox to\BoxBreadth{\strut\hfil (c,b)}\kern\scalable%
\hbox to\BoxBreadth{\strut\hfil (b,c)}\kern\scalable%
\hbox to\BoxBreadth{\strut\hfil (a,d)}\kern\scalable%
\hbox to\BoxBreadth{\strut\hfil (d,b)}\kern\scalable%
\hbox to\BoxBreadth{\strut\hfil (c,c)}\kern\scalable%
\hbox to\BoxBreadth{\strut\hfil (b,d)}\kern\scalable%
\hbox to\BoxBreadth{\strut\hfil (d,c)}\kern\scalable%
\hbox to\BoxBreadth{\strut\hfil (c,d)}\kern\scalable%
\hbox to\BoxBreadth{\strut\hfil (d,d)}}}%
\def\ColNames{\def\scalable{\interspacereduction}
\hbox{\rotatebox{90}{\strut (a,a)}\kern\scalable%
\rotatebox{90}{\strut (b,a)}\kern\scalable%
\rotatebox{90}{\strut (a,b)}\kern\scalable%
\rotatebox{90}{\strut (c,a)}\kern\scalable%
\rotatebox{90}{\strut (b,b)}\kern\scalable%
\rotatebox{90}{\strut (a,c)}\kern\scalable%
\rotatebox{90}{\strut (d,a)}\kern\scalable%
\rotatebox{90}{\strut (c,b)}\kern\scalable%
\rotatebox{90}{\strut (b,c)}\kern\scalable%
\rotatebox{90}{\strut (a,d)}\kern\scalable%
\rotatebox{90}{\strut (d,b)}\kern\scalable%
\rotatebox{90}{\strut (c,c)}\kern\scalable%
\rotatebox{90}{\strut (b,d)}\kern\scalable%
\rotatebox{90}{\strut (d,c)}\kern\scalable%
\rotatebox{90}{\strut (c,d)}\kern\scalable%
\rotatebox{90}{\strut (d,d)}\kern\scalable%
\kern0pt
}}%
\def\Matrix{\spmatrix{%
\noalign{\kern-2pt}
 {\CoefTrue}&\n&\n&\n&\n&\n&\n&\n&\n&\n&\n&\n&\n&\n&\n&\n\cr
 \n&\n&\n&\n&\n&\n&\n&\n&\n&\n&\n&\n&\n&\n&\n&\n\cr
 \n&\n&{\CoefTrue}&\n&\n&\n&\n&\n&\n&\n&\n&\n&\n&\n&\n&\n\cr
 \n&\n&\n&\n&\n&\n&\n&\n&\n&\n&\n&\n&\n&\n&\n&\n\cr
 \n&\n&\n&\n&\n&\n&\n&\n&\n&\n&\n&\n&\n&\n&\n&\n\cr
 \n&\n&\n&\n&\n&{\CoefTrue}&\n&\n&\n&\n&\n&\n&\n&\n&\n&\n\cr
 \n&\n&\n&\n&\n&\n&\n&\n&\n&\n&\n&\n&\n&\n&\n&\n\cr
 \n&\n&\n&\n&\n&\n&\n&\n&\n&\n&\n&\n&\n&\n&\n&\n\cr
 \n&\n&\n&\n&\n&\n&\n&\n&\n&\n&\n&\n&\n&\n&\n&\n\cr
 \n&\n&\n&\n&\n&\n&\n&\n&\n&{\CoefTrue}&\n&\n&\n&\n&\n&\n\cr
 \n&\n&\n&\n&\n&\n&\n&\n&\n&\n&\n&\n&\n&\n&\n&\n\cr
 \n&\n&\n&\n&\n&\n&\n&\n&\n&\n&\n&\n&\n&\n&\n&\n\cr
 \n&\n&\n&\n&\n&\n&\n&\n&\n&\n&\n&\n&\n&\n&\n&\n\cr
 \n&\n&\n&\n&\n&\n&\n&\n&\n&\n&\n&\n&\n&\n&\n&\n\cr
 \n&\n&\n&\n&\n&\n&\n&\n&\n&\n&\n&\n&\n&\n&\n&\n\cr
 \n&\n&\n&\n&\n&\n&\n&\n&\n&\n&\n&\n&\n&\n&\n&\n\cr
\noalign{\kern-2pt}}}%
\vbox{\setbox8=\hbox{$\RowNames\Matrix$}
\hbox to\wd8{\hfil$\ColNames$\kern\ColEntryShiftHoriz}\kern\ColEntryShiftVerti
\box8}}
{\footnotesize%
\BoxBreadth=0pt%
\setbox7=\hbox{a}%
\ifdim\wd7>\BoxBreadth\BoxBreadth=\wd7\fi%
\setbox7=\hbox{b}%
\ifdim\wd7>\BoxBreadth\BoxBreadth=\wd7\fi%
\setbox7=\hbox{c}%
\ifdim\wd7>\BoxBreadth\BoxBreadth=\wd7\fi%
\setbox7=\hbox{d}%
\ifdim\wd7>\BoxBreadth\BoxBreadth=\wd7\fi%
\def\RowNames{\def\scalable{\interspacereduction}
\vcenter{\offinterlineskip\baselineskip=\matrixskip%
\hbox to\BoxBreadth{\strut\hfil a}\kern\scalable%
\hbox to\BoxBreadth{\strut\hfil b}\kern\scalable%
\hbox to\BoxBreadth{\strut\hfil c}\kern\scalable%
\hbox to\BoxBreadth{\strut\hfil d}}}%
\def\ColNames{\def\scalable{\interspacereduction}
\hbox{\rotatebox{90}{\strut a}\kern\scalable%
\rotatebox{90}{\strut b}\kern\scalable%
\rotatebox{90}{\strut c}\kern\scalable%
\rotatebox{90}{\strut d}\kern\scalable%
\kern0pt
}}%
\def\Matrix{\spmatrix{%
\noalign{\kern-2pt}
 {\CoefTrue}&\n&\n&\n\cr
 {\CoefTrue}&\n&\n&\n\cr
 {\CoefTrue}&\n&\n&\n\cr
 {\CoefTrue}&\n&\n&\n\cr
\noalign{\kern-2pt}}}%
\vbox{\setbox8=\hbox{$\RowNames\Matrix$}
\hbox to\wd8{\hfil$\ColNames$\kern\ColEntryShiftHoriz}\kern\ColEntryShiftVerti
\box8}}
{\footnotesize%
\BoxBreadth=0pt%
\setbox7=\hbox{a}%
\ifdim\wd7>\BoxBreadth\BoxBreadth=\wd7\fi%
\setbox7=\hbox{b}%
\ifdim\wd7>\BoxBreadth\BoxBreadth=\wd7\fi%
\setbox7=\hbox{c}%
\ifdim\wd7>\BoxBreadth\BoxBreadth=\wd7\fi%
\setbox7=\hbox{d}%
\ifdim\wd7>\BoxBreadth\BoxBreadth=\wd7\fi%
\def\RowNames{\def\scalable{\interspacereduction}
\vcenter{\offinterlineskip\baselineskip=\matrixskip%
\hbox to\BoxBreadth{\strut\hfil a}\kern\scalable%
\hbox to\BoxBreadth{\strut\hfil b}\kern\scalable%
\hbox to\BoxBreadth{\strut\hfil c}\kern\scalable%
\hbox to\BoxBreadth{\strut\hfil d}}}%
\def\ColNames{\def\scalable{\interspacereduction}
\hbox{\rotatebox{90}{\strut a}\kern\scalable%
\rotatebox{90}{\strut b}\kern\scalable%
\rotatebox{90}{\strut c}\kern\scalable%
\rotatebox{90}{\strut d}\kern\scalable%
\kern0pt
}}%
\def\Matrix{\spmatrix{%
\noalign{\kern-2pt}
 {\CoefTrue}&\n&\n&\n\cr
 {\CoefTrue}&\n&\n&\n\cr
 {\CoefTrue}&\n&\n&\n\cr
 {\CoefTrue}&\n&\n&\n\cr
\noalign{\kern-2pt}}}%
\vbox{\setbox8=\hbox{$\RowNames\Matrix$}
\hbox to\wd8{\hfil$\ColNames$\kern\ColEntryShiftHoriz}\kern\ColEntryShiftVerti
\box8}}
\hfil}
\hbox to\textwidth{\hfil$\TarskiAdd\hfil\qquad\qquad\qquad\qquad\TarskiDelta\hfil\qquad\qquad\TarskiNeutrR\hfil\TarskiInversR\hfil$}
}}
{Non-existence of right-neutral elements}{FigHasRightNeutralUgly}

\noindent
Right- or left-neutral elements may exist or not.
In Fig.~\FigTwoNeutral\ we see what it means to be 
right-neutral: The corresponding two columns correspond to the row-inscriptions.

\Caption{\hbox{$\TarskiAdd=$ {\footnotesize%
\BoxBreadth=0pt%
\setbox7=\hbox{a}%
\ifdim\wd7>\BoxBreadth\BoxBreadth=\wd7\fi%
\setbox7=\hbox{b}%
\ifdim\wd7>\BoxBreadth\BoxBreadth=\wd7\fi%
\setbox7=\hbox{c}%
\ifdim\wd7>\BoxBreadth\BoxBreadth=\wd7\fi%
\setbox7=\hbox{d}%
\ifdim\wd7>\BoxBreadth\BoxBreadth=\wd7\fi%
\setbox7=\hbox{e}%
\ifdim\wd7>\BoxBreadth\BoxBreadth=\wd7\fi%
\setbox7=\hbox{f}%
\ifdim\wd7>\BoxBreadth\BoxBreadth=\wd7\fi%
\setbox7=\hbox{g}%
\ifdim\wd7>\BoxBreadth\BoxBreadth=\wd7\fi%
\def\RowNames{\def\scalable{0.45\baselineskip}
\vcenter{\offinterlineskip\baselineskip=\matrixskip%
\hbox to\BoxBreadth{\strut\hfil a}\kern\scalable%
\hbox to\BoxBreadth{\strut\hfil b}\kern\scalable%
\hbox to\BoxBreadth{\strut\hfil c}\kern\scalable%
\hbox to\BoxBreadth{\strut\hfil d}\kern\scalable%
\hbox to\BoxBreadth{\strut\hfil e}\kern\scalable%
\hbox to\BoxBreadth{\strut\hfil f}\kern\scalable%
\hbox to\BoxBreadth{\strut\hfil g}}}%
\def\ColNames{\def\scalable{0.17\baselineskip}
\hbox{\rotatebox{90}{\strut a}\kern\scalable%
\rotatebox{90}{\strut b}\kern\scalable%
\rotatebox{90}{\strut c}\kern\scalable%
\rotatebox{90}{\strut d}\kern\scalable%
\rotatebox{90}{\strut e}\kern\scalable%
\rotatebox{90}{\strut f}\kern\scalable%
\rotatebox{90}{\strut g}\kern\scalable%
\kern0pt
}}%
\def\Matrix{\spmatrix{%
\noalign{\kern-2pt}
 \hbox{\vbox to14pt{\vfil\hbox to14pt{\hfil 3\hfil}\vfil}}&\hbox{\vbox to14pt{\vfil\hbox to14pt{\hfil 2\hfil}\vfil}}&\hbox{\vbox to14pt{\vfil\hbox to14pt{\hfil 1\hfil}\vfil}}&\hbox{\vbox to14pt{\vfil\hbox to14pt{\hfil 4\hfil}\vfil}}&\hbox{\vbox to14pt{\vfil\hbox to14pt{\hfil 1\hfil}\vfil}}&\hbox{\vbox to14pt{\vfil\hbox to14pt{\hfil 6\hfil}\vfil}}&\hbox{\vbox to14pt{\vfil\hbox to14pt{\hfil 7\hfil}\vfil}}\cr
 \hbox{\vbox to14pt{\vfil\hbox to14pt{\hfil 1\hfil}\vfil}}&\hbox{\vbox to14pt{\vfil\hbox to14pt{\hfil 3\hfil}\vfil}}&\hbox{\vbox to14pt{\vfil\hbox to14pt{\hfil 2\hfil}\vfil}}&\hbox{\vbox to14pt{\vfil\hbox to14pt{\hfil 4\hfil}\vfil}}&\hbox{\vbox to14pt{\vfil\hbox to14pt{\hfil 2\hfil}\vfil}}&\hbox{\vbox to14pt{\vfil\hbox to14pt{\hfil 6\hfil}\vfil}}&\hbox{\vbox to14pt{\vfil\hbox to14pt{\hfil 7\hfil}\vfil}}\cr
 \hbox{\vbox to14pt{\vfil\hbox to14pt{\hfil 1\hfil}\vfil}}&\hbox{\vbox to14pt{\vfil\hbox to14pt{\hfil 2\hfil}\vfil}}&\hbox{\vbox to14pt{\vfil\hbox to14pt{\hfil 3\hfil}\vfil}}&\hbox{\vbox to14pt{\vfil\hbox to14pt{\hfil 4\hfil}\vfil}}&\hbox{\vbox to14pt{\vfil\hbox to14pt{\hfil 3\hfil}\vfil}}&\hbox{\vbox to14pt{\vfil\hbox to14pt{\hfil 6\hfil}\vfil}}&\hbox{\vbox to14pt{\vfil\hbox to14pt{\hfil 7\hfil}\vfil}}\cr
 \hbox{\vbox to14pt{\vfil\hbox to14pt{\hfil 1\hfil}\vfil}}&\hbox{\vbox to14pt{\vfil\hbox to14pt{\hfil 5\hfil}\vfil}}&\hbox{\vbox to14pt{\vfil\hbox to14pt{\hfil 4\hfil}\vfil}}&\hbox{\vbox to14pt{\vfil\hbox to14pt{\hfil 7\hfil}\vfil}}&\hbox{\vbox to14pt{\vfil\hbox to14pt{\hfil 4\hfil}\vfil}}&\hbox{\vbox to14pt{\vfil\hbox to14pt{\hfil 3\hfil}\vfil}}&\hbox{\vbox to14pt{\vfil\hbox to14pt{\hfil 6\hfil}\vfil}}\cr
 \hbox{\vbox to14pt{\vfil\hbox to14pt{\hfil 1\hfil}\vfil}}&\hbox{\vbox to14pt{\vfil\hbox to14pt{\hfil 2\hfil}\vfil}}&\hbox{\vbox to14pt{\vfil\hbox to14pt{\hfil 5\hfil}\vfil}}&\hbox{\vbox to14pt{\vfil\hbox to14pt{\hfil 4\hfil}\vfil}}&\hbox{\vbox to14pt{\vfil\hbox to14pt{\hfil 5\hfil}\vfil}}&\hbox{\vbox to14pt{\vfil\hbox to14pt{\hfil 6\hfil}\vfil}}&\hbox{\vbox to14pt{\vfil\hbox to14pt{\hfil 7\hfil}\vfil}}\cr
 \hbox{\vbox to14pt{\vfil\hbox to14pt{\hfil 1\hfil}\vfil}}&\hbox{\vbox to14pt{\vfil\hbox to14pt{\hfil 3\hfil}\vfil}}&\hbox{\vbox to14pt{\vfil\hbox to14pt{\hfil 6\hfil}\vfil}}&\hbox{\vbox to14pt{\vfil\hbox to14pt{\hfil 5\hfil}\vfil}}&\hbox{\vbox to14pt{\vfil\hbox to14pt{\hfil 6\hfil}\vfil}}&\hbox{\vbox to14pt{\vfil\hbox to14pt{\hfil 4\hfil}\vfil}}&\hbox{\vbox to14pt{\vfil\hbox to14pt{\hfil 7\hfil}\vfil}}\cr
 \hbox{\vbox to14pt{\vfil\hbox to14pt{\hfil 1\hfil}\vfil}}&\hbox{\vbox to14pt{\vfil\hbox to14pt{\hfil 2\hfil}\vfil}}&\hbox{\vbox to14pt{\vfil\hbox to14pt{\hfil 7\hfil}\vfil}}&\hbox{\vbox to14pt{\vfil\hbox to14pt{\hfil 4\hfil}\vfil}}&\hbox{\vbox to14pt{\vfil\hbox to14pt{\hfil 7\hfil}\vfil}}&\hbox{\vbox to14pt{\vfil\hbox to14pt{\hfil 6\hfil}\vfil}}&\hbox{\vbox to14pt{\vfil\hbox to14pt{\hfil 3\hfil}\vfil}}\cr
\noalign{\kern-2pt}}}
\vbox{\setbox8=\hbox{$\RowNames\Matrix$}
\hbox to\wd8{\hfil$\ColNames$\kern0.6\matrixskip}\kern0.1cm
\box8}}

\quad 

{\footnotesize%
\BoxBreadth=0pt%
\setbox7=\hbox{a}%
\ifdim\wd7>\BoxBreadth\BoxBreadth=\wd7\fi%
\setbox7=\hbox{b}%
\ifdim\wd7>\BoxBreadth\BoxBreadth=\wd7\fi%
\setbox7=\hbox{c}%
\ifdim\wd7>\BoxBreadth\BoxBreadth=\wd7\fi%
\setbox7=\hbox{d}%
\ifdim\wd7>\BoxBreadth\BoxBreadth=\wd7\fi%
\setbox7=\hbox{e}%
\ifdim\wd7>\BoxBreadth\BoxBreadth=\wd7\fi%
\setbox7=\hbox{f}%
\ifdim\wd7>\BoxBreadth\BoxBreadth=\wd7\fi%
\setbox7=\hbox{g}%
\ifdim\wd7>\BoxBreadth\BoxBreadth=\wd7\fi%
\def\RowNames{\vcenter{\offinterlineskip\baselineskip=\matrixskip%
\hbox to\BoxBreadth{\strut\hfil a}\kern\interspacereduction%
\hbox to\BoxBreadth{\strut\hfil b}\kern\interspacereduction%
\hbox to\BoxBreadth{\strut\hfil c}\kern\interspacereduction%
\hbox to\BoxBreadth{\strut\hfil d}\kern\interspacereduction%
\hbox to\BoxBreadth{\strut\hfil e}\kern\interspacereduction%
\hbox to\BoxBreadth{\strut\hfil f}\kern\interspacereduction%
\hbox to\BoxBreadth{\strut\hfil g}}}%
\def\MarkVect{\setbox2=\hbox{\strut}\matrixskip=\ht2\spmatrix{\noalign{\kern-2pt}%
\n\cr
\n\cr
{\CoefTrue}\cr
\n\cr
{\CoefTrue}\cr
\n\cr
\n\cr
\noalign{\kern-2pt}}%
}%
$\RowNames\MarkVect=\TarskiNull_{\rm right}$}

\quad {\footnotesize%
\BoxBreadth=0pt%
\setbox7=\hbox{a}%
\ifdim\wd7>\BoxBreadth\BoxBreadth=\wd7\fi%
\setbox7=\hbox{b}%
\ifdim\wd7>\BoxBreadth\BoxBreadth=\wd7\fi%
\setbox7=\hbox{c}%
\ifdim\wd7>\BoxBreadth\BoxBreadth=\wd7\fi%
\setbox7=\hbox{d}%
\ifdim\wd7>\BoxBreadth\BoxBreadth=\wd7\fi%
\setbox7=\hbox{e}%
\ifdim\wd7>\BoxBreadth\BoxBreadth=\wd7\fi%
\setbox7=\hbox{f}%
\ifdim\wd7>\BoxBreadth\BoxBreadth=\wd7\fi%
\setbox7=\hbox{g}%
\ifdim\wd7>\BoxBreadth\BoxBreadth=\wd7\fi%
\def\RowNames{\def\scalable{\interspacereduction}
\vcenter{\offinterlineskip\baselineskip=\matrixskip%
\hbox to\BoxBreadth{\strut\hfil a}\kern\scalable%
\hbox to\BoxBreadth{\strut\hfil b}\kern\scalable%
\hbox to\BoxBreadth{\strut\hfil c}\kern\scalable%
\hbox to\BoxBreadth{\strut\hfil d}\kern\scalable%
\hbox to\BoxBreadth{\strut\hfil e}\kern\scalable%
\hbox to\BoxBreadth{\strut\hfil f}\kern\scalable%
\hbox to\BoxBreadth{\strut\hfil g}}}%
\def\ColNames{\def\scalable{\interspacereduction}
\hbox{\rotatebox{90}{\strut a}\kern\scalable%
\rotatebox{90}{\strut b}\kern\scalable%
\rotatebox{90}{\strut c}\kern\scalable%
\rotatebox{90}{\strut d}\kern\scalable%
\rotatebox{90}{\strut e}\kern\scalable%
\rotatebox{90}{\strut f}\kern\scalable%
\rotatebox{90}{\strut g}\kern\scalable%
\kern0pt
}}%
\def\Matrix{\spmatrix{%
\noalign{\kern-2pt}
 {\CoefTrue}&\n&\n&\n&\n&\n&\n\cr
 \n&{\CoefTrue}&\n&\n&\n&\n&\n\cr
 \n&\n&{\CoefTrue}&\n&{\CoefTrue}&\n&\n\cr
 \n&{\CoefTrue}&\n&\n&\n&{\CoefTrue}&\n\cr
 \n&\n&{\CoefTrue}&\n&{\CoefTrue}&\n&\n\cr
 \n&{\CoefTrue}&\n&{\CoefTrue}&\n&\n&\n\cr
 \n&\n&\n&\n&\n&\n&{\CoefTrue}\cr
\noalign{\kern-2pt}}}%
\vbox{\setbox8=\hbox{$\RowNames\Matrix$}
\hbox to\wd8{\hfil$\ColNames$\kern\ColEntryShiftHoriz}\kern\ColEntryShiftVerti
\box8}}\quad 

{\footnotesize%
\BoxBreadth=0pt%
\setbox7=\hbox{a}%
\ifdim\wd7>\BoxBreadth\BoxBreadth=\wd7\fi%
\setbox7=\hbox{b}%
\ifdim\wd7>\BoxBreadth\BoxBreadth=\wd7\fi%
\setbox7=\hbox{c}%
\ifdim\wd7>\BoxBreadth\BoxBreadth=\wd7\fi%
\setbox7=\hbox{d}%
\ifdim\wd7>\BoxBreadth\BoxBreadth=\wd7\fi%
\setbox7=\hbox{e}%
\ifdim\wd7>\BoxBreadth\BoxBreadth=\wd7\fi%
\setbox7=\hbox{f}%
\ifdim\wd7>\BoxBreadth\BoxBreadth=\wd7\fi%
\setbox7=\hbox{g}%
\ifdim\wd7>\BoxBreadth\BoxBreadth=\wd7\fi%
\def\RowNames{\def\scalable{\interspacereduction}
\vcenter{\offinterlineskip\baselineskip=\matrixskip%
\hbox to\BoxBreadth{\strut\hfil a}\kern\scalable%
\hbox to\BoxBreadth{\strut\hfil b}\kern\scalable%
\hbox to\BoxBreadth{\strut\hfil c}\kern\scalable%
\hbox to\BoxBreadth{\strut\hfil d}\kern\scalable%
\hbox to\BoxBreadth{\strut\hfil e}\kern\scalable%
\hbox to\BoxBreadth{\strut\hfil f}\kern\scalable%
\hbox to\BoxBreadth{\strut\hfil g}}}%
\def\ColNames{\def\scalable{\interspacereduction}
\hbox{\rotatebox{90}{\strut a}\kern\scalable%
\rotatebox{90}{\strut b}\kern\scalable%
\rotatebox{90}{\strut c}\kern\scalable%
\rotatebox{90}{\strut d}\kern\scalable%
\rotatebox{90}{\strut e}\kern\scalable%
\rotatebox{90}{\strut f}\kern\scalable%
\rotatebox{90}{\strut g}\kern\scalable%
\kern0pt
}}%
\def\Matrix{\spmatrix{%
\noalign{\kern-2pt}
 \n&\n&\n&\n&\n&\n&\n\cr
 \n&\n&\n&\n&\n&\n&\n\cr
 \n&\n&\n&\n&\n&\n&\n\cr
 \n&{\CoefTrue}&\n&\n&\n&\n&\n\cr
 \n&\n&{\CoefTrue}&\n&{\CoefTrue}&\n&\n\cr
 \n&\n&\n&{\CoefTrue}&\n&\n&\n\cr
 \n&\n&\n&\n&\n&\n&\n\cr
\noalign{\kern-2pt}}}%
\vbox{\setbox8=\hbox{$\RowNames\Matrix$}
\hbox to\wd8{\hfil$\ColNames$\kern\ColEntryShiftHoriz}\kern\ColEntryShiftVerti
\box8}}

}}
{Binary map without left- but two right-neutrals $c,e$ and right-inverses wrt.~to $c$ and $e$}{FigTwoNeutral}

\noindent
A left-neutral element in analogy, gives rise to a row identical with the column numbering. From this fact it will become clear that there can be at most one point as neutral element $e$. The aforementioned transition to inverses

\smallskip
$i:=\pi\RELtraOP(\TarskiAdd\RELcompOP e\RELandOP\rho)$.

\smallskip
\noindent 
will then be a bijective mapping, which it was neither for $c$ nor for $e$ in Fig.~\FigTwoNeutral.

\enunc{}{Proposition}{}{LeftRightNeutral} For some binary mapping $\TarskiAdd $ we consider the left- as well as right-neutral element sets $\TarskiNeutrL, \TarskiNeutrR$. If both contain points $e_l,e_r$, these will be equal.

\Proof We apply the result obtained before in two directions 

\smallskip
$\TarskiAdd\RELtraOP\RELcompOP(\pi\RELcompOP Y\RELandOP\rho\RELcompOP e_r\RELcompOP\RELtop)
= 
\TarskiAdd\RELtraOP\RELcompOP(\pi\RELandOP\rho\RELcompOP e_r\RELcompOP\RELtop)\RELcompOP Y 
=
\TarskiAdd\RELtraOP\RELcompOP g\RELtraOP\RELcompOP Y 
=
\RELide\RELcompOP Y
=
Y
$,

\smallskip
\noindent 
and correspondingly

\smallskip
$\TarskiAdd\RELtraOP\RELcompOP(\pi\RELcompOP e_l\RELcompOP\RELtop\RELandOP\rho\RELcompOP Z)
= 
\TarskiAdd\RELtraOP\RELcompOP(\rho\RELandOP\pi\RELcompOP e_l\RELcompOP\RELtop)\RELcompOP Z 
=
\TarskiAdd\RELtraOP\RELcompOP f\RELtraOP\RELcompOP Z
=
\RELide\RELcompOP Z
=
Z
$.

\newpage
\noindent 
Therefore

\smallskip
$e_l\RELcompOP\RELtop
=
\TarskiAdd\RELtraOP\RELcompOP\StrictJoin{e_l\RELcompOP\RELtop}{e_r\RELcompOP\RELtop}
=
\TarskiAdd\RELtraOP\RELcompOP(\pi\RELcompOP e_l\RELcompOP\RELtop\RELandOP\rho\RELcompOP e_r\RELcompOP\RELtop)
=
e_r\RELcompOP\RELtop
$
\Bewende

\noindent
Should there exist more than one in either one of $n_l,n_r$ they will thus all be equal.

\Caption{\includegraphics[scale=0.5]{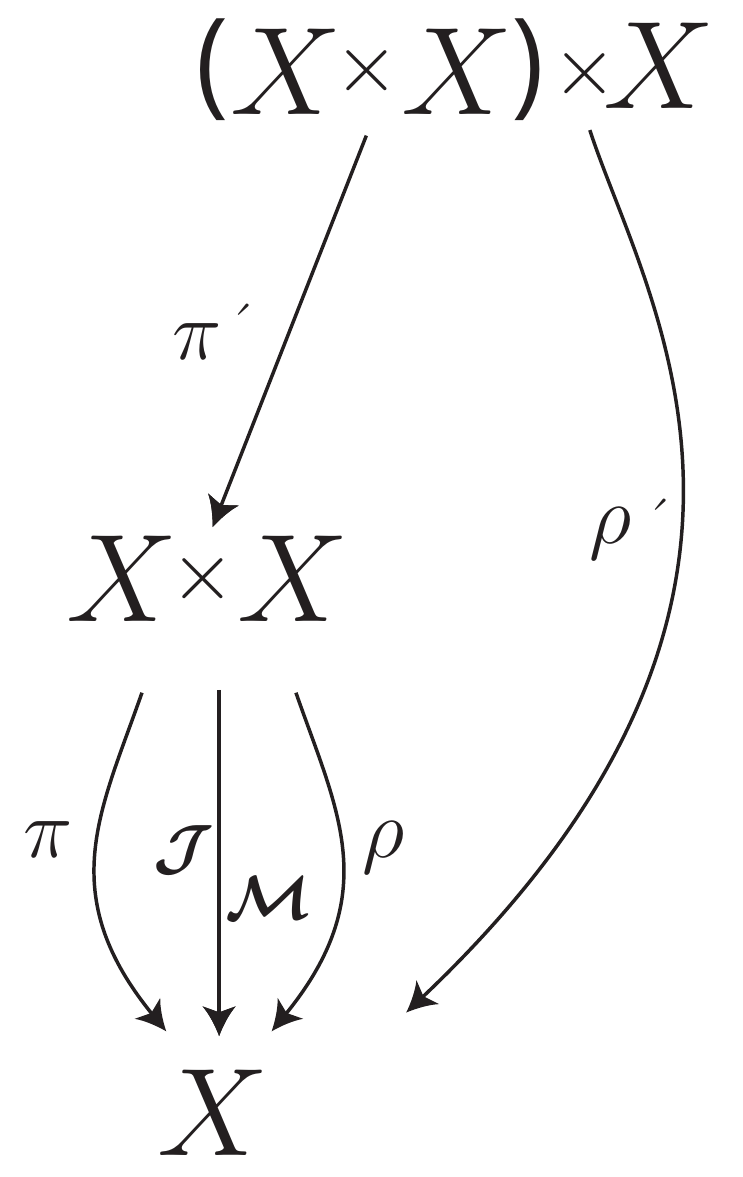}}
{Illustrating distributivity}{FigDistribut}

\def\JotHat{\hat{\liftedJoin}}

\noindent
Also the concept of distributivity\index{distributive} may be formulated relationally in case there are two binary mappings $\liftedJoin,\liftedMeet$, as, e.g., in a lattice the join and meet.

\enunc{}{Definition}{}{DefDistr} Given two binary mappings, we say that $\liftedJoin$ {\bf distributes} over $\liftedMeet$, when

\smallskip
$\StrictFork{\Kronecker{\pi}{\RELide}\RELcompOP\liftedJoin}{\Kronecker{\rho}{\RELide}\RELcompOP\liftedJoin}\RELcompOP\liftedMeet=\Kronecker{\liftedMeet}{\RELide}\RELcompOP\liftedJoin
$,

\smallskip
\noindent
or else, when $\JotHat\RELcompOP\liftedMeet=\Kronecker{\liftedMeet}{\RELide}\RELcompOP\liftedJoin$ as we will later slightly abbreviate.
\Bewende

\noindent
One might also demand in blown-up form resembling $(a\predor c)\predand(b\predor c)=(a\predand b)\predor c$

\smallskip
$\big[(\pi'\RELcompOP\pi\RELcompOP\pi\RELtraOP\RELandOP\rho'\RELcompOP\rho\RELtraOP)\RELcompOP\liftedJoin\RELcompOP\pi\RELtraOP\RELandOP
(\pi'\RELcompOP\rho\RELcompOP\pi\RELtraOP\RELandOP\rho'\RELcompOP\rho\RELtraOP)\RELcompOP\liftedJoin\RELcompOP\rho\RELtraOP\big]\RELcompOP\liftedMeet
=(\pi'\RELcompOP\liftedMeet\RELcompOP\pi\RELtraOP\RELandOP\rho'\RELcompOP\rho\RELtraOP)\RELcompOP\liftedJoin
$.


\chapter{\Boolean\ algebras\index{Boolean algebra@\Boolean\ algebra}}\label{SectBoolAlg}
\EnuncNo=0
\CaptionNo=0


\noindent
A note seems necessary concerning \Boolean\ algebras; here supported with visualization in a concrete example. The
peculiar recursive and fractal symmetries of these examples often give additional insight --- and have already triggered secretaries to stitch such patterns for a pot cloth.

\bigskip
\noindent
Most people work with subsets $U\subseteq X$, while we distinguish 
between a subset in this standard form and the corresponding element $e$ in the powerset, considered as a point.
The two are related via the membership relation $\varepsilon$ as shown in Fig.~\FigSubsetElement\ together with the powerset ordering $\Omega=\RELneg{\varepsilon\RELtraOP\RELcompOP\RELneg{\varepsilon}}$.

\bigskip
\noindent
Theoreticians frequently consider \Boolean\ algebras \lq\lq with signature
$\langle X,\,\cdot\,,+,-,\n,\CoefTrue\rangle$\rq\rq. Following their idea, we find on $X$ the operations $\RELandOP,\RELorOP,\RELneg{\phantom{X}},\RELbot,\RELtop$.

\Caption{$\vcenter{\hbox{\qquad\quad$U=\varepsilon\RELcompOP e\qquad e=\syqq{\varepsilon}{U}$} 
\kern0.5cm
\hbox{{\footnotesize%
\BoxBreadth=0pt%
\setbox7=\hbox{a}%
\ifdim\wd7>\BoxBreadth\BoxBreadth=\wd7\fi%
\setbox7=\hbox{b}%
\ifdim\wd7>\BoxBreadth\BoxBreadth=\wd7\fi%
\setbox7=\hbox{c}%
\ifdim\wd7>\BoxBreadth\BoxBreadth=\wd7\fi%
\setbox7=\hbox{d}%
\ifdim\wd7>\BoxBreadth\BoxBreadth=\wd7\fi%
\def\RowNames{\vcenter{\offinterlineskip\baselineskip=\matrixskip%
\hbox to\BoxBreadth{\strut\hfil a}\kern\interspacereduction%
\hbox to\BoxBreadth{\strut\hfil b}\kern\interspacereduction%
\hbox to\BoxBreadth{\strut\hfil c}\kern\interspacereduction%
\hbox to\BoxBreadth{\strut\hfil d}}}%
\def\ColNames{\hbox{\rotatebox{90}{\strut \{\}}\kern\interspacereduction%
\rotatebox{90}{\strut \{a\}}\kern\interspacereduction%
\rotatebox{90}{\strut \{b\}}\kern\interspacereduction%
\rotatebox{90}{\strut \{a,b\}}\kern\interspacereduction%
\rotatebox{90}{\strut \{c\}}\kern\interspacereduction%
\rotatebox{90}{\strut \{a,c\}}\kern\interspacereduction%
\rotatebox{90}{\strut \{b,c\}}\kern\interspacereduction%
\rotatebox{90}{\strut \{a,b,c\}}\kern\interspacereduction%
\rotatebox{90}{\strut \{d\}}\kern\interspacereduction%
\rotatebox{90}{\strut \{a,d\}}\kern\interspacereduction%
\rotatebox{90}{\strut \{b,d\}}\kern\interspacereduction%
\rotatebox{90}{\strut \{a,b,d\}}\kern\interspacereduction%
\rotatebox{90}{\strut \{c,d\}}\kern\interspacereduction%
\rotatebox{90}{\strut \{a,c,d\}}\kern\interspacereduction%
\rotatebox{90}{\strut \{b,c,d\}}\kern\interspacereduction%
\rotatebox{90}{\strut \{a,b,c,d\}}\kern\interspacereduction%
}}%
\def\Matrix{\spmatrix{%
\noalign{\kern-2pt}
 \n&{\CoefTrue}&\n&{\CoefTrue}&\n&{\CoefTrue}&\n&{\CoefTrue}&\n&{\CoefTrue}&\n&{\CoefTrue}&\n&{\CoefTrue}&\n&{\CoefTrue}\cr
 \n&\n&{\CoefTrue}&{\CoefTrue}&\n&\n&{\CoefTrue}&{\CoefTrue}&\n&\n&{\CoefTrue}&{\CoefTrue}&\n&\n&{\CoefTrue}&{\CoefTrue}\cr
 \n&\n&\n&\n&{\CoefTrue}&{\CoefTrue}&{\CoefTrue}&{\CoefTrue}&\n&\n&\n&\n&{\CoefTrue}&{\CoefTrue}&{\CoefTrue}&{\CoefTrue}\cr
 \n&\n&\n&\n&\n&\n&\n&\n&{\CoefTrue}&{\CoefTrue}&{\CoefTrue}&{\CoefTrue}&{\CoefTrue}&{\CoefTrue}&{\CoefTrue}&{\CoefTrue}\cr
\noalign{\kern-2pt}}}%
\vbox{\setbox8=\hbox{$\RowNames\Matrix$}
\hbox to\wd8{\hfil$\ColNames$\kern\ColEntryShiftHoriz}\kern\ColEntryShiftVerti
\box8}}
\kern-0.1cm
{$\footnotesize\def\MatrixU{\spmatrix{%
\noalign{\kern-2pt}
 \n\cr
 {\CoefTrue}\cr
 \n\cr
 {\CoefTrue}\cr
\noalign{\kern-2pt}}}
\MatrixU$}}
\kern0.5cm
\hbox{\footnotesize\kern0.41cm$\spmatrix{%
 \n&\n&\n&\n&\n&\n&\n&\n&\n&\n&{\CoefTrue}\n&\n&\n&\n&\n\cr}=e\RELtraOP$}}
\kern-0.2cm
{\footnotesize%
\BoxBreadth=0pt%
\setbox7=\hbox{\{\}}%
\ifdim\wd7>\BoxBreadth\BoxBreadth=\wd7\fi%
\setbox7=\hbox{\{a\}}%
\ifdim\wd7>\BoxBreadth\BoxBreadth=\wd7\fi%
\setbox7=\hbox{\{b\}}%
\ifdim\wd7>\BoxBreadth\BoxBreadth=\wd7\fi%
\setbox7=\hbox{\{a,b\}}%
\ifdim\wd7>\BoxBreadth\BoxBreadth=\wd7\fi%
\setbox7=\hbox{\{c\}}%
\ifdim\wd7>\BoxBreadth\BoxBreadth=\wd7\fi%
\setbox7=\hbox{\{a,c\}}%
\ifdim\wd7>\BoxBreadth\BoxBreadth=\wd7\fi%
\setbox7=\hbox{\{b,c\}}%
\ifdim\wd7>\BoxBreadth\BoxBreadth=\wd7\fi%
\setbox7=\hbox{\{a,b,c\}}%
\ifdim\wd7>\BoxBreadth\BoxBreadth=\wd7\fi%
\setbox7=\hbox{\{d\}}%
\ifdim\wd7>\BoxBreadth\BoxBreadth=\wd7\fi%
\setbox7=\hbox{\{a,d\}}%
\ifdim\wd7>\BoxBreadth\BoxBreadth=\wd7\fi%
\setbox7=\hbox{\{b,d\}}%
\ifdim\wd7>\BoxBreadth\BoxBreadth=\wd7\fi%
\setbox7=\hbox{\{a,b,d\}}%
\ifdim\wd7>\BoxBreadth\BoxBreadth=\wd7\fi%
\setbox7=\hbox{\{c,d\}}%
\ifdim\wd7>\BoxBreadth\BoxBreadth=\wd7\fi%
\setbox7=\hbox{\{a,c,d\}}%
\ifdim\wd7>\BoxBreadth\BoxBreadth=\wd7\fi%
\setbox7=\hbox{\{b,c,d\}}%
\ifdim\wd7>\BoxBreadth\BoxBreadth=\wd7\fi%
\setbox7=\hbox{\{a,b,c,d\}}%
\ifdim\wd7>\BoxBreadth\BoxBreadth=\wd7\fi%
\def\RowNames{\vcenter{\offinterlineskip\baselineskip=\matrixskip%
\hbox to\BoxBreadth{\strut\hfil \{\}}\kern\interspacereduction%
\hbox to\BoxBreadth{\strut\hfil \{a\}}\kern\interspacereduction%
\hbox to\BoxBreadth{\strut\hfil \{b\}}\kern\interspacereduction%
\hbox to\BoxBreadth{\strut\hfil \{a,b\}}\kern\interspacereduction%
\hbox to\BoxBreadth{\strut\hfil \{c\}}\kern\interspacereduction%
\hbox to\BoxBreadth{\strut\hfil \{a,c\}}\kern\interspacereduction%
\hbox to\BoxBreadth{\strut\hfil \{b,c\}}\kern\interspacereduction%
\hbox to\BoxBreadth{\strut\hfil \{a,b,c\}}\kern\interspacereduction%
\hbox to\BoxBreadth{\strut\hfil \{d\}}\kern\interspacereduction%
\hbox to\BoxBreadth{\strut\hfil \{a,d\}}\kern\interspacereduction%
\hbox to\BoxBreadth{\strut\hfil \{b,d\}}\kern\interspacereduction%
\hbox to\BoxBreadth{\strut\hfil \{a,b,d\}}\kern\interspacereduction%
\hbox to\BoxBreadth{\strut\hfil \{c,d\}}\kern\interspacereduction%
\hbox to\BoxBreadth{\strut\hfil \{a,c,d\}}\kern\interspacereduction%
\hbox to\BoxBreadth{\strut\hfil \{b,c,d\}}\kern\interspacereduction%
\hbox to\BoxBreadth{\strut\hfil \{a,b,c,d\}}}}%
\def\ColNames{\hbox{\rotatebox{90}{\strut \{\}}\kern\interspacereduction%
\rotatebox{90}{\strut \{a\}}\kern\interspacereduction%
\rotatebox{90}{\strut \{b\}}\kern\interspacereduction%
\rotatebox{90}{\strut \{a,b\}}\kern\interspacereduction%
\rotatebox{90}{\strut \{c\}}\kern\interspacereduction%
\rotatebox{90}{\strut \{a,c\}}\kern\interspacereduction%
\rotatebox{90}{\strut \{b,c\}}\kern\interspacereduction%
\rotatebox{90}{\strut \{a,b,c\}}\kern\interspacereduction%
\rotatebox{90}{\strut \{d\}}\kern\interspacereduction%
\rotatebox{90}{\strut \{a,d\}}\kern\interspacereduction%
\rotatebox{90}{\strut \{b,d\}}\kern\interspacereduction%
\rotatebox{90}{\strut \{a,b,d\}}\kern\interspacereduction%
\rotatebox{90}{\strut \{c,d\}}\kern\interspacereduction%
\rotatebox{90}{\strut \{a,c,d\}}\kern\interspacereduction%
\rotatebox{90}{\strut \{b,c,d\}}\kern\interspacereduction%
\rotatebox{90}{\strut \{a,b,c,d\}}\kern\interspacereduction%
}}%
\def\Matrix{\spmatrix{%
\noalign{\kern-2pt}
 {\CoefTrue}&{\CoefTrue}&{\CoefTrue}&{\CoefTrue}&{\CoefTrue}&{\CoefTrue}&{\CoefTrue}&{\CoefTrue}&{\CoefTrue}&{\CoefTrue}&{\CoefTrue}&{\CoefTrue}&{\CoefTrue}&{\CoefTrue}&{\CoefTrue}&{\CoefTrue}\cr
 \n&{\CoefTrue}&\n&{\CoefTrue}&\n&{\CoefTrue}&\n&{\CoefTrue}&\n&{\CoefTrue}&\n&{\CoefTrue}&\n&{\CoefTrue}&\n&{\CoefTrue}\cr
 \n&\n&{\CoefTrue}&{\CoefTrue}&\n&\n&{\CoefTrue}&{\CoefTrue}&\n&\n&{\CoefTrue}&{\CoefTrue}&\n&\n&{\CoefTrue}&{\CoefTrue}\cr
 \n&\n&\n&{\CoefTrue}&\n&\n&\n&{\CoefTrue}&\n&\n&\n&{\CoefTrue}&\n&\n&\n&{\CoefTrue}\cr
 \n&\n&\n&\n&{\CoefTrue}&{\CoefTrue}&{\CoefTrue}&{\CoefTrue}&\n&\n&\n&\n&{\CoefTrue}&{\CoefTrue}&{\CoefTrue}&{\CoefTrue}\cr
 \n&\n&\n&\n&\n&{\CoefTrue}&\n&{\CoefTrue}&\n&\n&\n&\n&\n&{\CoefTrue}&\n&{\CoefTrue}\cr
 \n&\n&\n&\n&\n&\n&{\CoefTrue}&{\CoefTrue}&\n&\n&\n&\n&\n&\n&{\CoefTrue}&{\CoefTrue}\cr
 \n&\n&\n&\n&\n&\n&\n&{\CoefTrue}&\n&\n&\n&\n&\n&\n&\n&{\CoefTrue}\cr
 \n&\n&\n&\n&\n&\n&\n&\n&{\CoefTrue}&{\CoefTrue}&{\CoefTrue}&{\CoefTrue}&{\CoefTrue}&{\CoefTrue}&{\CoefTrue}&{\CoefTrue}\cr
 \n&\n&\n&\n&\n&\n&\n&\n&\n&{\CoefTrue}&\n&{\CoefTrue}&\n&{\CoefTrue}&\n&{\CoefTrue}\cr
 \n&\n&\n&\n&\n&\n&\n&\n&\n&\n&{\CoefTrue}&{\CoefTrue}&\n&\n&{\CoefTrue}&{\CoefTrue}\cr
 \n&\n&\n&\n&\n&\n&\n&\n&\n&\n&\n&{\CoefTrue}&\n&\n&\n&{\CoefTrue}\cr
 \n&\n&\n&\n&\n&\n&\n&\n&\n&\n&\n&\n&{\CoefTrue}&{\CoefTrue}&{\CoefTrue}&{\CoefTrue}\cr
 \n&\n&\n&\n&\n&\n&\n&\n&\n&\n&\n&\n&\n&{\CoefTrue}&\n&{\CoefTrue}\cr
 \n&\n&\n&\n&\n&\n&\n&\n&\n&\n&\n&\n&\n&\n&{\CoefTrue}&{\CoefTrue}\cr
 \n&\n&\n&\n&\n&\n&\n&\n&\n&\n&\n&\n&\n&\n&\n&{\CoefTrue}\cr
\noalign{\kern-2pt}}}%
\vbox{\setbox8=\hbox{$\RowNames\Matrix$}
\hbox to\wd8{\hfil$\ColNames$\kern\ColEntryShiftHoriz}\kern\ColEntryShiftVerti
\box8}
\def\Matrixe{\spmatrix{%
\noalign{\kern-2pt}
 \n\cr
 \n\cr
 \n\cr
 \n\cr
 \n\cr
 \n\cr
 \n\cr
 \n\cr
 \n\cr
 \n\cr
 {\CoefTrue}\cr
 \n\cr
 \n\cr
 \n\cr
 \n\cr
 \n\cr
\noalign{\kern-2pt}}}
\Matrixe
}
$}
{Subset $U$ and corresponding point $e$ in the powerset via $\varepsilon,\Omega$}{FigSubsetElement}

\noindent
There is, however, a second \lq\lq lifted\rq\rq\ form, for which the elements are taken from $\PowTWO{X}$ with corresponding operations consisting of

\smallskip
$\liftedMeet,\quad\liftedJoin,\quad N,\quad(\RELneg{\varepsilon\RELtraOP\RELcompOP\RELtop}=)\syqq{\varepsilon}{\RELbot},\quad(\RELneg{\RELneg{\varepsilon}\RELtraOP\RELcompOP\RELtop}=)\syqq{\varepsilon}{\RELtop}
$,

\smallskip
\noindent
as defined below. Easiest to observe are the $0$-ary operators or elements $\RELneg{\varepsilon\RELtraOP\RELcompOP\RELtop}\approx\n,\RELneg{\RELneg{\varepsilon}\RELtraOP\RELcompOP\RELtop}\approx\CoefTrue$ for which obviously, looking at Fig.~\FigSubsetElement,

\smallskip
$\RELbot=\varepsilon\RELcompOP\RELneg{\varepsilon\RELtraOP\RELcompOP\RELtop}
=\varepsilon\RELcompOP\syqq{\varepsilon}{\RELbot}
$,
\quad
$\RELtop=\varepsilon\RELcompOP\RELneg{\RELneg{\varepsilon}\RELtraOP\RELcompOP\RELtop}
=\varepsilon\RELcompOP\syqq{\varepsilon}{\RELtop}$.

\bigskip
\noindent
Next we study the unary operator 

\smallskip
$N:=\syqq{\RELneg{\varepsilon}}{\varepsilon}
\qquad\qquad
\RELfromTO{N}{\PowTWO{X}}{\PowTWO{X}}
$,

\smallskip
\noindent
visualized in Fig.~\FigNegaDisjoint, for which we show in advance

\smallskip
$\RELneg{\varepsilon}\RELcompOP N
=
\RELneg{\varepsilon}\RELcompOP \syqq{\RELneg{\varepsilon}}{\varepsilon}
=
\varepsilon
$
\quad \quad \quad \quad 
$\varepsilon\RELcompOP N
=
\varepsilon\RELcompOP \syqq{\RELneg{\varepsilon}}{\varepsilon}
=
\varepsilon\RELcompOP \syqq{\varepsilon}{\RELneg{\varepsilon}}
=
\RELneg{\varepsilon}$

\smallskip
$\RELide\RELenthOP
\Omega=\RELneg{\varepsilon\RELtraOP\RELcompOP\RELneg{\varepsilon}} 
=\RELneg{\varepsilon\RELtraOP\RELcompOP\varepsilon\RELcompOP N} 
\quad\Longrightarrow\quad
N \RELenthOP\RELneg{\varepsilon\RELtraOP\RELcompOP\varepsilon} $.

\smallskip
\noindent
Multiplying a relation with $N$ from the left flips this relation upside/down, 
while multiplying from the right side flips it left/right. Sometimes, we have to apply $N$ to both sides of a pair, for which purpose we also introduce

\smallskip 
$\RELfromTO{{\cal N}
:=\Kronecker{N}{N}
=\pi\RELcompOP N\RELcompOP\pi\RELtraOP\RELandOP\rho\RELcompOP N\RELcompOP\rho\RELtraOP}
{\PowTWO{X}\times\PowTWO{X}}{\PowTWO{X}\times\PowTWO{X}}$.

\bigskip
\noindent
We identify here disjointness $\RELneg{\varepsilon\RELtraOP\RELcompOP\varepsilon}$ which is shown in Fig.~\FigNegaDisjoint. It looks as if the powerset ordering $\Omega$ of Fig.~\FigSubsetElement\ were rotated by an angle of $-90$ degrees, which may more mathematically be expressed as $\Omega\RELcompOP N=\RELneg{\varepsilon\RELtraOP\RELcompOP\varepsilon} $; this time flipping left/right.

\Caption{${\footnotesize%
\BoxBreadth=0pt%
\setbox7=\hbox{\{\}}%
\ifdim\wd7>\BoxBreadth\BoxBreadth=\wd7\fi%
\setbox7=\hbox{\{a\}}%
\ifdim\wd7>\BoxBreadth\BoxBreadth=\wd7\fi%
\setbox7=\hbox{\{b\}}%
\ifdim\wd7>\BoxBreadth\BoxBreadth=\wd7\fi%
\setbox7=\hbox{\{a,b\}}%
\ifdim\wd7>\BoxBreadth\BoxBreadth=\wd7\fi%
\setbox7=\hbox{\{c\}}%
\ifdim\wd7>\BoxBreadth\BoxBreadth=\wd7\fi%
\setbox7=\hbox{\{a,c\}}%
\ifdim\wd7>\BoxBreadth\BoxBreadth=\wd7\fi%
\setbox7=\hbox{\{b,c\}}%
\ifdim\wd7>\BoxBreadth\BoxBreadth=\wd7\fi%
\setbox7=\hbox{\{a,b,c\}}%
\ifdim\wd7>\BoxBreadth\BoxBreadth=\wd7\fi%
\setbox7=\hbox{\{d\}}%
\ifdim\wd7>\BoxBreadth\BoxBreadth=\wd7\fi%
\setbox7=\hbox{\{a,d\}}%
\ifdim\wd7>\BoxBreadth\BoxBreadth=\wd7\fi%
\setbox7=\hbox{\{b,d\}}%
\ifdim\wd7>\BoxBreadth\BoxBreadth=\wd7\fi%
\setbox7=\hbox{\{a,b,d\}}%
\ifdim\wd7>\BoxBreadth\BoxBreadth=\wd7\fi%
\setbox7=\hbox{\{c,d\}}%
\ifdim\wd7>\BoxBreadth\BoxBreadth=\wd7\fi%
\setbox7=\hbox{\{a,c,d\}}%
\ifdim\wd7>\BoxBreadth\BoxBreadth=\wd7\fi%
\setbox7=\hbox{\{b,c,d\}}%
\ifdim\wd7>\BoxBreadth\BoxBreadth=\wd7\fi%
\setbox7=\hbox{\{a,b,c,d\}}%
\ifdim\wd7>\BoxBreadth\BoxBreadth=\wd7\fi%
\def\RowNames{\vcenter{\offinterlineskip\baselineskip=\matrixskip%
\hbox to\BoxBreadth{\strut\hfil \{\}}\kern\interspacereduction%
\hbox to\BoxBreadth{\strut\hfil \{a\}}\kern\interspacereduction%
\hbox to\BoxBreadth{\strut\hfil \{b\}}\kern\interspacereduction%
\hbox to\BoxBreadth{\strut\hfil \{a,b\}}\kern\interspacereduction%
\hbox to\BoxBreadth{\strut\hfil \{c\}}\kern\interspacereduction%
\hbox to\BoxBreadth{\strut\hfil \{a,c\}}\kern\interspacereduction%
\hbox to\BoxBreadth{\strut\hfil \{b,c\}}\kern\interspacereduction%
\hbox to\BoxBreadth{\strut\hfil \{a,b,c\}}\kern\interspacereduction%
\hbox to\BoxBreadth{\strut\hfil \{d\}}\kern\interspacereduction%
\hbox to\BoxBreadth{\strut\hfil \{a,d\}}\kern\interspacereduction%
\hbox to\BoxBreadth{\strut\hfil \{b,d\}}\kern\interspacereduction%
\hbox to\BoxBreadth{\strut\hfil \{a,b,d\}}\kern\interspacereduction%
\hbox to\BoxBreadth{\strut\hfil \{c,d\}}\kern\interspacereduction%
\hbox to\BoxBreadth{\strut\hfil \{a,c,d\}}\kern\interspacereduction%
\hbox to\BoxBreadth{\strut\hfil \{b,c,d\}}\kern\interspacereduction%
\hbox to\BoxBreadth{\strut\hfil \{a,b,c,d\}}}}%
\def\ColNames{\hbox{\rotatebox{90}{\strut \{\}}\kern\interspacereduction%
\rotatebox{90}{\strut \{a\}}\kern\interspacereduction%
\rotatebox{90}{\strut \{b\}}\kern\interspacereduction%
\rotatebox{90}{\strut \{a,b\}}\kern\interspacereduction%
\rotatebox{90}{\strut \{c\}}\kern\interspacereduction%
\rotatebox{90}{\strut \{a,c\}}\kern\interspacereduction%
\rotatebox{90}{\strut \{b,c\}}\kern\interspacereduction%
\rotatebox{90}{\strut \{a,b,c\}}\kern\interspacereduction%
\rotatebox{90}{\strut \{d\}}\kern\interspacereduction%
\rotatebox{90}{\strut \{a,d\}}\kern\interspacereduction%
\rotatebox{90}{\strut \{b,d\}}\kern\interspacereduction%
\rotatebox{90}{\strut \{a,b,d\}}\kern\interspacereduction%
\rotatebox{90}{\strut \{c,d\}}\kern\interspacereduction%
\rotatebox{90}{\strut \{a,c,d\}}\kern\interspacereduction%
\rotatebox{90}{\strut \{b,c,d\}}\kern\interspacereduction%
\rotatebox{90}{\strut \{a,b,c,d\}}\kern\interspacereduction%
}}%
\def\Matrix{\spmatrix{%
\noalign{\kern-2pt}
 \n&\n&\n&\n&\n&\n&\n&\n&\n&\n&\n&\n&\n&\n&\n&{\CoefTrue}\cr
 \n&\n&\n&\n&\n&\n&\n&\n&\n&\n&\n&\n&\n&\n&{\CoefTrue}&\n\cr
 \n&\n&\n&\n&\n&\n&\n&\n&\n&\n&\n&\n&\n&{\CoefTrue}&\n&\n\cr
 \n&\n&\n&\n&\n&\n&\n&\n&\n&\n&\n&\n&{\CoefTrue}&\n&\n&\n\cr
 \n&\n&\n&\n&\n&\n&\n&\n&\n&\n&\n&{\CoefTrue}&\n&\n&\n&\n\cr
 \n&\n&\n&\n&\n&\n&\n&\n&\n&\n&{\CoefTrue}&\n&\n&\n&\n&\n\cr
 \n&\n&\n&\n&\n&\n&\n&\n&\n&{\CoefTrue}&\n&\n&\n&\n&\n&\n\cr
 \n&\n&\n&\n&\n&\n&\n&\n&{\CoefTrue}&\n&\n&\n&\n&\n&\n&\n\cr
 \n&\n&\n&\n&\n&\n&\n&{\CoefTrue}&\n&\n&\n&\n&\n&\n&\n&\n\cr
 \n&\n&\n&\n&\n&\n&{\CoefTrue}&\n&\n&\n&\n&\n&\n&\n&\n&\n\cr
 \n&\n&\n&\n&\n&{\CoefTrue}&\n&\n&\n&\n&\n&\n&\n&\n&\n&\n\cr
 \n&\n&\n&\n&{\CoefTrue}&\n&\n&\n&\n&\n&\n&\n&\n&\n&\n&\n\cr
 \n&\n&\n&{\CoefTrue}&\n&\n&\n&\n&\n&\n&\n&\n&\n&\n&\n&\n\cr
 \n&\n&{\CoefTrue}&\n&\n&\n&\n&\n&\n&\n&\n&\n&\n&\n&\n&\n\cr
 \n&{\CoefTrue}&\n&\n&\n&\n&\n&\n&\n&\n&\n&\n&\n&\n&\n&\n\cr
 {\CoefTrue}&\n&\n&\n&\n&\n&\n&\n&\n&\n&\n&\n&\n&\n&\n&\n\cr
\noalign{\kern-2pt}}}%
\vbox{\setbox8=\hbox{$\RowNames\Matrix$}
\hbox to\wd8{\hfil$\ColNames$\kern\ColEntryShiftHoriz}\kern\ColEntryShiftVerti
\box8}}
{\footnotesize%
\BoxBreadth=0pt%
\setbox7=\hbox{\{\}}%
\ifdim\wd7>\BoxBreadth\BoxBreadth=\wd7\fi%
\setbox7=\hbox{\{a\}}%
\ifdim\wd7>\BoxBreadth\BoxBreadth=\wd7\fi%
\setbox7=\hbox{\{b\}}%
\ifdim\wd7>\BoxBreadth\BoxBreadth=\wd7\fi%
\setbox7=\hbox{\{a,b\}}%
\ifdim\wd7>\BoxBreadth\BoxBreadth=\wd7\fi%
\setbox7=\hbox{\{c\}}%
\ifdim\wd7>\BoxBreadth\BoxBreadth=\wd7\fi%
\setbox7=\hbox{\{a,c\}}%
\ifdim\wd7>\BoxBreadth\BoxBreadth=\wd7\fi%
\setbox7=\hbox{\{b,c\}}%
\ifdim\wd7>\BoxBreadth\BoxBreadth=\wd7\fi%
\setbox7=\hbox{\{a,b,c\}}%
\ifdim\wd7>\BoxBreadth\BoxBreadth=\wd7\fi%
\setbox7=\hbox{\{d\}}%
\ifdim\wd7>\BoxBreadth\BoxBreadth=\wd7\fi%
\setbox7=\hbox{\{a,d\}}%
\ifdim\wd7>\BoxBreadth\BoxBreadth=\wd7\fi%
\setbox7=\hbox{\{b,d\}}%
\ifdim\wd7>\BoxBreadth\BoxBreadth=\wd7\fi%
\setbox7=\hbox{\{a,b,d\}}%
\ifdim\wd7>\BoxBreadth\BoxBreadth=\wd7\fi%
\setbox7=\hbox{\{c,d\}}%
\ifdim\wd7>\BoxBreadth\BoxBreadth=\wd7\fi%
\setbox7=\hbox{\{a,c,d\}}%
\ifdim\wd7>\BoxBreadth\BoxBreadth=\wd7\fi%
\setbox7=\hbox{\{b,c,d\}}%
\ifdim\wd7>\BoxBreadth\BoxBreadth=\wd7\fi%
\setbox7=\hbox{\{a,b,c,d\}}%
\ifdim\wd7>\BoxBreadth\BoxBreadth=\wd7\fi%
\def\RowNames{\vcenter{\offinterlineskip\baselineskip=\matrixskip%
\hbox to\BoxBreadth{\strut\hfil \{\}}\kern\interspacereduction%
\hbox to\BoxBreadth{\strut\hfil \{a\}}\kern\interspacereduction%
\hbox to\BoxBreadth{\strut\hfil \{b\}}\kern\interspacereduction%
\hbox to\BoxBreadth{\strut\hfil \{a,b\}}\kern\interspacereduction%
\hbox to\BoxBreadth{\strut\hfil \{c\}}\kern\interspacereduction%
\hbox to\BoxBreadth{\strut\hfil \{a,c\}}\kern\interspacereduction%
\hbox to\BoxBreadth{\strut\hfil \{b,c\}}\kern\interspacereduction%
\hbox to\BoxBreadth{\strut\hfil \{a,b,c\}}\kern\interspacereduction%
\hbox to\BoxBreadth{\strut\hfil \{d\}}\kern\interspacereduction%
\hbox to\BoxBreadth{\strut\hfil \{a,d\}}\kern\interspacereduction%
\hbox to\BoxBreadth{\strut\hfil \{b,d\}}\kern\interspacereduction%
\hbox to\BoxBreadth{\strut\hfil \{a,b,d\}}\kern\interspacereduction%
\hbox to\BoxBreadth{\strut\hfil \{c,d\}}\kern\interspacereduction%
\hbox to\BoxBreadth{\strut\hfil \{a,c,d\}}\kern\interspacereduction%
\hbox to\BoxBreadth{\strut\hfil \{b,c,d\}}\kern\interspacereduction%
\hbox to\BoxBreadth{\strut\hfil \{a,b,c,d\}}}}%
\def\ColNames{\hbox{\rotatebox{90}{\strut \{\}}\kern\interspacereduction%
\rotatebox{90}{\strut \{a\}}\kern\interspacereduction%
\rotatebox{90}{\strut \{b\}}\kern\interspacereduction%
\rotatebox{90}{\strut \{a,b\}}\kern\interspacereduction%
\rotatebox{90}{\strut \{c\}}\kern\interspacereduction%
\rotatebox{90}{\strut \{a,c\}}\kern\interspacereduction%
\rotatebox{90}{\strut \{b,c\}}\kern\interspacereduction%
\rotatebox{90}{\strut \{a,b,c\}}\kern\interspacereduction%
\rotatebox{90}{\strut \{d\}}\kern\interspacereduction%
\rotatebox{90}{\strut \{a,d\}}\kern\interspacereduction%
\rotatebox{90}{\strut \{b,d\}}\kern\interspacereduction%
\rotatebox{90}{\strut \{a,b,d\}}\kern\interspacereduction%
\rotatebox{90}{\strut \{c,d\}}\kern\interspacereduction%
\rotatebox{90}{\strut \{a,c,d\}}\kern\interspacereduction%
\rotatebox{90}{\strut \{b,c,d\}}\kern\interspacereduction%
\rotatebox{90}{\strut \{a,b,c,d\}}\kern\interspacereduction%
}}%
\def\Matrix{\spmatrix{%
\noalign{\kern-2pt}
 {\CoefTrue}&{\CoefTrue}&{\CoefTrue}&{\CoefTrue}&{\CoefTrue}&{\CoefTrue}&{\CoefTrue}&{\CoefTrue}&{\CoefTrue}&{\CoefTrue}&{\CoefTrue}&{\CoefTrue}&{\CoefTrue}&{\CoefTrue}&{\CoefTrue}&{\CoefTrue}\cr
 {\CoefTrue}&\n&{\CoefTrue}&\n&{\CoefTrue}&\n&{\CoefTrue}&\n&{\CoefTrue}&\n&{\CoefTrue}&\n&{\CoefTrue}&\n&{\CoefTrue}&\n\cr
 {\CoefTrue}&{\CoefTrue}&\n&\n&{\CoefTrue}&{\CoefTrue}&\n&\n&{\CoefTrue}&{\CoefTrue}&\n&\n&{\CoefTrue}&{\CoefTrue}&\n&\n\cr
 {\CoefTrue}&\n&\n&\n&{\CoefTrue}&\n&\n&\n&{\CoefTrue}&\n&\n&\n&{\CoefTrue}&\n&\n&\n\cr
 {\CoefTrue}&{\CoefTrue}&{\CoefTrue}&{\CoefTrue}&\n&\n&\n&\n&{\CoefTrue}&{\CoefTrue}&{\CoefTrue}&{\CoefTrue}&\n&\n&\n&\n\cr
 {\CoefTrue}&\n&{\CoefTrue}&\n&\n&\n&\n&\n&{\CoefTrue}&\n&{\CoefTrue}&\n&\n&\n&\n&\n\cr
 {\CoefTrue}&{\CoefTrue}&\n&\n&\n&\n&\n&\n&{\CoefTrue}&{\CoefTrue}&\n&\n&\n&\n&\n&\n\cr
 {\CoefTrue}&\n&\n&\n&\n&\n&\n&\n&{\CoefTrue}&\n&\n&\n&\n&\n&\n&\n\cr
 {\CoefTrue}&{\CoefTrue}&{\CoefTrue}&{\CoefTrue}&{\CoefTrue}&{\CoefTrue}&{\CoefTrue}&{\CoefTrue}&\n&\n&\n&\n&\n&\n&\n&\n\cr
 {\CoefTrue}&\n&{\CoefTrue}&\n&{\CoefTrue}&\n&{\CoefTrue}&\n&\n&\n&\n&\n&\n&\n&\n&\n\cr
 {\CoefTrue}&{\CoefTrue}&\n&\n&{\CoefTrue}&{\CoefTrue}&\n&\n&\n&\n&\n&\n&\n&\n&\n&\n\cr
 {\CoefTrue}&\n&\n&\n&{\CoefTrue}&\n&\n&\n&\n&\n&\n&\n&\n&\n&\n&\n\cr
 {\CoefTrue}&{\CoefTrue}&{\CoefTrue}&{\CoefTrue}&\n&\n&\n&\n&\n&\n&\n&\n&\n&\n&\n&\n\cr
 {\CoefTrue}&\n&{\CoefTrue}&\n&\n&\n&\n&\n&\n&\n&\n&\n&\n&\n&\n&\n\cr
 {\CoefTrue}&{\CoefTrue}&\n&\n&\n&\n&\n&\n&\n&\n&\n&\n&\n&\n&\n&\n\cr
 {\CoefTrue}&\n&\n&\n&\n&\n&\n&\n&\n&\n&\n&\n&\n&\n&\n&\n\cr
\noalign{\kern-2pt}}}%
\vbox{\setbox8=\hbox{$\RowNames\Matrix$}
\hbox to\wd8{\hfil$\ColNames$\kern\ColEntryShiftHoriz}\kern\ColEntryShiftVerti
\box8}}$}
{Negation $N$ and disjointness\index{disjointness} $\RELneg{\varepsilon\RELtraOP\RELcompOP\varepsilon}=\Omega\RELcompOP N
$ in the powerset}{FigNegaDisjoint}

\noindent
At last, we consider the binary operations meet $\liftedMeet$ and join $\liftedJoin$ which we mainly obtain
specializing the result of Prop.~\PropSumPowToPowProd\ to the case $X=Y$ and integrate them into the relational mechanism using the least upper, resp.~greatest lower, bound taken rowwise according to \cite{RelaMath2010} Prop.~9.10.

\Caption{\includegraphics[scale=0.4]{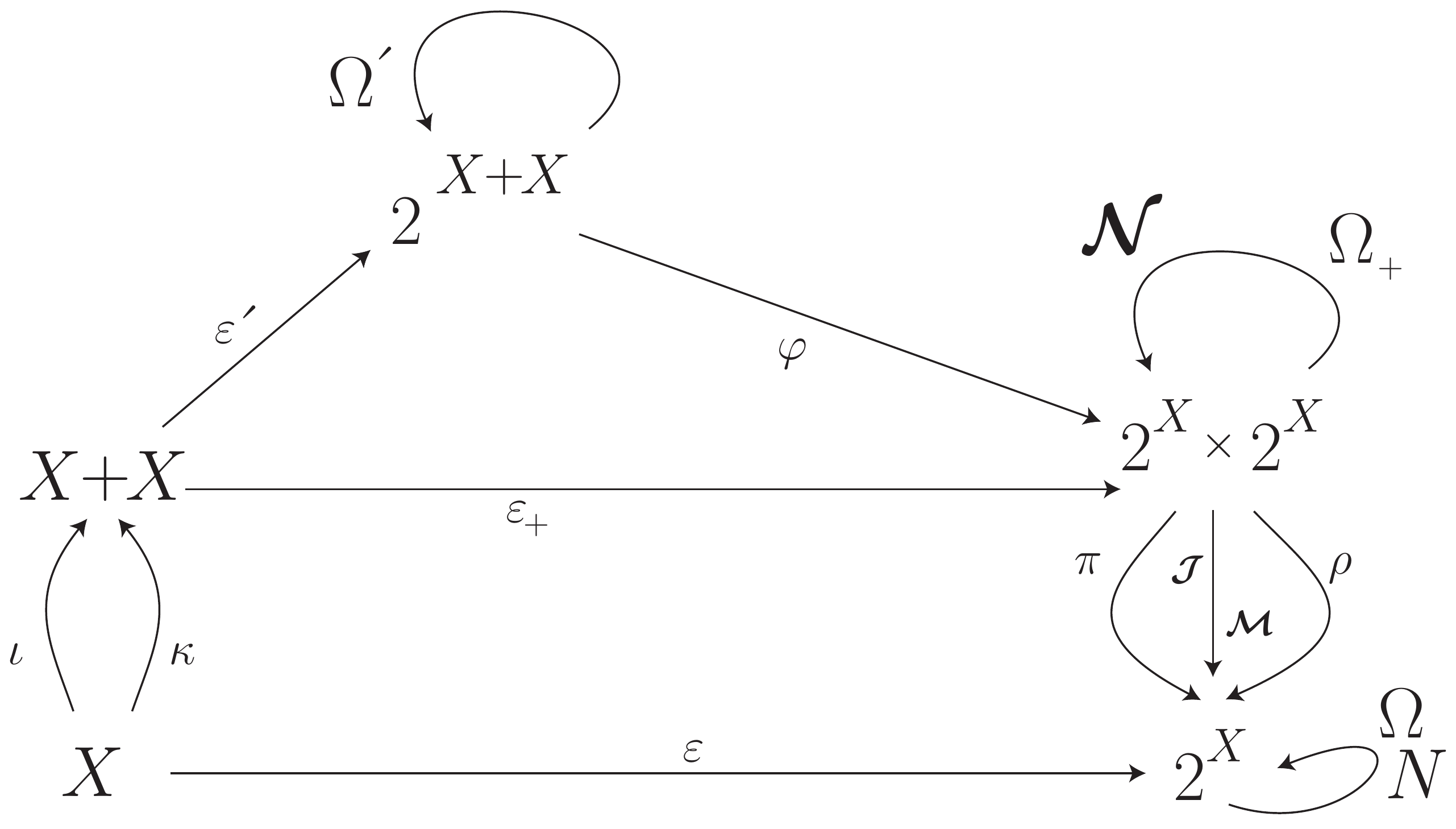}}
{Converting subsets of a sum to products of subsets with join $\liftedJoin$ and meet $\liftedMeet$}{FigSumPowToPowProdOnX}

\noindent
A first step is the investigation of the bijection $\varphi$ of Figs.~\FigSumPowToPowProdOnX\ and \FigSumPowToPowProdMatrix.
We show the relation indicating with $>$a, respectively a$<$ whether an element has been injected to the left or to the right. Only when restricting to somehow coherent visualizations of $\PowTWO{X+ X}$ and $\PowTWO{X}\times\PowTWO{X}$, this will show a \lq diagonal\rq.

\enunc{}{Proposition}{}{CorSumPowToPowProd} We assume the setting of Prop.~\PropSumPowToPowProd, however with $X=Y$, so that additional formulae may be formulated including join\index{join} and meet\index{meet}.

\begin{enumerate}[i)]
\item $\;\liftedJoin=\syqq{\iota\RELcompOP\varepsilon_+\RELorOP\kappa\RELcompOP\varepsilon_+}{\varepsilon}
=
\syqq{\varepsilon\RELcompOP\pi\RELtraOP\RELorOP\varepsilon\RELcompOP\rho\RELtraOP}{\varepsilon}
=
\lubR_\Omega(\pi\RELorOP\rho)
=
\syqq{\varepsilon\RELcompOP\lbrack\pi\RELorOP\rho\rbrack\RELtraOP}{\varepsilon}
$

\item[]$=\syqq{\StrictFork{\RELneg{\varepsilon}}{\RELneg{\varepsilon}}}{\RELneg{\varepsilon}}=
\syqq{\RELneg{\varepsilon}\RELcompOP\pi\RELtraOP\RELandOP\RELneg{\varepsilon}\RELcompOP\rho\RELtraOP}{\RELneg{\varepsilon}}
$

\item $\liftedMeet=\syqq{\iota\RELcompOP\varepsilon_+\RELandOP\kappa\RELcompOP\varepsilon_+}{\varepsilon}
=
\syqq{\varepsilon\RELcompOP\pi\RELtraOP\RELandOP\varepsilon\RELcompOP\rho\RELtraOP}{\varepsilon}
=
\glbR_ \Omega(\pi\RELorOP\rho)
=
\syqq{\RELneg{\varepsilon}\RELcompOP\lbrack\pi\RELorOP\rho\rbrack\RELtraOP}{\RELneg{\varepsilon}}
$

\item[]$=\syqq{\StrictFork{\varepsilon}{\varepsilon}}{\varepsilon}
$

\item $\;\varepsilon\RELcompOP\liftedJoin\RELtraOP=\iota\RELcompOP\varepsilon_+\RELorOP\kappa\RELcompOP\varepsilon_+
=
\varepsilon\RELcompOP\pi\RELtraOP\RELorOP\varepsilon\RELcompOP\rho\RELtraOP$

\item $
\varepsilon\RELcompOP\liftedMeet\RELtraOP=\iota\RELcompOP\varepsilon_+\RELandOP\kappa\RELcompOP\varepsilon_+
=
\varepsilon\RELcompOP\pi\RELtraOP\RELandOP\varepsilon\RELcompOP\rho\RELtraOP
=
\StrictFork{\varepsilon}{\varepsilon}$
\end{enumerate}

\kern-\baselineskip

\Caption{{\footnotesize%
\BoxBreadth=0pt%
\setbox7=\hbox{\{\}}%
\ifdim\wd7>\BoxBreadth\BoxBreadth=\wd7\fi%
\setbox7=\hbox{\{a$<$\}}%
\ifdim\wd7>\BoxBreadth\BoxBreadth=\wd7\fi%
\setbox7=\hbox{\{$>$a\}}%
\ifdim\wd7>\BoxBreadth\BoxBreadth=\wd7\fi%
\setbox7=\hbox{\{a$<$,$>$a\}}%
\ifdim\wd7>\BoxBreadth\BoxBreadth=\wd7\fi%
\setbox7=\hbox{\{b$<$\}}%
\ifdim\wd7>\BoxBreadth\BoxBreadth=\wd7\fi%
\setbox7=\hbox{\{a$<$,b$<$\}}%
\ifdim\wd7>\BoxBreadth\BoxBreadth=\wd7\fi%
\setbox7=\hbox{\{$>$a,b$<$\}}%
\ifdim\wd7>\BoxBreadth\BoxBreadth=\wd7\fi%
\setbox7=\hbox{\{a$<$,$>$a,b$<$\}}%
\ifdim\wd7>\BoxBreadth\BoxBreadth=\wd7\fi%
\setbox7=\hbox{\{$>$b\}}%
\ifdim\wd7>\BoxBreadth\BoxBreadth=\wd7\fi%
\setbox7=\hbox{\{a$<$,$>$b\}}%
\ifdim\wd7>\BoxBreadth\BoxBreadth=\wd7\fi%
\setbox7=\hbox{\{$>$a,$>$b\}}%
\ifdim\wd7>\BoxBreadth\BoxBreadth=\wd7\fi%
\setbox7=\hbox{\{a$<$,$>$a,$>$b\}}%
\ifdim\wd7>\BoxBreadth\BoxBreadth=\wd7\fi%
\setbox7=\hbox{\{b$<$,$>$b\}}%
\ifdim\wd7>\BoxBreadth\BoxBreadth=\wd7\fi%
\setbox7=\hbox{\{a$<$,b$<$,$>$b\}}%
\ifdim\wd7>\BoxBreadth\BoxBreadth=\wd7\fi%
\setbox7=\hbox{\{$>$a,b$<$,$>$b\}}%
\ifdim\wd7>\BoxBreadth\BoxBreadth=\wd7\fi%
\setbox7=\hbox{\{a$<$,$>$a,b$<$,$>$b\}}%
\ifdim\wd7>\BoxBreadth\BoxBreadth=\wd7\fi%
\def\RowNames{\vcenter{\offinterlineskip\baselineskip=\matrixskip%
\hbox to\BoxBreadth{\strut\hfil \{\}}\kern\interspacereduction%
\hbox to\BoxBreadth{\strut\hfil \{a$<$\}}\kern\interspacereduction%
\hbox to\BoxBreadth{\strut\hfil \{$>$a\}}\kern\interspacereduction%
\hbox to\BoxBreadth{\strut\hfil \{a$<$,$>$a\}}\kern\interspacereduction%
\hbox to\BoxBreadth{\strut\hfil \{b$<$\}}\kern\interspacereduction%
\hbox to\BoxBreadth{\strut\hfil \{a$<$,b$<$\}}\kern\interspacereduction%
\hbox to\BoxBreadth{\strut\hfil \{$>$a,b$<$\}}\kern\interspacereduction%
\hbox to\BoxBreadth{\strut\hfil \{a$<$,$>$a,b$<$\}}\kern\interspacereduction%
\hbox to\BoxBreadth{\strut\hfil \{$>$b\}}\kern\interspacereduction%
\hbox to\BoxBreadth{\strut\hfil \{a$<$,$>$b\}}\kern\interspacereduction%
\hbox to\BoxBreadth{\strut\hfil \{$>$a,$>$b\}}\kern\interspacereduction%
\hbox to\BoxBreadth{\strut\hfil \{a$<$,$>$a,$>$b\}}\kern\interspacereduction%
\hbox to\BoxBreadth{\strut\hfil \{b$<$,$>$b\}}\kern\interspacereduction%
\hbox to\BoxBreadth{\strut\hfil \{a$<$,b$<$,$>$b\}}\kern\interspacereduction%
\hbox to\BoxBreadth{\strut\hfil \{$>$a,b$<$,$>$b\}}\kern\interspacereduction%
\hbox to\BoxBreadth{\strut\hfil \{a$<$,$>$a,b$<$,$>$b\}}}}%
\def\ColNames{\hbox{\rotatebox{90}{\strut (\{\},\{\})}\kern\interspacereduction%
\rotatebox{90}{\strut (\{a\},\{\})}\kern\interspacereduction%
\rotatebox{90}{\strut (\{\},\{a\})}\kern\interspacereduction%
\rotatebox{90}{\strut (\{b\},\{\})}\kern\interspacereduction%
\rotatebox{90}{\strut (\{a\},\{a\})}\kern\interspacereduction%
\rotatebox{90}{\strut (\{\},\{b\})}\kern\interspacereduction%
\rotatebox{90}{\strut (\{a,b\},\{\})}\kern\interspacereduction%
\rotatebox{90}{\strut (\{b\},\{a\})}\kern\interspacereduction%
\rotatebox{90}{\strut (\{a\},\{b\})}\kern\interspacereduction%
\rotatebox{90}{\strut (\{\},\{a,b\})}\kern\interspacereduction%
\rotatebox{90}{\strut (\{a,b\},\{a\})}\kern\interspacereduction%
\rotatebox{90}{\strut (\{b\},\{b\})}\kern\interspacereduction%
\rotatebox{90}{\strut (\{a\},\{a,b\})}\kern\interspacereduction%
\rotatebox{90}{\strut (\{a,b\},\{b\})}\kern\interspacereduction%
\rotatebox{90}{\strut (\{b\},\{a,b\})}\kern\interspacereduction%
\rotatebox{90}{\strut (\{a,b\},\{a,b\})}\kern\interspacereduction%
}}%
\def\Matrix{\spmatrix{%
\noalign{\kern-2pt}
 {\CoefTrue}&\n&\n&\n&\n&\n&\n&\n&\n&\n&\n&\n&\n&\n&\n&\n\cr
 \n&{\CoefTrue}&\n&\n&\n&\n&\n&\n&\n&\n&\n&\n&\n&\n&\n&\n\cr
 \n&\n&{\CoefTrue}&\n&\n&\n&\n&\n&\n&\n&\n&\n&\n&\n&\n&\n\cr
 \n&\n&\n&\n&{\CoefTrue}&\n&\n&\n&\n&\n&\n&\n&\n&\n&\n&\n\cr
 \n&\n&\n&{\CoefTrue}&\n&\n&\n&\n&\n&\n&\n&\n&\n&\n&\n&\n\cr
 \n&\n&\n&\n&\n&\n&{\CoefTrue}&\n&\n&\n&\n&\n&\n&\n&\n&\n\cr
 \n&\n&\n&\n&\n&\n&\n&{\CoefTrue}&\n&\n&\n&\n&\n&\n&\n&\n\cr
 \n&\n&\n&\n&\n&\n&\n&\n&\n&\n&{\CoefTrue}&\n&\n&\n&\n&\n\cr
 \n&\n&\n&\n&\n&{\CoefTrue}&\n&\n&\n&\n&\n&\n&\n&\n&\n&\n\cr
 \n&\n&\n&\n&\n&\n&\n&\n&{\CoefTrue}&\n&\n&\n&\n&\n&\n&\n\cr
 \n&\n&\n&\n&\n&\n&\n&\n&\n&{\CoefTrue}&\n&\n&\n&\n&\n&\n\cr
 \n&\n&\n&\n&\n&\n&\n&\n&\n&\n&\n&\n&{\CoefTrue}&\n&\n&\n\cr
 \n&\n&\n&\n&\n&\n&\n&\n&\n&\n&\n&{\CoefTrue}&\n&\n&\n&\n\cr
 \n&\n&\n&\n&\n&\n&\n&\n&\n&\n&\n&\n&\n&{\CoefTrue}&\n&\n\cr
 \n&\n&\n&\n&\n&\n&\n&\n&\n&\n&\n&\n&\n&\n&{\CoefTrue}&\n\cr
 \n&\n&\n&\n&\n&\n&\n&\n&\n&\n&\n&\n&\n&\n&\n&{\CoefTrue}\cr
\noalign{\kern-2pt}}}%
\vbox{\setbox8=\hbox{$\RowNames\Matrix$}
\hbox to\wd8{\hfil$\ColNames$\kern\ColEntryShiftHoriz}\kern\ColEntryShiftVerti
\box8}}}
{Relation $\varphi$ converting subsets of a sum to products of subsets for $X:=\{{\rm a,b}\}$}{FigSumPowToPowProdMatrix}

\Proof i) We formulate the join $\liftedJoin$ as a least upper bound and recall Prop.~9.10\ of \cite{RelaMath2010}  

\smallskip
$\liftedJoin\RELtraOP
=
\lub_\Omega(\lbrack\pi\RELorOP\rho\rbrack\RELtraOP)
=
\syqq{\varepsilon}{\varepsilon\RELcompOP\lbrack\pi\RELorOP\rho\rbrack\RELtraOP}
=
\syqq{\varepsilon}{\varepsilon\RELcompOP\pi\RELtraOP\RELorOP\varepsilon\RELcompOP\rho\RELtraOP}
=
\syqq{\varepsilon}{\iota\RELcompOP\varepsilon_+\RELorOP\kappa\RELcompOP\varepsilon_+}
$

\bigskip
\noindent
ii) $\liftedMeet\RELtraOP
=
\glb_\Omega(\lbrack\pi\RELorOP\rho\rbrack\RELtraOP)
=
\syqq{\RELneg{\varepsilon}}{\RELneg{\varepsilon}\RELcompOP\lbrack\pi\RELorOP\rho\rbrack\RELtraOP}
=
\syqq{\varepsilon}{\varepsilon\RELcompOP\pi\RELtraOP\RELandOP\varepsilon\RELcompOP\rho\RELtraOP}
=
\syqq{\varepsilon}{\iota\RELcompOP\varepsilon_+\RELandOP\kappa\RELcompOP\varepsilon_+}
$

\bigskip

$\liftedMeet=\glbR_ \Omega(\pi\RELorOP\rho)$ by definition

$=\big\lbrack\glb_\Omega(\lbrack\pi\RELorOP\rho\rbrack\RELtraOP)\big\rbrack\RELtraOP$ by definition

$=\big\lbrack\syqq{\RELneg{\varepsilon}}{\RELneg{\varepsilon}\RELcompOP\lbrack\pi\RELorOP\rho\rbrack\RELtraOP}\big\rbrack\RELtraOP$ Prop.~9.10 of \cite{RelaMath2010}

$=\syqq{\RELneg{\varepsilon}\RELcompOP\pi\RELtraOP\RELorOP\RELneg{\varepsilon}\RELcompOP\rho\RELtraOP}{\RELneg{\varepsilon}}$ 

$=\syqq{\RELneg{\RELneg{\varepsilon}\RELcompOP\pi\RELtraOP\RELorOP\RELneg{\varepsilon}\RELcompOP\rho\RELtraOP}}{\varepsilon}$  Prop.~8.10.i of \cite{RelaMath2010}

$=\syqq{\RELneg{\RELneg{\varepsilon}\RELcompOP\pi\RELtraOP}\RELandOP\RELneg{\RELneg{\varepsilon}\RELcompOP\rho\RELtraOP}}{\varepsilon}$ 

$=\syqq{\RELneg{\RELneg{\varepsilon}}\RELcompOP\pi\RELtraOP\RELandOP\RELneg{\RELneg{\varepsilon}}\RELcompOP\rho\RELtraOP}{\varepsilon}$ 

$=\syqq{\varepsilon\RELcompOP\pi\RELtraOP\RELandOP\varepsilon\RELcompOP\rho\RELtraOP}{\varepsilon}$ 

\bigskip
$\liftedJoin=\lubR_ \Omega(\pi\RELorOP\rho)$ by definition

$=\big\lbrack\lub_\Omega(\lbrack\pi\RELorOP\rho\rbrack\RELtraOP)\big\rbrack\RELtraOP$ by definition

$=\big\lbrack\syqq{\varepsilon}{\varepsilon\RELcompOP\lbrack\pi\RELorOP\rho\rbrack\RELtraOP}\big\rbrack\RELtraOP$ Prop.~9.10 of \cite{RelaMath2010}

$=\syqq{\varepsilon\RELcompOP\pi\RELtraOP\RELorOP\varepsilon\RELcompOP\rho\RELtraOP}{\varepsilon}$

\bigskip
\noindent
iv) $\liftedMeet\RELcompOP\varepsilon\RELtraOP
=
\big\lbrack\varepsilon\RELcompOP\liftedMeet\RELtraOP\big\rbrack\RELtraOP
=
\big\lbrack\varepsilon\RELcompOP\syqq{\varepsilon}{\varepsilon\RELcompOP\pi\RELtraOP\RELandOP\varepsilon\RELcompOP\rho\RELtraOP}\big\rbrack\RELtraOP
=
\big\lbrack\varepsilon\RELcompOP\pi\RELtraOP\RELandOP\varepsilon\RELcompOP\rho\RELtraOP\big\rbrack\RELtraOP
=
\pi\RELcompOP\varepsilon\RELtraOP\RELandOP\rho\RELcompOP\varepsilon\RELtraOP
=
\StrictJoin{\varepsilon}{\varepsilon}
$

\bigskip
\noindent
iii) The proof for $\liftedJoin\RELcompOP\varepsilon\RELtraOP$ is established in a similar way.
\Bewende

\noindent
We convince us formally that $\liftedMeet$ is commutative:

\smallskip
$\TarskiSwitch\RELcompOP\liftedMeet
=
\TarskiSwitch\RELcompOP\syqq{\StrictFork{\varepsilon}{\varepsilon}}{\varepsilon}
$\quad by definition

$=
\syqq{\StrictFork{\varepsilon}{\varepsilon}\RELcompOP\,\TarskiSwitch}{\varepsilon}
$\quad since $\TarskiSwitch$ is a bijective mapping

$=
\syqq{\varepsilon\RELcompOP\pi\RELtraOP\RELcompOP\TarskiSwitch\RELandOP\varepsilon\RELcompOP\rho\RELtraOP\RELcompOP\TarskiSwitch}{\varepsilon}
$\quad 

$=
\syqq{\varepsilon\RELcompOP\rho\RELtraOP\RELandOP\varepsilon\RELcompOP\pi\RELtraOP}{\varepsilon}
$\quad 

$=
\syqq{\StrictFork{\varepsilon}{\varepsilon}}{\varepsilon}=\liftedMeet
$\quad

\bigskip
\noindent
A trivial remark is in order, namely that a pair with coinciding first and second component will have precisely this coinciding set as its meet, i.e.

\smallskip
$\pi\RELandOP\rho\RELenthOP\liftedMeet$\quad or\quad $\StrictJoin{\RELide}{\RELide}\RELenthOP\liftedMeet$.

\smallskip
\noindent
The proof can also be carried out in a fully formal way:

\smallskip
$\iff\quad\pi\RELandOP\rho
\RELenthOP
\syqq{\varepsilon\RELcompOP\pi\RELtraOP\RELandOP\varepsilon\RELcompOP\rho\RELtraOP}{\varepsilon}
$

$\iff\quad
\RELneg{\varepsilon\RELcompOP\pi\RELtraOP\RELandOP\varepsilon\RELcompOP\rho\RELtraOP}\RELtraOP\RELcompOP\varepsilon\RELorOP(\varepsilon\RELcompOP\pi\RELtraOP\RELandOP\varepsilon\RELcompOP\rho\RELtraOP)\RELtraOP\RELcompOP\RELneg{\varepsilon}
\RELenthOP
\RELneg{\pi}\RELorOP\RELneg{\rho}
$

$\iff\quad
(\pi\RELandOP\rho)\RELcompOP\varepsilon\RELtraOP
\RELenthOP
(\varepsilon\RELcompOP\pi\RELtraOP\RELandOP\varepsilon\RELcompOP\rho\RELtraOP)\RELtraOP
$
\quad and \quad 
$
(\pi\RELandOP\rho)\RELcompOP\RELneg{\varepsilon}\RELtraOP
\RELenthOP
\RELneg{\varepsilon\RELcompOP\pi\RELtraOP\RELandOP\varepsilon\RELcompOP\rho\RELtraOP}\RELtraOP
$

$\iff\quad
(\pi\RELandOP\rho)\RELcompOP\varepsilon\RELtraOP
\RELenthOP
\pi\RELcompOP\varepsilon\RELtraOP\RELandOP\rho\RELcompOP\varepsilon\RELtraOP
$
\quad and \quad 
$
(\pi\RELandOP\rho)\RELcompOP\RELneg{\varepsilon}\RELtraOP
\RELenthOP
\RELneg{\pi\RELcompOP\varepsilon\RELtraOP}\RELorOP\RELneg{\rho\RELcompOP\varepsilon\RELtraOP}
=
\pi\RELcompOP\RELneg{\varepsilon\RELtraOP}\RELorOP\rho\RELcompOP\RELneg{\varepsilon\RELtraOP}
$ which is true.

\bigskip
\noindent
Some other helpful formulae:

\enunc{}{Proposition}{}{PropPowerMeet} 

\begin{enumerate}[i)]
\item ${\cal N}\RELcompOP\pi=\pi\RELcompOP N,\quad {\cal N}\RELcompOP\rho=\rho\RELcompOP N,\qquad
{\cal N}\RELcompOP\liftedMeet=\liftedJoin\RELcompOP N,\quad{\cal N}\RELcompOP\liftedJoin=\liftedMeet\RELcompOP N$
\item $\liftedMeet\RELtraOP\RELcompOP\pi=\Omega$\qquad$\liftedMeet\RELtraOP\RELcompOP\rho=\Omega$
\item $\liftedJoin\RELtraOP\RELcompOP\pi=\Omega\RELtraOP$\qquad$\liftedJoin\RELtraOP\RELcompOP\rho=\Omega\RELtraOP$ 
\item $\liftedMeet\RELcompOP\Omega\RELtraOP=\pi\RELcompOP\Omega\RELtraOP\RELandOP\rho\RELcompOP\Omega\RELtraOP
=
\StrictJoin{\Omega\RELtraOP}{\Omega\RELtraOP}
\qquad
\liftedJoin\RELcompOP\Omega=\pi\RELcompOP\Omega\RELandOP\rho\RELcompOP\Omega
=
\StrictJoin{\Omega}{\Omega}
$
\item $(\varepsilon\RELcompOP\pi\RELtraOP\RELandOP\varepsilon\RELcompOP\rho\RELtraOP)\RELcompOP\liftedMeet=\varepsilon
\qquad
(\varepsilon\RELcompOP\pi\RELtraOP\RELorOP\varepsilon\RELcompOP\rho\RELtraOP)\RELcompOP\liftedJoin=\varepsilon$
\quad variant form\quad $\StrictFork{\varepsilon}{\varepsilon}\RELcompOP\liftedMeet=\varepsilon$

\item $\StrictFork{\varepsilon}{\varepsilon}\RELcompOP\Kronecker{\Omega}{\Omega}
=
\StrictFork{\varepsilon}{\varepsilon}
$
\item $\liftedMeet\RELtraOP\RELcompOP\syqq{(\varepsilon\RELcompOP\pi\RELtraOP\RELandOP\varepsilon\RELcompOP\rho\RELtraOP)}{X}
=
\syqq{(\varepsilon\RELcompOP\pi\RELtraOP\RELandOP\varepsilon\RELcompOP\rho\RELtraOP)\RELcompOP\liftedMeet}{X}
$
\item[] $\liftedMeet\RELtraOP\RELcompOP\syqq{\StrictFork{\varepsilon}{\varepsilon}}{X}
=
\syqq{\StrictFork{\varepsilon}{\varepsilon}\RELcompOP\liftedMeet}{X}
$
\item $\pi\RELcompOP\Omega\RELandOP\rho\RELenthOP\liftedJoin\qquad\rho\RELcompOP\Omega\RELandOP\pi\RELenthOP\liftedJoin$\quad or in variant form
\item[] $\StrictJoin{\Omega}{\RELide}\RELenthOP\liftedJoin\qquad\StrictJoin{\RELide}{\Omega}\RELenthOP\liftedJoin
$
\end{enumerate}

\Proof i) Since $N, {\cal N}$ are mappings, we may apply Prop.~\PropPiKrhoEquGeneral.ii to the first two and then proceed with, e.g.

\smallskip

${\cal N}\RELcompOP\liftedMeet\RELcompOP N
=
{\cal N}\RELcompOP
\syqq{\varepsilon\RELcompOP\pi\RELtraOP\RELandOP\varepsilon\RELcompOP\rho\RELtraOP}{\varepsilon}\RELcompOP N
=
\syqq{[\varepsilon\RELcompOP\pi\RELtraOP\RELandOP\varepsilon\RELcompOP\rho\RELtraOP]\RELcompOP{\cal N}\RELtraOP}{\varepsilon\RELcompOP N}
$

$=
\syqq{\varepsilon\RELcompOP\pi\RELtraOP\RELcompOP{\cal N}\RELtraOP\RELandOP\varepsilon\RELcompOP\rho\RELtraOP\RELcompOP{\cal N}\RELtraOP}{\RELneg{\varepsilon}}
=
\syqq{\varepsilon\RELcompOP\pi\RELtraOP\RELcompOP{\cal N}\RELandOP\varepsilon\RELcompOP\rho\RELtraOP\RELcompOP{\cal N}}{\RELneg{\varepsilon}}
$

$=
\syqq{\varepsilon\RELcompOP N\RELcompOP\pi\RELtraOP\RELandOP\varepsilon\RELcompOP N\RELcompOP\rho\RELtraOP}{\RELneg{\varepsilon}}
=
\syqq{\RELneg{\varepsilon}\RELcompOP\pi\RELtraOP\RELandOP\RELneg{\varepsilon}\RELcompOP\rho\RELtraOP}{\RELneg{\varepsilon}}
$

$
=
\syqq{\RELneg{\varepsilon\RELcompOP\pi\RELtraOP}\RELandOP\RELneg{\varepsilon\RELcompOP\rho\RELtraOP}}{\RELneg{\varepsilon}}
=
\syqq{\RELneg{\varepsilon\RELcompOP\pi\RELtraOP\RELorOP\varepsilon\RELcompOP\rho\RELtraOP}}{\RELneg{\varepsilon}}
=
\syqq{\varepsilon\RELcompOP\pi\RELtraOP\RELorOP\varepsilon\RELcompOP\rho\RELtraOP}{\varepsilon}
=
\liftedJoin
$

\bigskip
\noindent
ii) 
$\liftedMeet\RELtraOP
=
\syqq{\varepsilon}{\varepsilon\RELcompOP\pi\RELtraOP\RELandOP\varepsilon\RELcompOP\rho\RELtraOP}
=
\RELneg{\RELneg{\varepsilon}\RELtraOP\RELcompOP(\varepsilon\RELcompOP\pi\RELtraOP\RELandOP\varepsilon\RELcompOP\rho\RELtraOP)}\RELandOP\RELneg{\varepsilon\RELtraOP\RELcompOP\RELneg{\varepsilon\RELcompOP\pi\RELtraOP\RELandOP\varepsilon\RELcompOP\rho\RELtraOP}}
$

$
=
\RELneg{\RELneg{\varepsilon}\RELtraOP\RELcompOP(\varepsilon\RELcompOP\pi\RELtraOP\RELandOP\varepsilon\RELcompOP\rho\RELtraOP)}\RELandOP\RELneg{\varepsilon\RELtraOP\RELcompOP\RELneg{\varepsilon\RELcompOP\pi\RELtraOP}}\RELandOP\RELneg{\varepsilon\RELtraOP\RELcompOP\RELneg{\varepsilon\RELcompOP\rho\RELtraOP}}
=
\RELneg{\RELneg{\varepsilon}\RELtraOP\RELcompOP(\varepsilon\RELcompOP\pi\RELtraOP
\RELandOP\varepsilon\RELcompOP\rho\RELtraOP)}
\RELandOP
\Omega\RELcompOP\pi\RELtraOP
\RELandOP
\Omega\RELcompOP\rho\RELtraOP
$

\smallskip
\noindent
Now

\smallskip
$\liftedMeet\RELtraOP\RELcompOP\pi
=
\Big\lbrack\RELneg{\RELneg{\varepsilon}\RELtraOP\RELcompOP(\varepsilon\RELcompOP\pi\RELtraOP\RELandOP\varepsilon\RELcompOP\rho\RELtraOP)}\RELandOP\Omega\RELcompOP\pi\RELtraOP\RELandOP\Omega\RELcompOP\rho\RELtraOP\Big\rbrack\RELcompOP\pi
$

$
=
\Big\lbrack\Omega\RELcompOP\pi\RELtraOP\RELandOP\big\{\RELneg{\RELneg{\varepsilon}\RELtraOP\RELcompOP(\varepsilon\RELcompOP\pi\RELtraOP\RELandOP\varepsilon\RELcompOP\rho\RELtraOP)}\RELandOP\Omega\RELcompOP\rho\RELtraOP\big\}\Big\rbrack\RELcompOP\pi
=
\Omega\RELandOP\big\{\RELneg{\RELneg{\varepsilon}\RELtraOP\RELcompOP(\varepsilon\RELcompOP\pi\RELtraOP\RELandOP\varepsilon\RELcompOP\rho\RELtraOP)}\RELandOP\Omega\RELcompOP\rho\RELtraOP\big\}\RELcompOP\pi
=
\Omega\RELandOP\RELtop=\Omega
$

\smallskip
\noindent 
since

\smallskip
$\big\{\RELneg{\RELneg{\varepsilon}\RELtraOP\RELcompOP(\varepsilon\RELcompOP\pi\RELtraOP\RELandOP\varepsilon\RELcompOP\rho\RELtraOP)}\RELandOP\Omega\RELcompOP\rho\RELtraOP\big\}\RELcompOP\pi
\RELaboveOP
\big\{\RELneg{\RELneg{\varepsilon}\RELtraOP\RELcompOP\varepsilon\RELcompOP\rho\RELtraOP}\RELandOP\Omega\RELcompOP\rho\RELtraOP\big\}\RELcompOP\pi
=
\big\{\RELneg{\RELneg{\varepsilon}\RELtraOP\RELcompOP\varepsilon}\RELcompOP\rho\RELtraOP\RELandOP\Omega\RELcompOP\rho\RELtraOP\big\}\RELcompOP\pi
=
\big\{\RELneg{\RELneg{\varepsilon}\RELtraOP\RELcompOP\varepsilon}\RELandOP\Omega\big\}\RELcompOP\rho\RELtraOP\RELcompOP\pi
$

$=
\big\{\Omega\RELtraOP\RELandOP\Omega\big\}\RELcompOP\rho\RELtraOP\RELcompOP\pi
=
\RELide\RELcompOP\RELtop=\RELtop
$

\bigskip
\noindent
iii) $\liftedJoin\RELtraOP\RELcompOP\pi
=
N \RELcompOP\liftedMeet\RELtraOP\RELcompOP{\cal N}\RELcompOP\pi
=
N \RELcompOP\liftedMeet\RELtraOP\RELcompOP\pi\RELcompOP N
=
N \RELcompOP\Omega\RELcompOP N
=
N \RELcompOP\RELneg{\varepsilon\RELtraOP\RELcompOP\RELneg{\varepsilon}}\RELcompOP N
=
\RELneg{N \RELcompOP\varepsilon\RELtraOP\RELcompOP\RELneg{\varepsilon\RELcompOP N}}
=
\RELneg{\RELneg{\varepsilon}\RELtraOP\RELcompOP\RELneg{\RELneg{\varepsilon}}}
=
\RELneg{\RELneg{\varepsilon}\RELtraOP\RELcompOP\varepsilon}
=
\Omega\RELtraOP
$

\bigskip
\noindent
iv) From Prop.~\CorSumPowToPowProd.iv, we have $\liftedMeet\RELcompOP\varepsilon\RELtraOP
=
\pi\RELcompOP\varepsilon\RELtraOP\RELandOP\rho\RELcompOP\varepsilon\RELtraOP$. Negation and multiplication with $\varepsilon$ from the right side gives

\smallskip
$\RELneg{\liftedMeet\RELcompOP\varepsilon\RELtraOP}\RELcompOP\varepsilon
=
\RELneg{\pi\RELcompOP\varepsilon\RELtraOP}\RELcompOP\varepsilon\RELorOP\RELneg{\rho\RELcompOP\varepsilon\RELtraOP}\RELcompOP\varepsilon
$

\smallskip
$\iff\quad\RELneg{\RELneg{\liftedMeet\RELcompOP\varepsilon\RELtraOP}\RELcompOP\varepsilon}
=
\RELneg{\RELneg{\pi\RELcompOP\varepsilon\RELtraOP}\RELcompOP\varepsilon}\RELandOP\RELneg{\RELneg{\rho\RELcompOP\varepsilon\RELtraOP}\RELcompOP\varepsilon}
$

\smallskip
$\iff\quad\liftedMeet\RELcompOP\RELneg{\RELneg{\varepsilon\RELtraOP}\RELcompOP\varepsilon}
=
\pi\RELcompOP\RELneg{\RELneg{\varepsilon\RELtraOP}\RELcompOP\varepsilon}\RELandOP\rho\RELcompOP\RELneg{\RELneg{\varepsilon\RELtraOP}\RELcompOP\varepsilon}
$

\smallskip
$\iff\quad\liftedMeet\RELcompOP\Omega\RELtraOP
=
\pi\RELcompOP\Omega\RELtraOP\RELandOP\rho\RELcompOP\Omega\RELtraOP
$\quad meaning the intersection of lower cones

\bigskip
\noindent
Alternative proof:

\smallskip
$\Omega\RELcompOP\liftedMeet\RELtraOP
=
\Omega\RELcompOP\syqq{\varepsilon}{\StrictFork{\varepsilon}{\varepsilon}}
=
\RELneg{\varepsilon\RELtraOP\RELcompOP\RELneg{\varepsilon}}\RELcompOP\syqq{\varepsilon}{\StrictFork{\varepsilon}{\varepsilon}}
$

$=
\RELneg{\varepsilon\RELtraOP\RELcompOP\RELneg{\varepsilon\RELcompOP\syqq{\varepsilon}{\StrictFork{\varepsilon}{\varepsilon}}}}
$\quad since every $\syqq{\varepsilon}{\dots}$ is a transposed mapping

$=
\RELneg{\varepsilon\RELtraOP\RELcompOP\RELneg{\StrictFork{\varepsilon}{\varepsilon}}}
=\LeftResi{\varepsilon}{\StrictFork{\varepsilon}{\varepsilon}}
=\StrictFork{\LeftResi{\varepsilon}{\varepsilon}}{\LeftResi{\varepsilon}{\varepsilon}}
$\quad due to Prop.~\PropResiFork

$=
\StrictFork{\Omega}{\Omega}
$

\bigskip
\noindent
v) $(\varepsilon\RELcompOP\pi\RELtraOP\RELandOP\varepsilon\RELcompOP\rho\RELtraOP)
\RELcompOP\liftedMeet
=
(\varepsilon\RELcompOP\pi\RELtraOP\RELandOP\varepsilon\RELcompOP\rho\RELtraOP)
\RELcompOP\syqq{\varepsilon\RELcompOP\pi\RELtraOP\RELandOP\varepsilon\RELcompOP\rho\RELtraOP}{\varepsilon}
=
\varepsilon
$,

\smallskip
\noindent
since $\liftedMeet$ is surjective according to Prop.~\PropPiAndRhoAndMeetSurj.ii and \cite{RelaMath2010} 8.12.iii; for $\liftedJoin$ similarly.

\bigskip
\noindent
vi) The following is shown in two steps:

\smallskip
$\StrictFork{\varepsilon}{\varepsilon}\RELcompOP\Kronecker{\Omega}{\Omega}
=
(\varepsilon\RELcompOP\pi\RELtraOP\RELandOP\varepsilon\RELcompOP\rho\RELtraOP)
\RELcompOP
(\pi\RELcompOP\Omega\RELcompOP\pi\RELtraOP\RELandOP\rho\RELcompOP\Omega\RELcompOP\rho\RELtraOP)
$

$\RELenthOP
\varepsilon\RELcompOP\pi\RELtraOP\RELcompOP\pi\RELcompOP\Omega\RELcompOP\pi\RELtraOP\RELandOP\varepsilon\RELcompOP\rho\RELtraOP\RELcompOP\rho\RELcompOP\Omega\RELcompOP\rho\RELtraOP
$\quad isotony

$\RELenthOP
\varepsilon\RELcompOP\Omega\RELcompOP\pi\RELtraOP\RELandOP\varepsilon\RELcompOP\Omega\RELcompOP\rho\RELtraOP
$\quad $\pi,\rho$ are univalent

$=
\varepsilon\RELcompOP\pi\RELtraOP\RELandOP\varepsilon\RELcompOP\rho\RELtraOP
=
\StrictFork{\varepsilon}{\varepsilon}
$\quad since $\varepsilon\RELcompOP\Omega=\varepsilon$

\bigskip
\noindent
Short alternative proof:

\smallskip
$\StrictFork{\varepsilon}{\varepsilon}\RELcompOP\Kronecker{\Omega}{\Omega}
\RELenthOP
\StrictFork{\varepsilon\RELcompOP\Omega}{\varepsilon\RELcompOP\Omega}
=
\StrictFork{\Omega}{\Omega}
$\quad using Prop.~\PropForkMapKron.i

\smallskip
\noindent
On the other hand side

\smallskip
$\Kronecker{\Omega}{\Omega}
=
\pi\RELcompOP\Omega\RELcompOP\pi\RELtraOP\RELandOP\rho\RELcompOP\Omega\RELcompOP\rho\RELtraOP
\RELaboveOP
\pi\RELcompOP\pi\RELtraOP\RELandOP\rho\RELcompOP\rho\RELtraOP
=
\RELide
$,

\smallskip
\noindent
so that also

\smallskip
$\StrictFork{\varepsilon}{\varepsilon}\RELcompOP\Kronecker{\Omega}{\Omega}
\RELaboveOP
\StrictFork{\varepsilon}{\varepsilon}\RELcompOP\RELide
=
\StrictFork{\varepsilon}{\varepsilon}
$.

\bigskip
\noindent
vii) We apply Prop.~8.18 of \cite{RelaMath2010} and, therefore, prove just

\smallskip
$(\varepsilon\RELcompOP\pi\RELtraOP\RELandOP\varepsilon\RELcompOP\rho\RELtraOP)\RELcompOP\liftedMeet\RELcompOP\liftedMeet\RELtraOP
=
(\varepsilon\RELcompOP\pi\RELtraOP\RELandOP\varepsilon\RELcompOP\rho\RELtraOP)\RELcompOP\syqq{\varepsilon\RELcompOP\pi\RELtraOP\RELandOP\varepsilon\RELcompOP\rho\RELtraOP}{\varepsilon}\RELcompOP\liftedMeet\RELtraOP
=
\varepsilon\RELcompOP\liftedMeet\RELtraOP
=
(\varepsilon\RELcompOP\pi\RELtraOP\RELandOP\varepsilon\RELcompOP\rho\RELtraOP)$

\bigskip
\noindent
viii) $\pi\RELcompOP\Omega\RELandOP\rho\RELcompOP\RELide
=
\pi\RELcompOP\Omega\RELandOP\rho\RELcompOP(\Omega\RELandOP\Omega\RELtraOP)
=
\pi\RELcompOP\Omega\RELandOP\rho\RELcompOP\Omega\RELandOP\rho\RELcompOP\Omega\RELtraOP
$

$=
\liftedJoin\RELcompOP\Omega\RELandOP\rho\RELcompOP\Omega\RELtraOP
$\quad due to (iv)

$\RELenthOP
\liftedJoin\RELcompOP\Omega\RELandOP\liftedJoin\RELcompOP\Omega\RELtraOP
=
\liftedJoin\RELcompOP(\Omega\RELandOP\Omega\RELtraOP)
=
\liftedJoin
$\quad since $\rho\RELcompOP\Omega\RELtraOP\RELenthOP\liftedJoin\RELcompOP\Omega\RELtraOP
\;\iff\;
\liftedJoin\RELtraOP\RELcompOP\rho\RELcompOP\Omega\RELtraOP=\Omega\RELtraOP\RELcompOP\Omega\RELtraOP\RELenthOP\Omega\RELtraOP
$\Bewende

\bigskip
\noindent
Of course, the traditional reasoning with orderings, e.g., $a\leq c,a\leq d\;\Longrightarrow\; a\leq c\cap d$, assumes another shape.

\enunc{}{Proposition}{}{PropNonClassIntersection} i) For points $a,c,d$ we have

\smallskip
$\vcenter{\hbox{$a\RELenthOP\Omega\RELcompOP c$}\hbox{$a\RELenthOP\Omega\RELcompOP d$}}
\quad\Longrightarrow\quad
a\RELenthOP\Omega\RELcompOP\liftedMeet\RELtraOP\RELcompOP\StrictJoin{c}{d}
=
\StrictFork{\Omega}{\Omega}\RELcompOP\StrictJoin{c}{d}$

\bigskip
\noindent
ii) For points $b,c,d$ we have

\smallskip
$\vcenter{\hbox{$b\RELenthOP\Omega\RELtraOP\RELcompOP c$}\hbox{$b\RELenthOP\Omega\RELtraOP\RELcompOP d$}}
\quad\Longrightarrow\quad
b\RELenthOP
\Omega\RELtraOP\RELcompOP\liftedJoin\RELtraOP\RELcompOP\StrictJoin{c}{d}
=
\StrictFork{\Omega\RELtraOP}{\Omega\RELtraOP}\RELcompOP\StrictJoin{c}{d}$

\Proof i) $\liftedMeet\RELtraOP\RELcompOP\StrictJoin{c}{d}=
\syqq{\varepsilon}{(\varepsilon\RELcompOP\pi\RELtraOP\RELandOP\varepsilon\RELcompOP\rho\RELtraOP)}\RELcompOP(\pi\RELcompOP c\RELandOP\rho\RELcompOP d)
$\quad by definition

$=
\syqq{\varepsilon}{(\varepsilon\RELcompOP\pi\RELtraOP\RELandOP\varepsilon\RELcompOP\rho\RELtraOP)\RELcompOP(\pi\RELcompOP c\RELandOP\rho\RELcompOP  d)}
$\quad since $(\pi\RELcompOP c\RELandOP\rho\RELcompOP d)$ is a point

$=
\syqq{\varepsilon}{\varepsilon\RELcompOP\pi\RELtraOP\RELcompOP(\pi\RELcompOP c\RELandOP\rho\RELcompOP d)\RELandOP\varepsilon\RELcompOP\rho\RELtraOP\RELcompOP(\pi\RELcompOP c\RELandOP\rho\RELcompOP  d)}
$\quad again since $(\pi\RELcompOP c\RELandOP\rho\RELcompOP  d)$ is a point!

$=
\syqq{\varepsilon}{\varepsilon\RELcompOP(c\RELandOP\pi\RELtraOP\RELcompOP\rho\RELcompOP d)\RELandOP\varepsilon\RELcompOP(\rho\RELtraOP\RELcompOP\pi\RELcompOP c\RELandOP d)}
$

$=
\syqq{\varepsilon}{\varepsilon\RELcompOP(c\RELandOP\RELtop)\RELandOP\varepsilon\RELcompOP(\RELtop\RELandOP d)}
$

$=
\syqq{\varepsilon}{\varepsilon\RELcompOP c\RELandOP\varepsilon\RELcompOP d}=:s
$,\quad which is a point!

\bigskip
\noindent
Now, we may continue

\smallskip
$\Omega\RELcompOP\liftedMeet\RELtraOP\RELcompOP\StrictJoin{c}{d}
=
\Omega\RELcompOP s
=
\RELneg{\varepsilon\RELtraOP\RELcompOP\RELneg{\varepsilon}}\RELcompOP s
=
\RELneg{\varepsilon\RELtraOP\RELcompOP\RELneg{\varepsilon\RELcompOP s}}
=
\RELneg{\varepsilon\RELtraOP\RELcompOP\RELneg{\varepsilon\RELcompOP\syqq{\varepsilon}{\varepsilon\RELcompOP c\RELandOP\varepsilon\RELcompOP d}}}
=
\RELneg{\varepsilon\RELtraOP\RELcompOP\RELneg{\varepsilon\RELcompOP c\RELandOP\varepsilon\RELcompOP d}}
$

$=
\RELneg{\varepsilon\RELtraOP\RELcompOP(\RELneg{\varepsilon\RELcompOP c}\RELorOP\RELneg{\varepsilon\RELcompOP d)}}
=
\RELneg{\varepsilon\RELtraOP\RELcompOP\RELneg{\varepsilon\RELcompOP c}\RELorOP\varepsilon\RELtraOP\RELcompOP\RELneg{\varepsilon\RELcompOP d}}
=
\RELneg{\varepsilon\RELtraOP\RELcompOP\RELneg{\varepsilon\RELcompOP c}}\RELandOP\RELneg{\varepsilon\RELtraOP\RELcompOP\RELneg{\varepsilon\RELcompOP d}}
=
\RELneg{\varepsilon\RELtraOP\RELcompOP\RELneg{\varepsilon}}\RELcompOP c\RELandOP\RELneg{\varepsilon\RELtraOP\RELcompOP\RELneg{\varepsilon}}\RELcompOP d
=
\Omega\RELcompOP c\RELandOP\Omega\RELcompOP d
\RELaboveOP
a
$

\bigskip
\noindent
Short alternative proof:

\smallskip
$\Omega\RELcompOP\liftedMeet\RELtraOP\RELcompOP\StrictJoin{c}{d}
=
\StrictFork{\Omega}{\Omega}\RELcompOP\StrictJoin{c}{d}
=
\Omega\RELcompOP c\RELandOP\Omega\RELcompOP d
$\quad Prop.~\PropForkMapKron.iii

\bigskip
\noindent 
ii) is proved in a similar way.
\Bewende

\noindent
One will understand Prop.~\PropPowerMeet.iv when interpreting it with cone intersection: Lower cone of a meet means intersecting the lower cones of the projections. Upper cone of the join is the intersection of the upper cones of the projections. Prop.~\PropPowerMeet.i resembles the De Morgan rule\index{De Morgan rule}.

\enunc{}{Proposition}{}{PropPiAndRhoAndMeetSurj} Given any direct product with projections $\RELfromTO{\pi,\rho}{X\times X}{X}$, and meet- or join-forming $\liftedMeet,\liftedJoin$, 

\begin{enumerate}[i)]
\item the construct $p:=\pi\RELandOP\rho$ is univalent and surjective,
\item meet-forming $\liftedMeet$ and join-forming $\liftedJoin$ are surjective mappings,
\item concerning meet- and join-forming, $\liftedJoin$ {\bf distributes} over $\liftedMeet$,
\item meet-forming $\liftedMeet$ is a homomorphism and, even stronger, $\Kronecker{\Omega}{\Omega}\RELcompOP\liftedMeet=\liftedMeet\RELcompOP\Omega$.
\end{enumerate}
\Proof
i) We use that the direct product encompasses every pair and that projections are surjective before applying the Dedekind formula

\smallskip
$\RELide
=
\RELtop\RELandOP\RELide
=
\pi\RELtraOP\RELcompOP\rho\RELandOP\rho\RELtraOP\RELcompOP\rho
\RELenthOP
(\pi\RELtraOP\RELandOP\rho\RELtraOP\RELcompOP\rho\RELcompOP\rho\RELtraOP)\RELcompOP(\rho\RELandOP\pi\RELcompOP\rho\RELtraOP\RELcompOP\rho)
=
(\pi\RELtraOP\RELandOP\rho\RELtraOP)\RELcompOP(\rho\RELandOP\pi)
$

\bigskip
\noindent
ii) $\liftedMeet
=
\syqq{\varepsilon\RELcompOP\pi\RELtraOP\RELandOP\varepsilon\RELcompOP\rho\RELtraOP}{\varepsilon}
=
\RELneg{(\RELneg{\varepsilon\RELcompOP\pi\RELtraOP\RELandOP\varepsilon\RELcompOP\rho\RELtraOP})\RELtraOP\RELcompOP\varepsilon}
\RELandOP
\RELneg{(\varepsilon\RELcompOP\pi\RELtraOP\RELandOP\varepsilon\RELcompOP\rho\RELtraOP)\RELtraOP\RELcompOP\RELneg{\varepsilon}}
=:
A\RELandOP B
$

$A=
\RELneg{\RELneg{\pi\RELcompOP\varepsilon\RELtraOP}\RELcompOP\varepsilon\RELorOP\RELneg{\rho\RELcompOP\varepsilon\RELtraOP}\RELcompOP\varepsilon}
=
\RELneg{\RELneg{\pi\RELcompOP\varepsilon\RELtraOP}\RELcompOP\varepsilon}\RELandOP\RELneg{\RELneg{\rho\RELcompOP\varepsilon\RELtraOP}\RELcompOP\varepsilon}
=
\pi\RELcompOP\RELneg{\RELneg{\varepsilon\RELtraOP}\RELcompOP\varepsilon}\RELandOP\rho\RELcompOP\RELneg{\RELneg{\varepsilon\RELtraOP}\RELcompOP\varepsilon}
=
\pi\RELcompOP\Omega\RELtraOP\RELandOP\rho\RELcompOP\Omega\RELtraOP
$

$B
\RELaboveOP
\RELneg{(\pi\RELcompOP\varepsilon\RELtraOP\RELorOP\rho\RELcompOP\varepsilon\RELtraOP)\RELcompOP\RELneg{\varepsilon}}
=
\RELneg{\pi\RELcompOP\varepsilon\RELtraOP\RELcompOP\RELneg{\varepsilon}\RELorOP\rho\RELcompOP\varepsilon\RELtraOP\RELcompOP\RELneg{\varepsilon}}
=
\RELneg{\pi\RELcompOP\varepsilon\RELtraOP\RELcompOP\RELneg{\varepsilon}}\RELandOP\RELneg{\rho\RELcompOP\varepsilon\RELtraOP\RELcompOP\RELneg{\varepsilon}}
=
\pi\RELcompOP\RELneg{\varepsilon\RELtraOP\RELcompOP\RELneg{\varepsilon}}\RELandOP\rho\RELcompOP\RELneg{\varepsilon\RELtraOP\RELcompOP\RELneg{\varepsilon}}
=
\pi\RELcompOP\Omega\RELandOP\rho\RELcompOP\Omega
$

\smallskip
$\liftedMeet
=
A\RELandOP B
\RELaboveOP
(\pi\RELcompOP\Omega\RELtraOP\RELandOP\rho\RELcompOP\Omega\RELtraOP)
\RELandOP
(\pi\RELcompOP\Omega\RELandOP\rho\RELcompOP\Omega)
=
\pi\RELcompOP(\Omega\RELtraOP\RELandOP\Omega)
\RELandOP
\rho\RELcompOP(\Omega\RELtraOP\RELandOP\Omega)
=
\pi\RELcompOP\RELide
\RELandOP
\rho\RELcompOP\RELide
=
\pi\RELandOP\rho
$

\smallskip
\noindent
The latter is surjective owing to (i). The proof for $\liftedJoin$ is rather similar.

\bigskip
\noindent
iii) $\StrictFork{\Kronecker{\pi}{\RELide}\RELcompOP\,\liftedJoin}{\Kronecker{\rho}{\RELide}\RELcompOP\liftedJoin}\RELcompOP\liftedMeet
$\quad where the first factor is a mapping

$=
\StrictFork{\Kronecker{\pi}{\RELide}\RELcompOP\,\liftedJoin}{\Kronecker{\rho}{\RELide}\RELcompOP\liftedJoin}\RELcompOP\syqq{\StrictFork{\varepsilon}{\varepsilon}}{\varepsilon}
$\quad definition of $\liftedMeet$

$=\syqq{\StrictFork{\varepsilon}{\varepsilon}\RELcompOP\StrictFork{\Kronecker{\pi}{\RELide}\RELcompOP\,\liftedJoin}{\Kronecker{\rho}{\RELide}\RELcompOP\liftedJoin}\RELtraOP}{\varepsilon}
$

$=\syqq{\StrictFork{\varepsilon}{\varepsilon}\RELcompOP\StrictJoin{\liftedJoin\RELtraOP\RELcompOP\Kronecker{\pi\RELtraOP}{\RELide}}{\liftedJoin\RELtraOP\RELcompOP\Kronecker{\rho\RELtraOP}{\RELide}}}{\varepsilon}
$\quad transposed

$=\syqq{\varepsilon\RELcompOP\liftedJoin\RELtraOP\RELcompOP\Kronecker{\pi\RELtraOP}{\RELide}\RELandOP\,\,\varepsilon\RELcompOP\liftedJoin\RELtraOP\RELcompOP\Kronecker{\rho\RELtraOP}{\RELide}}{\varepsilon}
$\quad Prop.~\PropForkMapKron.i

$=\syqq{(\varepsilon\RELcompOP\pi\RELtraOP\RELorOP\varepsilon\RELcompOP\rho\RELtraOP)\RELcompOP\Kronecker{\pi\RELtraOP}{\RELide}\RELandOP\,\,(\varepsilon\RELcompOP\pi\RELtraOP\RELorOP\varepsilon\RELcompOP\rho\RELtraOP)\RELcompOP\Kronecker{\rho\RELtraOP}{\RELide}}{\varepsilon}
$\quad Prop.~\CorSumPowToPowProd.iii

$=\syqq{(\varepsilon\RELcompOP\pi\RELtraOP\RELcompOP\Kronecker{\pi\RELtraOP}{\RELide}\RELorOP\,\,\varepsilon\RELcompOP\rho\RELtraOP\RELcompOP\Kronecker{\pi\RELtraOP}{\RELide})\RELandOP(\varepsilon\RELcompOP\pi\RELtraOP\RELcompOP\Kronecker{\rho\RELtraOP}{\RELide}\RELorOP\,\,\varepsilon\RELcompOP\rho\RELtraOP\RELcompOP\Kronecker{\rho\RELtraOP}{\RELide})}{\varepsilon}
$

$=\syqq{(\varepsilon\RELcompOP\pi\RELtraOP\RELcompOP{\pi'}\RELtraOP\RELorOP\,\,\varepsilon\RELcompOP{\rho'}\RELtraOP)\RELandOP(\varepsilon\RELcompOP\rho\RELtraOP\RELcompOP{\pi'}\RELtraOP\RELorOP\,\,\varepsilon\RELcompOP{\rho'}\RELtraOP)}{\varepsilon}
$

$=\syqq{(\varepsilon\RELcompOP\pi\RELtraOP\RELcompOP{\pi'}\RELtraOP\RELandOP\varepsilon\RELcompOP\rho\RELtraOP\RELcompOP{\pi'}\RELtraOP)\RELorOP\varepsilon\RELcompOP{\rho'}\RELtraOP}{\varepsilon}
$

$=\syqq{(\varepsilon\RELcompOP\pi\RELtraOP\RELandOP\varepsilon\RELcompOP\rho\RELtraOP)\RELcompOP{\pi'}\RELtraOP\RELorOP\varepsilon\RELcompOP{\rho'}\RELtraOP}{\varepsilon}
$

$=
\syqq{\lbrack\varepsilon\RELcompOP\liftedMeet\RELtraOP\RELcompOP{\pi'}\RELtraOP
\RELorOP
\varepsilon\RELcompOP{\rho'}\RELtraOP\rbrack}{\varepsilon}
$

$=
\syqq{\lbrack\varepsilon\RELcompOP(\liftedMeet\RELtraOP\RELcompOP{\pi'}\RELtraOP\RELandOP\pi\RELtraOP\RELcompOP  \rho\RELcompOP{\rho'}\RELtraOP)
\RELorOP
\varepsilon\RELcompOP(\rho\RELtraOP\RELcompOP\pi\RELcompOP\liftedMeet\RELtraOP\RELcompOP{\pi'}\RELtraOP\RELandOP  {\rho'}\RELtraOP)\rbrack}{\varepsilon}
$

$=
\syqq{\lbrack\varepsilon\RELcompOP\pi\RELtraOP\RELcompOP  (\pi\RELcompOP\liftedMeet\RELtraOP\RELcompOP{\pi'}\RELtraOP\RELandOP  \rho\RELcompOP{\rho'}\RELtraOP)\RELorOP\varepsilon\RELcompOP\rho\RELtraOP\RELcompOP  (\pi\RELcompOP\liftedMeet\RELtraOP\RELcompOP{\pi'}\RELtraOP\RELandOP  \rho\RELcompOP{\rho'}\RELtraOP)\rbrack}{\varepsilon}
$

$=
\syqq{\lbrack\varepsilon\RELcompOP\pi\RELtraOP\RELorOP\varepsilon\RELcompOP\rho\RELtraOP\rbrack\RELcompOP  (\pi\RELcompOP\liftedMeet\RELtraOP\RELcompOP{\pi'}\RELtraOP\RELandOP  \rho\RELcompOP{\rho'}\RELtraOP)}{\varepsilon}
$

$=
\syqq{\varepsilon\RELcompOP\lbrack\pi\RELtraOP\RELorOP\rho\RELtraOP\rbrack\RELcompOP\Kronecker{\liftedMeet\RELtraOP}{\RELide}}{\varepsilon}
$

$=
\syqq{\varepsilon\RELcompOP\lbrack\pi\RELorOP\rho\rbrack\RELtraOP\RELcompOP\Kronecker{\liftedMeet}{\RELide}\RELtraOP}{\varepsilon}
$

$=
\Kronecker{\liftedMeet}{\RELide}\RELcompOP\syqq{\varepsilon\RELcompOP\lbrack\pi\RELorOP\rho\rbrack\RELtraOP}{\varepsilon}
$\quad since $\Kronecker{\liftedMeet}{\RELide}$ is a mapping

$=
\Kronecker{\liftedMeet}{\RELide}\RELcompOP\liftedJoin
$\quad Prop.~\CorSumPowToPowProd.i

\bigskip
\noindent
iv) \lq\lq$\RELenthOP$\rq\rq\ follows with shunting $\Kronecker{\Omega}{\Omega}\RELcompOP\liftedMeet\RELenthOP\liftedMeet\RELcompOP\Omega
\;\iff
\Kronecker{\Omega}{\Omega}
\RELenthOP
\liftedMeet\RELcompOP\Omega\RELcompOP\liftedMeet\RELtraOP\;$ from

\smallskip
$\liftedMeet\RELcompOP\Omega\RELcompOP\liftedMeet\RELtraOP
=
\RELneg{\liftedMeet\RELcompOP\varepsilon\RELtraOP\RELcompOP\RELneg{\varepsilon\RELcompOP\liftedMeet\RELtraOP}}
=
\RELneg{\StrictJoin{\varepsilon\RELtraOP}{\varepsilon\RELtraOP}\RELcompOP\RELneg{\StrictFork{\varepsilon}{\varepsilon}}}
=
\RELneg{\StrictFork{\varepsilon}{\varepsilon}\RELtraOP\RELcompOP\RELneg{\StrictFork{\varepsilon}{\varepsilon}}}
=
\LeftResi{\StrictFork{\varepsilon}{\varepsilon}} {\StrictFork{\varepsilon}{\varepsilon}}
$

$\RELaboveOP
\Kronecker{\LeftResi{\varepsilon}{\varepsilon}}{\,\LeftResi{\varepsilon}{\varepsilon}}
$\quad following Prop.~\PropForkMapKron.viii

$=
\Kronecker{\Omega}{\Omega}
$

\smallskip
\noindent
The other direction \lq\lq$\RELaboveOP$\rq\rq\ applies distributivity (iv):

\smallskip
$\liftedMeet\RELcompOP\Omega
=
{\pi'}\RELtraOP\RELcompOP\rho'\RELandOP\liftedMeet\RELcompOP\Omega
$\quad $\pi',\rho'$ form a direct product

$=
{\pi'}\RELtraOP\RELcompOP(\RELtop\RELandOP\rho')\RELandOP\liftedMeet\RELcompOP\Omega
$\quad from now on using the abbreviation of Def.~\DefDistr

$=
{\pi'}\RELtraOP\RELcompOP(\JotHat\RELcompOP{\pi'}\RELtraOP\RELcompOP\rho'\RELandOP\rho')\RELandOP\liftedMeet\RELcompOP\Omega
$\quad $\JotHat$ is total and $\pi',\rho'$ form a direct product

$=
{\pi'}\RELtraOP\RELcompOP(\JotHat\RELcompOP{\pi'}\RELtraOP\RELandOP\rho'\RELcompOP{\rho'}\RELtraOP)\RELcompOP\rho'\RELandOP\liftedMeet\RELcompOP\Omega
$\quad destroy and append

$=
\big[{\pi'}\RELtraOP\RELcompOP(\JotHat\RELcompOP{\pi'}\RELtraOP\RELandOP\rho'\RELcompOP{\rho'}\RELtraOP)\RELandOP\liftedMeet\RELcompOP\Omega\RELcompOP{\rho'}\RELtraOP\big]\RELcompOP\rho'
$\quad again destroy and append

$=
{\pi'}\RELtraOP\RELcompOP\big[\JotHat\RELcompOP{\pi'}\RELtraOP\RELandOP\rho'\RELcompOP{\rho'}\RELtraOP\RELandOP\pi'\RELcompOP\liftedMeet\RELcompOP\Omega\RELcompOP{\rho'}\RELtraOP\big]\RELcompOP\rho'
$\quad again destroy and append

$=
{\pi'}\RELtraOP\RELcompOP\big[\JotHat\RELcompOP{\pi'}\RELtraOP\RELandOP\Kronecker{\liftedMeet}{\RELide}\RELcompOP\StrictJoin{\Omega}{\RELide}\RELcompOP\,{\rho'}\RELtraOP\big]\RELcompOP\rho'
$\quad due to Prop.~\PropForkMapKron.iii

$\RELenthOP
{\pi'}\RELtraOP\RELcompOP\big[\JotHat\RELcompOP{\pi'}\RELtraOP\RELandOP\Kronecker{\liftedMeet}{\RELide}\RELcompOP\liftedJoin\RELcompOP\,{\rho'}\RELtraOP\big]\RELcompOP\rho'
$\quad due to Prop.~\PropPowerMeet.viii

$\RELenthOP
{\pi'}\RELtraOP\RELcompOP\big[\JotHat\RELcompOP{\pi'}\RELtraOP\RELandOP\StrictFork{\Kronecker{\pi}{\RELide}\RELcompOP\,\liftedJoin}{\Kronecker{\rho}{\RELide}\RELcompOP\,\liftedJoin}\RELcompOP\liftedMeet\RELcompOP\,{\rho'}\RELtraOP\big]\RELcompOP\rho'
$\quad due to Def.~\DefDistr, (iv)

$\RELenthOP
{\pi'}\RELtraOP\RELcompOP\big[\JotHat\RELcompOP{\pi'}\RELtraOP\RELandOP\JotHat\RELcompOP\liftedMeet\RELcompOP\,{\rho'}\RELtraOP\big]\RELcompOP\rho'
$\quad 

$\RELenthOP
{\pi'}\RELtraOP\RELcompOP\JotHat\RELcompOP\big[{\pi'}\RELtraOP\RELandOP\liftedMeet\RELcompOP\,{\rho'}\RELtraOP\big]\RELcompOP\rho'
$\quad 

$\RELenthOP
{\pi'}\RELtraOP\RELcompOP\JotHat\RELcompOP\big[{\pi'}\RELtraOP\RELcompOP\rho'\RELandOP\liftedMeet\big]
$\quad 

$\RELenthOP
{\pi'}\RELtraOP\RELcompOP\JotHat\RELcompOP\liftedMeet
$\quad 

$=
{\pi'}\RELtraOP\RELcompOP\StrictFork{\Kronecker{\pi}{\RELide}\RELcompOP\,\liftedJoin}{\Kronecker{\rho}{\RELide}\RELcompOP\liftedJoin}\RELcompOP\liftedMeet
$\quad expanded 

$\RELenthOP
\StrictFork{{\pi'}\RELtraOP\RELcompOP\Kronecker{\pi}{\RELide}\RELcompOP\,\liftedJoin}{{\pi'}\RELtraOP\RELcompOP\Kronecker{\rho}{\RELide}\RELcompOP\,\liftedJoin}\RELcompOP\liftedMeet
$\quad  

$=
\StrictFork{\pi\RELcompOP\pi\RELtraOP\RELcompOP\liftedJoin}{\rho\RELcompOP\pi\RELtraOP\RELcompOP\liftedJoin}\RELcompOP\liftedMeet
$\quad  

$=
\StrictFork{\pi\RELcompOP\Omega}{\rho\RELcompOP\Omega}\RELcompOP\liftedMeet
$\quad  

$=
\Kronecker{\Omega}{\Omega}\RELcompOP\,\liftedMeet
$\quad 
\Bewende

\noindent
When we proceed according to (iii) from a pair of sets to a set of possibly bigger ones and form their meet,
we might also first form the meet and then increase.

\Caption{\hbox{{\footnotesize%
\BoxBreadth=0pt%
\setbox7=\hbox{(\{\},\{\})}%
\ifdim\wd7>\BoxBreadth\BoxBreadth=\wd7\fi%
\setbox7=\hbox{(\{a\},\{\})}%
\ifdim\wd7>\BoxBreadth\BoxBreadth=\wd7\fi%
\setbox7=\hbox{(\{\},\{a\})}%
\ifdim\wd7>\BoxBreadth\BoxBreadth=\wd7\fi%
\setbox7=\hbox{(\{b\},\{\})}%
\ifdim\wd7>\BoxBreadth\BoxBreadth=\wd7\fi%
\setbox7=\hbox{(\{a\},\{a\})}%
\ifdim\wd7>\BoxBreadth\BoxBreadth=\wd7\fi%
\setbox7=\hbox{(\{\},\{b\})}%
\ifdim\wd7>\BoxBreadth\BoxBreadth=\wd7\fi%
\setbox7=\hbox{(\{a,b\},\{\})}%
\ifdim\wd7>\BoxBreadth\BoxBreadth=\wd7\fi%
\setbox7=\hbox{(\{b\},\{a\})}%
\ifdim\wd7>\BoxBreadth\BoxBreadth=\wd7\fi%
\setbox7=\hbox{(\{a\},\{b\})}%
\ifdim\wd7>\BoxBreadth\BoxBreadth=\wd7\fi%
\setbox7=\hbox{(\{\},\{a,b\})}%
\ifdim\wd7>\BoxBreadth\BoxBreadth=\wd7\fi%
\setbox7=\hbox{(\{c\},\{\})}%
\ifdim\wd7>\BoxBreadth\BoxBreadth=\wd7\fi%
\setbox7=\hbox{(\{a,b\},\{a\})}%
\ifdim\wd7>\BoxBreadth\BoxBreadth=\wd7\fi%
\setbox7=\hbox{(\{b\},\{b\})}%
\ifdim\wd7>\BoxBreadth\BoxBreadth=\wd7\fi%
\setbox7=\hbox{(\{a\},\{a,b\})}%
\ifdim\wd7>\BoxBreadth\BoxBreadth=\wd7\fi%
\setbox7=\hbox{(\{\},\{c\})}%
\ifdim\wd7>\BoxBreadth\BoxBreadth=\wd7\fi%
\setbox7=\hbox{(\{a,c\},\{\})}%
\ifdim\wd7>\BoxBreadth\BoxBreadth=\wd7\fi%
\setbox7=\hbox{(\{c\},\{a\})}%
\ifdim\wd7>\BoxBreadth\BoxBreadth=\wd7\fi%
\setbox7=\hbox{(\{a,b\},\{b\})}%
\ifdim\wd7>\BoxBreadth\BoxBreadth=\wd7\fi%
\setbox7=\hbox{(\{b\},\{a,b\})}%
\ifdim\wd7>\BoxBreadth\BoxBreadth=\wd7\fi%
\setbox7=\hbox{(\{a\},\{c\})}%
\ifdim\wd7>\BoxBreadth\BoxBreadth=\wd7\fi%
\setbox7=\hbox{(\{\},\{a,c\})}%
\ifdim\wd7>\BoxBreadth\BoxBreadth=\wd7\fi%
\setbox7=\hbox{(\{b,c\},\{\})}%
\ifdim\wd7>\BoxBreadth\BoxBreadth=\wd7\fi%
\setbox7=\hbox{(\{a,c\},\{a\})}%
\ifdim\wd7>\BoxBreadth\BoxBreadth=\wd7\fi%
\setbox7=\hbox{(\{c\},\{b\})}%
\ifdim\wd7>\BoxBreadth\BoxBreadth=\wd7\fi%
\setbox7=\hbox{(\{a,b\},\{a,b\})}%
\ifdim\wd7>\BoxBreadth\BoxBreadth=\wd7\fi%
\setbox7=\hbox{(\{b\},\{c\})}%
\ifdim\wd7>\BoxBreadth\BoxBreadth=\wd7\fi%
\setbox7=\hbox{(\{a\},\{a,c\})}%
\ifdim\wd7>\BoxBreadth\BoxBreadth=\wd7\fi%
\setbox7=\hbox{(\{\},\{b,c\})}%
\ifdim\wd7>\BoxBreadth\BoxBreadth=\wd7\fi%
\setbox7=\hbox{(\{a,b,c\},\{\})}%
\ifdim\wd7>\BoxBreadth\BoxBreadth=\wd7\fi%
\setbox7=\hbox{(\{b,c\},\{a\})}%
\ifdim\wd7>\BoxBreadth\BoxBreadth=\wd7\fi%
\setbox7=\hbox{(\{a,c\},\{b\})}%
\ifdim\wd7>\BoxBreadth\BoxBreadth=\wd7\fi%
\setbox7=\hbox{(\{c\},\{a,b\})}%
\ifdim\wd7>\BoxBreadth\BoxBreadth=\wd7\fi%
\setbox7=\hbox{(\{a,b\},\{c\})}%
\ifdim\wd7>\BoxBreadth\BoxBreadth=\wd7\fi%
\setbox7=\hbox{(\{b\},\{a,c\})}%
\ifdim\wd7>\BoxBreadth\BoxBreadth=\wd7\fi%
\setbox7=\hbox{(\{a\},\{b,c\})}%
\ifdim\wd7>\BoxBreadth\BoxBreadth=\wd7\fi%
\setbox7=\hbox{(\{\},\{a,b,c\})}%
\ifdim\wd7>\BoxBreadth\BoxBreadth=\wd7\fi%
\setbox7=\hbox{(\{d\},\{\})}%
\def\RowNames{\vcenter{\offinterlineskip\baselineskip=\matrixskip%
\hbox to\BoxBreadth{\strut\hfil (\{\},\{\})}\kern\interspacereduction%
\hbox to\BoxBreadth{\strut\hfil (\{a\},\{\})}\kern\interspacereduction%
\hbox to\BoxBreadth{\strut\hfil (\{\},\{a\})}\kern\interspacereduction%
\hbox to\BoxBreadth{\strut\hfil (\{b\},\{\})}\kern\interspacereduction%
\hbox to\BoxBreadth{\strut\hfil (\{a\},\{a\})}\kern\interspacereduction%
\hbox to\BoxBreadth{\strut\hfil (\{\},\{b\})}\kern\interspacereduction%
\hbox to\BoxBreadth{\strut\hfil (\{a,b\},\{\})}\kern\interspacereduction%
\hbox to\BoxBreadth{\strut\hfil (\{b\},\{a\})}\kern\interspacereduction%
\hbox to\BoxBreadth{\strut\hfil (\{a\},\{b\})}\kern\interspacereduction%
\hbox to\BoxBreadth{\strut\hfil (\{\},\{a,b\})}\kern\interspacereduction%
\hbox to\BoxBreadth{\strut\hfil (\{c\},\{\})}\kern\interspacereduction%
\hbox to\BoxBreadth{\strut\hfil (\{a,b\},\{a\})}\kern\interspacereduction%
\hbox to\BoxBreadth{\strut\hfil (\{b\},\{b\})}\kern\interspacereduction%
\hbox to\BoxBreadth{\strut\hfil (\{a\},\{a,b\})}\kern\interspacereduction%
\hbox to\BoxBreadth{\strut\hfil (\{\},\{c\})}\kern\interspacereduction%
\hbox to\BoxBreadth{\strut\hfil (\{a,c\},\{\})}\kern\interspacereduction%
\hbox to\BoxBreadth{\strut\hfil (\{c\},\{a\})}\kern\interspacereduction%
\hbox to\BoxBreadth{\strut\hfil (\{a,b\},\{b\})}\kern\interspacereduction%
\hbox to\BoxBreadth{\strut\hfil (\{b\},\{a,b\})}\kern\interspacereduction%
\hbox to\BoxBreadth{\strut\hfil (\{a\},\{c\})}\kern\interspacereduction%
\hbox to\BoxBreadth{\strut\hfil (\{\},\{a,c\})}\kern\interspacereduction%
\hbox to\BoxBreadth{\strut\hfil (\{b,c\},\{\})}\kern\interspacereduction%
\hbox to\BoxBreadth{\strut\hfil (\{a,c\},\{a\})}\kern\interspacereduction%
\hbox to\BoxBreadth{\strut\hfil (\{c\},\{b\})}\kern\interspacereduction%
\hbox to\BoxBreadth{\strut\hfil (\{a,b\},\{a,b\})}\kern\interspacereduction%
\hbox to\BoxBreadth{\strut\hfil (\{b\},\{c\})}\kern\interspacereduction%
\hbox to\BoxBreadth{\strut\hfil (\{a\},\{a,c\})}\kern\interspacereduction%
\hbox to\BoxBreadth{\strut\hfil (\{\},\{b,c\})}\kern\interspacereduction%
\hbox to\BoxBreadth{\strut\hfil (\{a,b,c\},\{\})}\kern\interspacereduction%
\hbox to\BoxBreadth{\strut\hfil (\{b,c\},\{a\})}\kern\interspacereduction%
\hbox to\BoxBreadth{\strut\hfil (\{a,c\},\{b\})}\kern\interspacereduction%
\hbox to\BoxBreadth{\strut\hfil (\{c\},\{a,b\})}\kern\interspacereduction%
\hbox to\BoxBreadth{\strut\hfil (\{a,b\},\{c\})}\kern\interspacereduction%
\hbox to\BoxBreadth{\strut\hfil (\{b\},\{a,c\})}\kern\interspacereduction%
\hbox to\BoxBreadth{\strut\hfil (\{a\},\{b,c\})}\kern\interspacereduction%
\hbox to\BoxBreadth{\strut\hfil (\{\},\{a,b,c\})}\kern\interspacereduction%
\hbox to\BoxBreadth{\strut\hfil (\{d\},\{\})}}}%
\def\ColNames{\hbox{\rotatebox{90}{\strut \{\}}\kern\interspacereduction%
\rotatebox{90}{\strut \{a\}}\kern\interspacereduction%
\rotatebox{90}{\strut \{b\}}\kern\interspacereduction%
\rotatebox{90}{\strut \{a,b\}}\kern\interspacereduction%
\rotatebox{90}{\strut \{c\}}\kern\interspacereduction%
\rotatebox{90}{\strut \{a,c\}}\kern\interspacereduction%
\rotatebox{90}{\strut \{b,c\}}\kern\interspacereduction%
\rotatebox{90}{\strut \{a,b,c\}}\kern\interspacereduction%
\rotatebox{90}{\strut \{d\}}\kern\interspacereduction%
\rotatebox{90}{\strut \{a,d\}}\kern\interspacereduction%
\rotatebox{90}{\strut \{b,d\}}\kern\interspacereduction%
\rotatebox{90}{\strut \{a,b,d\}}\kern\interspacereduction%
\rotatebox{90}{\strut \{c,d\}}\kern\interspacereduction%
\rotatebox{90}{\strut \{a,c,d\}}\kern\interspacereduction%
\rotatebox{90}{\strut \{b,c,d\}}\kern\interspacereduction%
\rotatebox{90}{\strut \{a,b,c,d\}}\kern\interspacereduction%
}}%
\def\Matrix{\spmatrix{%
\noalign{\kern-2pt}
 {\CoefTrue}&\n&\n&\n&\n&\n&\n&\n&\n&\n&\n&\n&\n&\n&\n&\n\cr
 \n&{\CoefTrue}&\n&\n&\n&\n&\n&\n&\n&\n&\n&\n&\n&\n&\n&\n\cr
 \n&{\CoefTrue}&\n&\n&\n&\n&\n&\n&\n&\n&\n&\n&\n&\n&\n&\n\cr
 \n&\n&{\CoefTrue}&\n&\n&\n&\n&\n&\n&\n&\n&\n&\n&\n&\n&\n\cr
 \n&{\CoefTrue}&\n&\n&\n&\n&\n&\n&\n&\n&\n&\n&\n&\n&\n&\n\cr
 \n&\n&{\CoefTrue}&\n&\n&\n&\n&\n&\n&\n&\n&\n&\n&\n&\n&\n\cr
 \n&\n&\n&{\CoefTrue}&\n&\n&\n&\n&\n&\n&\n&\n&\n&\n&\n&\n\cr
 \n&\n&\n&{\CoefTrue}&\n&\n&\n&\n&\n&\n&\n&\n&\n&\n&\n&\n\cr
 \n&\n&\n&{\CoefTrue}&\n&\n&\n&\n&\n&\n&\n&\n&\n&\n&\n&\n\cr
 \n&\n&\n&{\CoefTrue}&\n&\n&\n&\n&\n&\n&\n&\n&\n&\n&\n&\n\cr
 \n&\n&\n&\n&{\CoefTrue}&\n&\n&\n&\n&\n&\n&\n&\n&\n&\n&\n\cr
 \n&\n&\n&{\CoefTrue}&\n&\n&\n&\n&\n&\n&\n&\n&\n&\n&\n&\n\cr
 \n&\n&{\CoefTrue}&\n&\n&\n&\n&\n&\n&\n&\n&\n&\n&\n&\n&\n\cr
 \n&\n&\n&{\CoefTrue}&\n&\n&\n&\n&\n&\n&\n&\n&\n&\n&\n&\n\cr
 \n&\n&\n&\n&{\CoefTrue}&\n&\n&\n&\n&\n&\n&\n&\n&\n&\n&\n\cr
 \n&\n&\n&\n&\n&{\CoefTrue}&\n&\n&\n&\n&\n&\n&\n&\n&\n&\n\cr
 \n&\n&\n&\n&\n&{\CoefTrue}&\n&\n&\n&\n&\n&\n&\n&\n&\n&\n\cr
 \n&\n&\n&{\CoefTrue}&\n&\n&\n&\n&\n&\n&\n&\n&\n&\n&\n&\n\cr
 \n&\n&\n&{\CoefTrue}&\n&\n&\n&\n&\n&\n&\n&\n&\n&\n&\n&\n\cr
 \n&\n&\n&\n&\n&{\CoefTrue}&\n&\n&\n&\n&\n&\n&\n&\n&\n&\n\cr
 \n&\n&\n&\n&\n&{\CoefTrue}&\n&\n&\n&\n&\n&\n&\n&\n&\n&\n\cr
 \n&\n&\n&\n&\n&\n&{\CoefTrue}&\n&\n&\n&\n&\n&\n&\n&\n&\n\cr
 \n&\n&\n&\n&\n&{\CoefTrue}&\n&\n&\n&\n&\n&\n&\n&\n&\n&\n\cr
 \n&\n&\n&\n&\n&\n&{\CoefTrue}&\n&\n&\n&\n&\n&\n&\n&\n&\n\cr
 \n&\n&\n&{\CoefTrue}&\n&\n&\n&\n&\n&\n&\n&\n&\n&\n&\n&\n\cr
 \n&\n&\n&\n&\n&\n&{\CoefTrue}&\n&\n&\n&\n&\n&\n&\n&\n&\n\cr
 \n&\n&\n&\n&\n&{\CoefTrue}&\n&\n&\n&\n&\n&\n&\n&\n&\n&\n\cr
 \n&\n&\n&\n&\n&\n&{\CoefTrue}&\n&\n&\n&\n&\n&\n&\n&\n&\n\cr
 \n&\n&\n&\n&\n&\n&\n&{\CoefTrue}&\n&\n&\n&\n&\n&\n&\n&\n\cr
 \n&\n&\n&\n&\n&\n&\n&{\CoefTrue}&\n&\n&\n&\n&\n&\n&\n&\n\cr
 \n&\n&\n&\n&\n&\n&\n&{\CoefTrue}&\n&\n&\n&\n&\n&\n&\n&\n\cr
 \n&\n&\n&\n&\n&\n&\n&{\CoefTrue}&\n&\n&\n&\n&\n&\n&\n&\n\cr
 \n&\n&\n&\n&\n&\n&\n&{\CoefTrue}&\n&\n&\n&\n&\n&\n&\n&\n\cr
 \n&\n&\n&\n&\n&\n&\n&{\CoefTrue}&\n&\n&\n&\n&\n&\n&\n&\n\cr
 \n&\n&\n&\n&\n&\n&\n&{\CoefTrue}&\n&\n&\n&\n&\n&\n&\n&\n\cr
 \n&\n&\n&\n&\n&\n&\n&{\CoefTrue}&\n&\n&\n&\n&\n&\n&\n&\n\cr
 \n&\n&\n&\n&\n&\n&\n&\n&{\CoefTrue}&\n&\n&\n&\n&\n&\n&\n\cr
\noalign{\kern-2pt}}}%
\vbox{\setbox8=\hbox{$\RowNames\Matrix$}
\hbox to\wd8{\hfil$\ColNames$\kern\ColEntryShiftHoriz}\kern\ColEntryShiftVerti
\box8}}
\quad
{\footnotesize%
\BoxBreadth=0pt%
\setbox7=\hbox{(\{\},\{\})}%
\ifdim\wd7>\BoxBreadth\BoxBreadth=\wd7\fi%
\setbox7=\hbox{(\{a\},\{\})}%
\ifdim\wd7>\BoxBreadth\BoxBreadth=\wd7\fi%
\setbox7=\hbox{(\{\},\{a\})}%
\ifdim\wd7>\BoxBreadth\BoxBreadth=\wd7\fi%
\setbox7=\hbox{(\{b\},\{\})}%
\ifdim\wd7>\BoxBreadth\BoxBreadth=\wd7\fi%
\setbox7=\hbox{(\{a\},\{a\})}%
\ifdim\wd7>\BoxBreadth\BoxBreadth=\wd7\fi%
\setbox7=\hbox{(\{\},\{b\})}%
\ifdim\wd7>\BoxBreadth\BoxBreadth=\wd7\fi%
\setbox7=\hbox{(\{a,b\},\{\})}%
\ifdim\wd7>\BoxBreadth\BoxBreadth=\wd7\fi%
\setbox7=\hbox{(\{b\},\{a\})}%
\ifdim\wd7>\BoxBreadth\BoxBreadth=\wd7\fi%
\setbox7=\hbox{(\{a\},\{b\})}%
\ifdim\wd7>\BoxBreadth\BoxBreadth=\wd7\fi%
\setbox7=\hbox{(\{\},\{a,b\})}%
\ifdim\wd7>\BoxBreadth\BoxBreadth=\wd7\fi%
\setbox7=\hbox{(\{c\},\{\})}%
\ifdim\wd7>\BoxBreadth\BoxBreadth=\wd7\fi%
\setbox7=\hbox{(\{a,b\},\{a\})}%
\ifdim\wd7>\BoxBreadth\BoxBreadth=\wd7\fi%
\setbox7=\hbox{(\{b\},\{b\})}%
\ifdim\wd7>\BoxBreadth\BoxBreadth=\wd7\fi%
\setbox7=\hbox{(\{a\},\{a,b\})}%
\ifdim\wd7>\BoxBreadth\BoxBreadth=\wd7\fi%
\setbox7=\hbox{(\{\},\{c\})}%
\ifdim\wd7>\BoxBreadth\BoxBreadth=\wd7\fi%
\setbox7=\hbox{(\{a,c\},\{\})}%
\ifdim\wd7>\BoxBreadth\BoxBreadth=\wd7\fi%
\setbox7=\hbox{(\{c\},\{a\})}%
\ifdim\wd7>\BoxBreadth\BoxBreadth=\wd7\fi%
\setbox7=\hbox{(\{a,b\},\{b\})}%
\ifdim\wd7>\BoxBreadth\BoxBreadth=\wd7\fi%
\setbox7=\hbox{(\{b\},\{a,b\})}%
\ifdim\wd7>\BoxBreadth\BoxBreadth=\wd7\fi%
\setbox7=\hbox{(\{a\},\{c\})}%
\ifdim\wd7>\BoxBreadth\BoxBreadth=\wd7\fi%
\setbox7=\hbox{(\{\},\{a,c\})}%
\ifdim\wd7>\BoxBreadth\BoxBreadth=\wd7\fi%
\setbox7=\hbox{(\{b,c\},\{\})}%
\ifdim\wd7>\BoxBreadth\BoxBreadth=\wd7\fi%
\setbox7=\hbox{(\{a,c\},\{a\})}%
\ifdim\wd7>\BoxBreadth\BoxBreadth=\wd7\fi%
\setbox7=\hbox{(\{c\},\{b\})}%
\ifdim\wd7>\BoxBreadth\BoxBreadth=\wd7\fi%
\setbox7=\hbox{(\{a,b\},\{a,b\})}%
\ifdim\wd7>\BoxBreadth\BoxBreadth=\wd7\fi%
\setbox7=\hbox{(\{b\},\{c\})}%
\ifdim\wd7>\BoxBreadth\BoxBreadth=\wd7\fi%
\setbox7=\hbox{(\{a\},\{a,c\})}%
\ifdim\wd7>\BoxBreadth\BoxBreadth=\wd7\fi%
\setbox7=\hbox{(\{\},\{b,c\})}%
\ifdim\wd7>\BoxBreadth\BoxBreadth=\wd7\fi%
\setbox7=\hbox{(\{a,b,c\},\{\})}%
\ifdim\wd7>\BoxBreadth\BoxBreadth=\wd7\fi%
\setbox7=\hbox{(\{b,c\},\{a\})}%
\ifdim\wd7>\BoxBreadth\BoxBreadth=\wd7\fi%
\setbox7=\hbox{(\{a,c\},\{b\})}%
\ifdim\wd7>\BoxBreadth\BoxBreadth=\wd7\fi%
\setbox7=\hbox{(\{c\},\{a,b\})}%
\ifdim\wd7>\BoxBreadth\BoxBreadth=\wd7\fi%
\setbox7=\hbox{(\{a,b\},\{c\})}%
\ifdim\wd7>\BoxBreadth\BoxBreadth=\wd7\fi%
\setbox7=\hbox{(\{b\},\{a,c\})}%
\ifdim\wd7>\BoxBreadth\BoxBreadth=\wd7\fi%
\setbox7=\hbox{(\{a\},\{b,c\})}%
\ifdim\wd7>\BoxBreadth\BoxBreadth=\wd7\fi%
\setbox7=\hbox{(\{\},\{a,b,c\})}%
\ifdim\wd7>\BoxBreadth\BoxBreadth=\wd7\fi%
\setbox7=\hbox{(\{d\},\{\})}%
\def\RowNames{\vcenter{\offinterlineskip\baselineskip=\matrixskip%
\hbox to\BoxBreadth{\strut\hfil (\{\},\{\})}\kern\interspacereduction%
\hbox to\BoxBreadth{\strut\hfil (\{a\},\{\})}\kern\interspacereduction%
\hbox to\BoxBreadth{\strut\hfil (\{\},\{a\})}\kern\interspacereduction%
\hbox to\BoxBreadth{\strut\hfil (\{b\},\{\})}\kern\interspacereduction%
\hbox to\BoxBreadth{\strut\hfil (\{a\},\{a\})}\kern\interspacereduction%
\hbox to\BoxBreadth{\strut\hfil (\{\},\{b\})}\kern\interspacereduction%
\hbox to\BoxBreadth{\strut\hfil (\{a,b\},\{\})}\kern\interspacereduction%
\hbox to\BoxBreadth{\strut\hfil (\{b\},\{a\})}\kern\interspacereduction%
\hbox to\BoxBreadth{\strut\hfil (\{a\},\{b\})}\kern\interspacereduction%
\hbox to\BoxBreadth{\strut\hfil (\{\},\{a,b\})}\kern\interspacereduction%
\hbox to\BoxBreadth{\strut\hfil (\{c\},\{\})}\kern\interspacereduction%
\hbox to\BoxBreadth{\strut\hfil (\{a,b\},\{a\})}\kern\interspacereduction%
\hbox to\BoxBreadth{\strut\hfil (\{b\},\{b\})}\kern\interspacereduction%
\hbox to\BoxBreadth{\strut\hfil (\{a\},\{a,b\})}\kern\interspacereduction%
\hbox to\BoxBreadth{\strut\hfil (\{\},\{c\})}\kern\interspacereduction%
\hbox to\BoxBreadth{\strut\hfil (\{a,c\},\{\})}\kern\interspacereduction%
\hbox to\BoxBreadth{\strut\hfil (\{c\},\{a\})}\kern\interspacereduction%
\hbox to\BoxBreadth{\strut\hfil (\{a,b\},\{b\})}\kern\interspacereduction%
\hbox to\BoxBreadth{\strut\hfil (\{b\},\{a,b\})}\kern\interspacereduction%
\hbox to\BoxBreadth{\strut\hfil (\{a\},\{c\})}\kern\interspacereduction%
\hbox to\BoxBreadth{\strut\hfil (\{\},\{a,c\})}\kern\interspacereduction%
\hbox to\BoxBreadth{\strut\hfil (\{b,c\},\{\})}\kern\interspacereduction%
\hbox to\BoxBreadth{\strut\hfil (\{a,c\},\{a\})}\kern\interspacereduction%
\hbox to\BoxBreadth{\strut\hfil (\{c\},\{b\})}\kern\interspacereduction%
\hbox to\BoxBreadth{\strut\hfil (\{a,b\},\{a,b\})}\kern\interspacereduction%
\hbox to\BoxBreadth{\strut\hfil (\{b\},\{c\})}\kern\interspacereduction%
\hbox to\BoxBreadth{\strut\hfil (\{a\},\{a,c\})}\kern\interspacereduction%
\hbox to\BoxBreadth{\strut\hfil (\{\},\{b,c\})}\kern\interspacereduction%
\hbox to\BoxBreadth{\strut\hfil (\{a,b,c\},\{\})}\kern\interspacereduction%
\hbox to\BoxBreadth{\strut\hfil (\{b,c\},\{a\})}\kern\interspacereduction%
\hbox to\BoxBreadth{\strut\hfil (\{a,c\},\{b\})}\kern\interspacereduction%
\hbox to\BoxBreadth{\strut\hfil (\{c\},\{a,b\})}\kern\interspacereduction%
\hbox to\BoxBreadth{\strut\hfil (\{a,b\},\{c\})}\kern\interspacereduction%
\hbox to\BoxBreadth{\strut\hfil (\{b\},\{a,c\})}\kern\interspacereduction%
\hbox to\BoxBreadth{\strut\hfil (\{a\},\{b,c\})}\kern\interspacereduction%
\hbox to\BoxBreadth{\strut\hfil (\{\},\{a,b,c\})}\kern\interspacereduction%
\hbox to\BoxBreadth{\strut\hfil (\{d\},\{\})}}}%
\def\ColNames{\hbox{\rotatebox{90}{\strut \{\}}\kern\interspacereduction%
\rotatebox{90}{\strut \{a\}}\kern\interspacereduction%
\rotatebox{90}{\strut \{b\}}\kern\interspacereduction%
\rotatebox{90}{\strut \{a,b\}}\kern\interspacereduction%
\rotatebox{90}{\strut \{c\}}\kern\interspacereduction%
\rotatebox{90}{\strut \{a,c\}}\kern\interspacereduction%
\rotatebox{90}{\strut \{b,c\}}\kern\interspacereduction%
\rotatebox{90}{\strut \{a,b,c\}}\kern\interspacereduction%
\rotatebox{90}{\strut \{d\}}\kern\interspacereduction%
\rotatebox{90}{\strut \{a,d\}}\kern\interspacereduction%
\rotatebox{90}{\strut \{b,d\}}\kern\interspacereduction%
\rotatebox{90}{\strut \{a,b,d\}}\kern\interspacereduction%
\rotatebox{90}{\strut \{c,d\}}\kern\interspacereduction%
\rotatebox{90}{\strut \{a,c,d\}}\kern\interspacereduction%
\rotatebox{90}{\strut \{b,c,d\}}\kern\interspacereduction%
\rotatebox{90}{\strut \{a,b,c,d\}}\kern\interspacereduction%
}}%
\def\Matrix{\spmatrix{%
\noalign{\kern-2pt}
 {\CoefTrue}&\n&\n&\n&\n&\n&\n&\n&\n&\n&\n&\n&\n&\n&\n&\n\cr
 {\CoefTrue}&\n&\n&\n&\n&\n&\n&\n&\n&\n&\n&\n&\n&\n&\n&\n\cr
 {\CoefTrue}&\n&\n&\n&\n&\n&\n&\n&\n&\n&\n&\n&\n&\n&\n&\n\cr
 {\CoefTrue}&\n&\n&\n&\n&\n&\n&\n&\n&\n&\n&\n&\n&\n&\n&\n\cr
 \n&{\CoefTrue}&\n&\n&\n&\n&\n&\n&\n&\n&\n&\n&\n&\n&\n&\n\cr
 {\CoefTrue}&\n&\n&\n&\n&\n&\n&\n&\n&\n&\n&\n&\n&\n&\n&\n\cr
 {\CoefTrue}&\n&\n&\n&\n&\n&\n&\n&\n&\n&\n&\n&\n&\n&\n&\n\cr
 {\CoefTrue}&\n&\n&\n&\n&\n&\n&\n&\n&\n&\n&\n&\n&\n&\n&\n\cr
 {\CoefTrue}&\n&\n&\n&\n&\n&\n&\n&\n&\n&\n&\n&\n&\n&\n&\n\cr
 {\CoefTrue}&\n&\n&\n&\n&\n&\n&\n&\n&\n&\n&\n&\n&\n&\n&\n\cr
 {\CoefTrue}&\n&\n&\n&\n&\n&\n&\n&\n&\n&\n&\n&\n&\n&\n&\n\cr
 \n&{\CoefTrue}&\n&\n&\n&\n&\n&\n&\n&\n&\n&\n&\n&\n&\n&\n\cr
 \n&\n&{\CoefTrue}&\n&\n&\n&\n&\n&\n&\n&\n&\n&\n&\n&\n&\n\cr
 \n&{\CoefTrue}&\n&\n&\n&\n&\n&\n&\n&\n&\n&\n&\n&\n&\n&\n\cr
 {\CoefTrue}&\n&\n&\n&\n&\n&\n&\n&\n&\n&\n&\n&\n&\n&\n&\n\cr
 {\CoefTrue}&\n&\n&\n&\n&\n&\n&\n&\n&\n&\n&\n&\n&\n&\n&\n\cr
 {\CoefTrue}&\n&\n&\n&\n&\n&\n&\n&\n&\n&\n&\n&\n&\n&\n&\n\cr
 \n&\n&{\CoefTrue}&\n&\n&\n&\n&\n&\n&\n&\n&\n&\n&\n&\n&\n\cr
 \n&\n&{\CoefTrue}&\n&\n&\n&\n&\n&\n&\n&\n&\n&\n&\n&\n&\n\cr
 {\CoefTrue}&\n&\n&\n&\n&\n&\n&\n&\n&\n&\n&\n&\n&\n&\n&\n\cr
 {\CoefTrue}&\n&\n&\n&\n&\n&\n&\n&\n&\n&\n&\n&\n&\n&\n&\n\cr
 {\CoefTrue}&\n&\n&\n&\n&\n&\n&\n&\n&\n&\n&\n&\n&\n&\n&\n\cr
 \n&{\CoefTrue}&\n&\n&\n&\n&\n&\n&\n&\n&\n&\n&\n&\n&\n&\n\cr
 {\CoefTrue}&\n&\n&\n&\n&\n&\n&\n&\n&\n&\n&\n&\n&\n&\n&\n\cr
 \n&\n&\n&{\CoefTrue}&\n&\n&\n&\n&\n&\n&\n&\n&\n&\n&\n&\n\cr
 {\CoefTrue}&\n&\n&\n&\n&\n&\n&\n&\n&\n&\n&\n&\n&\n&\n&\n\cr
 \n&{\CoefTrue}&\n&\n&\n&\n&\n&\n&\n&\n&\n&\n&\n&\n&\n&\n\cr
 {\CoefTrue}&\n&\n&\n&\n&\n&\n&\n&\n&\n&\n&\n&\n&\n&\n&\n\cr
 {\CoefTrue}&\n&\n&\n&\n&\n&\n&\n&\n&\n&\n&\n&\n&\n&\n&\n\cr
 {\CoefTrue}&\n&\n&\n&\n&\n&\n&\n&\n&\n&\n&\n&\n&\n&\n&\n\cr
 {\CoefTrue}&\n&\n&\n&\n&\n&\n&\n&\n&\n&\n&\n&\n&\n&\n&\n\cr
 {\CoefTrue}&\n&\n&\n&\n&\n&\n&\n&\n&\n&\n&\n&\n&\n&\n&\n\cr
 {\CoefTrue}&\n&\n&\n&\n&\n&\n&\n&\n&\n&\n&\n&\n&\n&\n&\n\cr
 {\CoefTrue}&\n&\n&\n&\n&\n&\n&\n&\n&\n&\n&\n&\n&\n&\n&\n\cr
 {\CoefTrue}&\n&\n&\n&\n&\n&\n&\n&\n&\n&\n&\n&\n&\n&\n&\n\cr
 {\CoefTrue}&\n&\n&\n&\n&\n&\n&\n&\n&\n&\n&\n&\n&\n&\n&\n\cr
 {\CoefTrue}&\n&\n&\n&\n&\n&\n&\n&\n&\n&\n&\n&\n&\n&\n&\n\cr
\noalign{\kern-2pt}}}%
\vbox{\setbox8=\hbox{$
\Matrix$}
\hbox to\wd8{\hfil$\ColNames$\kern\ColEntryShiftHoriz}\kern\ColEntryShiftVerti
\box8}}
}
}
{The initial ones of 256 rows of the relations $\RELfromTO{\liftedJoin,\liftedMeet}{\PowTWO{X}\times\PowTWO{X}}{\PowTWO{X}}$}{FigJoinMeet}

\bigskip
\noindent
One will identify the commutative law in (ii,iii), where it is expressed that the collection of results doesn't change when starting from the first as opposed to the second component. The other laws may be found later in Prop.~\PropAbsorbAssoc.

\Caption{{\footnotesize%
\BoxBreadth=0pt%
\setbox7=\hbox{\{\}}%
\ifdim\wd7>\BoxBreadth\BoxBreadth=\wd7\fi%
\setbox7=\hbox{\{a\}}%
\ifdim\wd7>\BoxBreadth\BoxBreadth=\wd7\fi%
\setbox7=\hbox{\{b\}}%
\ifdim\wd7>\BoxBreadth\BoxBreadth=\wd7\fi%
\setbox7=\hbox{\{a,b\}}%
\ifdim\wd7>\BoxBreadth\BoxBreadth=\wd7\fi%
\setbox7=\hbox{\{c\}}%
\ifdim\wd7>\BoxBreadth\BoxBreadth=\wd7\fi%
\setbox7=\hbox{\{a,c\}}%
\ifdim\wd7>\BoxBreadth\BoxBreadth=\wd7\fi%
\setbox7=\hbox{\{b,c\}}%
\ifdim\wd7>\BoxBreadth\BoxBreadth=\wd7\fi%
\setbox7=\hbox{abc}%
\ifdim\wd7>\BoxBreadth\BoxBreadth=\wd7\fi%
\setbox7=\hbox{\{d\}}%
\ifdim\wd7>\BoxBreadth\BoxBreadth=\wd7\fi%
\setbox7=\hbox{\{a,d\}}%
\ifdim\wd7>\BoxBreadth\BoxBreadth=\wd7\fi%
\setbox7=\hbox{\{b,d\}}%
\ifdim\wd7>\BoxBreadth\BoxBreadth=\wd7\fi%
\setbox7=\hbox{abd}%
\ifdim\wd7>\BoxBreadth\BoxBreadth=\wd7\fi%
\setbox7=\hbox{\{c,d\}}%
\ifdim\wd7>\BoxBreadth\BoxBreadth=\wd7\fi%
\setbox7=\hbox{acd}%
\ifdim\wd7>\BoxBreadth\BoxBreadth=\wd7\fi%
\setbox7=\hbox{bcd}%
\ifdim\wd7>\BoxBreadth\BoxBreadth=\wd7\fi%
\setbox7=\hbox{all}%
\ifdim\wd7>\BoxBreadth\BoxBreadth=\wd7\fi%
%
\def\RowNames{\vcenter{\offinterlineskip\baselineskip=\matrixskip%
\hbox to1.7000987\BoxBreadth{\strut\hfil \{\}}\kern0.53234cm%
\hbox to1.7000987\BoxBreadth{\strut\hfil \{a\}}\kern0.53234cm%
\hbox to1.7000987\BoxBreadth{\strut\hfil \{b\}}\kern0.53234cm%
\hbox to1.7000987\BoxBreadth{\strut\hfil \{a,b\}}\kern0.53234cm%
\hbox to1.7000987\BoxBreadth{\strut\hfil \{c\}}\kern0.53234cm%
\hbox to1.7000987\BoxBreadth{\strut\hfil \{a,c\}}\kern0.53234cm%
\hbox to1.7000987\BoxBreadth{\strut\hfil \{b,c\}}\kern0.53234cm%
\hbox to1.7000987\BoxBreadth{\strut\hfil abc}\kern0.53234cm%
\hbox to1.7000987\BoxBreadth{\strut\hfil \{d\}}\kern0.53234cm%
\hbox to1.7000987\BoxBreadth{\strut\hfil \{a,d\}}\kern0.53234cm%
\hbox to1.7000987\BoxBreadth{\strut\hfil \{b,d\}}\kern0.53234cm%
\hbox to1.7000987\BoxBreadth{\strut\hfil abd}\kern0.53234cm%
\hbox to1.7000987\BoxBreadth{\strut\hfil \{c,d\}}\kern0.53234cm%
\hbox to1.7000987\BoxBreadth{\strut\hfil acd}\kern0.53234cm%
\hbox to1.7000987\BoxBreadth{\strut\hfil bcd}\kern0.53234cm%
\hbox to1.7000987\BoxBreadth{\strut\hfil \{a,b,c,d\}}}}%
\def\ColNames{\hbox{\rotatebox{90}{\strut \{\}}\kern0.409001cm%
\rotatebox{90}{\strut \{a\}}\kern0.409001cm%
\rotatebox{90}{\strut \{b\}}\kern0.409001cm%
\rotatebox{90}{\strut \{a,b\}}\kern0.409001cm%
\rotatebox{90}{\strut \{c\}}\kern0.409001cm%
\rotatebox{90}{\strut \{a,c\}}\kern0.409001cm%
\rotatebox{90}{\strut \{b,c\}}\kern0.409001cm%
\rotatebox{90}{\strut abc}\kern0.409001cm%
\rotatebox{90}{\strut \{d\}}\kern0.409001cm%
\rotatebox{90}{\strut \{a,d\}}\kern0.409001cm%
\rotatebox{90}{\strut \{b,d\}}\kern0.409001cm%
\rotatebox{90}{\strut abd}\kern0.409001cm%
\rotatebox{90}{\strut \{c,d\}}\kern0.409001cm%
\rotatebox{90}{\strut acd}\kern0.409001cm%
\rotatebox{90}{\strut bcd}\kern0.409001cm%
\rotatebox{90}{\strut all}\kern0.409001cm%
}}%
\def\Matrix{\spmatrix{%
\noalign{\kern-2pt}
 \hbox{\vbox to\BoxBreadth{\vfil\hbox to\BoxBreadth{\hfil \{\}\hfil}\vfil}}&\hbox{\vbox to\BoxBreadth{\vfil\hbox to\BoxBreadth{\hfil \{a\}\hfil}\vfil}}&\hbox{\vbox to\BoxBreadth{\vfil\hbox to\BoxBreadth{\hfil \{b\}\hfil}\vfil}}&\hbox{\vbox to\BoxBreadth{\vfil\hbox to\BoxBreadth{\hfil \{a,b\}\hfil}\vfil}}&\hbox{\vbox to\BoxBreadth{\vfil\hbox to\BoxBreadth{\hfil \{c\}\hfil}\vfil}}&\hbox{\vbox to\BoxBreadth{\vfil\hbox to\BoxBreadth{\hfil \{a,c\}\hfil}\vfil}}&\hbox{\vbox to\BoxBreadth{\vfil\hbox to\BoxBreadth{\hfil \{b,c\}\hfil}\vfil}}&\hbox{\vbox to\BoxBreadth{\vfil\hbox to\BoxBreadth{\hfil abc\hfil}\vfil}}&\hbox{\vbox to\BoxBreadth{\vfil\hbox to\BoxBreadth{\hfil \{d\}\hfil}\vfil}}&\hbox{\vbox to\BoxBreadth{\vfil\hbox to\BoxBreadth{\hfil \{a,d\}\hfil}\vfil}}&\hbox{\vbox to\BoxBreadth{\vfil\hbox to\BoxBreadth{\hfil \{b,d\}\hfil}\vfil}}&\hbox{\vbox to\BoxBreadth{\vfil\hbox to\BoxBreadth{\hfil abd\hfil}\vfil}}&\hbox{\vbox to\BoxBreadth{\vfil\hbox to\BoxBreadth{\hfil \{c,d\}\hfil}\vfil}}&\hbox{\vbox to\BoxBreadth{\vfil\hbox to\BoxBreadth{\hfil acd\hfil}\vfil}}&\hbox{\vbox to\BoxBreadth{\vfil\hbox to\BoxBreadth{\hfil bcd\hfil}\vfil}}&\hbox{\vbox to\BoxBreadth{\vfil\hbox to\BoxBreadth{\hfil all\hfil}\vfil}}\cr
 \hbox{\vbox to\BoxBreadth{\vfil\hbox to\BoxBreadth{\hfil \{a\}\hfil}\vfil}}&\hbox{\vbox to\BoxBreadth{\vfil\hbox to\BoxBreadth{\hfil \{a\}\hfil}\vfil}}&\hbox{\vbox to\BoxBreadth{\vfil\hbox to\BoxBreadth{\hfil \{a,b\}\hfil}\vfil}}&\hbox{\vbox to\BoxBreadth{\vfil\hbox to\BoxBreadth{\hfil \{a,b\}\hfil}\vfil}}&\hbox{\vbox to\BoxBreadth{\vfil\hbox to\BoxBreadth{\hfil \{a,c\}\hfil}\vfil}}&\hbox{\vbox to\BoxBreadth{\vfil\hbox to\BoxBreadth{\hfil \{a,c\}\hfil}\vfil}}&\hbox{\vbox to\BoxBreadth{\vfil\hbox to\BoxBreadth{\hfil abc\hfil}\vfil}}&\hbox{\vbox to\BoxBreadth{\vfil\hbox to\BoxBreadth{\hfil abc\hfil}\vfil}}&\hbox{\vbox to\BoxBreadth{\vfil\hbox to\BoxBreadth{\hfil \{a,d\}\hfil}\vfil}}&\hbox{\vbox to\BoxBreadth{\vfil\hbox to\BoxBreadth{\hfil \{a,d\}\hfil}\vfil}}&\hbox{\vbox to\BoxBreadth{\vfil\hbox to\BoxBreadth{\hfil abd\hfil}\vfil}}&\hbox{\vbox to\BoxBreadth{\vfil\hbox to\BoxBreadth{\hfil abd\hfil}\vfil}}&\hbox{\vbox to\BoxBreadth{\vfil\hbox to\BoxBreadth{\hfil acd\hfil}\vfil}}&\hbox{\vbox to\BoxBreadth{\vfil\hbox to\BoxBreadth{\hfil acd\hfil}\vfil}}&\hbox{\vbox to\BoxBreadth{\vfil\hbox to\BoxBreadth{\hfil all\hfil}\vfil}}&\hbox{\vbox to\BoxBreadth{\vfil\hbox to\BoxBreadth{\hfil all\hfil}\vfil}}\cr
 \hbox{\vbox to\BoxBreadth{\vfil\hbox to\BoxBreadth{\hfil \{b\}\hfil}\vfil}}&\hbox{\vbox to\BoxBreadth{\vfil\hbox to\BoxBreadth{\hfil \{a,b\}\hfil}\vfil}}&\hbox{\vbox to\BoxBreadth{\vfil\hbox to\BoxBreadth{\hfil \{b\}\hfil}\vfil}}&\hbox{\vbox to\BoxBreadth{\vfil\hbox to\BoxBreadth{\hfil \{a,b\}\hfil}\vfil}}&\hbox{\vbox to\BoxBreadth{\vfil\hbox to\BoxBreadth{\hfil \{b,c\}\hfil}\vfil}}&\hbox{\vbox to\BoxBreadth{\vfil\hbox to\BoxBreadth{\hfil abc\hfil}\vfil}}&\hbox{\vbox to\BoxBreadth{\vfil\hbox to\BoxBreadth{\hfil \{b,c\}\hfil}\vfil}}&\hbox{\vbox to\BoxBreadth{\vfil\hbox to\BoxBreadth{\hfil abc\hfil}\vfil}}&\hbox{\vbox to\BoxBreadth{\vfil\hbox to\BoxBreadth{\hfil \{b,d\}\hfil}\vfil}}&\hbox{\vbox to\BoxBreadth{\vfil\hbox to\BoxBreadth{\hfil abd\hfil}\vfil}}&\hbox{\vbox to\BoxBreadth{\vfil\hbox to\BoxBreadth{\hfil \{b,d\}\hfil}\vfil}}&\hbox{\vbox to\BoxBreadth{\vfil\hbox to\BoxBreadth{\hfil abd\hfil}\vfil}}&\hbox{\vbox to\BoxBreadth{\vfil\hbox to\BoxBreadth{\hfil bcd\hfil}\vfil}}&\hbox{\vbox to\BoxBreadth{\vfil\hbox to\BoxBreadth{\hfil all\hfil}\vfil}}&\hbox{\vbox to\BoxBreadth{\vfil\hbox to\BoxBreadth{\hfil bcd\hfil}\vfil}}&\hbox{\vbox to\BoxBreadth{\vfil\hbox to\BoxBreadth{\hfil all\hfil}\vfil}}\cr
 \hbox{\vbox to\BoxBreadth{\vfil\hbox to\BoxBreadth{\hfil \{a,b\}\hfil}\vfil}}&\hbox{\vbox to\BoxBreadth{\vfil\hbox to\BoxBreadth{\hfil \{a,b\}\hfil}\vfil}}&\hbox{\vbox to\BoxBreadth{\vfil\hbox to\BoxBreadth{\hfil \{a,b\}\hfil}\vfil}}&\hbox{\vbox to\BoxBreadth{\vfil\hbox to\BoxBreadth{\hfil \{a,b\}\hfil}\vfil}}&\hbox{\vbox to\BoxBreadth{\vfil\hbox to\BoxBreadth{\hfil abc\hfil}\vfil}}&\hbox{\vbox to\BoxBreadth{\vfil\hbox to\BoxBreadth{\hfil abc\hfil}\vfil}}&\hbox{\vbox to\BoxBreadth{\vfil\hbox to\BoxBreadth{\hfil abc\hfil}\vfil}}&\hbox{\vbox to\BoxBreadth{\vfil\hbox to\BoxBreadth{\hfil abc\hfil}\vfil}}&\hbox{\vbox to\BoxBreadth{\vfil\hbox to\BoxBreadth{\hfil abd\hfil}\vfil}}&\hbox{\vbox to\BoxBreadth{\vfil\hbox to\BoxBreadth{\hfil abd\hfil}\vfil}}&\hbox{\vbox to\BoxBreadth{\vfil\hbox to\BoxBreadth{\hfil abd\hfil}\vfil}}&\hbox{\vbox to\BoxBreadth{\vfil\hbox to\BoxBreadth{\hfil abd\hfil}\vfil}}&\hbox{\vbox to\BoxBreadth{\vfil\hbox to\BoxBreadth{\hfil all\hfil}\vfil}}&\hbox{\vbox to\BoxBreadth{\vfil\hbox to\BoxBreadth{\hfil all\hfil}\vfil}}&\hbox{\vbox to\BoxBreadth{\vfil\hbox to\BoxBreadth{\hfil all\hfil}\vfil}}&\hbox{\vbox to\BoxBreadth{\vfil\hbox to\BoxBreadth{\hfil all\hfil}\vfil}}\cr
 \hbox{\vbox to\BoxBreadth{\vfil\hbox to\BoxBreadth{\hfil \{c\}\hfil}\vfil}}&\hbox{\vbox to\BoxBreadth{\vfil\hbox to\BoxBreadth{\hfil \{a,c\}\hfil}\vfil}}&\hbox{\vbox to\BoxBreadth{\vfil\hbox to\BoxBreadth{\hfil \{b,c\}\hfil}\vfil}}&\hbox{\vbox to\BoxBreadth{\vfil\hbox to\BoxBreadth{\hfil abc\hfil}\vfil}}&\hbox{\vbox to\BoxBreadth{\vfil\hbox to\BoxBreadth{\hfil \{c\}\hfil}\vfil}}&\hbox{\vbox to\BoxBreadth{\vfil\hbox to\BoxBreadth{\hfil \{a,c\}\hfil}\vfil}}&\hbox{\vbox to\BoxBreadth{\vfil\hbox to\BoxBreadth{\hfil \{b,c\}\hfil}\vfil}}&\hbox{\vbox to\BoxBreadth{\vfil\hbox to\BoxBreadth{\hfil abc\hfil}\vfil}}&\hbox{\vbox to\BoxBreadth{\vfil\hbox to\BoxBreadth{\hfil \{c,d\}\hfil}\vfil}}&\hbox{\vbox to\BoxBreadth{\vfil\hbox to\BoxBreadth{\hfil acd\hfil}\vfil}}&\hbox{\vbox to\BoxBreadth{\vfil\hbox to\BoxBreadth{\hfil bcd\hfil}\vfil}}&\hbox{\vbox to\BoxBreadth{\vfil\hbox to\BoxBreadth{\hfil all\hfil}\vfil}}&\hbox{\vbox to\BoxBreadth{\vfil\hbox to\BoxBreadth{\hfil \{c,d\}\hfil}\vfil}}&\hbox{\vbox to\BoxBreadth{\vfil\hbox to\BoxBreadth{\hfil acd\hfil}\vfil}}&\hbox{\vbox to\BoxBreadth{\vfil\hbox to\BoxBreadth{\hfil bcd\hfil}\vfil}}&\hbox{\vbox to\BoxBreadth{\vfil\hbox to\BoxBreadth{\hfil all\hfil}\vfil}}\cr
 \hbox{\vbox to\BoxBreadth{\vfil\hbox to\BoxBreadth{\hfil \{a,c\}\hfil}\vfil}}&\hbox{\vbox to\BoxBreadth{\vfil\hbox to\BoxBreadth{\hfil \{a,c\}\hfil}\vfil}}&\hbox{\vbox to\BoxBreadth{\vfil\hbox to\BoxBreadth{\hfil abc\hfil}\vfil}}&\hbox{\vbox to\BoxBreadth{\vfil\hbox to\BoxBreadth{\hfil abc\hfil}\vfil}}&\hbox{\vbox to\BoxBreadth{\vfil\hbox to\BoxBreadth{\hfil \{a,c\}\hfil}\vfil}}&\hbox{\vbox to\BoxBreadth{\vfil\hbox to\BoxBreadth{\hfil \{a,c\}\hfil}\vfil}}&\hbox{\vbox to\BoxBreadth{\vfil\hbox to\BoxBreadth{\hfil abc\hfil}\vfil}}&\hbox{\vbox to\BoxBreadth{\vfil\hbox to\BoxBreadth{\hfil abc\hfil}\vfil}}&\hbox{\vbox to\BoxBreadth{\vfil\hbox to\BoxBreadth{\hfil acd\hfil}\vfil}}&\hbox{\vbox to\BoxBreadth{\vfil\hbox to\BoxBreadth{\hfil acd\hfil}\vfil}}&\hbox{\vbox to\BoxBreadth{\vfil\hbox to\BoxBreadth{\hfil all\hfil}\vfil}}&\hbox{\vbox to\BoxBreadth{\vfil\hbox to\BoxBreadth{\hfil all\hfil}\vfil}}&\hbox{\vbox to\BoxBreadth{\vfil\hbox to\BoxBreadth{\hfil acd\hfil}\vfil}}&\hbox{\vbox to\BoxBreadth{\vfil\hbox to\BoxBreadth{\hfil acd\hfil}\vfil}}&\hbox{\vbox to\BoxBreadth{\vfil\hbox to\BoxBreadth{\hfil all\hfil}\vfil}}&\hbox{\vbox to\BoxBreadth{\vfil\hbox to\BoxBreadth{\hfil all\hfil}\vfil}}\cr
 \hbox{\vbox to\BoxBreadth{\vfil\hbox to\BoxBreadth{\hfil \{b,c\}\hfil}\vfil}}&\hbox{\vbox to\BoxBreadth{\vfil\hbox to\BoxBreadth{\hfil abc\hfil}\vfil}}&\hbox{\vbox to\BoxBreadth{\vfil\hbox to\BoxBreadth{\hfil \{b,c\}\hfil}\vfil}}&\hbox{\vbox to\BoxBreadth{\vfil\hbox to\BoxBreadth{\hfil abc\hfil}\vfil}}&\hbox{\vbox to\BoxBreadth{\vfil\hbox to\BoxBreadth{\hfil \{b,c\}\hfil}\vfil}}&\hbox{\vbox to\BoxBreadth{\vfil\hbox to\BoxBreadth{\hfil abc\hfil}\vfil}}&\hbox{\vbox to\BoxBreadth{\vfil\hbox to\BoxBreadth{\hfil \{b,c\}\hfil}\vfil}}&\hbox{\vbox to\BoxBreadth{\vfil\hbox to\BoxBreadth{\hfil abc\hfil}\vfil}}&\hbox{\vbox to\BoxBreadth{\vfil\hbox to\BoxBreadth{\hfil bcd\hfil}\vfil}}&\hbox{\vbox to\BoxBreadth{\vfil\hbox to\BoxBreadth{\hfil all\hfil}\vfil}}&\hbox{\vbox to\BoxBreadth{\vfil\hbox to\BoxBreadth{\hfil bcd\hfil}\vfil}}&\hbox{\vbox to\BoxBreadth{\vfil\hbox to\BoxBreadth{\hfil all\hfil}\vfil}}&\hbox{\vbox to\BoxBreadth{\vfil\hbox to\BoxBreadth{\hfil bcd\hfil}\vfil}}&\hbox{\vbox to\BoxBreadth{\vfil\hbox to\BoxBreadth{\hfil all\hfil}\vfil}}&\hbox{\vbox to\BoxBreadth{\vfil\hbox to\BoxBreadth{\hfil bcd\hfil}\vfil}}&\hbox{\vbox to\BoxBreadth{\vfil\hbox to\BoxBreadth{\hfil all\hfil}\vfil}}\cr
 \hbox{\vbox to\BoxBreadth{\vfil\hbox to\BoxBreadth{\hfil abc\hfil}\vfil}}&\hbox{\vbox to\BoxBreadth{\vfil\hbox to\BoxBreadth{\hfil abc\hfil}\vfil}}&\hbox{\vbox to\BoxBreadth{\vfil\hbox to\BoxBreadth{\hfil abc\hfil}\vfil}}&\hbox{\vbox to\BoxBreadth{\vfil\hbox to\BoxBreadth{\hfil abc\hfil}\vfil}}&\hbox{\vbox to\BoxBreadth{\vfil\hbox to\BoxBreadth{\hfil abc\hfil}\vfil}}&\hbox{\vbox to\BoxBreadth{\vfil\hbox to\BoxBreadth{\hfil abc\hfil}\vfil}}&\hbox{\vbox to\BoxBreadth{\vfil\hbox to\BoxBreadth{\hfil abc\hfil}\vfil}}&\hbox{\vbox to\BoxBreadth{\vfil\hbox to\BoxBreadth{\hfil abc\hfil}\vfil}}&\hbox{\vbox to\BoxBreadth{\vfil\hbox to\BoxBreadth{\hfil all\hfil}\vfil}}&\hbox{\vbox to\BoxBreadth{\vfil\hbox to\BoxBreadth{\hfil all\hfil}\vfil}}&\hbox{\vbox to\BoxBreadth{\vfil\hbox to\BoxBreadth{\hfil all\hfil}\vfil}}&\hbox{\vbox to\BoxBreadth{\vfil\hbox to\BoxBreadth{\hfil all\hfil}\vfil}}&\hbox{\vbox to\BoxBreadth{\vfil\hbox to\BoxBreadth{\hfil all\hfil}\vfil}}&\hbox{\vbox to\BoxBreadth{\vfil\hbox to\BoxBreadth{\hfil all\hfil}\vfil}}&\hbox{\vbox to\BoxBreadth{\vfil\hbox to\BoxBreadth{\hfil all\hfil}\vfil}}&\hbox{\vbox to\BoxBreadth{\vfil\hbox to\BoxBreadth{\hfil all\hfil}\vfil}}\cr
 \hbox{\vbox to\BoxBreadth{\vfil\hbox to\BoxBreadth{\hfil \{d\}\hfil}\vfil}}&\hbox{\vbox to\BoxBreadth{\vfil\hbox to\BoxBreadth{\hfil \{a,d\}\hfil}\vfil}}&\hbox{\vbox to\BoxBreadth{\vfil\hbox to\BoxBreadth{\hfil \{b,d\}\hfil}\vfil}}&\hbox{\vbox to\BoxBreadth{\vfil\hbox to\BoxBreadth{\hfil abd\hfil}\vfil}}&\hbox{\vbox to\BoxBreadth{\vfil\hbox to\BoxBreadth{\hfil \{c,d\}\hfil}\vfil}}&\hbox{\vbox to\BoxBreadth{\vfil\hbox to\BoxBreadth{\hfil acd\hfil}\vfil}}&\hbox{\vbox to\BoxBreadth{\vfil\hbox to\BoxBreadth{\hfil bcd\hfil}\vfil}}&\hbox{\vbox to\BoxBreadth{\vfil\hbox to\BoxBreadth{\hfil all\hfil}\vfil}}&\hbox{\vbox to\BoxBreadth{\vfil\hbox to\BoxBreadth{\hfil \{d\}\hfil}\vfil}}&\hbox{\vbox to\BoxBreadth{\vfil\hbox to\BoxBreadth{\hfil \{a,d\}\hfil}\vfil}}&\hbox{\vbox to\BoxBreadth{\vfil\hbox to\BoxBreadth{\hfil \{b,d\}\hfil}\vfil}}&\hbox{\vbox to\BoxBreadth{\vfil\hbox to\BoxBreadth{\hfil abd\hfil}\vfil}}&\hbox{\vbox to\BoxBreadth{\vfil\hbox to\BoxBreadth{\hfil \{c,d\}\hfil}\vfil}}&\hbox{\vbox to\BoxBreadth{\vfil\hbox to\BoxBreadth{\hfil acd\hfil}\vfil}}&\hbox{\vbox to\BoxBreadth{\vfil\hbox to\BoxBreadth{\hfil bcd\hfil}\vfil}}&\hbox{\vbox to\BoxBreadth{\vfil\hbox to\BoxBreadth{\hfil all\hfil}\vfil}}\cr
 \hbox{\vbox to\BoxBreadth{\vfil\hbox to\BoxBreadth{\hfil \{a,d\}\hfil}\vfil}}&\hbox{\vbox to\BoxBreadth{\vfil\hbox to\BoxBreadth{\hfil \{a,d\}\hfil}\vfil}}&\hbox{\vbox to\BoxBreadth{\vfil\hbox to\BoxBreadth{\hfil abd\hfil}\vfil}}&\hbox{\vbox to\BoxBreadth{\vfil\hbox to\BoxBreadth{\hfil abd\hfil}\vfil}}&\hbox{\vbox to\BoxBreadth{\vfil\hbox to\BoxBreadth{\hfil acd\hfil}\vfil}}&\hbox{\vbox to\BoxBreadth{\vfil\hbox to\BoxBreadth{\hfil acd\hfil}\vfil}}&\hbox{\vbox to\BoxBreadth{\vfil\hbox to\BoxBreadth{\hfil all\hfil}\vfil}}&\hbox{\vbox to\BoxBreadth{\vfil\hbox to\BoxBreadth{\hfil all\hfil}\vfil}}&\hbox{\vbox to\BoxBreadth{\vfil\hbox to\BoxBreadth{\hfil \{a,d\}\hfil}\vfil}}&\hbox{\vbox to\BoxBreadth{\vfil\hbox to\BoxBreadth{\hfil \{a,d\}\hfil}\vfil}}&\hbox{\vbox to\BoxBreadth{\vfil\hbox to\BoxBreadth{\hfil abd\hfil}\vfil}}&\hbox{\vbox to\BoxBreadth{\vfil\hbox to\BoxBreadth{\hfil abd\hfil}\vfil}}&\hbox{\vbox to\BoxBreadth{\vfil\hbox to\BoxBreadth{\hfil acd\hfil}\vfil}}&\hbox{\vbox to\BoxBreadth{\vfil\hbox to\BoxBreadth{\hfil acd\hfil}\vfil}}&\hbox{\vbox to\BoxBreadth{\vfil\hbox to\BoxBreadth{\hfil all\hfil}\vfil}}&\hbox{\vbox to\BoxBreadth{\vfil\hbox to\BoxBreadth{\hfil all\hfil}\vfil}}\cr
 \hbox{\vbox to\BoxBreadth{\vfil\hbox to\BoxBreadth{\hfil \{b,d\}\hfil}\vfil}}&\hbox{\vbox to\BoxBreadth{\vfil\hbox to\BoxBreadth{\hfil abd\hfil}\vfil}}&\hbox{\vbox to\BoxBreadth{\vfil\hbox to\BoxBreadth{\hfil \{b,d\}\hfil}\vfil}}&\hbox{\vbox to\BoxBreadth{\vfil\hbox to\BoxBreadth{\hfil abd\hfil}\vfil}}&\hbox{\vbox to\BoxBreadth{\vfil\hbox to\BoxBreadth{\hfil bcd\hfil}\vfil}}&\hbox{\vbox to\BoxBreadth{\vfil\hbox to\BoxBreadth{\hfil all\hfil}\vfil}}&\hbox{\vbox to\BoxBreadth{\vfil\hbox to\BoxBreadth{\hfil bcd\hfil}\vfil}}&\hbox{\vbox to\BoxBreadth{\vfil\hbox to\BoxBreadth{\hfil all\hfil}\vfil}}&\hbox{\vbox to\BoxBreadth{\vfil\hbox to\BoxBreadth{\hfil \{b,d\}\hfil}\vfil}}&\hbox{\vbox to\BoxBreadth{\vfil\hbox to\BoxBreadth{\hfil abd\hfil}\vfil}}&\hbox{\vbox to\BoxBreadth{\vfil\hbox to\BoxBreadth{\hfil \{b,d\}\hfil}\vfil}}&\hbox{\vbox to\BoxBreadth{\vfil\hbox to\BoxBreadth{\hfil abd\hfil}\vfil}}&\hbox{\vbox to\BoxBreadth{\vfil\hbox to\BoxBreadth{\hfil bcd\hfil}\vfil}}&\hbox{\vbox to\BoxBreadth{\vfil\hbox to\BoxBreadth{\hfil all\hfil}\vfil}}&\hbox{\vbox to\BoxBreadth{\vfil\hbox to\BoxBreadth{\hfil bcd\hfil}\vfil}}&\hbox{\vbox to\BoxBreadth{\vfil\hbox to\BoxBreadth{\hfil all\hfil}\vfil}}\cr
 \hbox{\vbox to\BoxBreadth{\vfil\hbox to\BoxBreadth{\hfil abd\hfil}\vfil}}&\hbox{\vbox to\BoxBreadth{\vfil\hbox to\BoxBreadth{\hfil abd\hfil}\vfil}}&\hbox{\vbox to\BoxBreadth{\vfil\hbox to\BoxBreadth{\hfil abd\hfil}\vfil}}&\hbox{\vbox to\BoxBreadth{\vfil\hbox to\BoxBreadth{\hfil abd\hfil}\vfil}}&\hbox{\vbox to\BoxBreadth{\vfil\hbox to\BoxBreadth{\hfil all\hfil}\vfil}}&\hbox{\vbox to\BoxBreadth{\vfil\hbox to\BoxBreadth{\hfil all\hfil}\vfil}}&\hbox{\vbox to\BoxBreadth{\vfil\hbox to\BoxBreadth{\hfil all\hfil}\vfil}}&\hbox{\vbox to\BoxBreadth{\vfil\hbox to\BoxBreadth{\hfil all\hfil}\vfil}}&\hbox{\vbox to\BoxBreadth{\vfil\hbox to\BoxBreadth{\hfil abd\hfil}\vfil}}&\hbox{\vbox to\BoxBreadth{\vfil\hbox to\BoxBreadth{\hfil abd\hfil}\vfil}}&\hbox{\vbox to\BoxBreadth{\vfil\hbox to\BoxBreadth{\hfil abd\hfil}\vfil}}&\hbox{\vbox to\BoxBreadth{\vfil\hbox to\BoxBreadth{\hfil acd\hfil}\vfil}}&\hbox{\vbox to\BoxBreadth{\vfil\hbox to\BoxBreadth{\hfil all\hfil}\vfil}}&\hbox{\vbox to\BoxBreadth{\vfil\hbox to\BoxBreadth{\hfil all\hfil}\vfil}}&\hbox{\vbox to\BoxBreadth{\vfil\hbox to\BoxBreadth{\hfil all\hfil}\vfil}}&\hbox{\vbox to\BoxBreadth{\vfil\hbox to\BoxBreadth{\hfil all\hfil}\vfil}}\cr
 \hbox{\vbox to\BoxBreadth{\vfil\hbox to\BoxBreadth{\hfil \{c,d\}\hfil}\vfil}}&\hbox{\vbox to\BoxBreadth{\vfil\hbox to\BoxBreadth{\hfil acd\hfil}\vfil}}&\hbox{\vbox to\BoxBreadth{\vfil\hbox to\BoxBreadth{\hfil bcd\hfil}\vfil}}&\hbox{\vbox to\BoxBreadth{\vfil\hbox to\BoxBreadth{\hfil all\hfil}\vfil}}&\hbox{\vbox to\BoxBreadth{\vfil\hbox to\BoxBreadth{\hfil \{c,d\}\hfil}\vfil}}&\hbox{\vbox to\BoxBreadth{\vfil\hbox to\BoxBreadth{\hfil acd\hfil}\vfil}}&\hbox{\vbox to\BoxBreadth{\vfil\hbox to\BoxBreadth{\hfil bcd\hfil}\vfil}}&\hbox{\vbox to\BoxBreadth{\vfil\hbox to\BoxBreadth{\hfil all\hfil}\vfil}}&\hbox{\vbox to\BoxBreadth{\vfil\hbox to\BoxBreadth{\hfil \{c,d\}\hfil}\vfil}}&\hbox{\vbox to\BoxBreadth{\vfil\hbox to\BoxBreadth{\hfil acd\hfil}\vfil}}&\hbox{\vbox to\BoxBreadth{\vfil\hbox to\BoxBreadth{\hfil bcd\hfil}\vfil}}&\hbox{\vbox to\BoxBreadth{\vfil\hbox to\BoxBreadth{\hfil all\hfil}\vfil}}&\hbox{\vbox to\BoxBreadth{\vfil\hbox to\BoxBreadth{\hfil \{c,d\}\hfil}\vfil}}&\hbox{\vbox to\BoxBreadth{\vfil\hbox to\BoxBreadth{\hfil acd\hfil}\vfil}}&\hbox{\vbox to\BoxBreadth{\vfil\hbox to\BoxBreadth{\hfil bcd\hfil}\vfil}}&\hbox{\vbox to\BoxBreadth{\vfil\hbox to\BoxBreadth{\hfil all\hfil}\vfil}}\cr
 \hbox{\vbox to\BoxBreadth{\vfil\hbox to\BoxBreadth{\hfil acd\hfil}\vfil}}&\hbox{\vbox to\BoxBreadth{\vfil\hbox to\BoxBreadth{\hfil acd\hfil}\vfil}}&\hbox{\vbox to\BoxBreadth{\vfil\hbox to\BoxBreadth{\hfil all\hfil}\vfil}}&\hbox{\vbox to\BoxBreadth{\vfil\hbox to\BoxBreadth{\hfil all\hfil}\vfil}}&\hbox{\vbox to\BoxBreadth{\vfil\hbox to\BoxBreadth{\hfil acd\hfil}\vfil}}&\hbox{\vbox to\BoxBreadth{\vfil\hbox to\BoxBreadth{\hfil acd\hfil}\vfil}}&\hbox{\vbox to\BoxBreadth{\vfil\hbox to\BoxBreadth{\hfil all\hfil}\vfil}}&\hbox{\vbox to\BoxBreadth{\vfil\hbox to\BoxBreadth{\hfil all\hfil}\vfil}}&\hbox{\vbox to\BoxBreadth{\vfil\hbox to\BoxBreadth{\hfil acd\hfil}\vfil}}&\hbox{\vbox to\BoxBreadth{\vfil\hbox to\BoxBreadth{\hfil acd\hfil}\vfil}}&\hbox{\vbox to\BoxBreadth{\vfil\hbox to\BoxBreadth{\hfil all\hfil}\vfil}}&\hbox{\vbox to\BoxBreadth{\vfil\hbox to\BoxBreadth{\hfil all\hfil}\vfil}}&\hbox{\vbox to\BoxBreadth{\vfil\hbox to\BoxBreadth{\hfil acd\hfil}\vfil}}&\hbox{\vbox to\BoxBreadth{\vfil\hbox to\BoxBreadth{\hfil acd\hfil}\vfil}}&\hbox{\vbox to\BoxBreadth{\vfil\hbox to\BoxBreadth{\hfil all\hfil}\vfil}}&\hbox{\vbox to\BoxBreadth{\vfil\hbox to\BoxBreadth{\hfil all\hfil}\vfil}}\cr
 \hbox{\vbox to\BoxBreadth{\vfil\hbox to\BoxBreadth{\hfil bcd\hfil}\vfil}}&\hbox{\vbox to\BoxBreadth{\vfil\hbox to\BoxBreadth{\hfil all\hfil}\vfil}}&\hbox{\vbox to\BoxBreadth{\vfil\hbox to\BoxBreadth{\hfil bcd\hfil}\vfil}}&\hbox{\vbox to\BoxBreadth{\vfil\hbox to\BoxBreadth{\hfil all\hfil}\vfil}}&\hbox{\vbox to\BoxBreadth{\vfil\hbox to\BoxBreadth{\hfil bcd\hfil}\vfil}}&\hbox{\vbox to\BoxBreadth{\vfil\hbox to\BoxBreadth{\hfil all\hfil}\vfil}}&\hbox{\vbox to\BoxBreadth{\vfil\hbox to\BoxBreadth{\hfil bcd\hfil}\vfil}}&\hbox{\vbox to\BoxBreadth{\vfil\hbox to\BoxBreadth{\hfil all\hfil}\vfil}}&\hbox{\vbox to\BoxBreadth{\vfil\hbox to\BoxBreadth{\hfil bcd\hfil}\vfil}}&\hbox{\vbox to\BoxBreadth{\vfil\hbox to\BoxBreadth{\hfil all\hfil}\vfil}}&\hbox{\vbox to\BoxBreadth{\vfil\hbox to\BoxBreadth{\hfil bcd\hfil}\vfil}}&\hbox{\vbox to\BoxBreadth{\vfil\hbox to\BoxBreadth{\hfil all\hfil}\vfil}}&\hbox{\vbox to\BoxBreadth{\vfil\hbox to\BoxBreadth{\hfil bcd\hfil}\vfil}}&\hbox{\vbox to\BoxBreadth{\vfil\hbox to\BoxBreadth{\hfil all\hfil}\vfil}}&\hbox{\vbox to\BoxBreadth{\vfil\hbox to\BoxBreadth{\hfil bcd\hfil}\vfil}}&\hbox{\vbox to\BoxBreadth{\vfil\hbox to\BoxBreadth{\hfil all\hfil}\vfil}}\cr
 \hbox{\vbox to\BoxBreadth{\vfil\hbox to\BoxBreadth{\hfil all\hfil}\vfil}}&\hbox{\vbox to\BoxBreadth{\vfil\hbox to\BoxBreadth{\hfil all\hfil}\vfil}}&\hbox{\vbox to\BoxBreadth{\vfil\hbox to\BoxBreadth{\hfil all\hfil}\vfil}}&\hbox{\vbox to\BoxBreadth{\vfil\hbox to\BoxBreadth{\hfil all\hfil}\vfil}}&\hbox{\vbox to\BoxBreadth{\vfil\hbox to\BoxBreadth{\hfil all\hfil}\vfil}}&\hbox{\vbox to\BoxBreadth{\vfil\hbox to\BoxBreadth{\hfil all\hfil}\vfil}}&\hbox{\vbox to\BoxBreadth{\vfil\hbox to\BoxBreadth{\hfil all\hfil}\vfil}}&\hbox{\vbox to\BoxBreadth{\vfil\hbox to\BoxBreadth{\hfil all\hfil}\vfil}}&\hbox{\vbox to\BoxBreadth{\vfil\hbox to\BoxBreadth{\hfil all\hfil}\vfil}}&\hbox{\vbox to\BoxBreadth{\vfil\hbox to\BoxBreadth{\hfil all\hfil}\vfil}}&\hbox{\vbox to\BoxBreadth{\vfil\hbox to\BoxBreadth{\hfil all\hfil}\vfil}}&\hbox{\vbox to\BoxBreadth{\vfil\hbox to\BoxBreadth{\hfil all\hfil}\vfil}}&\hbox{\vbox to\BoxBreadth{\vfil\hbox to\BoxBreadth{\hfil all\hfil}\vfil}}&\hbox{\vbox to\BoxBreadth{\vfil\hbox to\BoxBreadth{\hfil all\hfil}\vfil}}&\hbox{\vbox to\BoxBreadth{\vfil\hbox to\BoxBreadth{\hfil all\hfil}\vfil}}&\hbox{\vbox to\BoxBreadth{\vfil\hbox to\BoxBreadth{\hfil all\hfil}\vfil}}\cr
\noalign{\kern-2pt}}}
\vbox{\setbox8=\hbox{$\RowNames\Matrix$}
\hbox to\wd8{\hfil$\ColNames$\kern0.2\matrixskip}
\box8}}}
{$\liftedJoin$ as function table $\liftedJoin\in\big\lbrack\PowTWO{X}\big\rbrack\sp{\PowTWO{X}\times\PowTWO{X}}\!$; abbreviated notation for 3- and 4-element sets}{FigJoinMeetA}

\enunc{}{Proposition}{}{PropAbsorbAssoc} $\liftedJoin,\liftedMeet$ satisfy

\begin{enumerate}[i)]
\item $\big\lbrack\pi\RELtraOP\RELandOP\rho\RELtraOP\RELcompOP\liftedMeet\RELcompOP\rho\RELtraOP\big\rbrack\RELcompOP\liftedJoin
=
\RELide,
\quad
\big\lbrack\pi\RELtraOP\RELandOP\rho\RELtraOP\RELcompOP\liftedJoin\RELcompOP\rho\RELtraOP\big\rbrack\RELcompOP\liftedMeet
=
\RELide$,\quad i.e., the absorption\index{absorption} laws

\item $\Kronecker{\liftedMeet}{\RELide}\RELcompOP\liftedMeet=T\RELcompOP\Kronecker{\RELide}{\liftedMeet}\RELcompOP\liftedMeet$\quad i.e., the associative law, where 
\item[]\quad $\!\!\RELfromTO{T}{(X\times X)\times X}{X\times(X\times X)}$ is the brace rearrangement bijection of Def.~\DefBinOp. 

\end{enumerate}

\Proof i) We start the proof of \lq\lq$\RELenthOP$\rq\rq\ with Prop.~\PropPowerMeet.i, Prop.~\CorSumPowToPowProd.i and shunting.

\smallskip
$\big\lbrack\pi\RELtraOP\RELandOP\rho\RELtraOP\RELcompOP\liftedMeet\RELcompOP\rho\RELtraOP\big\rbrack\RELcompOP\liftedJoin
=
\big\lbrack\pi\RELtraOP\RELandOP\Omega\RELtraOP\RELcompOP\rho\RELtraOP\big\rbrack\RELcompOP\liftedJoin
\RELenthOP
\RELide
\quad\iff\quad
\pi\RELtraOP\RELandOP\Omega\RELtraOP\RELcompOP\rho\RELtraOP
\RELenthOP
\liftedJoin\RELtraOP
$

$
\iff\quad
\RELneg{\liftedJoin}
\RELenthOP
\RELneg{\pi}
\RELorOP
\RELneg{\rho\RELcompOP\Omega}
\quad\iff\quad
\RELneg{\varepsilon\RELcompOP\pi\RELtraOP\RELorOP\varepsilon\RELcompOP\rho\RELtraOP}\RELtraOP\RELcompOP\varepsilon
\RELorOP
\big\lbrack\varepsilon\RELcompOP\pi\RELtraOP\RELorOP\varepsilon\RELcompOP\rho\RELtraOP\big\rbrack\RELtraOP\RELcompOP\RELneg{\varepsilon}
\RELenthOP
\RELneg{\pi}
\RELorOP
\rho\RELcompOP\varepsilon\RELtraOP\RELcompOP\RELneg{\varepsilon}
$

\smallskip
\noindent
The first term is contained in $\RELneg{\pi}$, because

\smallskip
$(\pi\RELcompOP\RELneg{\varepsilon}\RELtraOP\RELandOP\rho\RELcompOP\RELneg{\varepsilon}\RELtraOP)\RELcompOP\varepsilon
\RELenthOP
\RELneg{\pi\RELcompOP\varepsilon\RELtraOP}\RELcompOP\varepsilon
\RELenthOP
\RELneg{\pi}
$

\smallskip
\noindent
The second term is also contained in $\RELneg{\pi}$, owing to univalency of $\pi$

\smallskip
$\pi\RELcompOP\varepsilon\RELtraOP\RELcompOP\RELneg{\varepsilon}
\RELenthOP
\RELneg{\pi}
\quad\iff\quad
\varepsilon\RELcompOP\pi\RELtraOP\RELcompOP\pi
\RELenthOP
\varepsilon
$

\smallskip
\noindent
Finally, the third term is equal to the right-most one. This was the proof of containment only; but this suffices because the total (see Prop.~\PropPiAndRhoAndMeetSurj.ii) term $\big\lbrack\pi\RELtraOP\RELandOP\rho\RELtraOP\RELcompOP\liftedMeet\RELcompOP\rho\RELtraOP\big\rbrack\RELcompOP\liftedJoin$ contained in the univalent $\RELide$, so that both must be equal.

\bigskip
\noindent
ii) $T\RELcompOP\Kronecker{\RELide}{\liftedMeet}\RELcompOP\liftedMeet
=
T\RELcompOP\Kronecker{\RELide}{\liftedMeet}\RELcompOP\syqq{\StrictFork{\varepsilon}{\varepsilon}}{\varepsilon}
$\quad Prop.~\CorSumPowToPowProd.ii

$
=
T\RELcompOP\syqq{\StrictFork{\varepsilon}{\varepsilon}\RELcompOP\Kronecker{\RELide}{\liftedMeet}\RELtraOP}{\varepsilon}
$\quad since $\Kronecker{\RELide}{\liftedMeet}$ is a mapping

$
=
T\RELcompOP\syqq{\StrictFork{\varepsilon}{\varepsilon}\RELcompOP\Kronecker{\RELide}{\liftedMeet\RELtraOP}}{\varepsilon}
$\quad transposed

$
=
T\RELcompOP\syqq{\StrictFork{\varepsilon}{\varepsilon\RELcompOP\liftedMeet\RELtraOP}}{\varepsilon}
$\quad Prop.~\PropForkMapKron.iii

$
=
T\RELcompOP\syqq{\StrictFork{\varepsilon}{\StrictFork{\varepsilon}{\varepsilon}}}{\varepsilon}
$\quad Prop.~\CorSumPowToPowProd.iv

$
=
\StrictFork{\pi'\RELcompOP\pi}{\Kronecker{\rho}{\RELide}}\RELcompOP\syqq{\StrictFork{\varepsilon}{\StrictFork{\varepsilon}{\varepsilon}}}{\varepsilon}
$\quad expanding $T$ according to Def.~\DefBinOp.iii

$
=
\syqq{\StrictFork{\varepsilon}{\StrictFork{\varepsilon}{\varepsilon}}\RELcompOP\StrictFork{\pi'\RELcompOP\pi}{\Kronecker{\rho}{\RELide}}\RELtraOP}{\varepsilon}
$\quad $T$ is a map

$
=
\syqq{\StrictFork{\varepsilon}{\StrictFork{\varepsilon}{\varepsilon}}\RELcompOP\StrictJoin{\pi\RELtraOP\RELcompOP{\pi'}\RELtraOP}{\Kronecker{\rho\RELtraOP}{\RELide}}}{\varepsilon}
$\quad transposed

$
=
\syqq{\varepsilon\RELcompOP\pi\RELtraOP\RELcompOP{\pi'}\RELtraOP\RELandOP\StrictFork{\varepsilon}{\varepsilon}\RELcompOP\Kronecker{\rho\RELtraOP}{\RELide}}{\varepsilon}
$\quad Prop.~\PropForkMapKronZierer

$
=
\syqq{\varepsilon\RELcompOP\pi\RELtraOP\RELcompOP{\pi'}\RELtraOP\RELandOP\StrictFork{\varepsilon\RELcompOP\rho\RELtraOP}{\varepsilon}}{\varepsilon}
$\quad Prop.~\PropForkMapKron.iii 

$
=
\syqq{\varepsilon\RELcompOP\pi\RELtraOP\RELcompOP{\pi'}\RELtraOP\RELandOP\varepsilon\RELcompOP\rho\RELtraOP\RELcompOP{\pi'}\RELtraOP\RELandOP\varepsilon\RELcompOP{\rho'}\RELtraOP}{\varepsilon}
$

$
=
\syqq{(\varepsilon\RELcompOP\pi\RELtraOP\RELandOP\varepsilon\RELcompOP\rho\RELtraOP)\RELcompOP{\pi'}\RELtraOP\RELandOP\varepsilon\RELcompOP{\rho'}\RELtraOP}{\varepsilon}
$

$
=
\syqq{\StrictFork{\varepsilon}{\varepsilon}\RELcompOP\,{\pi'}\RELtraOP\RELandOP\varepsilon\RELcompOP{\rho'}\RELtraOP}{\varepsilon}
$

$=
\syqq{\StrictFork{\StrictFork{\varepsilon}{\varepsilon}}{\varepsilon}}{\varepsilon}
$\quad

$=
\syqq{\StrictFork{\varepsilon\RELcompOP\liftedMeet\RELtraOP}{\varepsilon}}{\varepsilon}
$\quad Prop.~\CorSumPowToPowProd.iv

$=
\syqq{\StrictFork{\varepsilon}{\varepsilon}\RELcompOP\Kronecker{\liftedMeet\RELtraOP}{\RELide}}{\varepsilon}
$

$=
\syqq{\StrictFork{\varepsilon}{\varepsilon}\RELcompOP\Kronecker{\liftedMeet}{\RELide}\RELtraOP}{\varepsilon}
$

$=
\Kronecker{\liftedMeet}{\RELide}\RELcompOP\,\syqq{\StrictFork{\varepsilon}{\varepsilon}}{\varepsilon}
$

$=
\Kronecker{\liftedMeet}{\RELide}\RELcompOP\,\liftedMeet
$
\Bewende


\chapter{Concluding Remarks\label{SectConclusion}}
\ExerciseNo=0
\EnuncNo=0
\CaptionNo=0


\noindent
These additions have already been broadly applied, not least in studies of relational topology. The relational language {\sc TituRel} 
reflecting all these ideas in functional programming style has made it possible to successfully explore discrete topologies, concepts of nearness, proximity that have been studied by logicians.

\bigskip
\noindent
These investigations further support our firm creed:
Mankind seems hardly capable of handling intellectually more than {\it linear\/} situations!

%
%

\begin{theindex}

  \item absorption, \hyperpage{42}
  \item addition theorem, \hyperpage{22}, \hyperpage{24}
  \item associative, \hyperpage{27}
  \item atom, \hyperpage{6}

  \indexspace

  \item binary operation, \hyperpage{25}
  \item Boolean\ algebra, \hyperpage{33}

  \indexspace

  \item cancellation, \hyperpage{3, 4}
  \item commutative, \hyperpage{27}

  \indexspace

  \item De Morgan rule, \hyperpage{39}
  \item Dedekind rule, \hyperpage{2}
  \item Descartes, Ren\Acut e, \hyperpage{2}
  \item direct
    \subitem product, \hyperpage{16}
    \subitem sum, \hyperpage{16}
  \item disjointness, \hyperpage{35}
  \item distributive, \hyperpage{33}

  \indexspace

  \item existential image, \hyperpage{8}

  \indexspace

  \item fork operator, \hyperpage{17}

  \indexspace

  \item injection, \hyperpage{16}
  \item invariant element, \hyperpage{30}
  \item inverse image, \hyperpage{9}

  \indexspace

  \item join, \hyperpage{35}
  \item join operator, \hyperpage{17}

  \indexspace

  \item Kronecker operator, \hyperpage{17}
  \item Kronecker product, \hyperpage{17}

  \indexspace

  \item left-invertible, \hyperpage{29}
  \item left-neutral element, \hyperpage{30}

  \indexspace

  \item meet, \hyperpage{35}
  \item membership relation, \hyperpage{5}

  \indexspace

  \item natural projection, \hyperpage{7}
  \item neutral element, \hyperpage{30}

  \indexspace

  \item observability, \hyperpage{19}

  \indexspace

  \item point, \hyperpage{25}
  \item Point Axiom, \hyperpage{18}
  \item power relator, \hyperpage{12}
  \item product, direct, \hyperpage{16}
  \item projection, \hyperpage{16}

  \indexspace

  \item quotient, \hyperpage{7}

  \indexspace

  \item residual, \hyperpage{3}
  \item residue cancellation, \hyperpage{3}
  \item right-invertible, \hyperpage{29}
  \item right-neutral element, \hyperpage{30}

  \indexspace

  \item Schr\"oder equivalences, \hyperpage{2}
  \item shunting, \hyperpage{2}
  \item simulation, \hyperpage{9}
  \item singleton injection, \hyperpage{6}
  \item sum, direct, \hyperpage{16}
  \item symmetric quotient, \hyperpage{4}

  \indexspace

  \item {\sc TituRel}, \hyperpage{14}

  \indexspace

  \item unsharpness, \hyperpage{17}

  \indexspace

  \item vectorization, \hyperpage{13}

\end{theindex}

\end{document}